\documentclass[12pt]{article} 
\usepackage{graphicx}
\usepackage{color}
\usepackage{rotating}
\usepackage{amsmath}
\usepackage{amssymb}
\usepackage{float}
\usepackage{url}
\usepackage{ulem} 

\usepackage{array,arydshln}

\usepackage{epstopdf}

\usepackage{enumerate}
\usepackage{enumitem}
\usepackage{xr}

\newcommand\bx {\mathbf x}

\newcommand\bA {\mathbf A}

\newcommand\bX {\mathbf X}

\newcommand\indica {\mathbb{I}}

\newcommand\wy {\widehat{y}}

\newcommand\wF {\widehat{F}}

\newcommand\wZ {\widehat{Z}}

\newcommand\itF {{\mathcal{F}}}
\newcommand\itG {{\mathcal{G}}}

\newcommand\itK {{\mathcal{K}}}

\newcommand\itM {{\mathcal{M}}}

\newcommand\itS {{\mathcal{S}}}

\newcommand\itV {{\mathcal{V}}}

\newcommand\balfa {\mbox{\boldmath $\alpha$}}
\newcommand\balfach {\mbox{\scriptsize\boldmath $\alpha$}}

\newcommand\bbe {\mbox{\boldmath $\beta$}}

\newcommand\bthe {\mbox{\boldmath $\theta$}}

\newcommand\bthech {\mbox{\scriptsize\boldmath $\theta$}}

\newcommand\bxi {\mbox{\boldmath $\xi$}}

\newcommand\weps {\widehat{\epsilon}}

\newcommand\wmu {\widehat{\mu}}

\newcommand\wbthe {\widehat{\bthe}}
\newcommand\wbthech {\mbox{\scriptsize$\wbthe$}} 

\newcommand\wpi {\widehat{\pi}}

\newcommand\wbxi  {\widehat{\bxi}}

\def\real{\mathbb{R}}

\def\qu{\mathbb{Q}}

\newcommand{\esp}{\mathbb{E}}
\newcommand{\prob}{\mathbb{P}}

\newcommand{\convpp}{ \buildrel{a.s.}\over\longrightarrow}

\newcommand{\convprob  }{ \buildrel{p}\over\longrightarrow}

\newcommand{\trasp}{^{\mbox{\footnotesize \sc t}}}

\newcommand\bcero {{\bf{0}}}

\def\dst{\displaystyle}

\def\ROC {{\mbox{ROC}}} 
\def\AUC {{\mbox{AUC}}} 
\def\wROC {{\widehat{\mbox{ROC}}}} 
\def\wAUC {{\widehat{\mbox{AUC}}}}

\newcommand{\identidad}{\mbox{\bf I}}

\newcommand\noi{\noindent}
\def\dst{\displaystyle}

\def\square{\ifmmode\sqr\else{$\sqr$}\fi}
\def\sqr{\vcenter{
         \hrule height.1mm
         \hbox{\vrule width.1mm height2.2mm\kern2.18mm
\vrule width.1mm}
         \hrule height.1mm}}

\newcommand{\simp}{\mbox{\scriptsize \sc simp}}

\newcommand\conv {\mbox{\scriptsize \sc conv}}

\newcommand\kernel {\mbox{\scriptsize \sc ker}}

\newcommand\ipw {\mbox{\scriptsize \sc ipw}}
\newcommand\emp {\mbox{\scriptsize \sc emp}}

\usepackage[utf8]{inputenc}

\begin{document}

\title{ Estimators for covariate-adjusted ROC curves with missing  biomarkers values} 
\author{Ana M. Bianco$^1$, Graciela Boente$^1$,\\ 
Wenceslao Gonz\'alez--Manteiga$^2$ and Ana P\'erez--Gonz\'alez$^3$\\
$^1$ Universidad de Buenos Aires and CONICET\\
$^2$ Universidad de Santiago de Compostela\\
$^3$ Universidad de Vigo
}
\date{}
\maketitle

\small 
\begin{abstract}
 In this paper, we present three estimators of the $\ROC$ curve when missing observations arise among the biomarkers. 
Two of the procedures assume that we have covariates that allow to estimate the propensity and the estimators are obtained using an inverse probability weighting method or a smoothed version of it. The other one  assumes that the covariates are related to the biomarkers through a regression model which enables us to construct convolution--based estimators of the distribution and quantile functions. Consistency results are obtained under mild conditions. Through  a numerical study we evaluate the finite sample performance of the different  proposals. A real data set is also analysed.
\end{abstract}

\normalsize

\section{Introduction}{\label{intro}}
 
 The Receiver Operating Characteristic (ROC) curves are useful in statistical procedures such as classification or discrimination, where  we typically have a set of individuals or items assigned to one of two  classes on the basis of disposable information of that individual. The use  of ROC curves has become more and more popular in medicine from the early 60's (see Gon{\c{c}}alves \textsl{et al.}, 2014, for a historical note).
 In fact, ROC curves are a very well known technique in medical studies where a continuous variable or biomarker  is used to diagnose a disease or to evaluate its evolution.
 However, assignations are not perfect and may lead to classification errors. In fact, during the assignment procedure some classification errors may occur,  in the sense that an individual or object may be allocated into the wrong class. At this point, ROC curves become an interesting strategy either to evaluate the quality of a given assignment rule or to compare two available procedures. 

To be more precise, let us fix some notation. Assume that we are in a medical decision scenario and that   we deal with two populations, identified as diseased and healthy. Besides, assume that a continuous score (or marker or diagnostic variable), $Y$, is considered for the assignment procedure in the classification purpose and that the rule is based on a cut--off value $c$. Thus, according to this assignment rule, an individual is classified as diseased if $Y \ge c$ and as healthy when $Y < c$. Let $F_{D}$ be the distribution of the marker on the diseased population and $F_{H}$ the distribution of $Y$ in the healthy one. Henceforth, for practical reasons, we denote  as $Y_D \sim F_D$ the marker in the diseased population and $Y_{H} \sim F_{H}$ the score in the healthy one.  Without loss of generality, we will assume that $Y_D$ is stochastically greater than  $Y_{H}$, that is,  $\prob(Y_{D} \le c) \le \prob(Y_{H} \le c)$ for all  $c$.
 Typically,   some individuals belonging to the diseased population may present a marker $Y_D < c$ and  in that case,  would be classified as healthy, corresponding to the false negative cases. In contrast, among healthy individuals, there may be cases  with a score $Y_H \ge c$ that would be assigned to the diseased population, corresponding to the false positive cases. It is clear that the classification errors depend on the threshold $c$. Therefore, it becomes of interest to study the triplets  $\{(c,1-F_{H}(c),1-F_{D}(c)),\quad c \in \real \}$,
which describe a geometrical object called ROC curve, that reflects the discriminatory capability of the marker. More precisely,   associated with each threshold value  $c$  is the probability of a true positive (sensitivity) and the probability of a true negative (specificity) result. If we take into account that the sensitivity of a test corresponds to the probability of a true positive $1-F_{D}(c)$, while the specificity to that of a true negative, that is, to $F_{H}(c)$, the ROC curve corresponds to the plot of the sensitivity  versus  $1 $ minus the specificity for all possible  cut--off values $c$. A re-parametrization of this curve in terms of the false positive rate, $p=1-F_{H}(c)$,  is usually considered, leading to  
$\{\left(p, 1-F_{D}\left(F_{H}^{-1}\left(1-p\right)\right)\right), \quad p \in (0,1) \}\, $
and therefore, to
$$\ROC(p)= 1-F_{D}\left(F_{H}^{-1}\left(1-p\right)\right), \quad p \in (0,1) \,. $$
In this  way, the ROC curve is a complete picture of the performance of the assignment procedure over all the possible threshold values. Nevertheless, different summary measures of the ROC curve are useful to sum up particular features of the curve. One of the most popular indices is the \textsl{area under the curve} (AUC), which is computed as
$$\AUC= \int_0^1 \ROC(p) dp \, .$$
This summary index takes values between 0 and 1 corresponding to low and high discriminatory capability, respectively,  while the interesting  cases  is when $AUC>0.5$.
After a change of variable and a bit of algebra, it is easy to see that
$\AUC= \prob(Y_{D} > Y_{H})$,
where $Y_{D}$ and $Y_{H}$ are the markers of two independent individuals randomly chosen from the diseased and healthy populations, respectively. Hence,  another interpretation of the area under the curve is that it measures the probability that the markers of a randomly chosen pair of subjects selected from the diseased and healthy population, were in the correct order. For that reason, values of $\AUC$ close to 1 suggest a high diagnostic accuracy of the   marker.

An extensively used model is the bi-normal model, which assumes that in both populations the marker is normally distributed. In this case, the distributions are characterized   by their parameters, i.e., the means $\mu_{D}$ and $\mu_{H}$ and the standard deviations $\sigma_{D}$ and $\sigma_{H}$. The resulting ROC curve can be written as
$$\ROC(p)= \Phi\left(\dfrac{\mu_{D}-\mu_{H}}{\sigma_{D}}+ \dfrac{\sigma_{H}}{\sigma_{D}} \Phi^{-1}(p)\right)\,,$$
while 
$\AUC= \Phi\left((\mu_{D}-\mu_{H})/ \sqrt{\sigma_{D}^2+ \sigma_{H}^2}\right)$.

In this paper, we face the problem of estimating the ROC curve when missing biomarkers arise in one or both populations. In fact, there are occasions where   the diagnostic variables can not be observed for all the individuals leading to  missing observations. One of such situations has been described in Long \textsl{et al.} (2011a)  in which the biomarker are missing but auxiliary variables with prediction capability may be fully recorded.  
This context of missing data must be distinguish from the setting of verification bias  where the biomarker is always obtained and a gold standard biomarker  is obtained only for some subjects. 

Different perspectives were given to estimate  the ROC curve when missing biomarkers arise. On one hand, under a   completely at random model for dropouts, Liu and Zhao (2012), An(2012), Yan and Zhao(2015)   impute the missing observations using either hot deck imputation or by assuming a parametric model and selecting random samples from a maximum likelihood estimator. On the other hand, when covariates are present and under a missing at random model,    Li and Ning (2015) use weighted or fully augmented weighted estimating equations to estimate a covariate-specific time-dependent roc curve with missing  biomarkers.  Similarly,  Long \textsl{et al.} (2011a) use auxiliary covariates to propose estimators of the AUC using an inversely probability weighing procedure combined with  doubly robust techniques. Besides, Long \textsl{et al.}(2011b) propose a multiple imputation procedure  to estimate the AUC and the ROC curve.

In this paper, we assume that the covariates are related to the biomarkers through a regression model which enables us to construct convolution--based estimators of the distribution and quantile functions. These convolution estimators are then used to construct estimators of the ROC curves. We also define an estimator which uses the covariates only to estimate the propensity and which extends the estimator defined in Pulit (2016) to the case of missing biomarkers. Section \ref{sec:proposal} reviews some approaches when all the observations are available. The estimating procedures when missing biomakers arise are described in Section \ref{sec:proposalmiss}. Some consistency results 
are given in Section \ref{sec:consist}. Finally, a numerical study carried out to evaluate the finite sample performance of the proposals is described in Section \ref{sec:monte}.  Proofs are relegated to the Appendix.

\section{The estimators}{\label{sec:proposal}}
\subsection{The case of complete observations}{\label{sec:proposalnomiss}}
Several estimators of the ROC curve and the area under the curve have been proposed.  Pepe  (2003)  and Krzanowski and Hand (2009) give a deep insight of different statistical aspects of their estimation, we also refer to Pardo-Fern\'andez \textsl{et al}. (2014)  for an overview on this topic. 

In particular, estimators of the ROC curve can be obtained by plugging--in appropriate estimators of the marginal distributions $F_{D}$ and $F_{H}$.  
 For instance, the empirical distribution function can be used to merge marginal distribution estimators into the definition of the ROC curve. More precisely, let us assume that for $i=D,H$, we have  data sets $ y_{i,j} $, $1\le j \le n_i$, then the empirical distribution estimator of the marginal distribution function $F_{i}$, $i=D, H$, is defined as
\begin{equation*}
\wF_{i}(y)=\frac{1}{n_i} \sum_{j=1}^{n_i}  \indica_{(-\infty,y]}(y_{i,j})\;,
\end{equation*}
Hence, estimators of the ROC curve can be defined as
\begin{equation}
\wROC_{\emp}(p)= 1-\wF_{D }\left(\wF_{H }^{-1}\left(1-p\right)\right), \quad p \in (0,1) \,,
\label{ROCempir}
\end{equation}
while an estimator of the area under the curve can be defined as $\wAUC_{\emp} = \int_0^1 \wROC_{\emp} (p) dp $ and approximated as $(1/N)\sum_{\ell=1}^N \wROC_{\emp}(p_\ell)$ with $\{p_{\ell}\}_{1\le \ell \le N}$ an equidistant grid over $(0,1)$.

As noted in Pulit (2016), the ROC curve can be written as the distribution function of $Z = 1 - F_H(y_D)$. Based on this property, that author proposed to estimate the ROC curve by means of the pseudo--observations $\wZ_j= 1-\wF_{H }(y_{D,j})$, $1\le j\le n_D$, using a kernel approach. The estimator defined in Pulit (2016) equals
\begin{equation}
\wROC_{\kernel}(p)= \frac{1}{n_D} \sum_{j=1}^{n_D} \itK\left(\frac{p-\wZ_j}{h}\right)
\label{ROCpulit}
\end{equation}
 where $\itK(t)=\int_{-\infty}^t K(u)du$ with $K$ a continuous symmetric density function with support $[-1, 1]$, $h$ is the smoothing parameter. A possible choice for $h$ is suggested in Pulit (2016).

\subsection{The situation of missing biomarkers}{\label{sec:proposalmiss}}
As in other settings, when the biomarker value is missing for
some observations, the ROC analysis based solely on the complete cases may be  biased. Some authors such as Long \textsl{et al.} (2011a,b) have investigated proposals based on multiple imputation of the missing biomarkers.  However, when ignorable missing biomarkers arise and the practitioner has some covariates with predictive capability on the missingness, the inverse probability weighted estimators is a common approach. Let us assume that for $i=D, H$, we have incomplete data sets $\left(y_{i,j},
\bx_{i,j}\trasp, \delta_{i,j}\right)$, $1\le j \le n_i$ where  the binary indicator  $\delta_{i,j}=1$ if
$y_{i,j}$ is observed and $\delta_{i,j}=0$ if $y_{i,j}$ is missing and the covariates $\bx_{i,j} \in \real^{d_i}$ allow to predict drop--outs. Furthermore,  assume that the responses are missing at random (MAR), that is,  we assume an ignorable missing mechanism such that the binary variables  and  the responses are conditionally independent given the covariates, i.e.,
\begin{equation}
\prob\left(\delta_{i,j}=1\vert (y_{i,j},\bx_{i,j}) \right)=\prob\left(\delta_{i,j}=1\vert
\bx_{i,j} \right)=\pi_i\left(\bx_{i,j}\right)\, .
\label{delta}
\end{equation}
 The  inverse probability weighting (\textsc{ipw}) estimation approach introduced in Horvitz and Thompson (1952), reduces bias by  weighting each observation according to the inverse of the estimated probability of dropouts. The \textsc{ipw} estimator of the marginal distribution function $F_{i}$, $i=D, H$, is defined as
\begin{equation*}
\wF_{i,\ipw}(y)=\frac{1}{\dst\sum_{\ell=1}^{n_i}\frac{\delta_{i,\ell}}{\wpi_i(\bx_{i,\ell})}} \sum_{j=1}^{n_i}\frac{\delta_{i,j}}{\wpi_i(\bx_{i,j})} \indica_{(-\infty,y]}(y_{i,j}) =\sum_{j=1}^{n_i} \tau_{i,j} \indica_{(-\infty,y]}(y_{i,j})\;,
\label{Qws}
\end{equation*}
Note that when there are no missing data, the estimator $\wF_{i,\ipw}$ reduces to $\wF_{i,\emp}$. Uniform strong consistency results for $\wF_{i,\ipw}$, under mild assumptions, are derived in Bianco \textsl{et al.} (2010) when \eqref{delta} holds. In contrast, under a non--ignorable missing setting, Ding and Tang (2018) obtained the pointwise asymptotic distribution of $\sqrt{n}\left(\wF_{i,\ipw}(y)-F_{i}(y)\right)$ for a family of kernel--based propensity score estimators.

Estimators of the ROC curve based on an inverse probability weighting  can be defined as
\begin{equation}
\wROC_{\ipw}(p)= 1-\wF_{D, \ipw}\left(\wF_{H, \ipw}^{-1}\left(1-p\right)\right), \quad p \in (0,1) \,,
\label{ROCIPW}
\end{equation}
while an estimator of the area under the curve can be defined as $\wAUC_{\ipw}= \int_0^1 \wROC_{\ipw}(p) dp $ and approximated as $(1/N)\sum_{\ell=1}^N \wROC_{\ipw}(p_\ell)$ with $\{p_{\ell}\}_{1\le \ell \le N}$ an equidistant grid over $(0,1)$.

The estimator defined in Pulit (2016) can also be extended to the case of missing biomarkers as follows. First of all, for each $1\le j\le n_D$ such that  $\delta_{D, j}=1$, define pseudo--observations $\wZ_j= 1-\wF_{H, \ipw}(y_{D,j})$. We propose to estimate the ROC curve   using a kernel approach combined with inverse probability weighting, that is,  
\begin{equation}
\wROC_{\kernel}(p)= \frac{1}{\dst\sum_{\ell=1}^{n_D}\frac{\delta_{D,\ell}}{\wpi_D(\bx_{D,\ell})}} \sum_{j=1}^{n_D} \frac{\delta_{D,j}}{\wpi_D(\bx_{D,j})} \itK\left(\frac{p-\wZ_j}{h}\right)
\label{ROCpulitmis}
\end{equation}
 where as above $\itK(t)=\int_{-\infty}^t K(u)du$ with $K$ a continuous symmetric density function with support $[-1, 1]$ and $h$ is the smoothing parameter.

When the practitioner records covariates $\bx_{i,j}$ with predictive capability for  $y_{i,j}$ and when missing data arise among  $y_{i,j}$,  using an approach related to that considered in M\"uller (2009) for linear functionals,  a different estimator of the marginal distribution function may be obtained using the regression model  and the fact that $F_y$ is the convolution of the errors and  the regression function distributions. This procedure can also be used in the complete data framework. It is worth mentioning that the covariates used to fit the biomarkers do not need to be the same as those involved when estimating the drop-out probability. For notation simplicity, we will assume that  the same set of explanatory variables is considered.

In this framework, one assumes that, for $i=D, H$,
\begin{equation}
y_{i,j}=\mu_i(\bx_{i,j})+\epsilon_{i,j}\qquad 1\le j\le n_i \;, \label{regresion}
\end{equation}
 where  the error $\epsilon_{i,j}$ are independent,  independent of $\bx_{i,j}$ and $\mu_i$ denote the regression functions. Denote as  
  $F_{i,\epsilon}$ and  $F_{i,\mu_i}$ the distribution functions   of the errors $\epsilon_{i,j}$ and  of the true regression function $\mu_i(\bx_{i,j})$, respectively. Using the convolution property, i.e.,   $F_i=F_{i,\epsilon}*F_{i,\mu_i}$, a consistent estimator for $F_i$ can be obtained plugging--in consistent estimators $\wF_{i,\epsilon}$ and $\wF_{i,\mu_i}$ of $F_{i,\epsilon}$ and $F_{i,\mu_i}$, respectively.

More precisely, let $\wmu_i(\bx)$ be a consistent estimator of $\mu_i(\bx)$. This consistent estimation can be accomplished in different ways according to  the model structure assumed on the regression function which may be  parametric, nonparametric or semiparametric.   Define
\begin{eqnarray*}
 {\wF}_{i,\mu_i} (u)&= &  \frac 1{n_i}\sum_{j=1}^{n_i}   \indica_{(-\infty,u]}(\wmu_i(\bx_{i,j}))  \,.
 \end{eqnarray*}

When $\delta_{i,j } =1$, the residuals can be effectively predicted as $\weps_{i,j}=y_{i,j}-\wmu_i(\bx_{i,j})$, so that an estimator of $F_{i,\epsilon}$ can be computed as
$$ 
\wF_{i,\epsilon}(e)=\frac 1{ \sum_{\ell=1}^{n_i} \delta_{i,\ell}} \sum_{j=1}^{n_i} \delta_{i,j}\indica_{(-\infty,e]}(\weps_{i,j})=   \sum_{j=1}^{n_i} \kappa_{i,j} \indica_{(-\infty,e]}(\weps_{i,j})\,, $$ 
with  $\kappa_{i,j}=\delta_{i,j}/\sum_{\ell=1}^{n_i} \delta_{i,\ell}$.
The convolution--based estimator of  $F_i$ is then defined as 
$\wF_{i, \conv} = \wF_{i,\epsilon} * \wF_{i,\mu_i} $ and is a weighted empirical distribution since it can be written as 
$$\wF_{i, \conv} (y)= \frac 1{n_i} \sum_{j=1}^{n_i}   \sum_{\ell=1}^{n_i}  \kappa_{i,j} \indica_{(-\infty,y]}(\wy_{i,j\ell})$$ 
  where $\wy_{i,j\ell} = \wmu_i(\bx_{i,\ell}) + \weps_{i,j}$, for $    j \in \{\delta_{i,\ell}=1\}  $. Note that for complete data sets, that is when no missing biomarkers are present in the $i-$th population,   $\delta_{i,j } =1$ for all the observations so the estimator reduces to 
  $$\wF_{i, \conv} (y)= \frac 1{{n_i}^2} \sum_{j=1}^{n_i}   \sum_{\ell=1}^{n_i}    \indica_{(-\infty,y]}(\wy_{i,j\ell})$$ 
  where $\wy_{i,j\ell} = \wmu_i(\bx_{i,\ell}) + \weps_{i,j}$ for all $j$.

The convolution--based estimators of the ROC curve are defined as in \eqref{ROCIPW}, but plugging--in the convolution--based estimators of the distribution and quantile functions, that is,  
\begin{equation}
\wROC_{\conv}(p)= 1-\wF_{D, \conv}(\wF_{H, \conv}^{-1}(1-p))), \quad p \in (0,1) \,,
\label{ROCCONV}
\end{equation}
and $\wAUC_{\conv}= \int_0^1 \wROC_{\conv}(p) dp $.

\section{Consistency results}{\label{sec:consist}}
In this section, we will derive uniform consistency results for the ROC curve estimators defined in Section \ref{sec:proposalmiss}. We begin, by stating the results for $\wROC_{\ipw}$ and $\wROC_{\conv}$. From now on, $\itS_i$ stands for the support of $\bx_{i,1}$, $i= D, H$.

\begin{enumerate}[label = \textbf{C\arabic*}]
 \item\label{ass:C1} $\dst\inf_{\bx\in \itS_i} \pi_i(\bx)=A_i>0$, for $i= D, H$.
\item\label{ass:C2} $\dst\sup_{\bx\in \itS_i} | \wpi_i(\bx)-  \pi_i(\bx) | \convpp 0$, for $i= D, H$.
\item\label{ass:C3} $\dst\sup_{\bx\in \itK_i} | \wmu_i(\bx)-  \mu_i(\bx) | \convpp 0$, for $i= D, H$, for any  compact set $\itK_i\in \real^{d_i}$.
\item\label{ass:C4} $F_H : \real\to  (0, 1)$ has an associated density $f_H$ such that $f_H(y)>0$, for all $y\in \real$.
\item\label{ass:C5} $F_D : \real\to  (0, 1)$ has an associated bounded density $f_D$.

\end{enumerate}

The following result is a direct consequence of Theorem  4.1 in Bianco \textsl{et al.} (2010) and Theorem 3.2 in Bianco \textsl{et al.} (2019), see also Theorem 1 in Sued and Yohai (2013) for the  situation of a  parametric regression model.

\noi \textbf{Proposition \ref{sec:consist}.1.} \textsl{Let  $\left(y_{i,j},
\bx_{i,j}\trasp, \delta_{i,j}\right)$, $1\le j \le n_i$, $i=D, H$, be such that  (\ref{delta}) hold.  
\begin{itemize}
\item[a)] Under \ref{ass:C1} and \ref{ass:C2}, we have that $\|\wF_{i,\ipw}-F_i\|_{\infty}\convpp 0$.
\item[b)] Assume that model \eqref{regresion} holds. Under \ref{ass:C3}, we have that $\|\wF_{i, \conv} -F_i\|_{\infty}\convpp 0$.
\end{itemize}
}

From Proposition \ref{sec:consist}.1 and the continuity of the quantile functionals when \ref{ass:C4} holds, we get the following result for the healthy subjects.

\noi \textbf{Proposition \ref{sec:consist}.2.} \textsl{Let  $\left(y_{H,j},
\bx_{H,j}\trasp, \delta_{H,j}\right)$, $1\le j \le n_H$,  be such that  (\ref{delta}) holds.  
\begin{itemize}
\item[a)] Under \ref{ass:C1}, \ref{ass:C2} and \ref{ass:C4}, we have that 
\begin{itemize}
\item[i)] $\wF_{H,\ipw}^{-1}(p) \convpp F_H^{-1}(p)$, for each $0<p<1$.
\end{itemize}  
\item[b)] Assume that model \eqref{regresion} holds. Under \ref{ass:C3} and \ref{ass:C4}, we have that 
\begin{itemize}
\item[i)] $\wF_{H,\conv}^{-1}(p) \convpp F_H^{-1}(p)$, for each $0<p<1$.
\end{itemize}  
\end{itemize}
}


We then get the following   result for the ROC curve estimator.

\noi \textbf{Theorem \ref{sec:consist}.1.} \textsl{Let  $\left(y_{i,j},
\bx_{i,j}\trasp, \delta_{i,j}\right)$, $1\le j \le n_i$, $i=D, H$, be such that  (\ref{delta})  is verified.  Assume that \ref{ass:C4} and \ref{ass:C5} hold.
\begin{itemize}
\item[a)] If in addition \ref{ass:C1} and \ref{ass:C2}  are satisfied,  we have that $\sup_{0<p<1} \left|\wROC_{\ipw}(p)-\ROC(p)\right| \convpp 0$.
\item[b)] Furthermore,  under the regression model \eqref{regresion}, if \ref{ass:C3} holds,  we have that  \linebreak $ \sup_{0<p<1}\left|\wROC_{\conv}(p)-\ROC(p)\right| \convpp 0$.
\end{itemize}
}

In order to obtain point--wise weakly consistency results for the smoothed estimator $\wROC_{\kernel}$ defined in \eqref{ROCpulitmis}, we will need the following additional assumptions
\begin{enumerate}[label = \textbf{C\arabic*}]
 \setcounter{enumi}{5}
 \item\label{ass:C6}  $K$ is bounded, continuously differentiable with bounded derivative and with support on $[-1,1]$.
\item\label{ass:C7} $n_D/(n_D+n_H)\to \tau$ with $0<\tau<1$.
\item\label{ass:C8}  $h\to 0$ and $n_D h^2\to \infty$.
 \end{enumerate}
 Furthermore,  in the next condition  we will assume a parametric model for the  propensity in the healthy population and that a root-$n$ estimate of the unknown parameter exists. An example of such situation is when the propensity is modelled using a logistic regression model for which assumption \ref{ass:C10} is satisfied when second moments exist for the covariate $\bx_H$.
\begin{enumerate}[label = \textbf{C\arabic*}]
 \setcounter{enumi}{8}
 \item\label{ass:C10} For the healthy population, the missingness probability is given by $\pi_{H}(\bx)=G_{H}(\bx, \bthe_{H})$ where $\bthe_{H}\in \real^{q_{H}}$ and is such that
\begin{enumerate}
  \item[a)]$G_{H}(\bx,  \bthe)$ is twice continuously differentiable with respect to $\bthe$. We will denote by $G_{H}^{\prime}(\bx, \bthe)$ and $G_{H}^{\prime\prime}(\bx ,\bthe)$ the gradient and Hessian matrix of $G_{H}(\bx,  \bthe)$, respectively. 
  \item[b)] $\esp\left\{\left\|G_{H}^{\prime }(\bx_{H} ,\bthe_H)\right\|/G_{H}(\bx_H,  \bthe)\right\}$  exists and is bounded in   a neighbourhood $\itV$ of $\bthe_{H}$. Moreover, $\sup_{\bthech \in \itV, \balfach \in \itV}\esp\left\{\lambda_1\left(G_{H}^{\prime\prime}(\bx_{H} ,\balfa)\right)/G_{H}(\bx_H,  \bthe)\right\}<\infty$, where for a   symmetric matrix $\bA$,  $\lambda_1(\bA)$  stands for the largest eigenvalue of $\bA$.
\item[c)]  The family of functions ${\itG}_H=\{1/G_H(\bx,  \bthe): \bthe\in \real^{q_H}\}$ satisfies the uniform--entropy condition, that is, 
$$\int_{0}^{\infty} \sup_{\qu} \sqrt{\log N(\epsilon, {\itG}_H, L^2(\qu))}\; d\epsilon<\infty\,,$$
where, for any class of function $\itF$, $N\left(\epsilon,{\itF},L^2(\qu)\right)$ stands for the covering number of the class $\itF$ with respect to $L^2(\qu)$ and $\qu$ stands  for any finitely discrete probability.
 \end{enumerate}
\end{enumerate}

\noi \textbf{Theorem \ref{sec:consist}.2.} \textsl{Let  $\left(y_{i,j},
\bx_{i,j}\trasp, \delta_{i,j}\right)$, $1\le j \le n_i$, $i=D, H$, be such that  (\ref{delta})   is satisfied.  Assume that \ref{ass:C1}, \ref{ass:C2}, \ref{ass:C4} to \ref{ass:C10} hold and that $\wpi_{H}(\bx )=G_{H}(\bx , \wbthe_{H})$, where $\sqrt{n_{H}}(  \wbthe_{H} -  \bthe_{H} )=O_{\prob}(1)$.
Then, we have that $\wROC_{\kernel}(p)\convprob \ROC(p)$.}
 
 According to Theorem \ref{sec:consist}.2, we should use a $\sqrt{n}-$consistent estimator of the parameter $\bthe_{H}$ of the propensity model. In the aforementioned situation of the logistic regression fit, we could employ the maximum likelihood estimator which satisfies this requirement under regularity conditions (see Fahrmeir and Kaufmann, 1985).  
 
It is worth mentioning that when no missing biomarkers arise, Theorem 1 in Pulit (2016), allow to conclude that, under \ref{ass:C4} to \ref{ass:C8}, the kernel--based estimator of the ROC curve is point--wise weakly consistent.

\section{Monte Carlo study}{\label{sec:monte}}

In this section, we summarize the results of a simulation study conducted to study the small sample performance of the proposal.    In all cases, we    generate $Nrep=1000$ datasets  of equal size $n_{D}=n_{H}=n$. The quantiles $p$ were chosen over an equidistant grid, ${\cal G}_p$ between 0 and 1  of length $N_p=99$. Several summary measures were considered to evaluate the performance of the estimators.  
 To provide a global measure of discrepancy over samples, for each replication,  we compute the mean over replications of the following summary measures which  quantify the  global  mismatch  between the estimated $\ROC$ curve, denoted $\wROC(p)$,  and the true one, $\ROC(p)$,
\begin{itemize}
\item the  mean squared error given by 
$$MSE=\frac{1}{ N_p}  \sum_{j=1}^{N_p} \left(\wROC(p_j) -{\ROC} (p_j)\right)^2 \, ,$$
\item  a measure inspired on the Kolmogorov distance calculated as
$$KS=  \sup_{p_j \in {\cal G}_p} \left| \wROC(p_j) - \ROC (p_j) \right| \, .$$
\end{itemize}
 For the estimators of the $\AUC$, we have computed the bias and the mean squared error ($MSE$) over replications as well as the mean relative bias ($RB$) defined as
$$RB= \mathop{\mbox{mean}}_{1\le \ell\le Nrep} \frac{\left|\wAUC_\ell- \AUC\right|}{\AUC}\,,$$
with $\wAUC_\ell$ the estimate obtained at the $\ell-$th replication and $\AUC$ corresponds to the true area under the curve for the current situation. This measure has been used in Long \textsl{et al.} (2011b) and adapts for the size of the $\AUC$.  

\subsection{Numerical study  for data sets without missing biomarkers}{\label{sec:dim2-caso2}}
We consider   homoscedastic  regression models for each populations. More precisely, we assume that
\begin{eqnarray}
Y_{D,i} &=& 2+ \bX_{D,i}\trasp \bbe_D+  \frac{1}{3} \epsilon_{D,i} \label{trued}\\
Y_{H,i} &=&  0.5+   \bX_{H,i}\trasp \bbe_H + \sqrt{\frac{24}{9}} \epsilon_{H,i} \label{trueh}\; ,
\end{eqnarray}
where $\bbe_D=(4,20)\trasp$, $\bbe_H=(\sqrt{17/2},20)\trasp$. For all $i=1,\dots,n$ $\epsilon_{j,i} \sim N(0,1)$ are independent and independent from   $\bX_{j,i} \sim N(\bcero,\identidad/9)$, for $j=D,H$,  where $\identidad$ denotes the identity matrix. Besides, the sample from one population was generated independently from the other one. The choice of the parameters ensure that  $Y_D$ is stochastically greater than  $Y_{H}$. Under this model the true area under the curve equals  $\AUC=0.56196$. 
 
The results corresponding to the estimators defined using the empirical distribution function, $\wROC_{\emp}$, those defined in Pulit (2016) denoted $\wROC_{\kernel}$ and those using a convolution approach labelled $\wROC_{\conv}$  are reported in Table  \ref{tab:dim2_caso2_sum_00},  for sample sizes   $n=20, 50$ and $100$. For the kernel--based estimator,    the local bandwidth suggested by Pulit (2016), i.e.,   $ h(p)= c_{n_D} \sqrt{({5 \,p\, (1-p)})/({ 2 n_H}) }$, with $c_{n_D} = 1 + 1.8\, n_D^{-1/5}$  was considered.

Besides, the situation in which the correct model is adjusted, we consider different settings where the regression is misspecified to analyse the sensitivity of the convolution based estimator. We considered a misspecification in which   the intercept is omitted and two situations in which the model is estimated as a linear regression one depending only on the first component of the covariates. In the first case, we fit the model assuming that it depends linearly on $X_{D,j,1}$ or $X_{H,j,1}$, while in the second one we consider 
a misspecified regression depending on $X_{D,j,1}^2$ or $X_{H,j,1}^2$. Tables \ref{tab:dim2_caso2_noint_00} to \ref{tab:dim2_caso2__solox1cuad_00} report the summary measures in these three settings.

Functionals boxplots   introduced by  Sun and Genton (2011)   are    useful to visualize a collection of curves. 
The area in purple represents the 50\% inner band of curves, the dotted red lines correspond to outlying curves, the black line indicates the central (deepest) function, while  we add a green line in the plot that corresponds to the  true  $\ROC$ curve.  Figure  \ref{fig:fbx_dim2_caso2_00} presents the functional boxplots of the estimated ROC curves for the situation in which the regression is correctly specified. Taking into account that the convolution based method depends on the fit of the regression model, Figures \ref{fig:fbx_dim2_caso2_00_nointer} to \ref{fig:fbx_dim2_caso2_00_solox1cuad} display the boxplots of $\wROC_{\conv}$ for different sample sizes when the model is correctly fitted and under  misspecification of the regression function.

The ROC curve estimators obtained using the estimator proposed by Pulit (2016) show their advantage over the other competitors both when considering the mean squared error or the measure based on the Kolmogorov distance. When the regression function is properly fitted, the convolution--based estimators outperform  $\wROC_{\emp}$. Note that when $n=20$ the $MSE$ of  $\wROC_{\conv}$ is a 40\% larger than that of  $\wROC_{\kernel}$, while for $n=100$ it shows only a 5\% increase, while the $\wROC_{\emp}$  have values of $MSE$ more than a 10\% larger than those obtained with the convolution method for all sample sizes.  

Figure  \ref{fig:fbx_dim2_caso2_00} illustrates the smoothness of the kernel--based estimators $\wROC_{\kernel}$.  
To compare the smoothness of the estimators, we have computed the mean over replications of the following measure that gives an approximation of $\int_0^1  \left(\wROC^{\prime}(p)\right)^2 dp$ 
$$SM(\wROC)=\frac 1{d^2} \frac{1}{ N_p-1}  \sum_{j=2}^{N_p}\left( \wROC (p_j)-\wROC (p_{j-1}) \right)^2$$
where $p_1\le p_2\le \dots \le p_{N_p}$ are the values of the   equidistant grid  ${\cal G}_p$ and $d=p_j-p_{j-1}$ is the spacing. For the considered situation, we have that $SM(\wROC_{\emp})=2.74$ while $SM(\wROC_{\conv})=1.34$. Besides, for the kernel smoothing procedure defined in Pulit (2016), we have  $SM(\wROC_{\kernel})= 1.12$, while the smoothness of the true curve equals $SM(\ROC)=1.04$.  The obtained results mean  that, as expected,  the convolution indeed smooths the estimation, but less than the kernel smoother.

Regarding the estimation of the area under the curve, again the kernel--based estimators outperform the other two in mean squared error and when considering the mean relative bias. However, when looking at the bias, the estimator $\wAUC_{\emp}$ has a smaller bias than the other two for $n=20$ and $n=50$, while for $n=100$ its bias is larger than that of the convolution based proposal.

Quite surprisingly the considered regression  misspecification settings  do not seem to affect the convolution--based estimators of the ROC curve. Both the summary measures and the functional boxplots of  $\wROC_{\conv}$ displayed in Figures \ref{fig:fbx_dim2_caso2_00_nointer} to \ref{fig:fbx_dim2_caso2_00_solox1cuad} remain quite stable when an incorrect model is fitted.

 In summary, from this first numerical experiment, we have that both, the smoothness of $\wROC_{\kernel}$ and its performance under the smaller sample sizes, make this estimator as a good competitor. 

\footnotesize
   
\begin{table}[ht!]
\centering
\setlength{\tabcolsep}{4pt}
\renewcommand{\arraystretch}{1}
\begin{tabular}{c|rrr|rrr|rrr|}
  \hline
  & \multicolumn{3}{c|}{$n=20$} & \multicolumn{3}{c|}{$n=50$} & \multicolumn{3}{c|}{$n=100$}\\
  \hline  
 $1000\times$ & $\wROC_{\ipw} $ & $\wROC_{\kernel}$ &  $\wROC_{\conv} $ &  $\wROC_{\ipw} $ & $\wROC_{\kernel}$ &  $\wROC_{\conv} $ &  $\wROC_{\ipw} $ & $\wROC_{\kernel}$ &  $\wROC_{\conv} $\\
  \hline

$MSE$ & 16.28 & 10.25 & 14.78 & 6.45 & 4.93 & 5.77 & 3.21 & 2.67 & 2.83  \\ 
  $KS$&  247.81 & 135.12 & 210.63 & 158.63 & 103.86 & 130.36 & 112.74 & 82.48 & 92.10 \\ 
   \hline
 
 $1000\times$ & $\wAUC_{\ipw}$ & $\wAUC_{\kernel}$ &  $\wAUC_{\conv}$ & $\wAUC_{\ipw}$ & $\wAUC_{\kernel}$ &  $\wAUC_{\conv}$ & $\wAUC_{\ipw}$ & $\wAUC_{\kernel}$ &  $\wAUC_{\conv}$\\
 \hline 
  Bias  & -0.71 & -8.10 & -2.58 & 0.66 & -2.89 & -0.97 & 1.62 & -0.93 & 0.01 \\ 
 $RB$ & 130.48 & 120.71 & 130.50 & 80.23 & 78.11 & 79.85 & 55.64 & 55.00 & 55.59 \\ 
$MSE$ & 8.43 & 7.20 & 8.43 & 3.27 & 3.10 & 3.24 & 1.54 & 1.50 & 1.53 \\ 
   \hline
\end{tabular}
 \caption{\label{tab:dim2_caso2_sum_00}Summary measures for ROC and AUC when $\bX\in \real^2$.} 
\end{table}


\begin{table}[ht!]
\centering
\setlength{\tabcolsep}{4pt}
\renewcommand{\arraystretch}{0.9}
\begin{tabular}{c|rrr|rrr|rrr|}
  \hline
 $1000\times$ & $\wROC_{\ipw} $ & $\wROC_{\kernel}$ &  $\wROC_{\conv} $ &  $\wROC_{\ipw} $ & $\wROC_{\kernel}$ &  $\wROC_{\conv} $ &  $\wROC_{\ipw} $ & $\wROC_{\kernel}$ &  $\wROC_{\conv} $\\
  \hline
    & \multicolumn{3}{c|}{$n=20$} & \multicolumn{3}{c|}{$n=50$} & \multicolumn{3}{c|}{$n=100$}\\
  \hline
 & \multicolumn{9}{c|}{(a)}\\
 \hline
  $MSE$ & 16.28 & 10.25 & 14.38 & 6.45 & 4.93 & 5.73 & 3.21 & 2.67 & 2.81 \\ 
 $KS$ & 247.81 & 135.12 & 202.53 & 158.63 & 103.86 & 127.76 & 112.74 & 82.48 & 91.15 \\ 
    \hline 
   & \multicolumn{9}{c|}{(b)}\\
 \hline
    
 $MSE$ & 16.28 & 10.25 & 14.33 & 6.45 & 4.93 & 5.70 & 3.21 & 2.67 & 2.79 \\ 
  $KS$ & 247.81 & 135.12 & 202.27 & 158.63 & 103.86 & 127.40 & 112.74 & 82.48 & 90.80 \\ 
  \hline
    & \multicolumn{9}{c|}{(c)}\\
 \hline
$MSE$  & 16.28 & 10.25 & 14.84 & 6.45 & 4.93 & 5.79 & 3.21 & 2.67 & 2.84 \\ 
  $KS$ & 247.81 & 135.12 & 211.14 & 158.63 & 103.86 & 130.73 & 112.74 & 82.48 & 92.43 \\ 
  \hline

  $1000\times$ & $\wAUC_{\ipw}$ & $\wAUC_{\kernel}$ &  $\wAUC_{\conv}$ & $\wAUC_{\ipw}$ & $\wAUC_{\kernel}$ &  $\wAUC_{\conv}$& $\wAUC_{\ipw}$ & $\wAUC_{\kernel}$ &  $\wAUC_{\conv}$ \\
 \hline
    & \multicolumn{3}{c|}{$n=20$} & \multicolumn{3}{c|}{$n=50$} & \multicolumn{3}{c|}{$n=100$}\\
   \hline
   & \multicolumn{9}{c|}{(a)}\\
 \hline 
 Bias & -0.71 & -8.10 & -4.04 & 0.66 & -2.89 & -1.51 & 1.62 & -0.93 & -0.33 \\ 
$RB$ & 130.48 & 120.71 & 128.86 & 80.23 & 78.11 & 79.07 & 55.64 & 55.00 & 54.92 \\ 
 $MSE$ & 8.43 & 7.20 & 8.23 & 3.27 & 3.10 & 3.18 & 1.54 & 1.50 & 1.49 \\    \hline
& \multicolumn{9}{c|}{(b)}\\
 \hline
Bias & -0.71 & -8.10 & -4.49 & 0.66 & -2.89 & -1.70 & 1.62 & -0.93 & -0.42 \\ 
  $RB$ & 130.48 & 120.71 & 128.63 & 80.23 & 78.11 & 78.87 & 55.64 & 55.00 & 54.73 \\ 
   $MSE$ & 8.43 & 7.20 & 8.21 & 3.27 & 3.10 & 3.17 & 1.54 & 1.50 & 1.48 \\ 
   \hline
 & \multicolumn{9}{c|}{(c)}\\
 \hline
  Bias & -0.71 & -8.10 & -2.15 & 0.66 & -2.89 & -0.78 & 1.62 & -0.93 & 0.09 \\ 
 $RB$ & 130.48 & 120.71 & 130.74 & 80.23 & 78.11 & 80.08 & 55.64 & 55.00 & 55.78 \\ 
$MSE$  & 8.43 & 7.20 & 8.46 & 3.27 & 3.10 & 3.26 & 1.54 & 1.50 & 1.54 \\ 
   \hline
 \end{tabular}
\caption{\label{tab:dim2_caso2_noint_00}Summary measures for the ROC curve and the  AUC when $\bX\in \real^2$ under  misspecification of the regression model which is assumed to be a linear one without intercept: (a) in both populations, (b) only in the diseased population and (c) only in the  healthy one.} 
\end{table}

\begin{table}[ht!]
\centering
\setlength{\tabcolsep}{4pt}
\renewcommand{\arraystretch}{0.9}
\begin{tabular}{c|rrr|rrr|rrr|}
  \hline
  $1000\times$ & $\wROC_{\ipw} $ & $\wROC_{\kernel}$ &  $\wROC_{\conv} $ &  $\wROC_{\ipw} $ & $\wROC_{\kernel}$ &  $\wROC_{\conv} $ &  $\wROC_{\ipw} $ & $\wROC_{\kernel}$ &  $\wROC_{\conv} $\\
  \hline
    & \multicolumn{3}{c|}{$n=20$} & \multicolumn{3}{c|}{$n=50$} & \multicolumn{3}{c|}{$n=100$}\\
  \hline
 & \multicolumn{9}{c|}{(a)}\\
\hline
$MSE$ & 16.28 & 10.25 & 13.99 & 6.45 & 4.93 & 5.62 & 3.21 & 2.67 & 2.79 \\ 
 $KS$ & 247.81 & 135.12 & 195.62 & 158.63 & 103.86 & 123.96 & 112.74 & 82.48 & 88.26 \\ 
   \hline
    & \multicolumn{9}{c|}{(b)}\\
\hline
$MSE$ & 16.28 & 10.25 & 13.86 & 6.45 & 4.93 & 5.36 & 3.21 & 2.67 & 2.65 \\ 
  $KS$ & 247.81 & 135.12 & 189.65 & 158.63 & 103.86 & 116.36 & 112.74 & 82.48 & 81.88 \\ 
   \hline
 
 & \multicolumn{9}{c|}{(c)}\\
\hline
$MSE$ & 16.28 & 10.25 & 14.93 & 6.45 & 4.93 & 6.03 & 3.21 & 2.67 & 2.97 \\ 
 $KS$ & 247.81 & 135.12 & 215.38 & 158.63 & 103.86 & 136.37 & 112.74 & 82.48 & 96.88 \\ 
   \hline
 
  $1000\times$ & $\wAUC_{\ipw}$ & $\wAUC_{\kernel}$ &  $\wAUC_{\conv}$ & $\wAUC_{\ipw}$ & $\wAUC_{\kernel}$ &  $\wAUC_{\conv}$ & $\wAUC_{\ipw}$ & $\wAUC_{\kernel}$ &  $\wAUC_{\conv}$\\
 \hline
    & \multicolumn{3}{c|}{$n=20$} & \multicolumn{3}{c|}{$n=50$} & \multicolumn{3}{c|}{$n=100$}\\
   \hline
   & \multicolumn{9}{c|}{(a)}\\
 \hline 
 
Bias & -0.71 & -8.10 & -2.41 & 0.66 & -2.89 & -0.59 & 1.62 & -0.93 & 0.03 \\ 
$RB$ & 130.48 & 120.71 & 130.50 & 80.23 & 78.11 & 80.38 & 55.64 & 55.00 & 55.69 \\ 
$MSE$ & 8.43 & 7.20 & 8.42 & 3.27 & 3.10 & 3.27 & 1.54 & 1.50 & 1.54 \\ 
   \hline
 
    & \multicolumn{9}{c|}{(b)}\\
\hline
Bias & -0.71 & -8.10 & -2.91 & 0.66 & -2.89 & -1.01 & 1.62 & -0.93 & 0.05 \\ 
$RB$ & 130.48 & 120.71 & 130.25 & 80.23 & 78.11 & 79.51 & 55.64 & 55.00 & 55.48 \\ 
$MSE$ & 8.43 & 7.20 & 8.43 & 3.27 & 3.10 & 3.22 & 1.54 & 1.50 & 1.53 \\ 
     \hline
    & \multicolumn{9}{c|}{(c)}\\
\hline
Bias & -0.71 & -8.10 & -2.19 & 0.66 & -2.89 & -0.65 & 1.62 & -0.93 & 0.03 \\ 
$RB$ & 130.48 & 120.71 & 130.73 & 80.23 & 78.11 & 80.82 & 55.64 & 55.00 & 55.68 \\ 
$MSE$ & 8.43 & 7.20 & 8.43 & 3.27 & 3.10 & 3.30 & 1.54 & 1.50 & 1.54 \\ 
\hline
\end{tabular}
\caption{\label{tab:dim2_caso2_solox1_00}Summary measures for the ROC curve and the  AUC when $\bX\in \real^2$ under misspecification of the regression model which is assumed to be a linear model depending only on the first component of $\bX$: (a) in both populations, (b) only in the diseased population and (c) only in the  healthy one.} 
\end{table}

\begin{table}[ht!]
\centering
\setlength{\tabcolsep}{4pt}
\renewcommand{\arraystretch}{0.9}
\begin{tabular}{c|rrr|rrr|rrr|}
  \hline
  $1000\times$ & $\wROC_{\ipw} $ & $\wROC_{\kernel}$ &  $\wROC_{\conv} $ &  $\wROC_{\ipw} $ & $\wROC_{\kernel}$ &  $\wROC_{\conv} $ &  $\wROC_{\ipw} $ & $\wROC_{\kernel}$ &  $\wROC_{\conv} $\\
  \hline
    & \multicolumn{3}{c|}{$n=20$} & \multicolumn{3}{c|}{$n=50$} & \multicolumn{3}{c|}{$n=100$}\\
  \hline
 & \multicolumn{9}{c|}{(a)}\\
 \hline
 $MSE$  & 16.28 & 10.25 & 14.84 & 6.45 & 4.93 & 5.99 & 3.21 & 2.67 & 3.04 \\ 
$KS$ & 247.81 & 135.12 & 214.31 & 158.63 & 103.86 & 140.33 & 112.74 & 82.48 & 103.00 \\ 
  \hline
  
   & \multicolumn{9}{c|}{(b)}\\
    \hline
  $MSE$  & 16.28 & 10.25 & 14.38 & 6.45 & 4.93 & 5.64 & 3.21 & 2.67 & 2.81 \\ 
$KS$ & 247.81 & 135.12 & 201.88 & 158.63 & 103.86 & 128.58 & 112.74 & 82.48 & 92.86 \\ 
 \hline
  & \multicolumn{9}{c|}{(c)}\\
  \hline
 
$MSE$  & 16.28 & 10.25 & 15.23 & 6.45 & 4.93 & 6.13 & 3.21 & 2.67 & 3.06 \\ 
$KS$  & 247.81 & 135.12 & 221.32 & 158.63 & 103.86 & 142.32 & 112.74 & 82.48 & 102.57 \\ 
   \hline
 
  $1000\times$ & $\wAUC_{\ipw}$ & $\wAUC_{\kernel}$ &  $\wAUC_{\conv}$ & $\wAUC_{\ipw}$ & $\wAUC_{\kernel}$ &  $\wAUC_{\conv}$ & $\wAUC_{\ipw}$ & $\wAUC_{\kernel}$ &  $\wAUC_{\conv}$\\
 \hline
    & \multicolumn{3}{c|}{$n=20$} & \multicolumn{3}{c|}{$n=50$} & \multicolumn{3}{c|}{$n=100$}\\
  \hline
   & \multicolumn{9}{c|}{(a)}\\
 \hline 
Bias & -0.71 & -8.10 & -2.41 & 0.66 & -2.89 & -0.75 & 1.62 & -0.93 & 0.02 \\ 
$RB$ & 130.48 & 120.71 & 130.92 & 80.23 & 78.11 & 80.24 & 55.64 & 55.00 & 55.84 \\ 
$MSE$ & 8.43 & 7.20 & 8.46 & 3.27 & 3.10 & 3.27 & 1.54 & 1.50 & 1.54 \\ 
    \hline
   & \multicolumn{9}{c|}{(b)}\\
 \hline 
 Bias & -0.71 & -8.10 & -2.50 & 0.66 & -2.89 & -1.06 & 1.62 & -0.93 & 0.01 \\ 
$RB$ & 130.48 & 120.71 & 130.83 & 80.23 & 78.11 & 79.71 & 55.64 & 55.00 & 55.64 \\ 
 $MSE$  & 8.43 & 7.20 & 8.48 & 3.27 & 3.10 & 3.24 & 1.54 & 1.50 & 1.53 \\ 
   \hline

 & \multicolumn{9}{c|}{(c)}\\
 \hline 
 Bias & -0.71 & -8.10 & -2.34 & 0.66 & -2.89 & -0.70 & 1.62 & -0.93 & 0.01 \\ 
$RB$ & 130.48 & 120.71 & 130.68 & 80.23 & 78.11 & 80.39 & 55.64 & 55.00 & 55.80 \\ 
$MSE$ & 8.43 & 7.20 & 8.43 & 3.27 & 3.10 & 3.28 & 1.54 & 1.50 & 1.54 \\ 

  \hline
\end{tabular}
\caption{\label{tab:dim2_caso2__solox1cuad_00}Summary measures for the ROC curve and the AUC  when $\bX\in \real^2$ under misspecification of the regression model which is assumed to be a linear model depending only on the square of the first component of $\bX$: (a) in both populations, (b) only in the diseased population and (c) only in the  healthy one. } 
\end{table}

\begin{figure}[ht!]
 \begin{center}
 \footnotesize
 \renewcommand{\arraystretch}{0.4}
 \newcolumntype{M}{>{\centering\arraybackslash}m{\dimexpr.1 \linewidth-1\tabcolsep}}
   \newcolumntype{G}{>{\centering\arraybackslash}m{\dimexpr.3\linewidth-1\tabcolsep}}
\begin{tabular}{M G G G}\\
& $\wROC_{\emp}$ & $\wROC_{\kernel}$ & $\wROC_{\conv}$\\
$n=20$ & \includegraphics[scale=0.3]{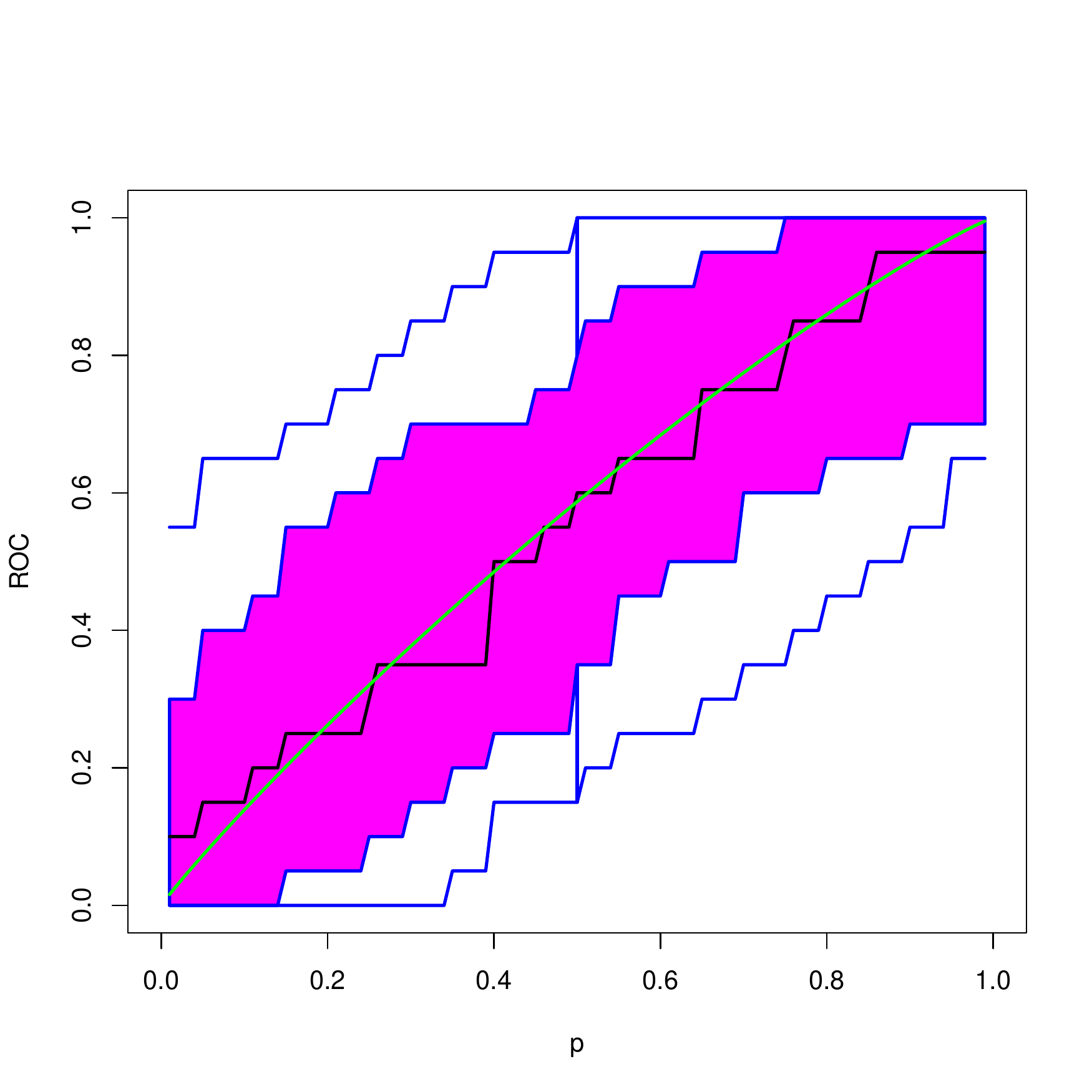}  
 & \includegraphics[scale=0.3]{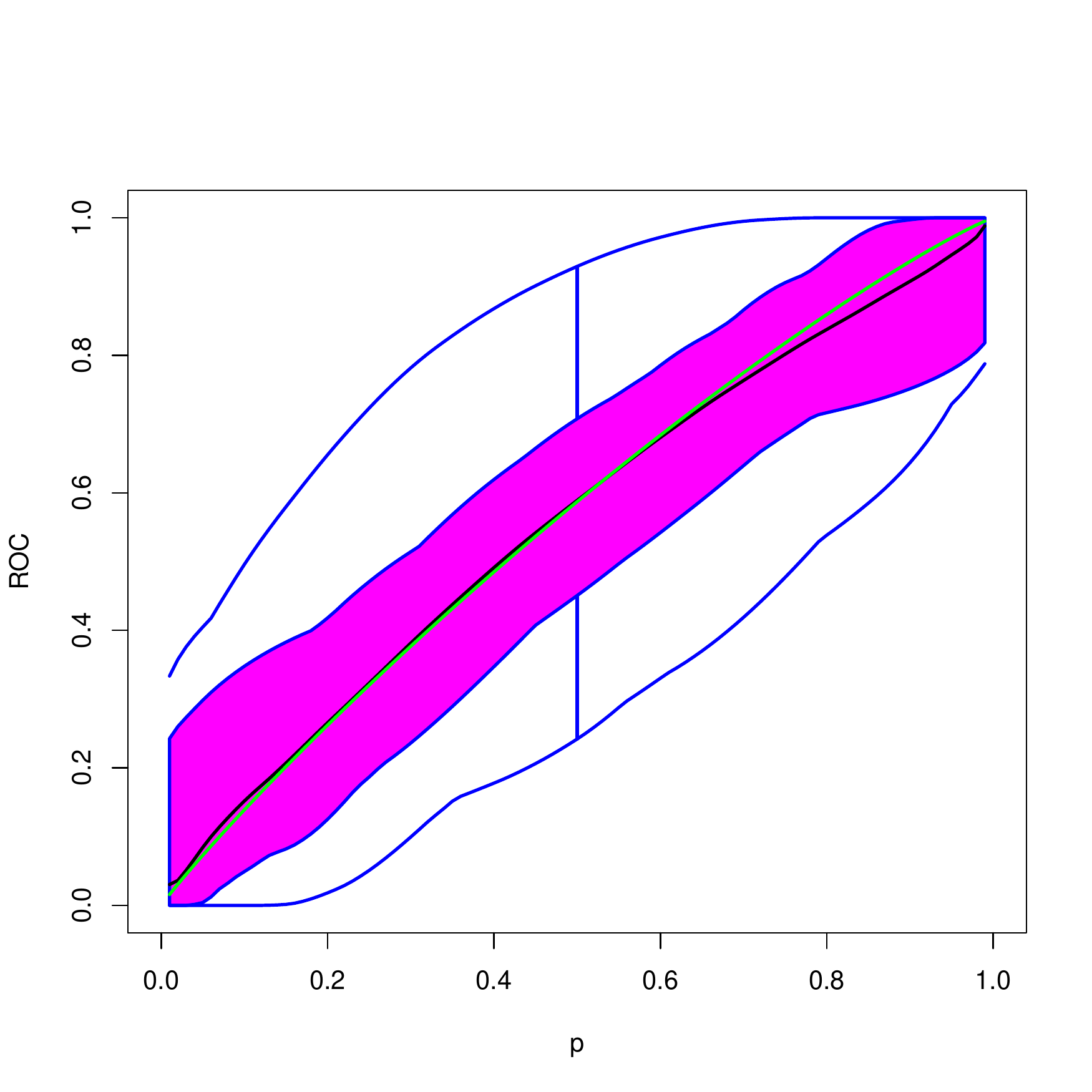}  
 &  \includegraphics[scale=0.3]{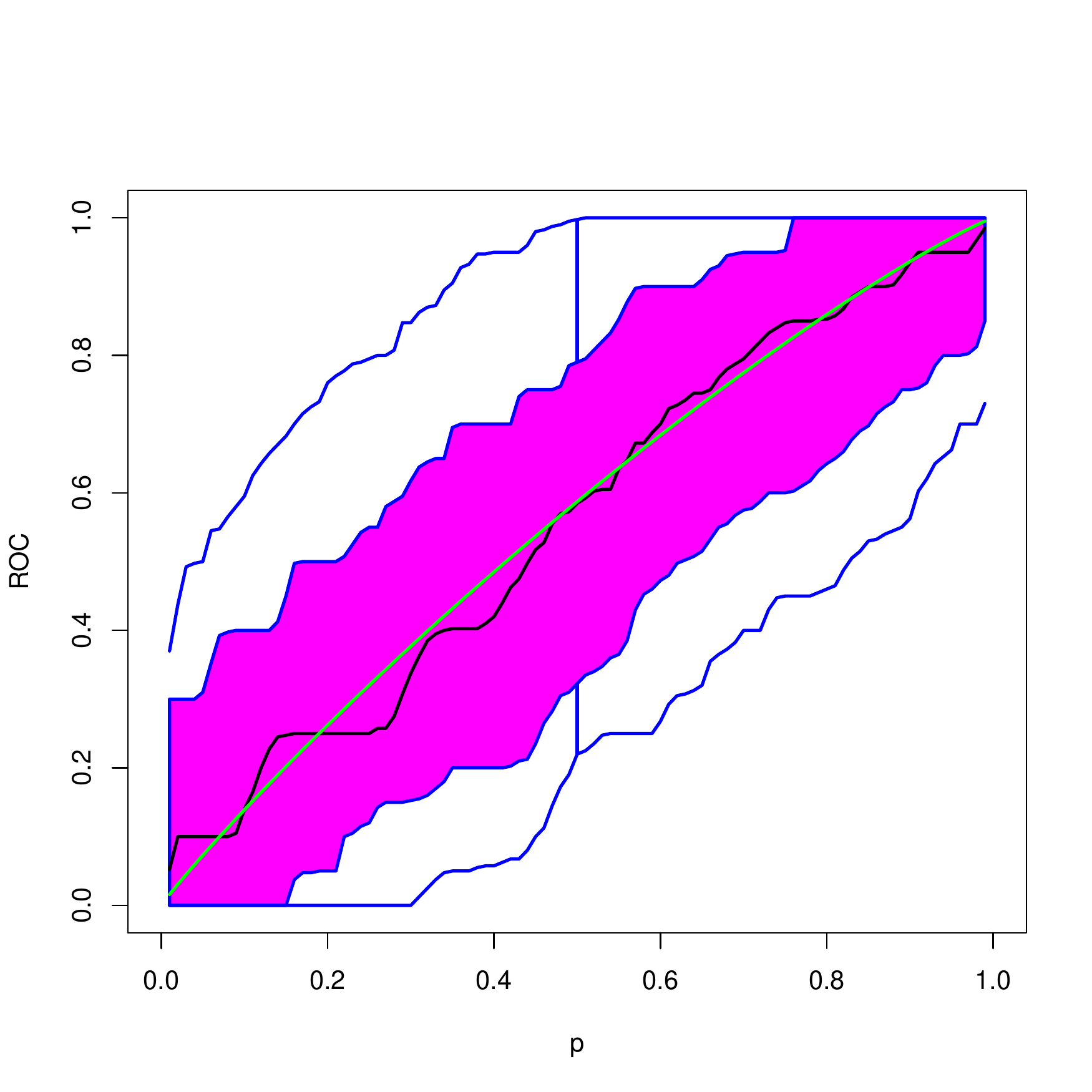}  
\\
$n=50$ & \includegraphics[scale=0.3]{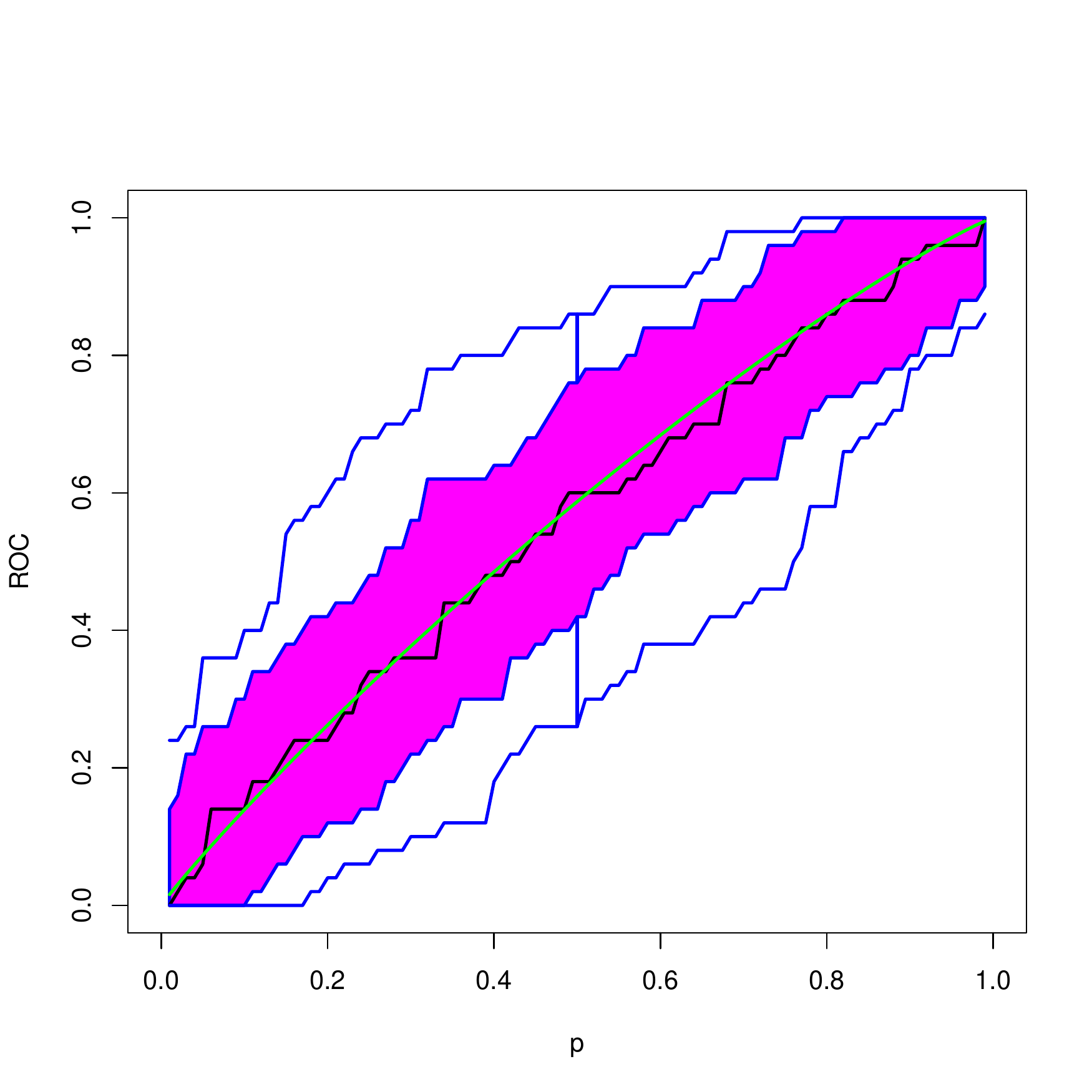}  
 & \includegraphics[scale=0.3]{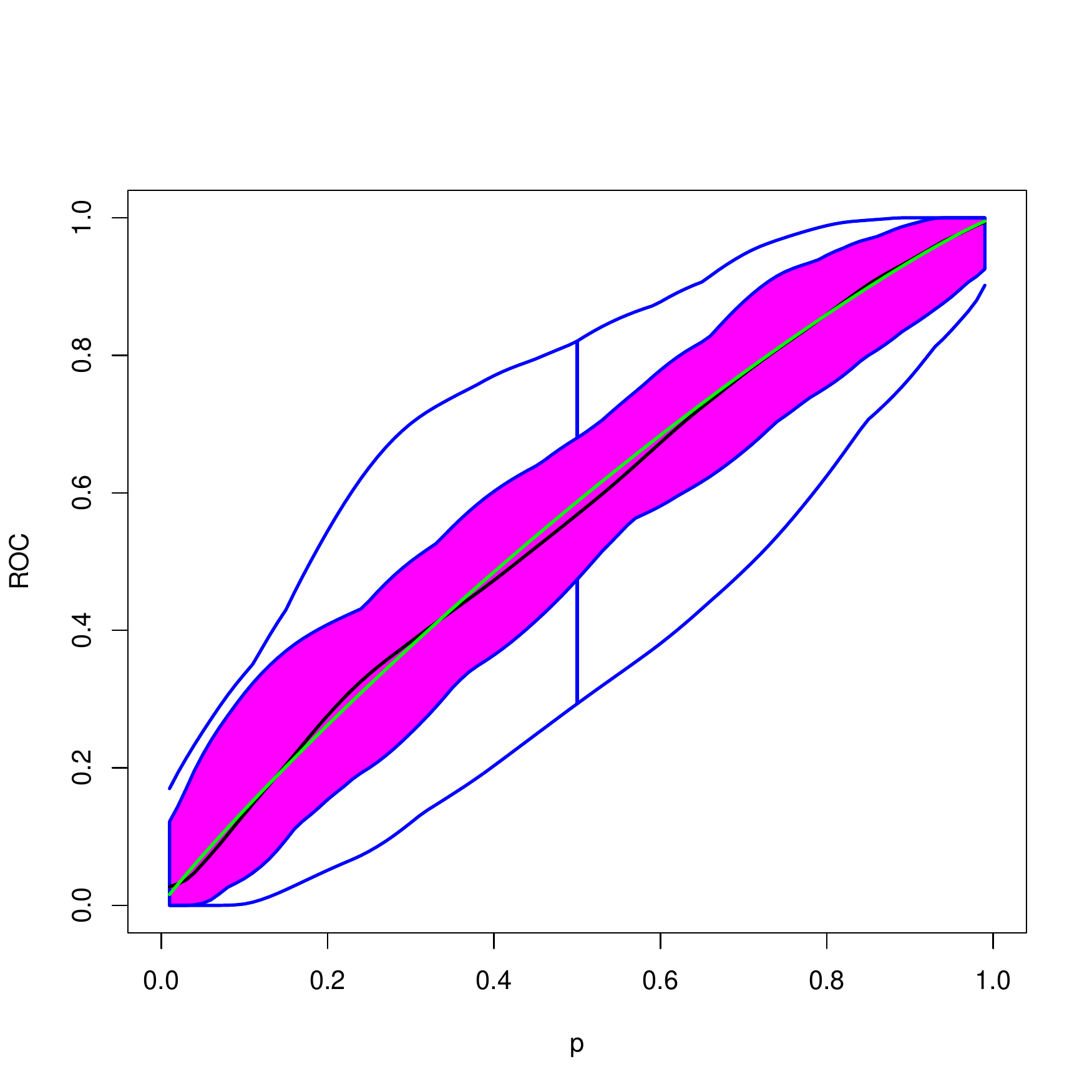}  
 &  \includegraphics[scale=0.3]{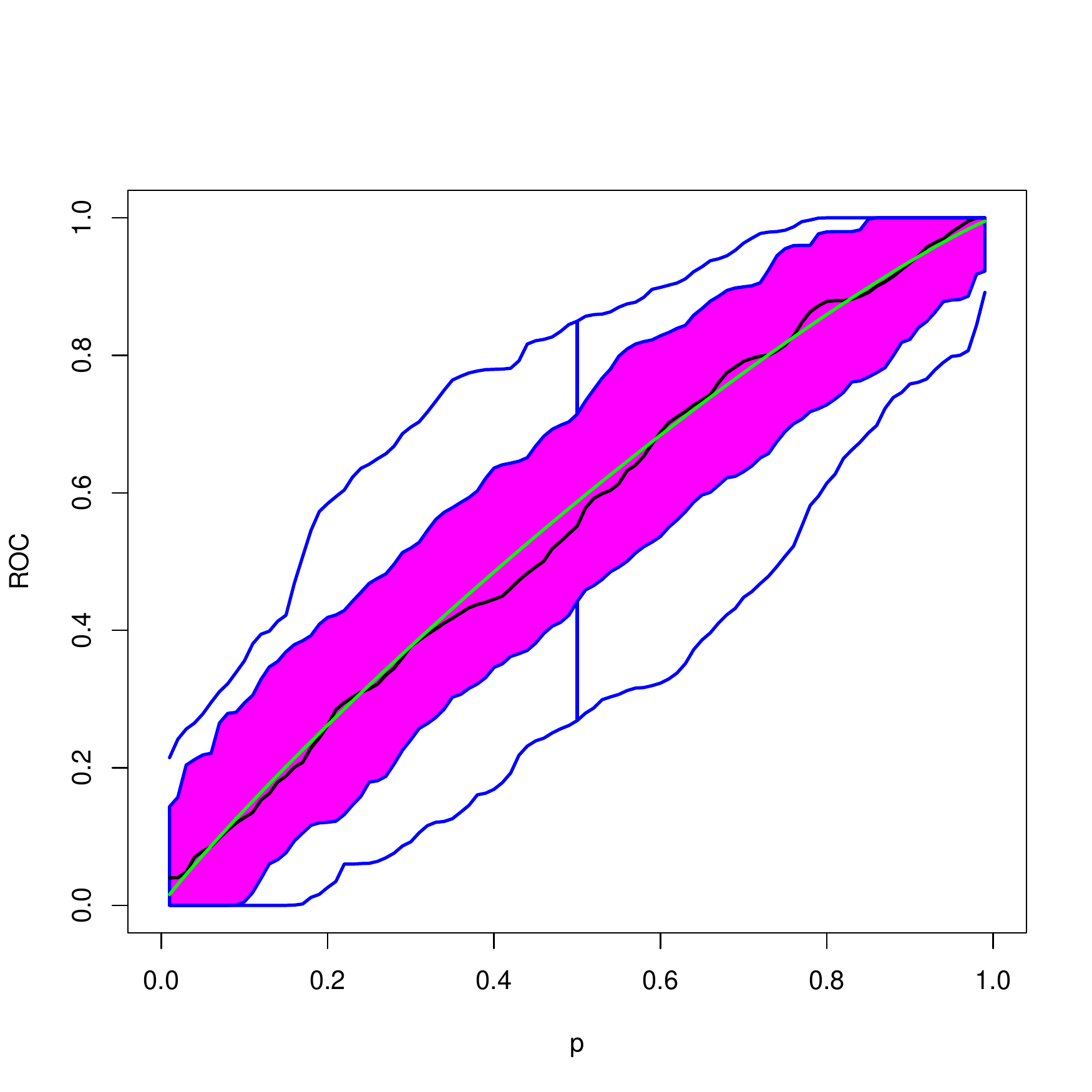}  
\\
 $n=100$ & \includegraphics[scale=0.3]{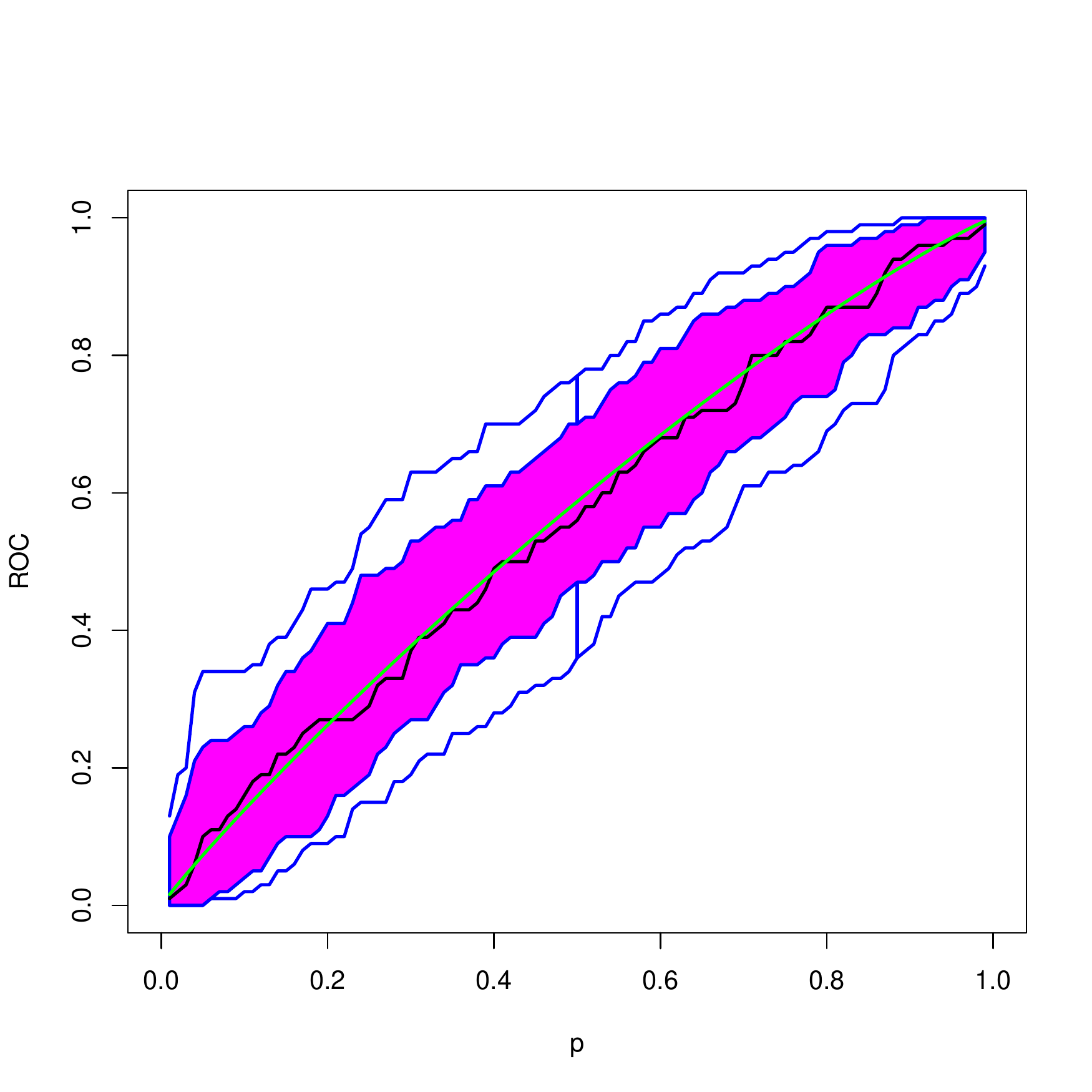}  
 & \includegraphics[scale=0.3]{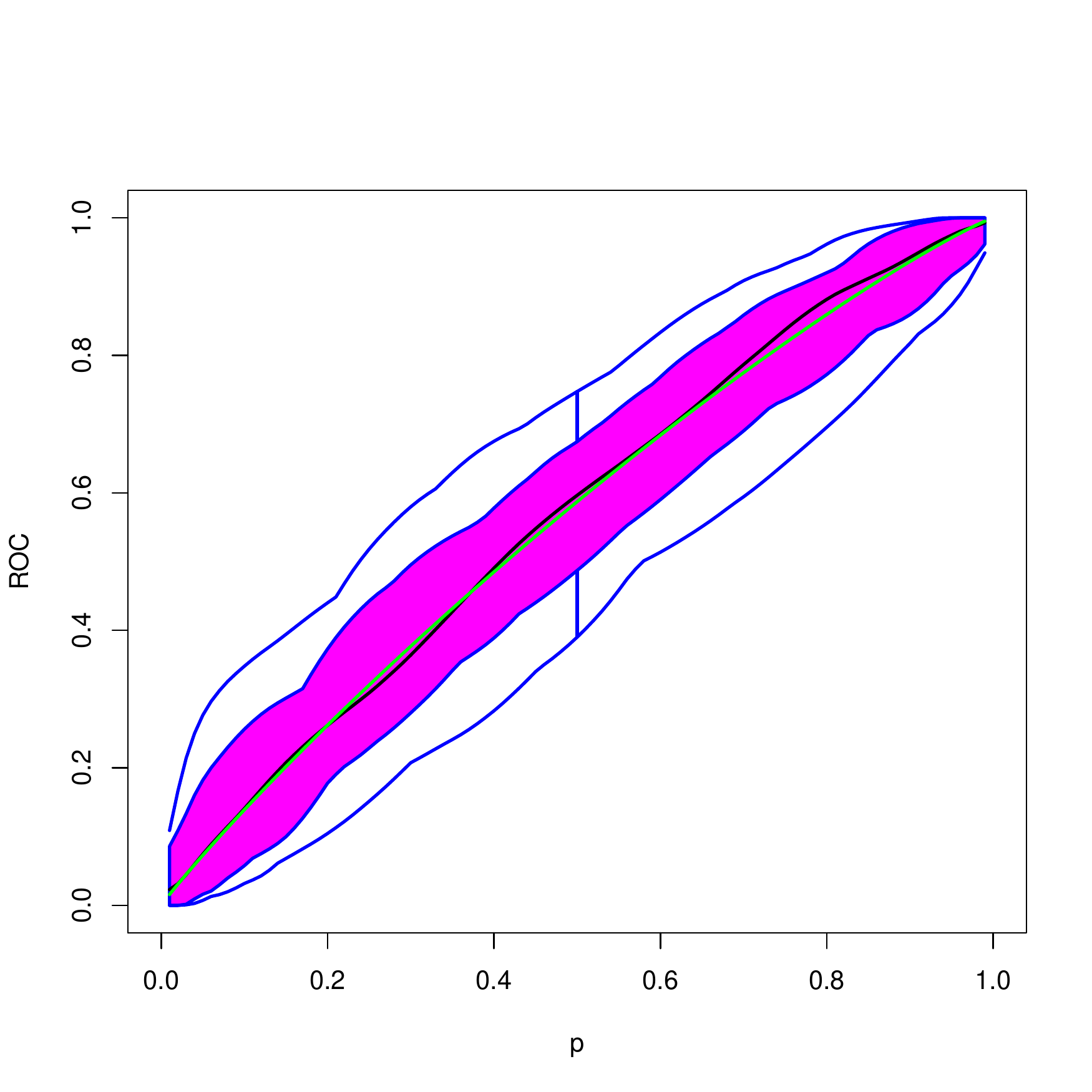}  
 &  \includegraphics[scale=0.3]{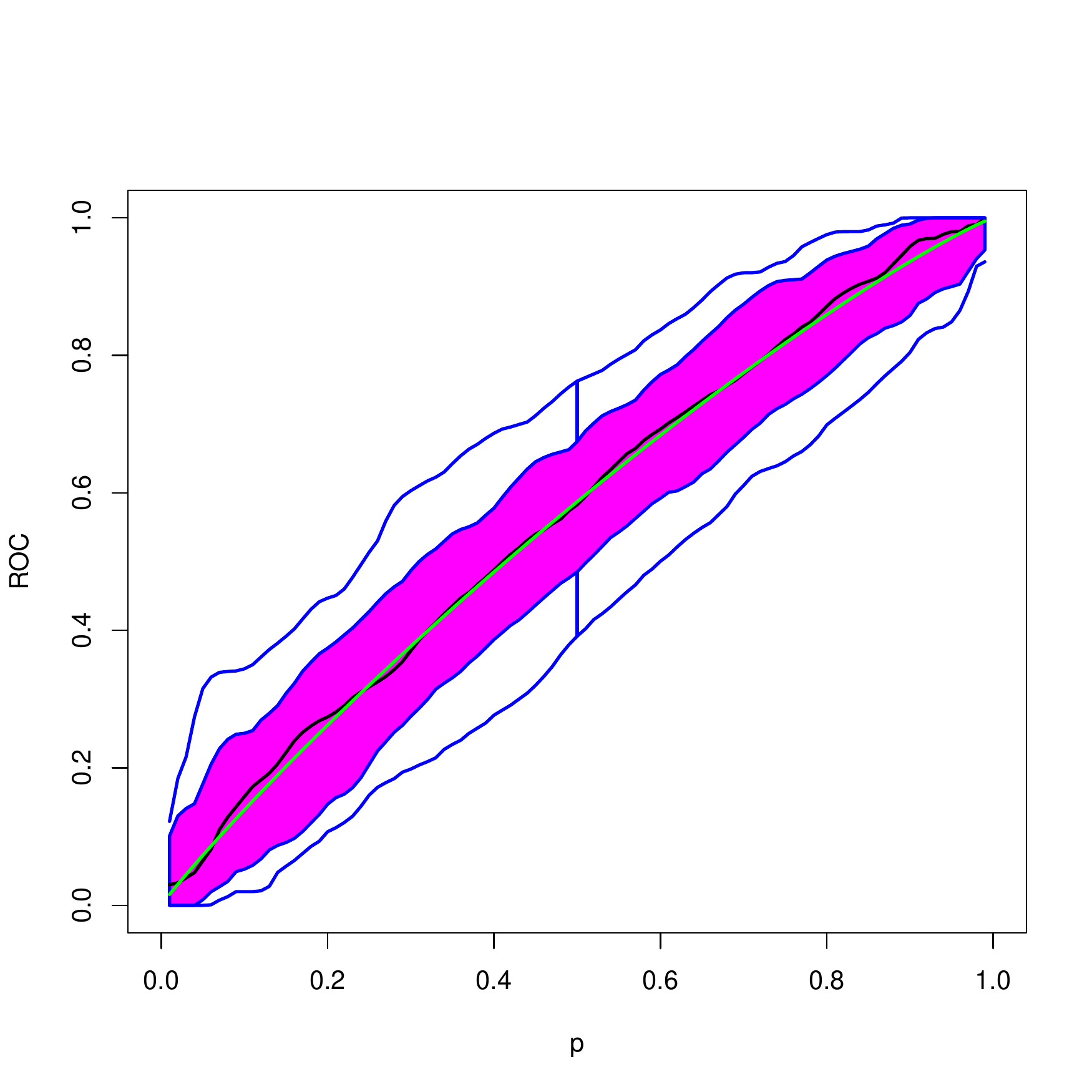}  
 \end{tabular}
\vskip-0.1in  
\caption{\label{fig:fbx_dim2_caso2_00}\small Functional boxplots of  $\wROC(p)$   for $\bX\in \real^2$ and $n_D=n_H=n=20,50$ and 100. The green line corresponds to the true $ROC(p)$ and the dotted red lines to the outlying curves detected by the functional boxplot.}
\end{center} 
\end{figure}
\normalsize


 
\begin{figure}[ht!]
 \begin{center}
 \footnotesize
 \renewcommand{\arraystretch}{0.4}
 \newcolumntype{M}{>{\centering\arraybackslash}m{\dimexpr.1\linewidth-1\tabcolsep}}
   \newcolumntype{G}{>{\centering\arraybackslash}m{\dimexpr.3\linewidth-1\tabcolsep}}
\begin{tabular}{MG G G}\\
 & $n=20$ & $n=50$ & $n=100$\\[-0.1in]
(a) 
&  \includegraphics[scale=0.25]{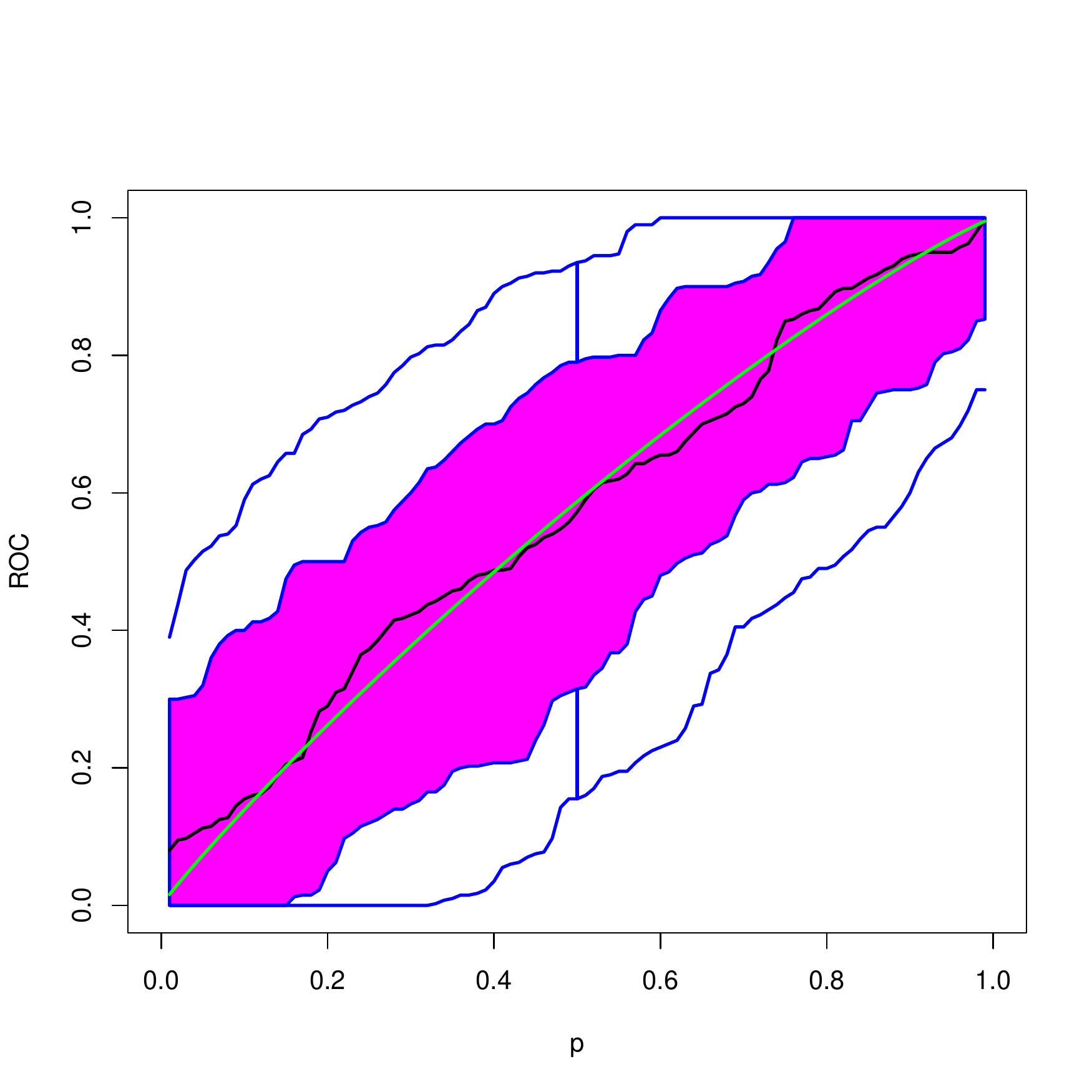} 
&  \includegraphics[scale=0.25]{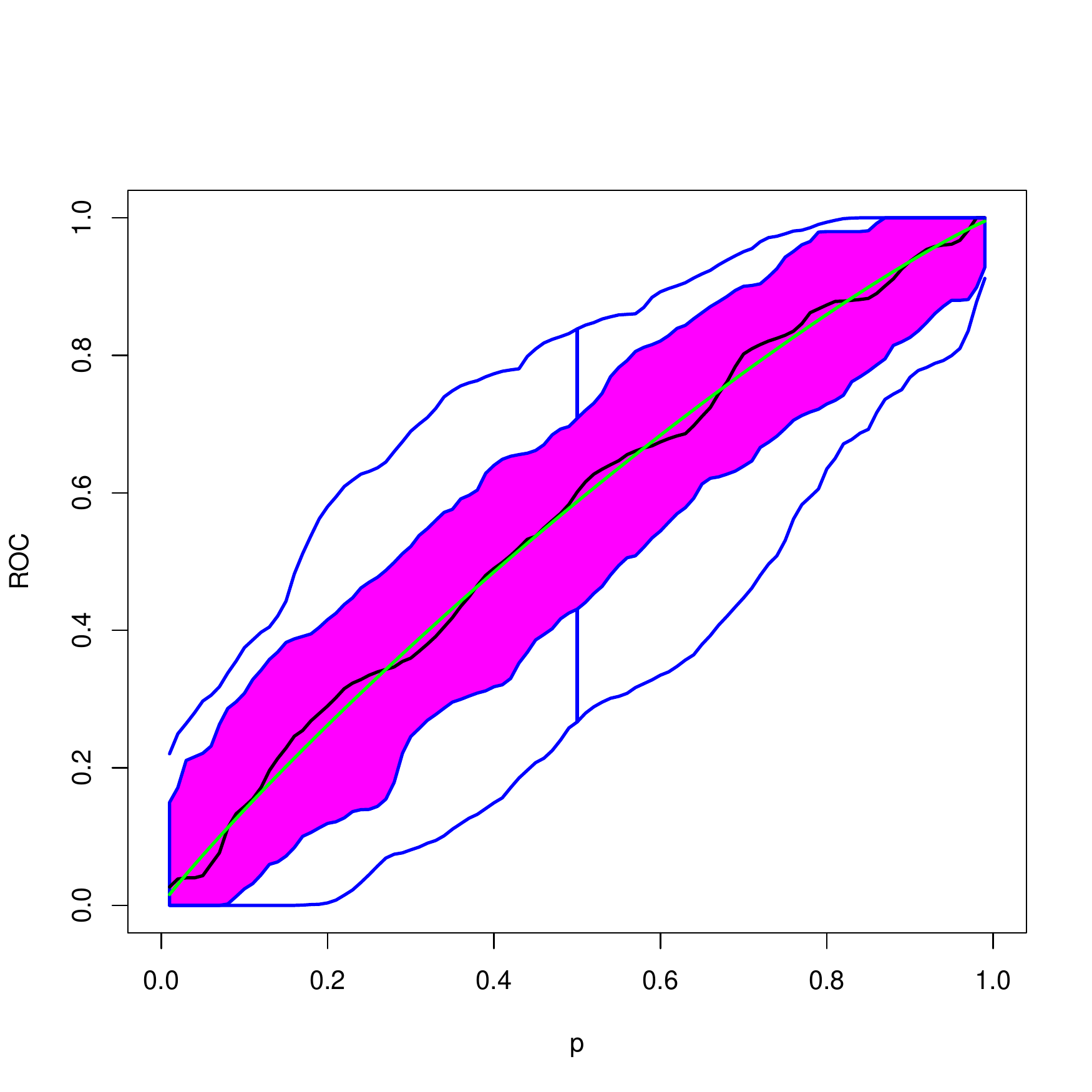}  
&  \includegraphics[scale=0.25]{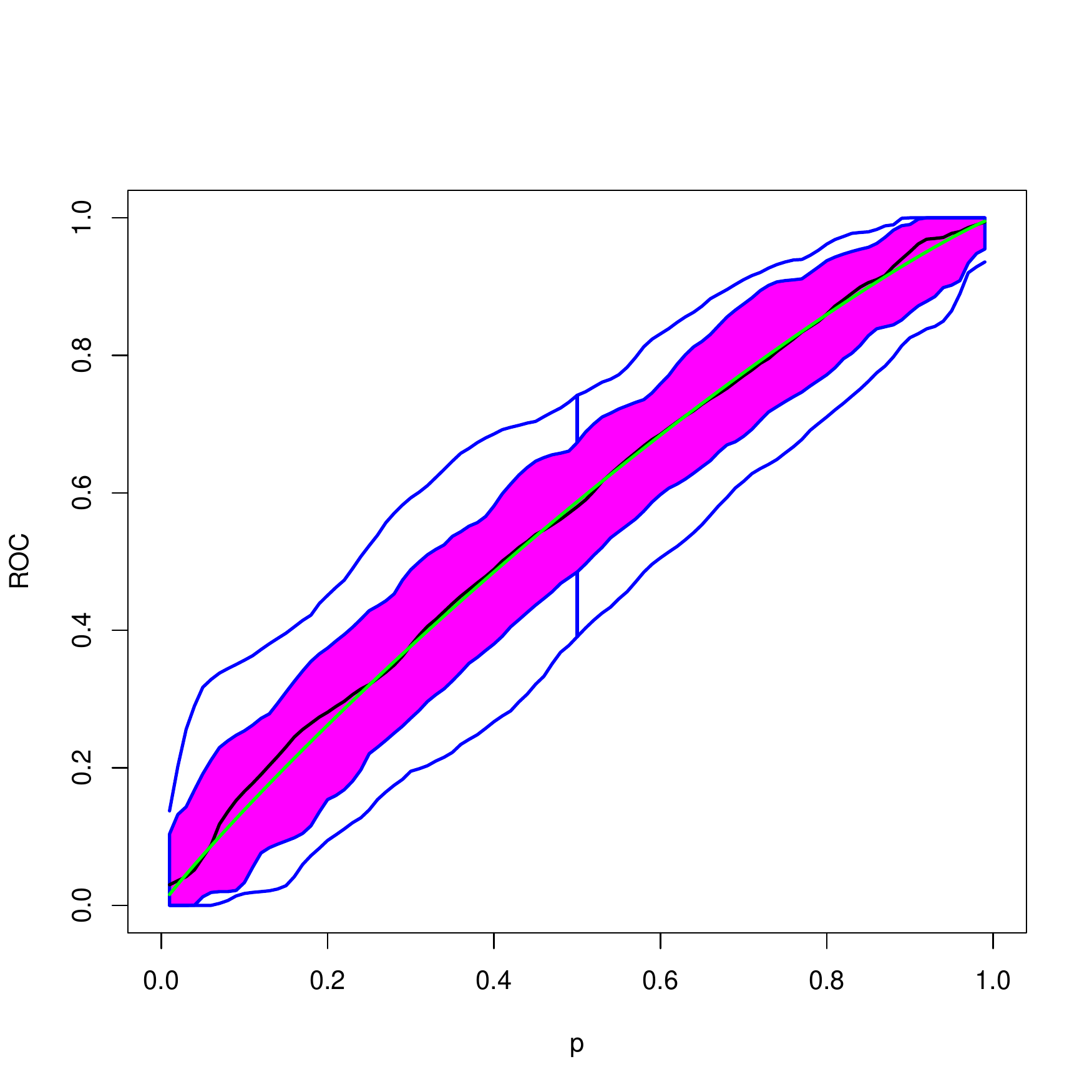}  \\[-0.1in]
(b)    
 & \includegraphics[scale=0.25]{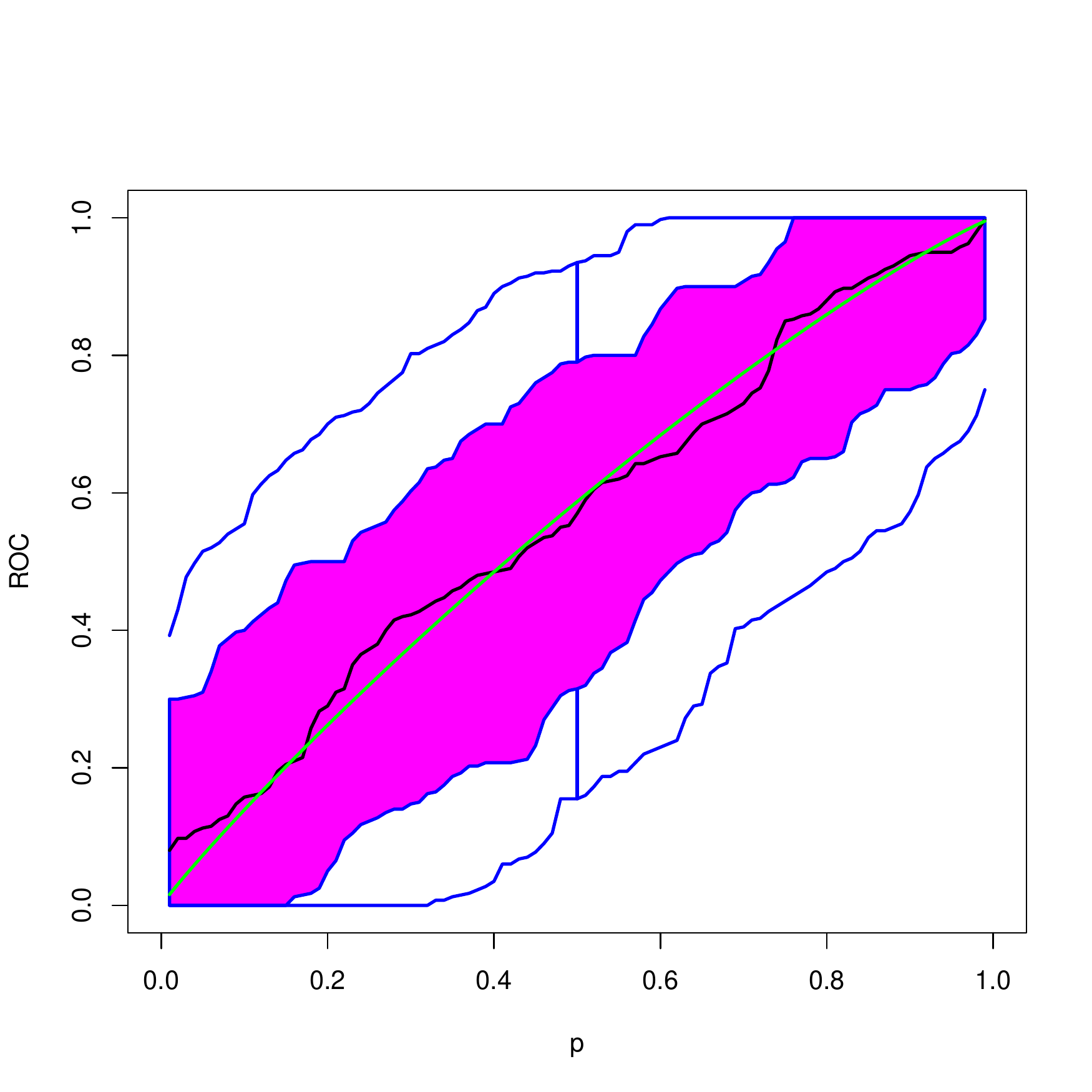}
 & \includegraphics[scale=0.25]{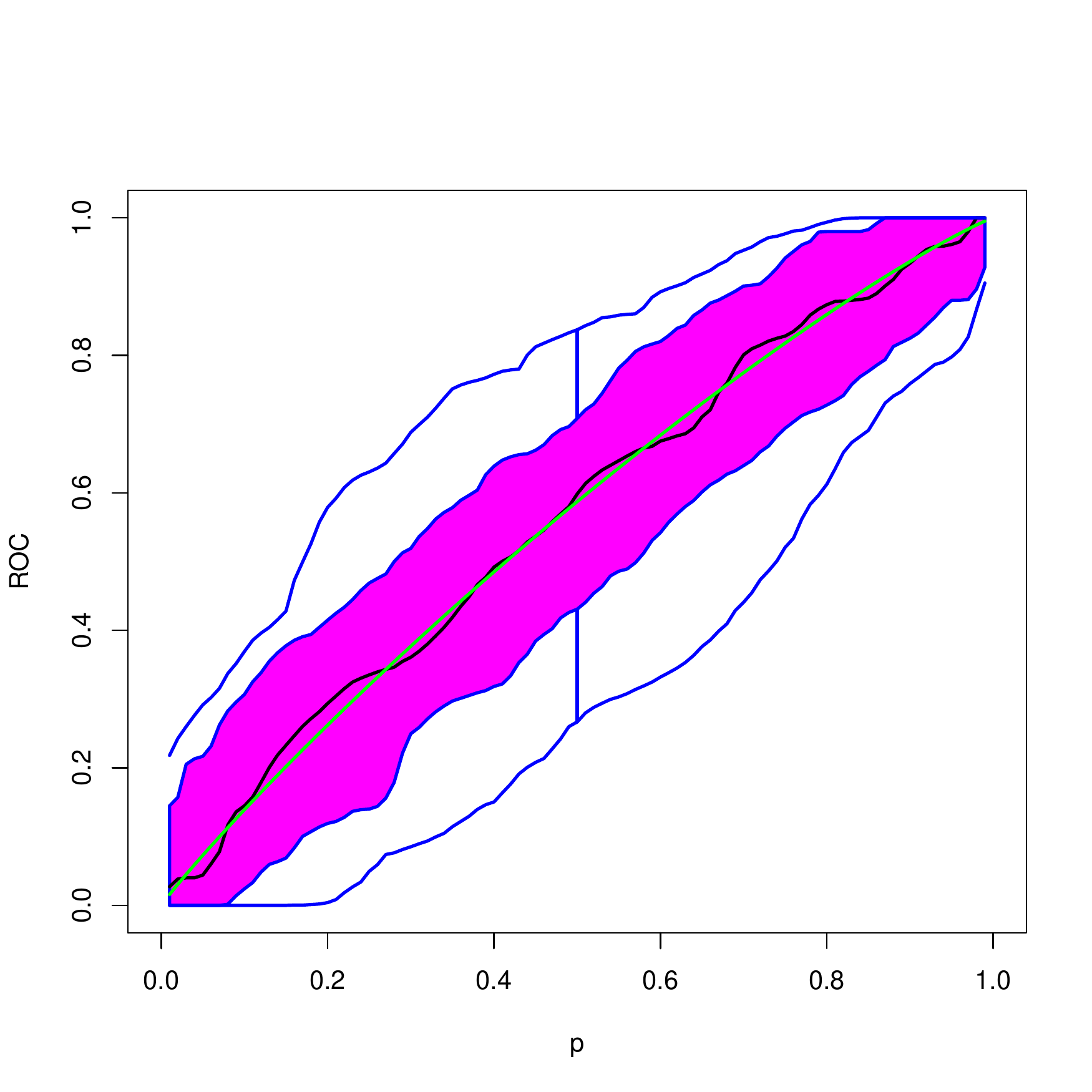}
 & \includegraphics[scale=0.25]{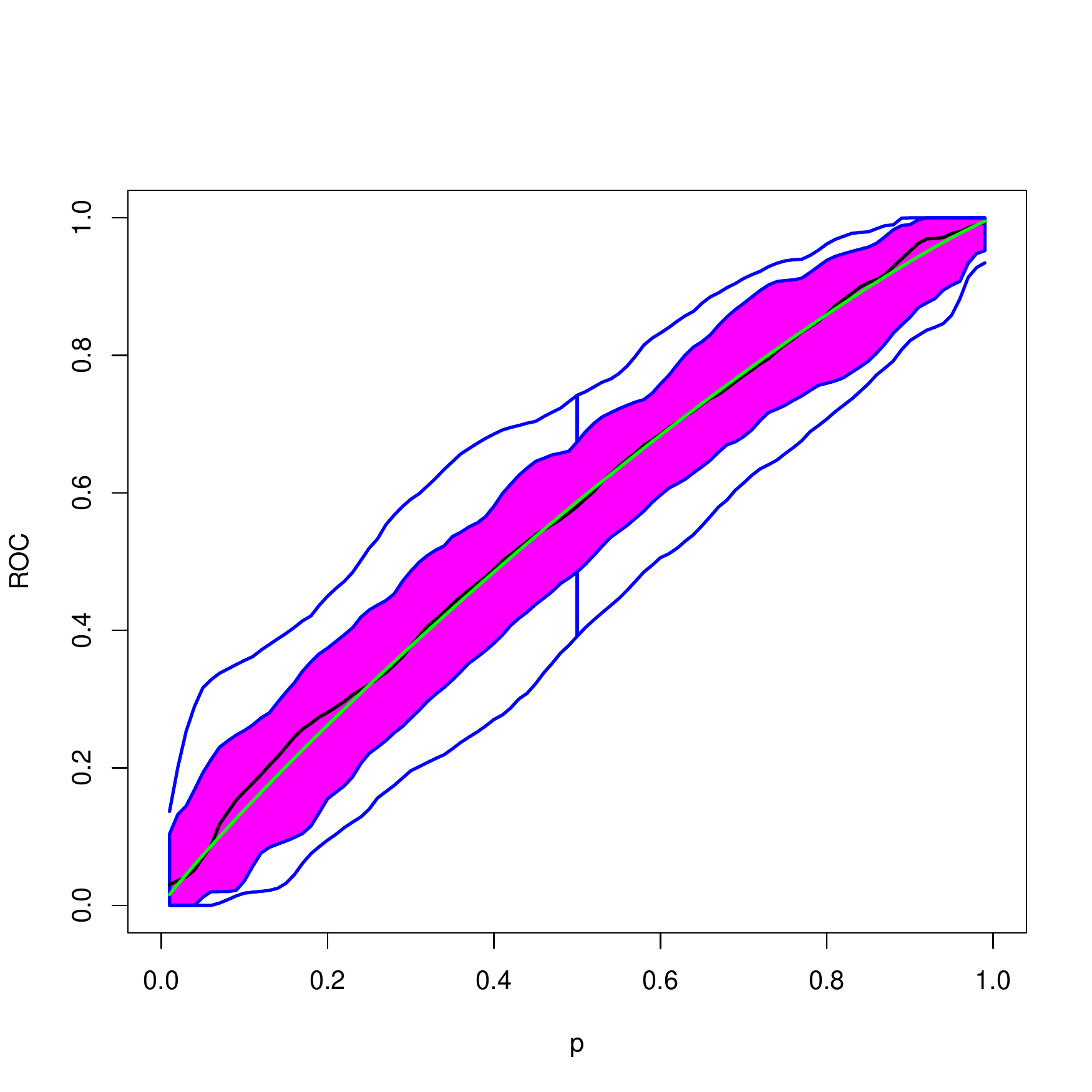} \\[-0.1in]
(c)  
 & \includegraphics[scale=0.25]{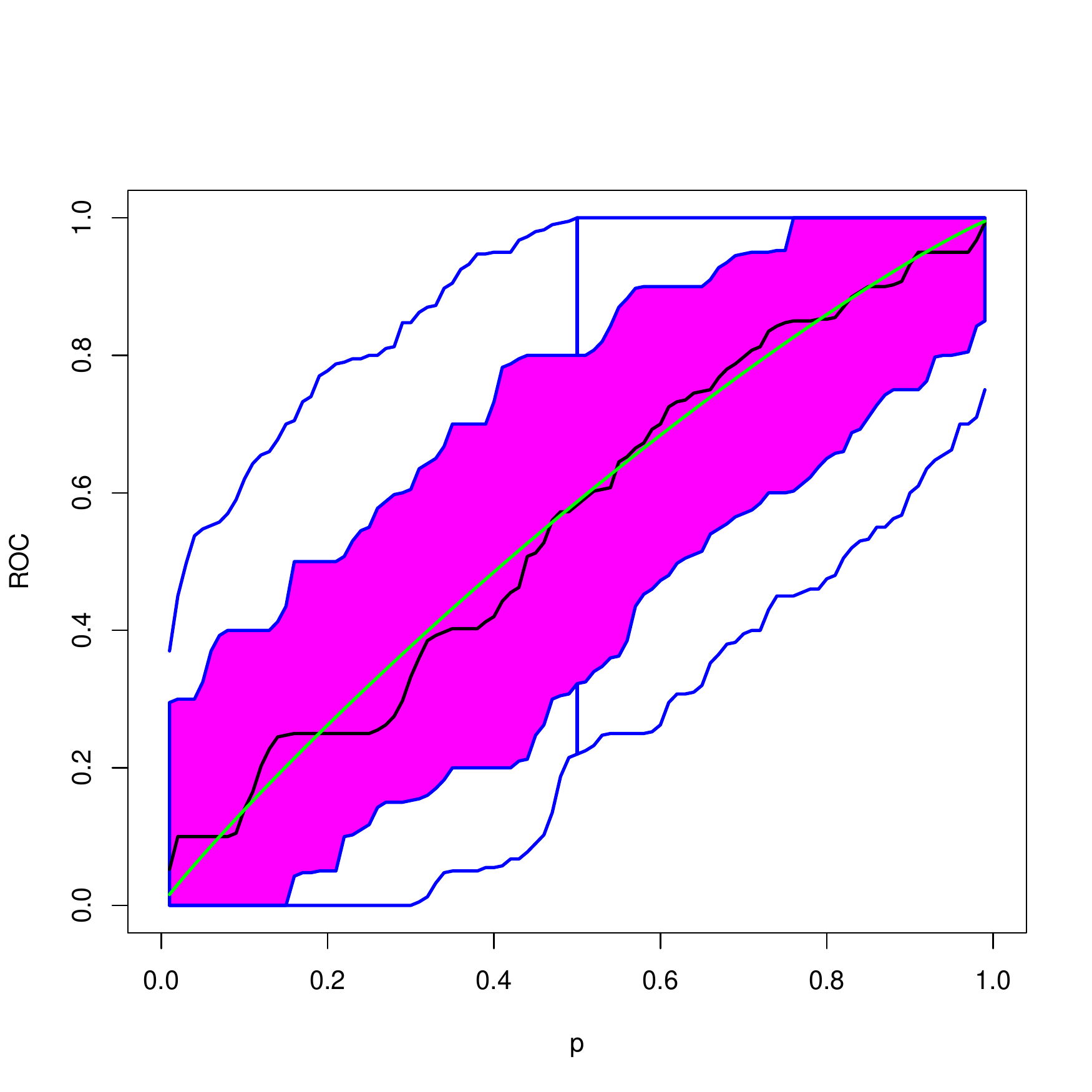}
 & \includegraphics[scale=0.25]{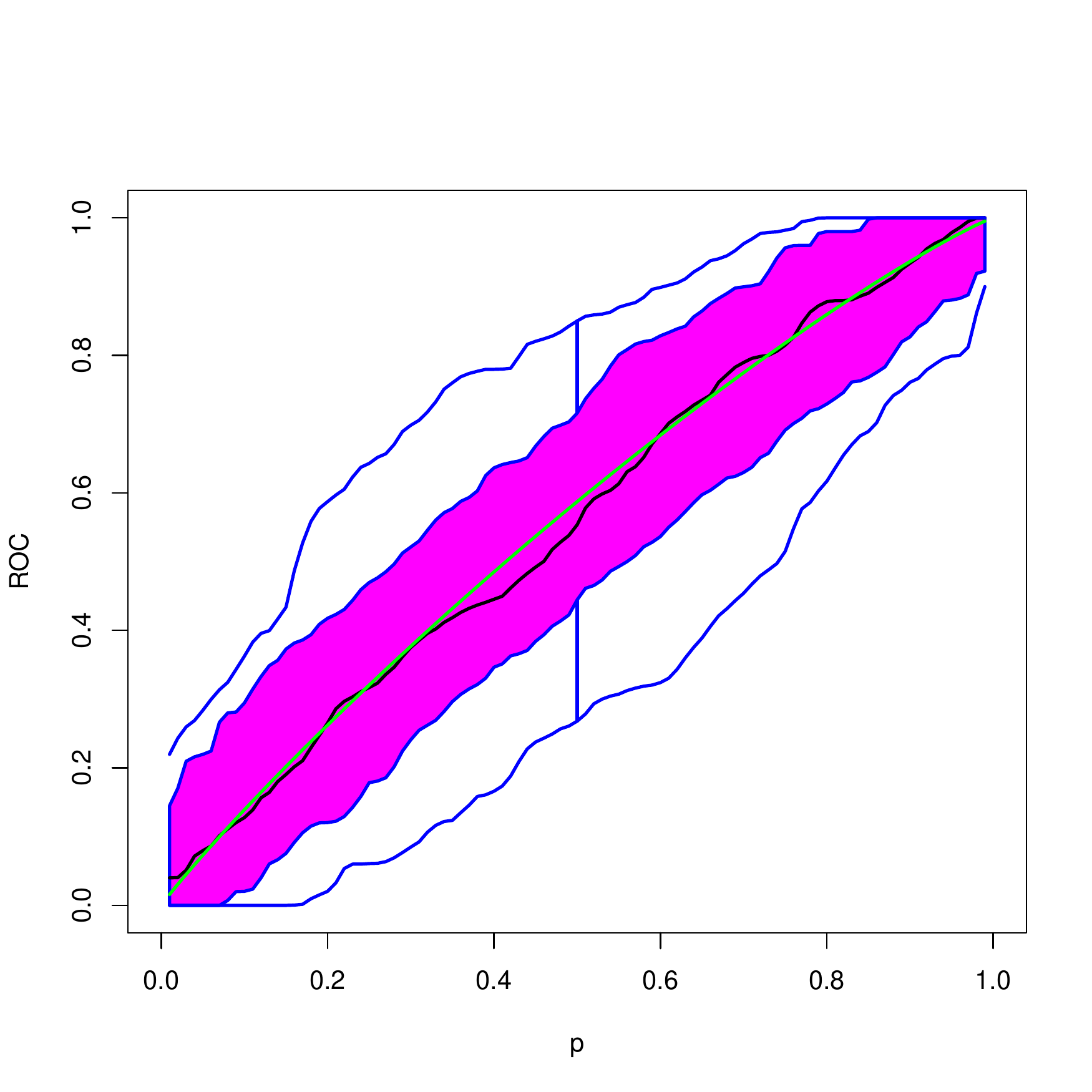}
 & \includegraphics[scale=0.25]{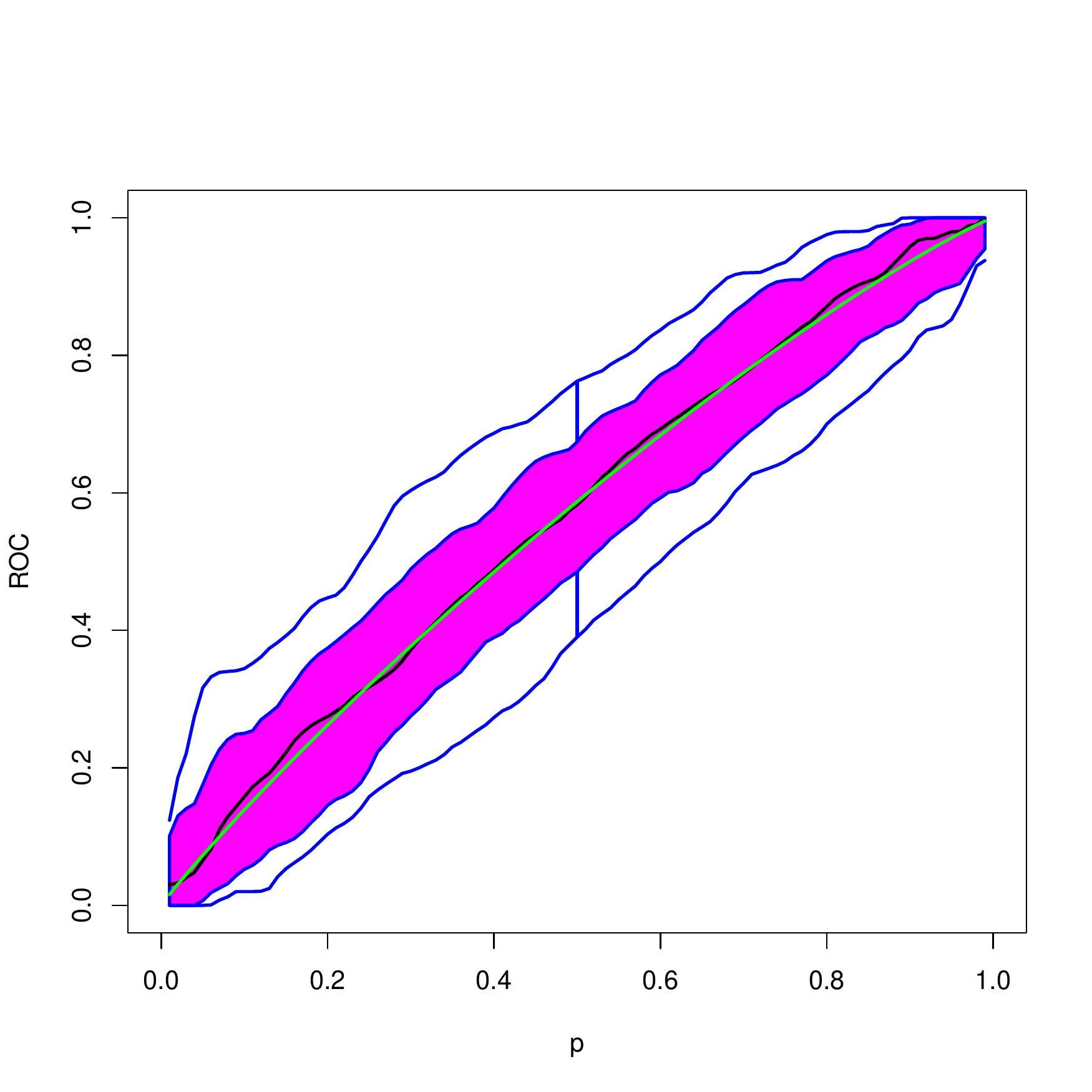} \\[-0.1in]
 (d)  
  &  \includegraphics[scale=0.25]{ROC_MUL_n20_H_preal_D_preallineallineal_mis_H_0_mis_D_0_dim2_caso2.pdf}  
  &  \includegraphics[scale=0.25]{ROC_MUL_n50_H_preal_D_preallineallineal_mis_H_0_mis_D_0_dim2_caso2.pdf}  
  &  \includegraphics[scale=0.25]{ROC_MUL_n100_H_preal_D_preallineallineal_mis_H_0_mis_D_0_dim2_caso2.pdf}  
 \end{tabular}
\vskip-0.1in  
\caption{\label{fig:fbx_dim2_caso2_00_nointer}\small Functional boxplots of  $\wROC_{\conv}(p)$   for  $\bX\in \real^2$, when the correct model is fitted (d) and under misspecification of the regression which is assumed to be a linear one without intercept: (a) in both populations, (b) only in the diseased population and (c) only in the  healthy one. The green line corresponds to the true $ROC(p)$ and the dotted red lines to the outlying curves detected by the functional boxplot.}
\end{center} 
\end{figure}


\begin{figure}[ht!]
 \begin{center}
 \footnotesize
 \renewcommand{\arraystretch}{0.4}
 \newcolumntype{M}{>{\centering\arraybackslash}m{\dimexpr.1\linewidth-1\tabcolsep}}
   \newcolumntype{G}{>{\centering\arraybackslash}m{\dimexpr.3\linewidth-1\tabcolsep}}
\begin{tabular}{MG G G}\\
  & $n=20$ & $n=50$ & $n=100$\\[-0.1in]
(a) 
 &  \includegraphics[scale=0.25]{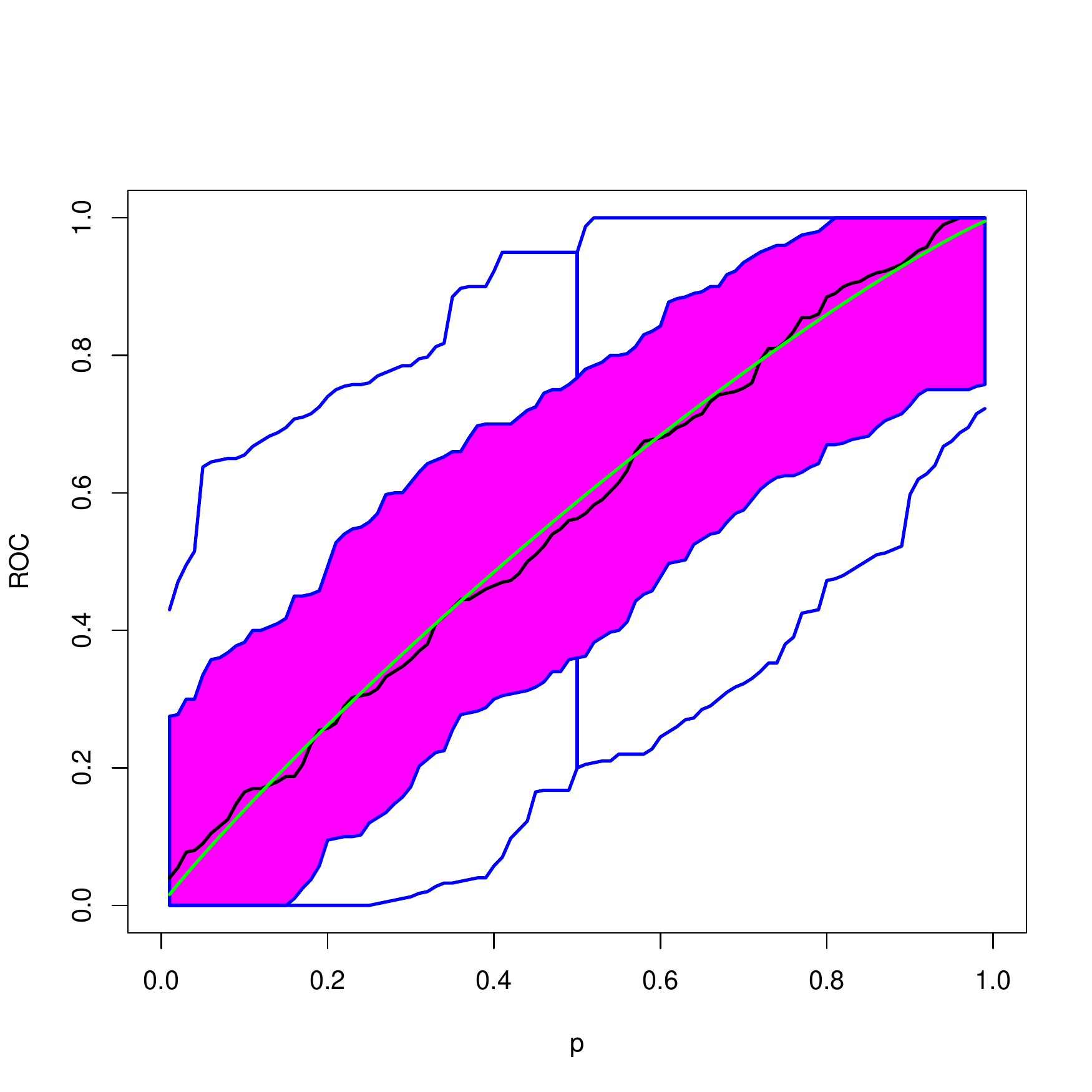} 
  &  \includegraphics[scale=0.25]{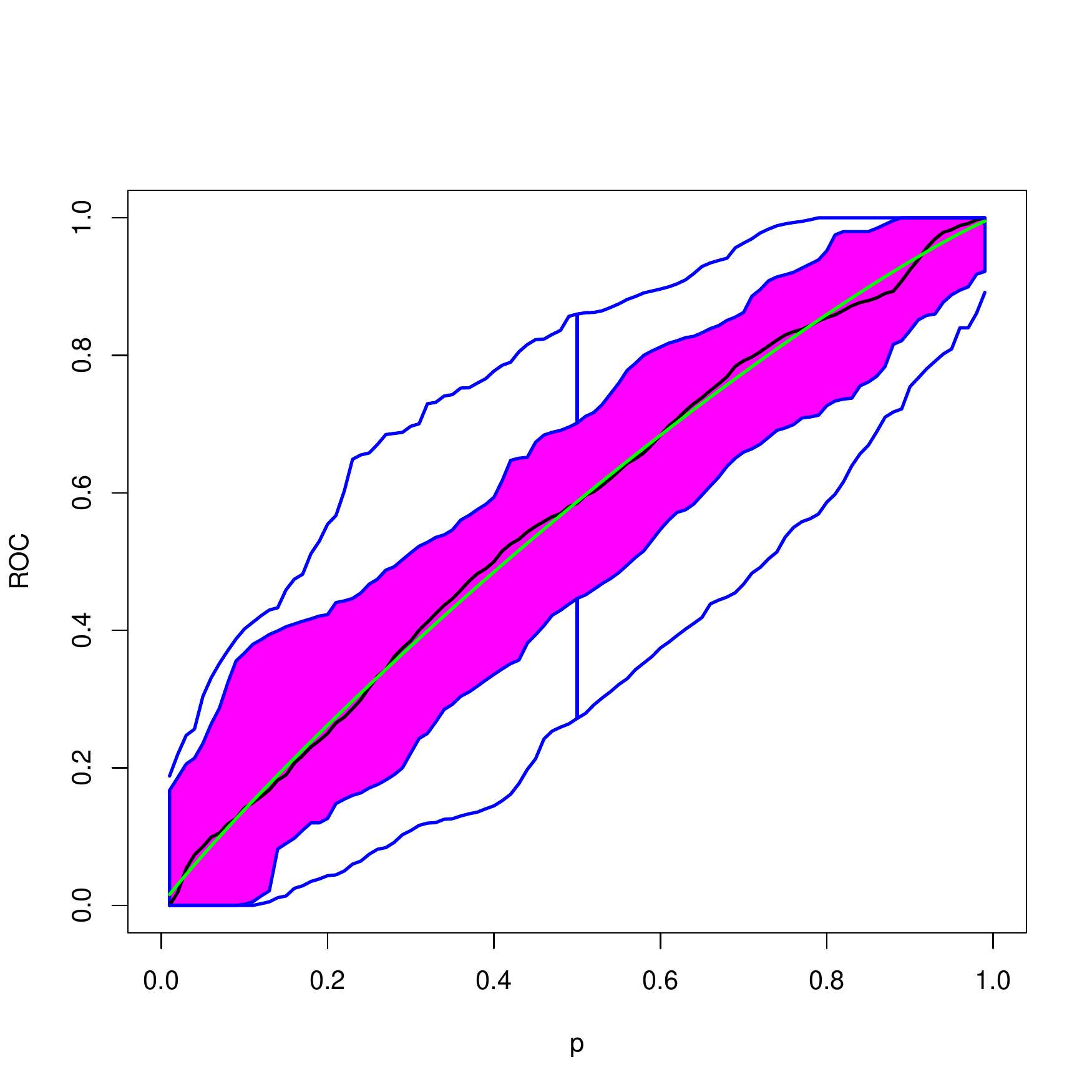}   
 &  \includegraphics[scale=0.25]{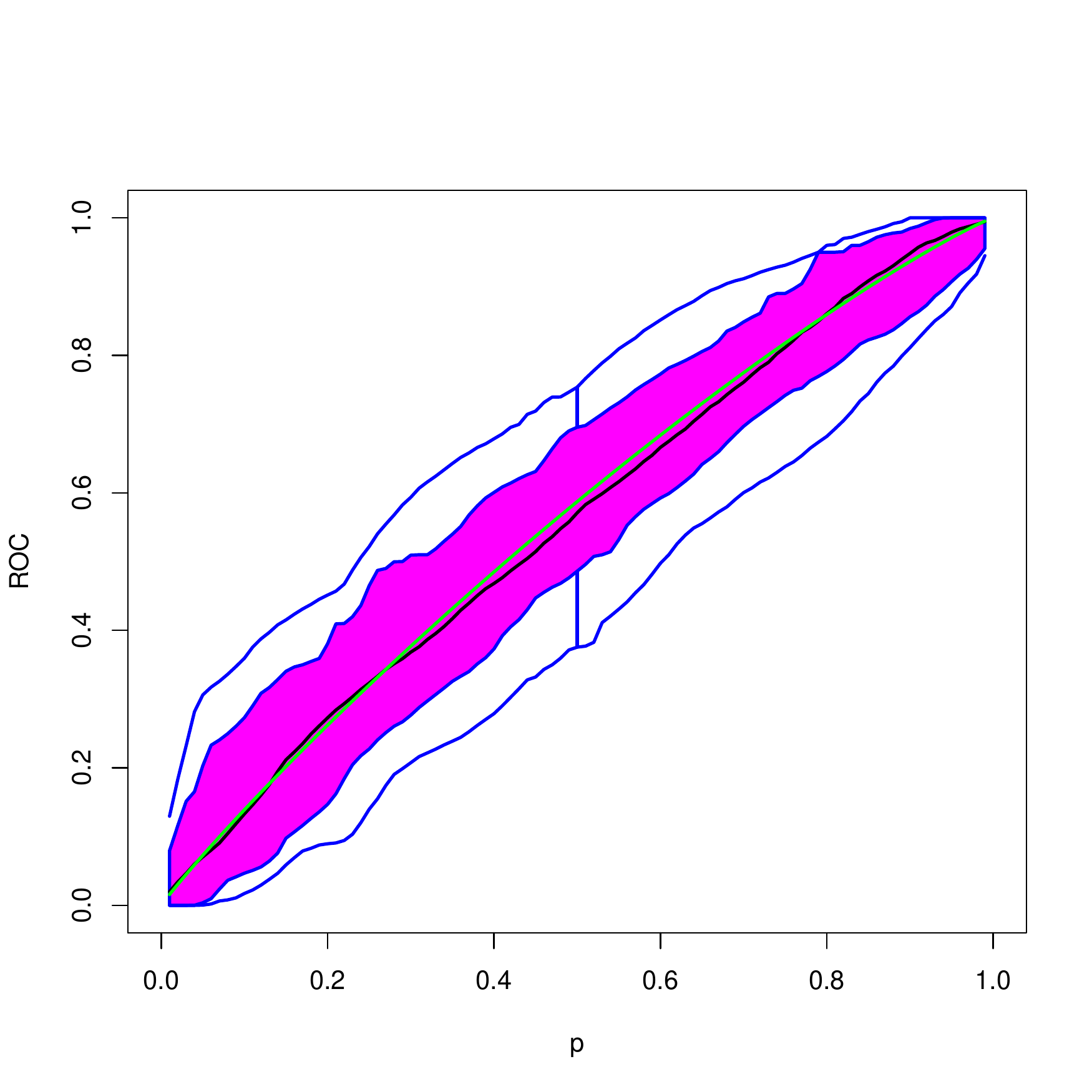}  \\[-0.1in]
(b)  
 & \includegraphics[scale=0.25]{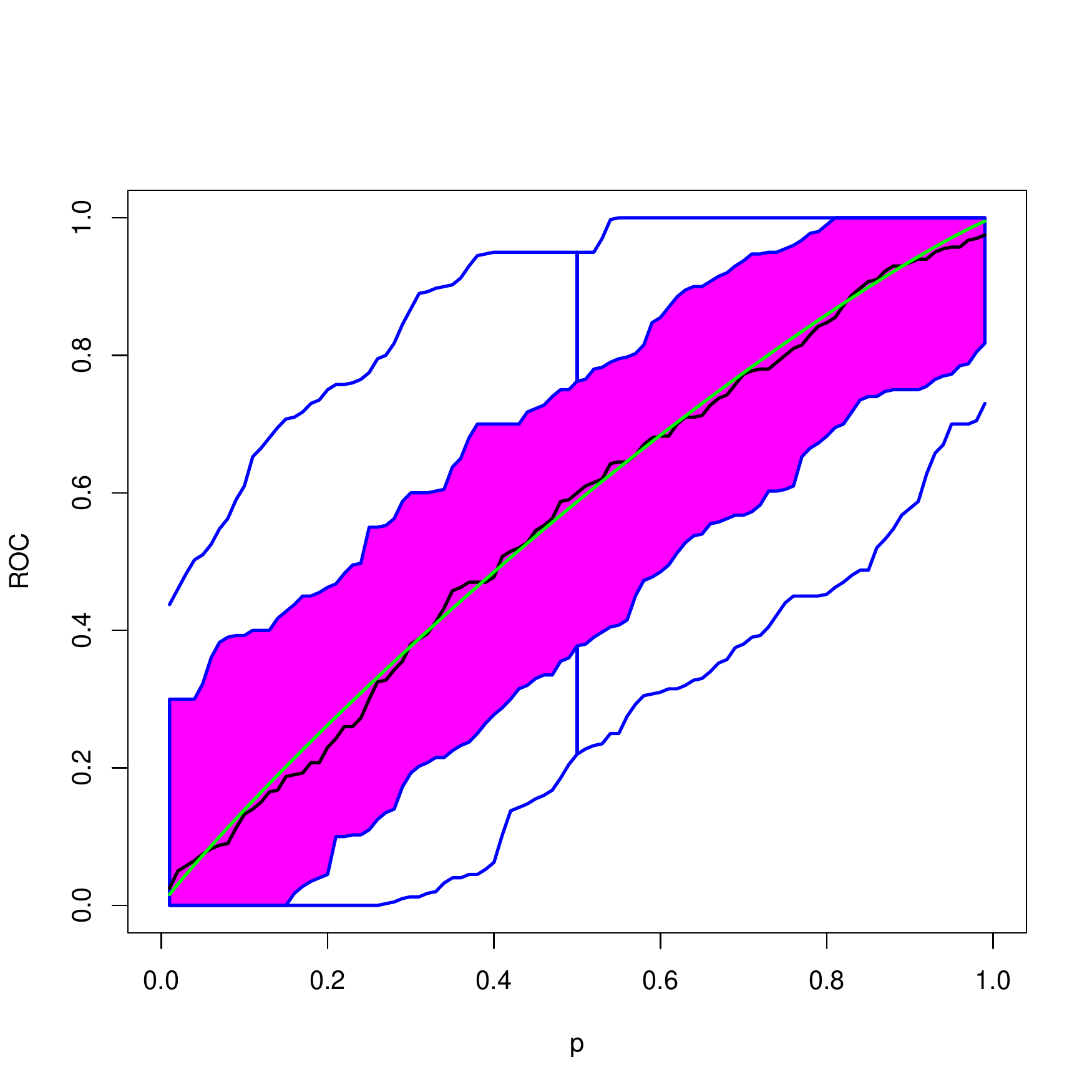}
 & \includegraphics[scale=0.25]{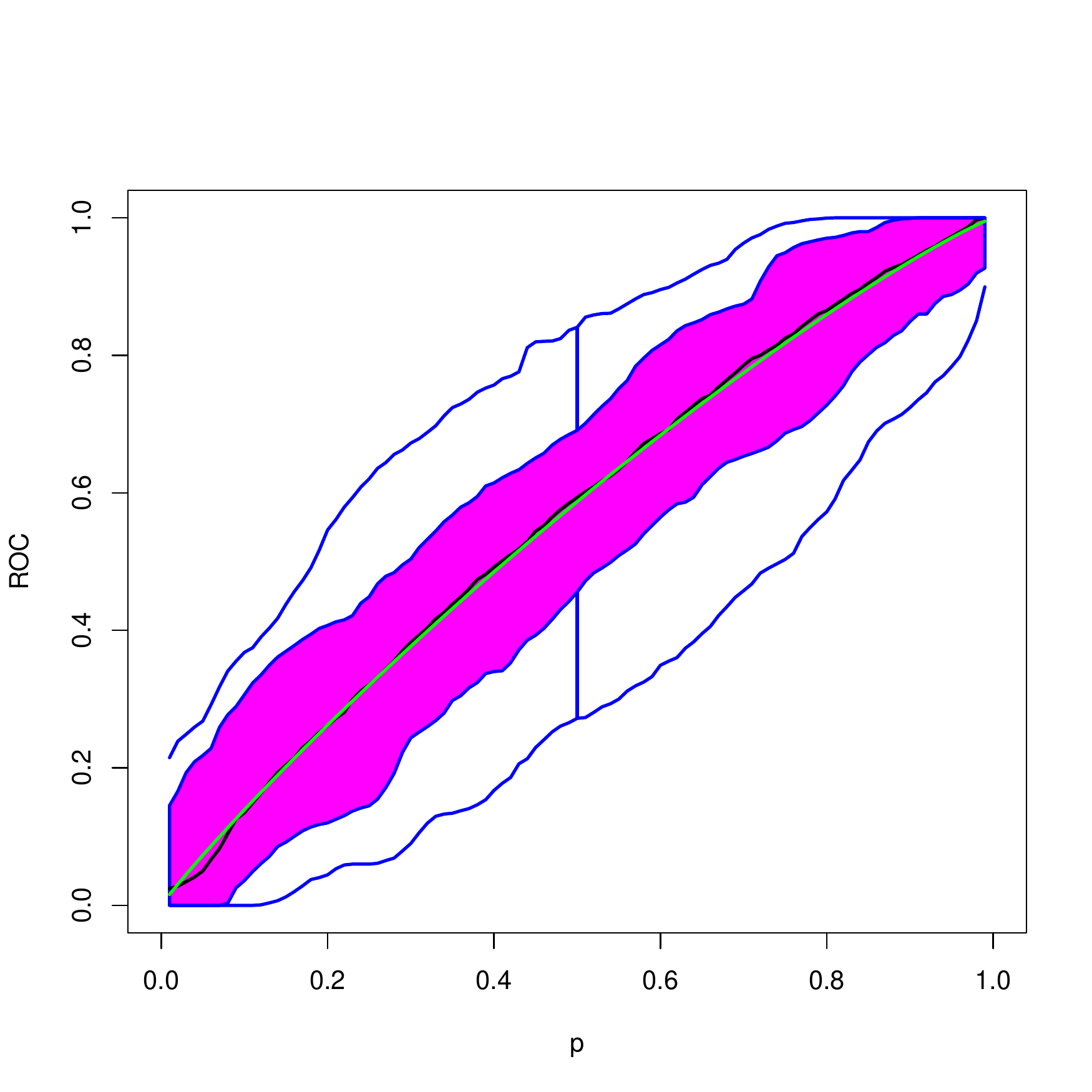}
 & \includegraphics[scale=0.25]{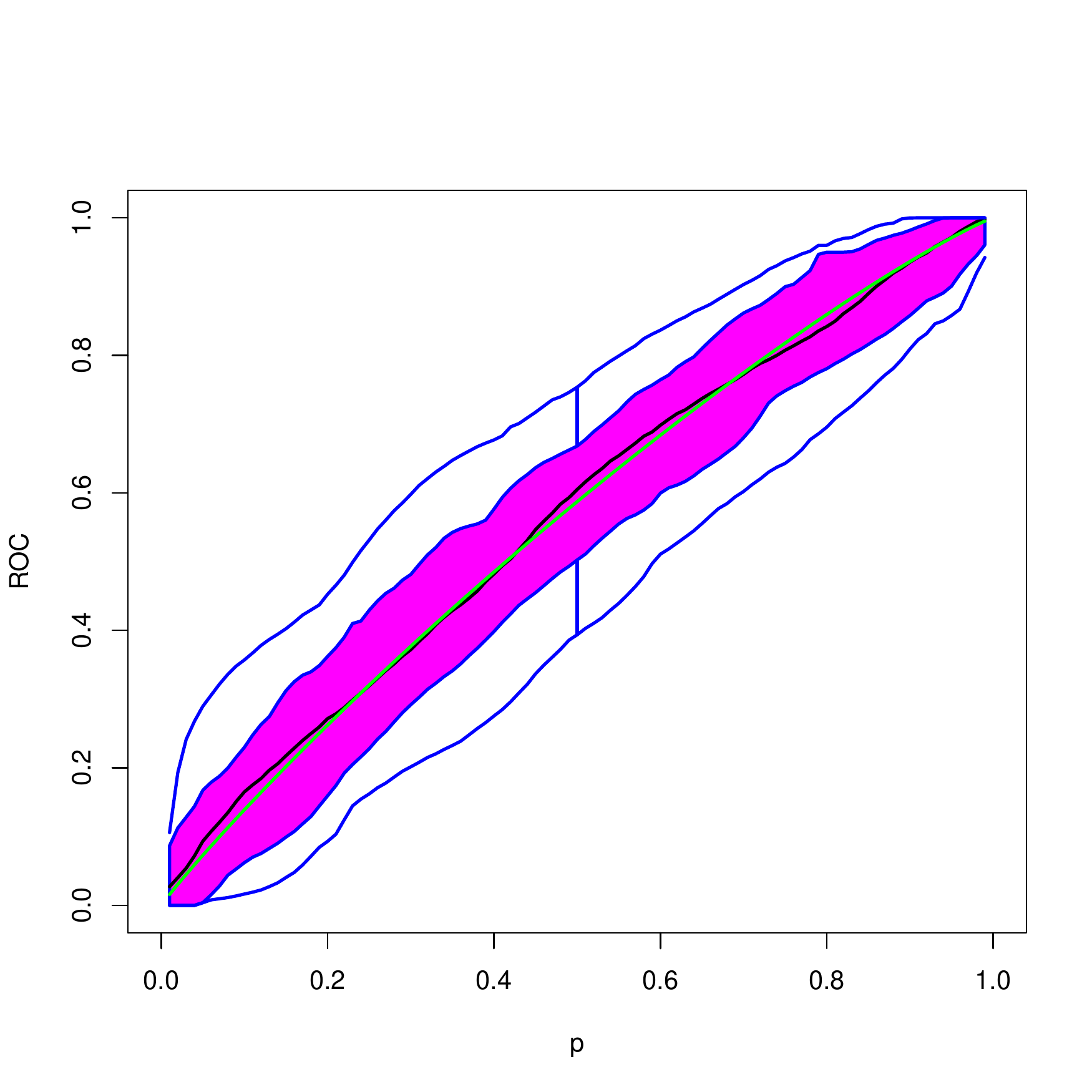} \\[-0.1in]
(c)
& \includegraphics[scale=0.25]{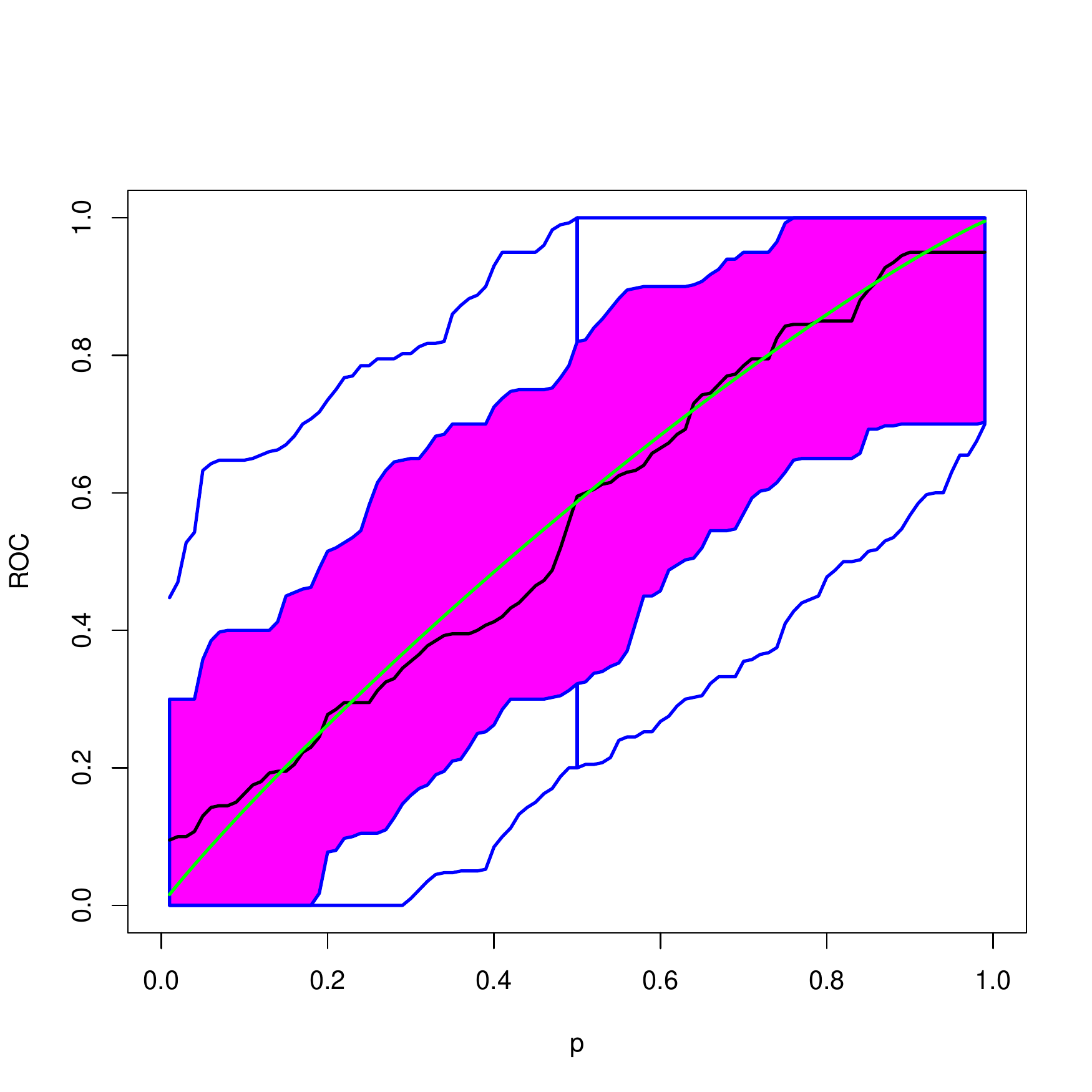} 
& \includegraphics[scale=0.25]{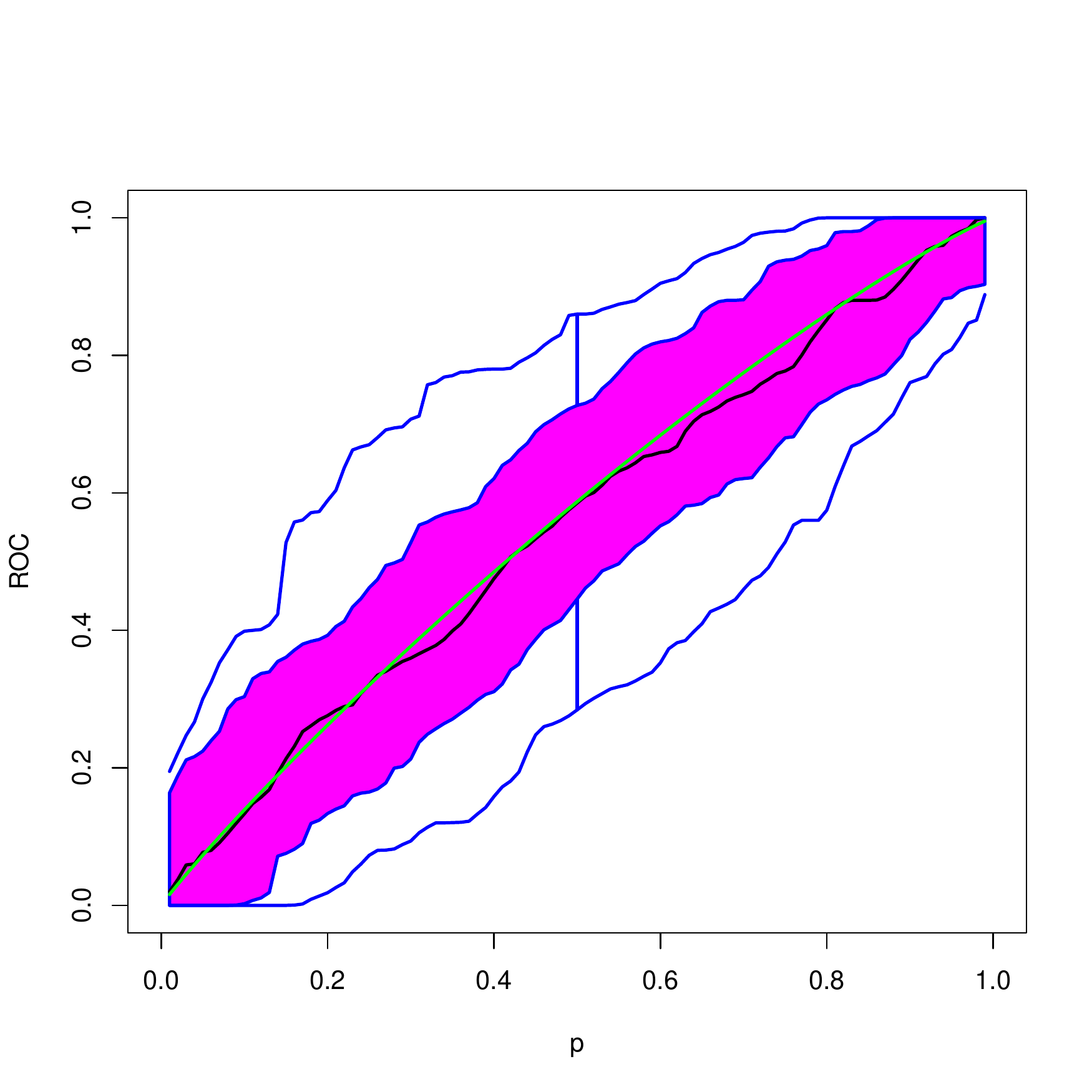}  
& \includegraphics[scale=0.25]{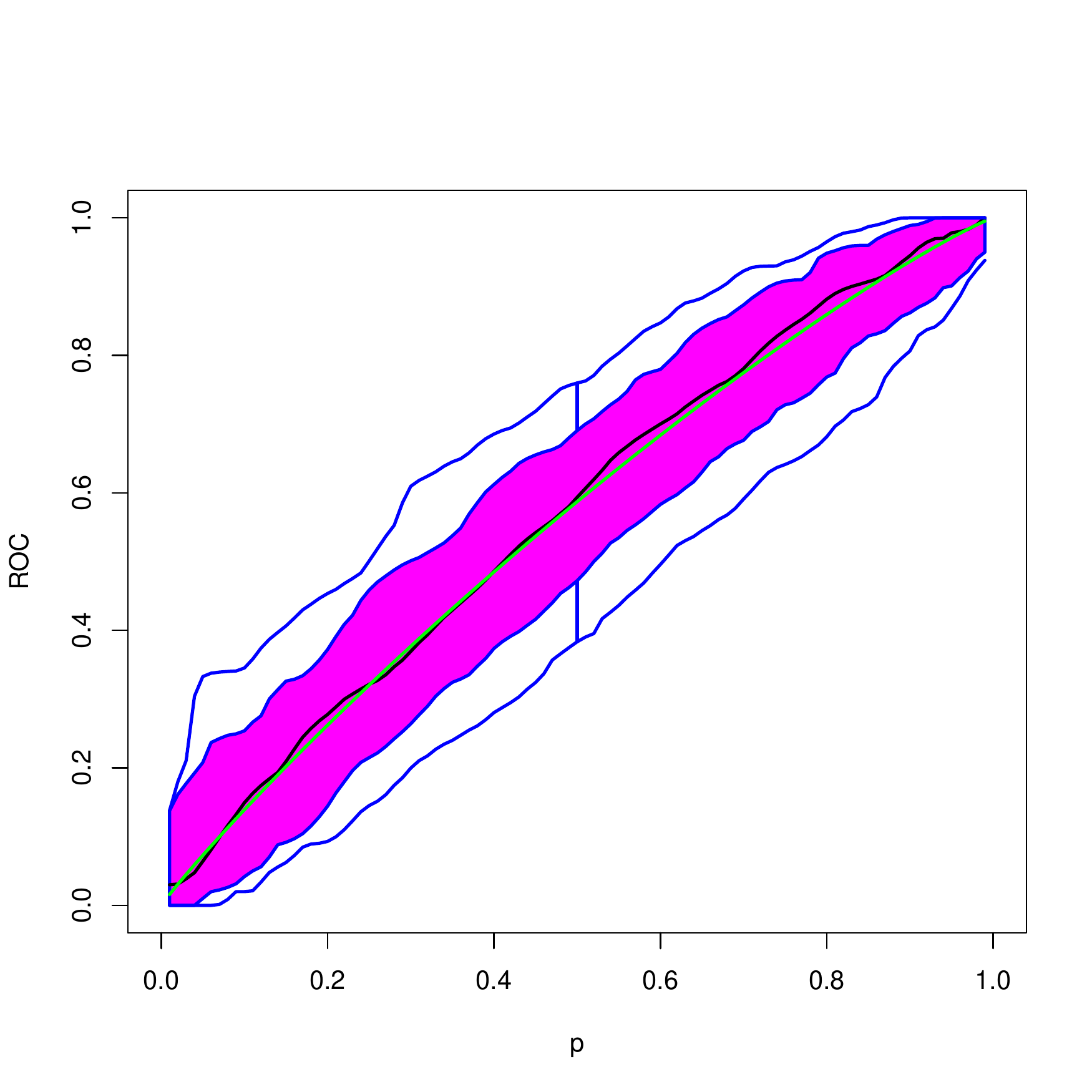} \\[-0.1in]
 (d)  
  &  \includegraphics[scale=0.25]{ROC_MUL_n20_H_preal_D_preallineallineal_mis_H_0_mis_D_0_dim2_caso2.pdf}  
  &  \includegraphics[scale=0.25]{ROC_MUL_n50_H_preal_D_preallineallineal_mis_H_0_mis_D_0_dim2_caso2.pdf}  
  &  \includegraphics[scale=0.25]{ROC_MUL_n100_H_preal_D_preallineallineal_mis_H_0_mis_D_0_dim2_caso2.pdf}  
 \end{tabular}
\vskip-0.1in  
\caption{\label{fig:fbx_dim2_caso2_00_solox1}\small Functional boxplots of  $\wROC_{\conv}(p)$   and $\bX\in \real^2$, when the correct model is fitted (d) and   under misspecification of the regression which is assumed to be a linear model depending only on the first component of $\bX$: (a) in both populations, (b) only in the diseased population and (c) only in the  healthy one.}
\end{center} 
\end{figure}



\begin{figure}[ht!]
 \begin{center}
 \footnotesize
 \renewcommand{\arraystretch}{0.4}
 \newcolumntype{M}{>{\centering\arraybackslash}m{\dimexpr.1\linewidth-1\tabcolsep}}
   \newcolumntype{G}{>{\centering\arraybackslash}m{\dimexpr.3\linewidth-1\tabcolsep}}
\begin{tabular}{MG G G}\\
  & $n=20$ & $n=50$ & $n=100$\\[-0.1in]
(a) 
 &  \includegraphics[scale=0.25]{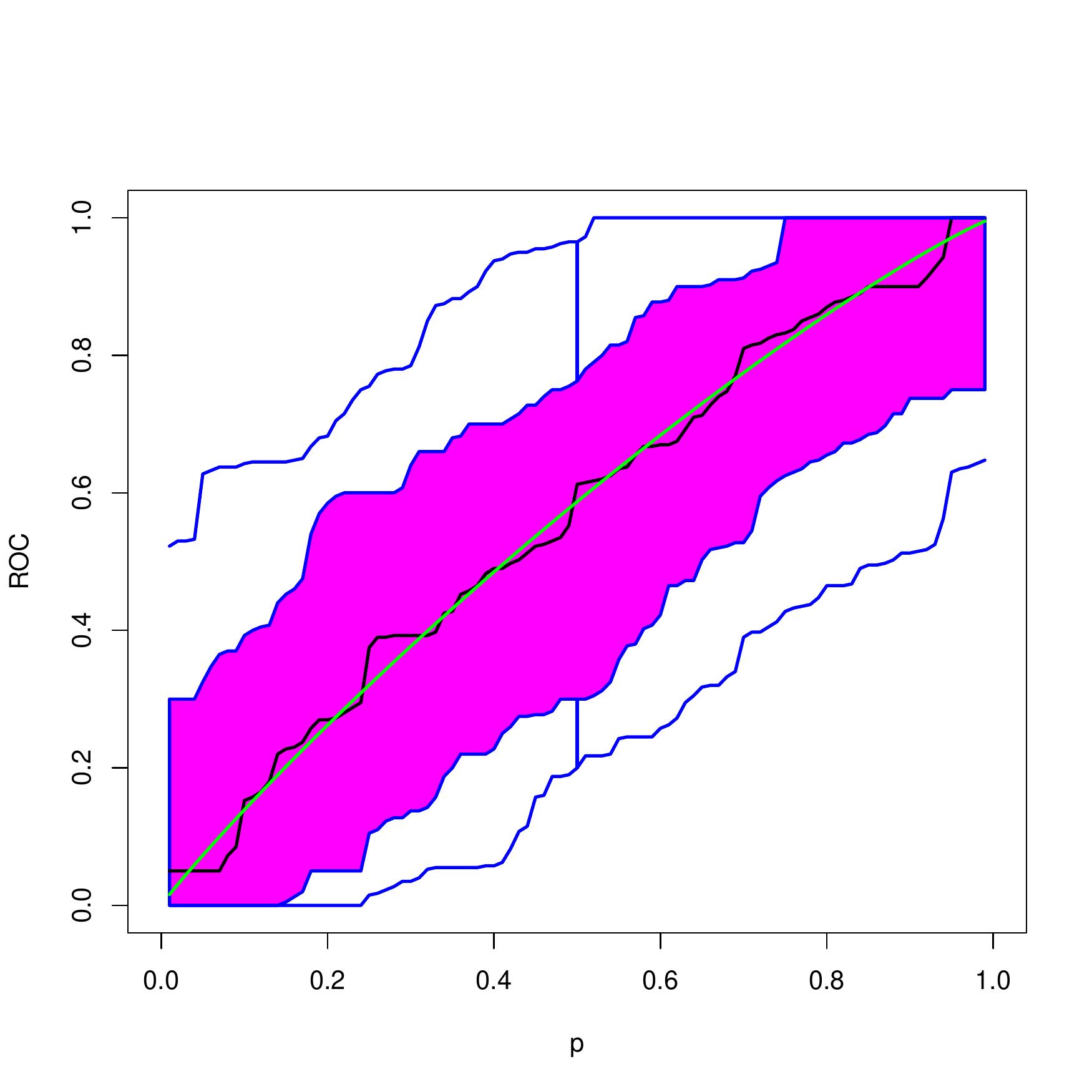} 
  &  \includegraphics[scale=0.25]{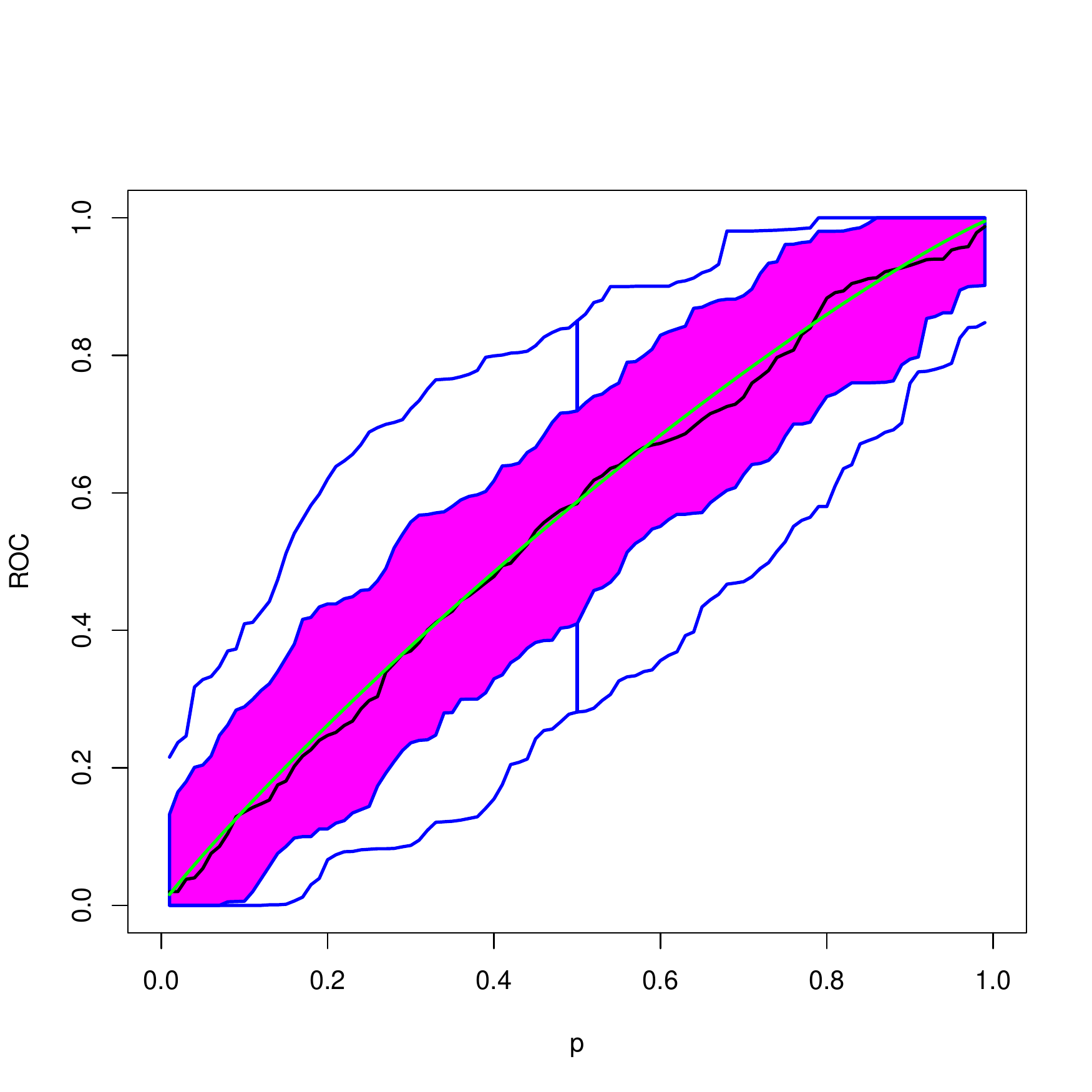} 
 &  \includegraphics[scale=0.25]{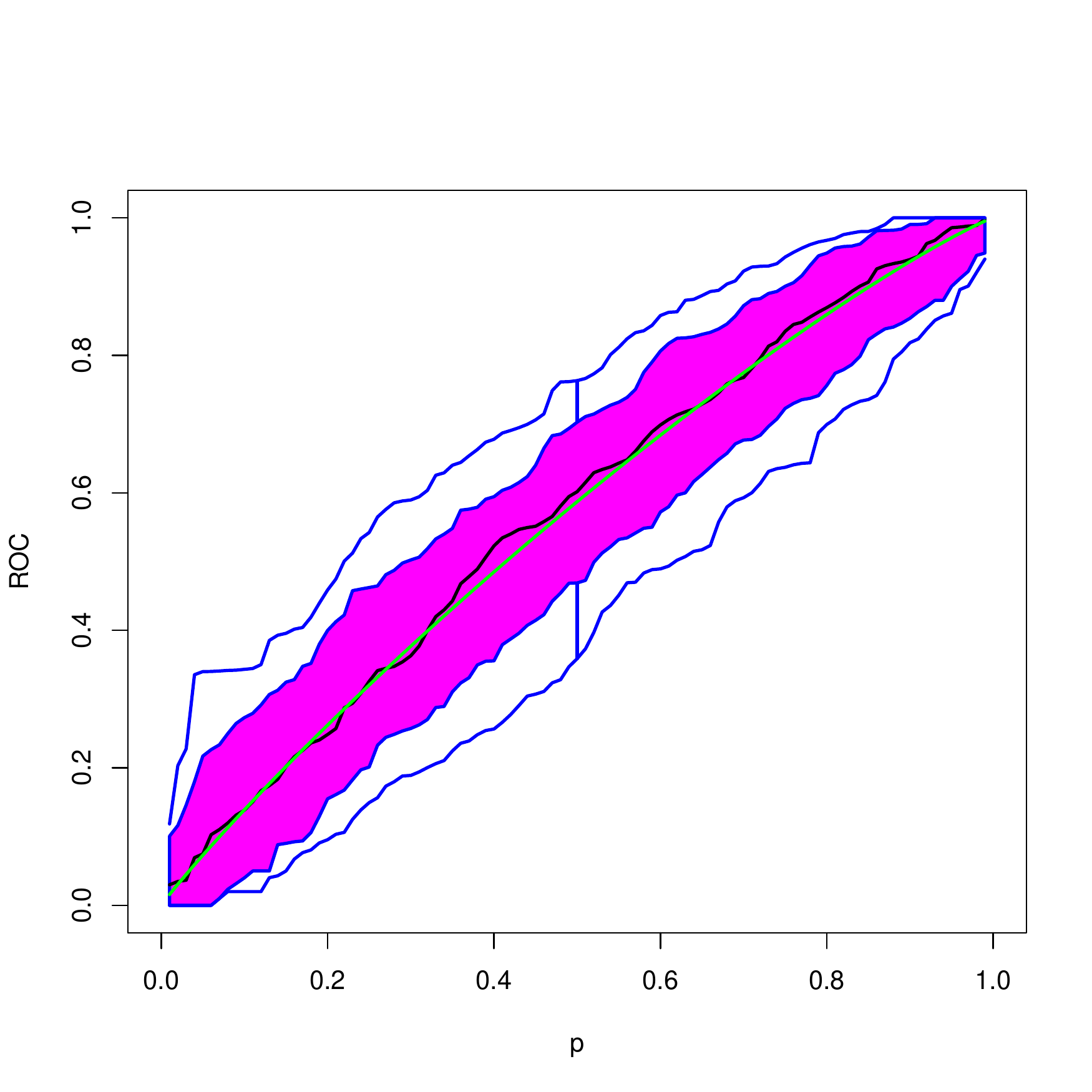}  \\[-0.1in]
(b)     
 & \includegraphics[scale=0.25]{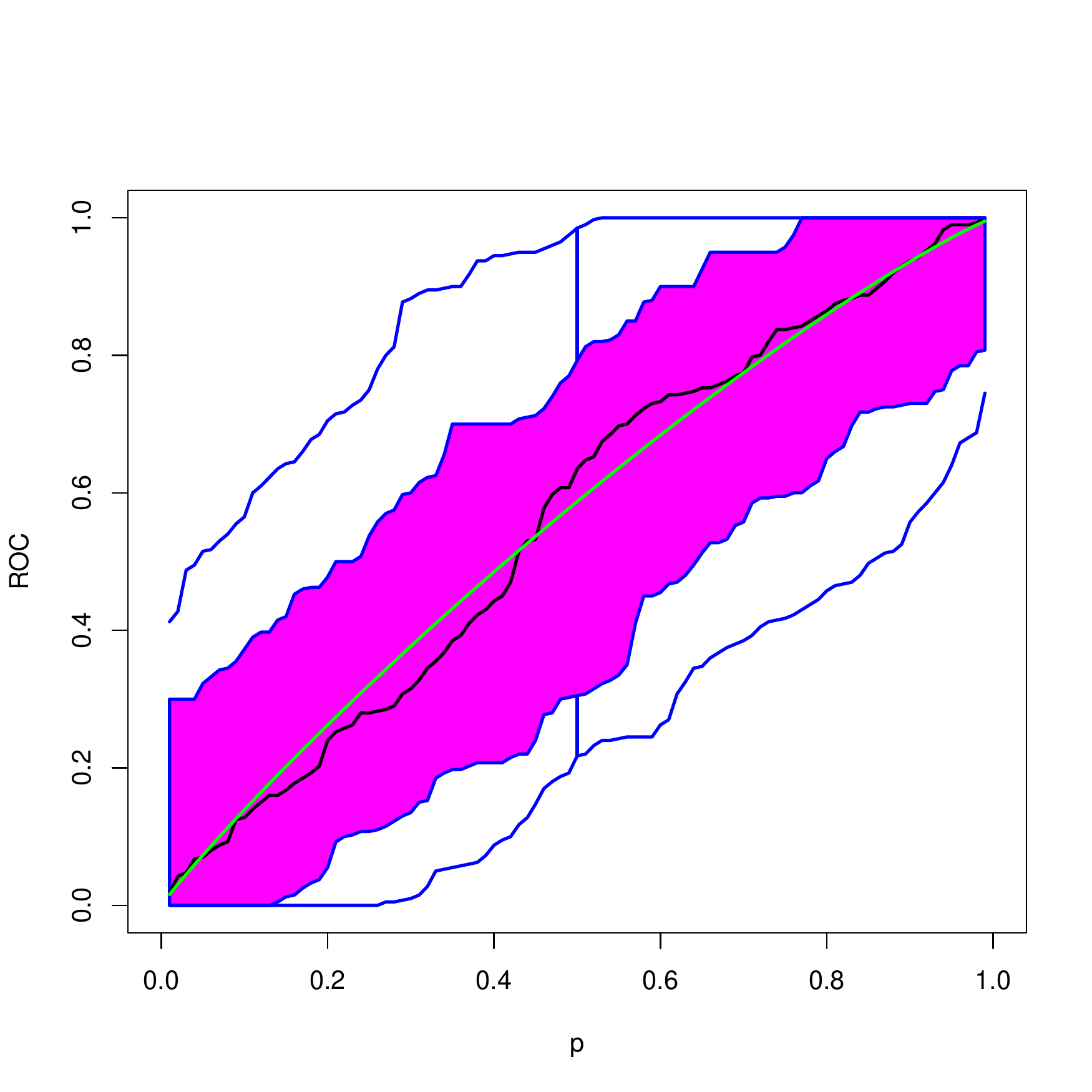} 
  & \includegraphics[scale=0.25]{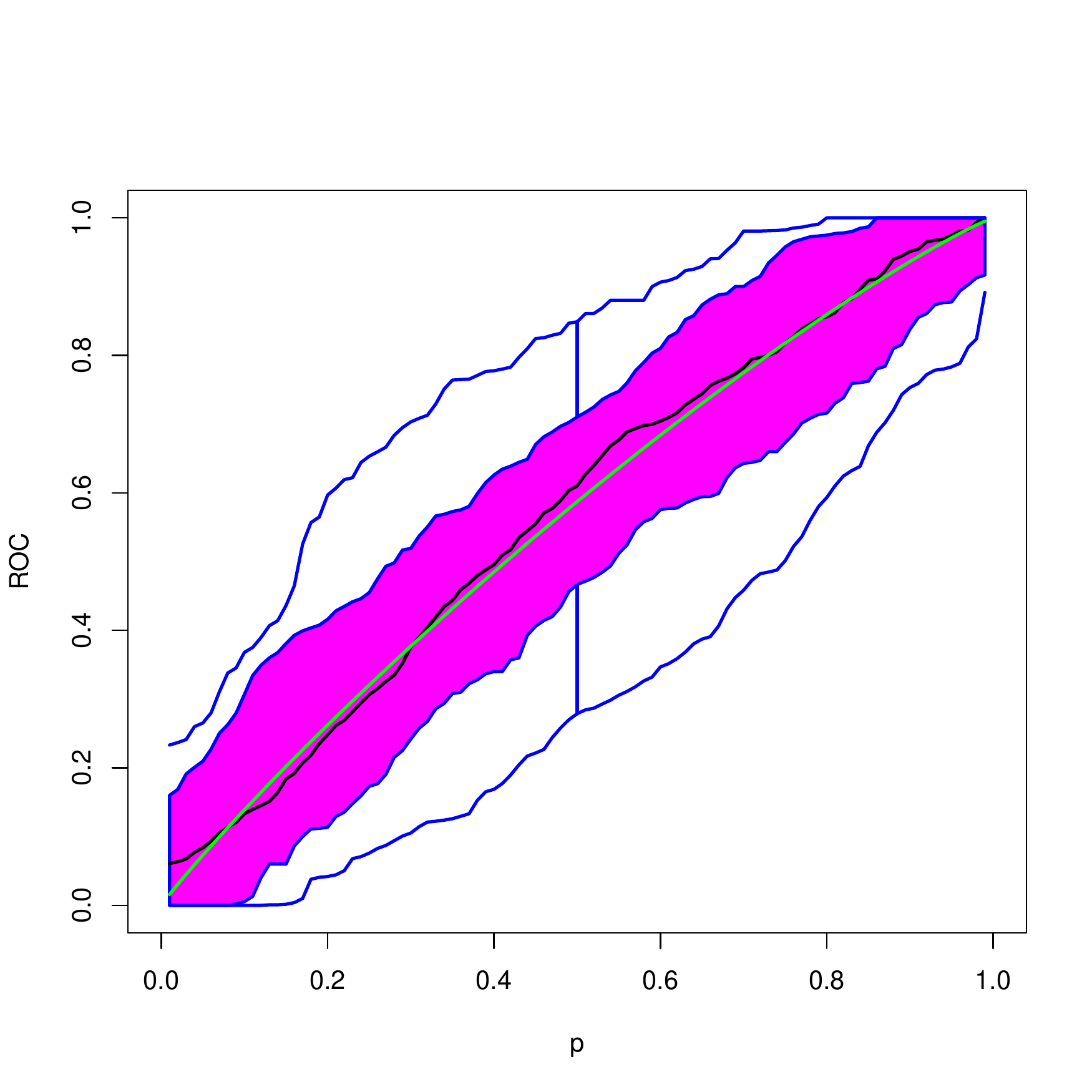} 
   & \includegraphics[scale=0.25]{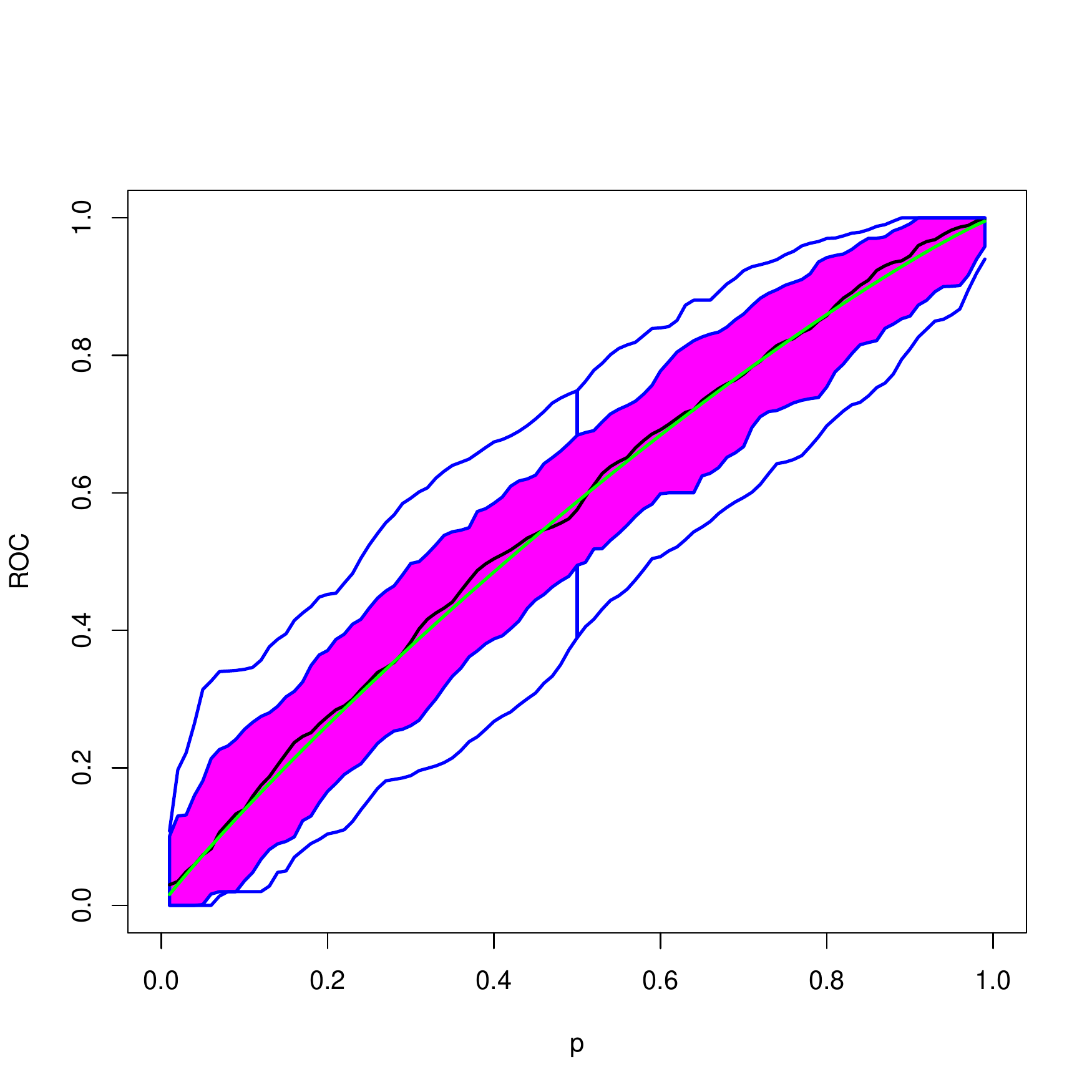} \\[-0.1in]
(c) 
 & \includegraphics[scale=0.25]{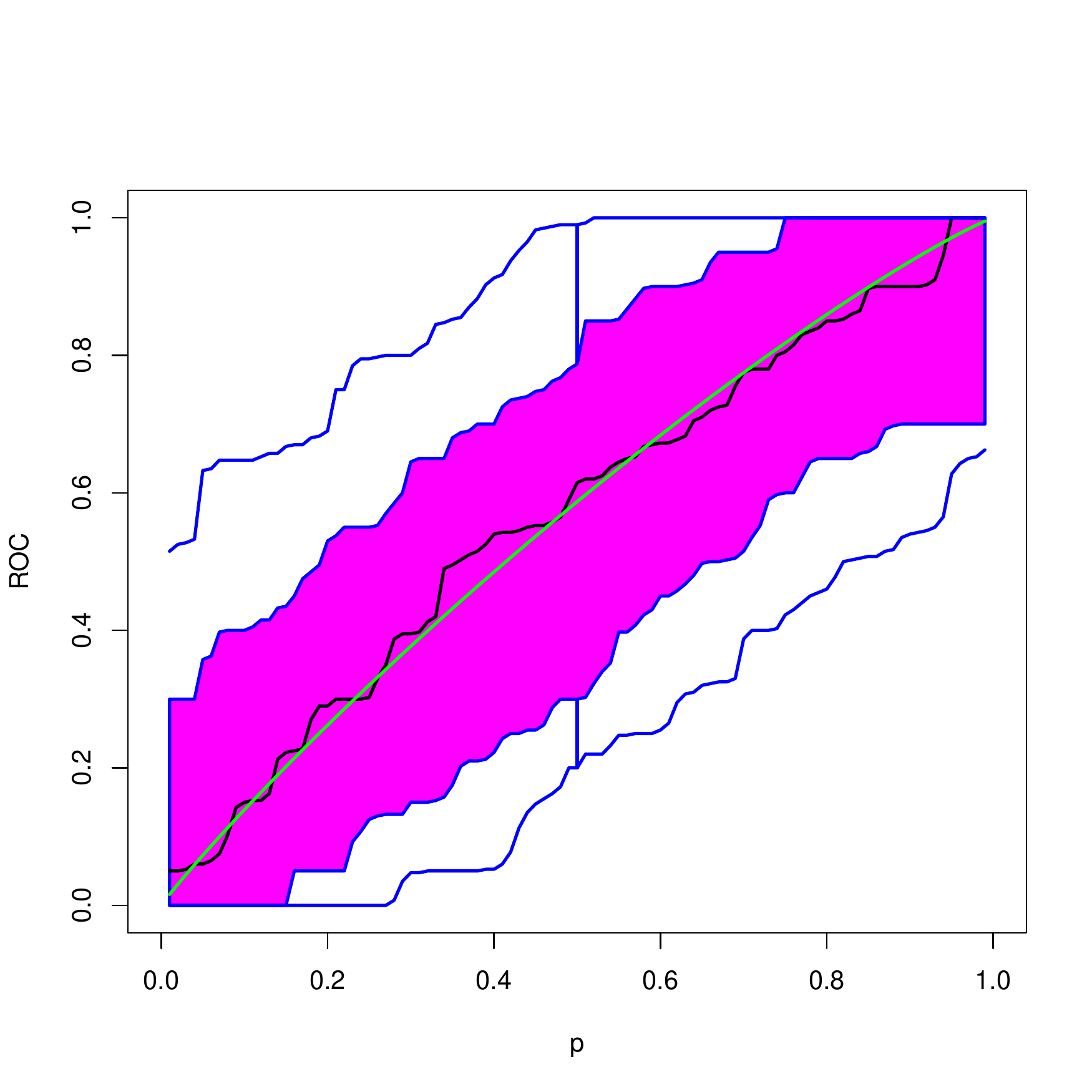}
  & \includegraphics[scale=0.25]{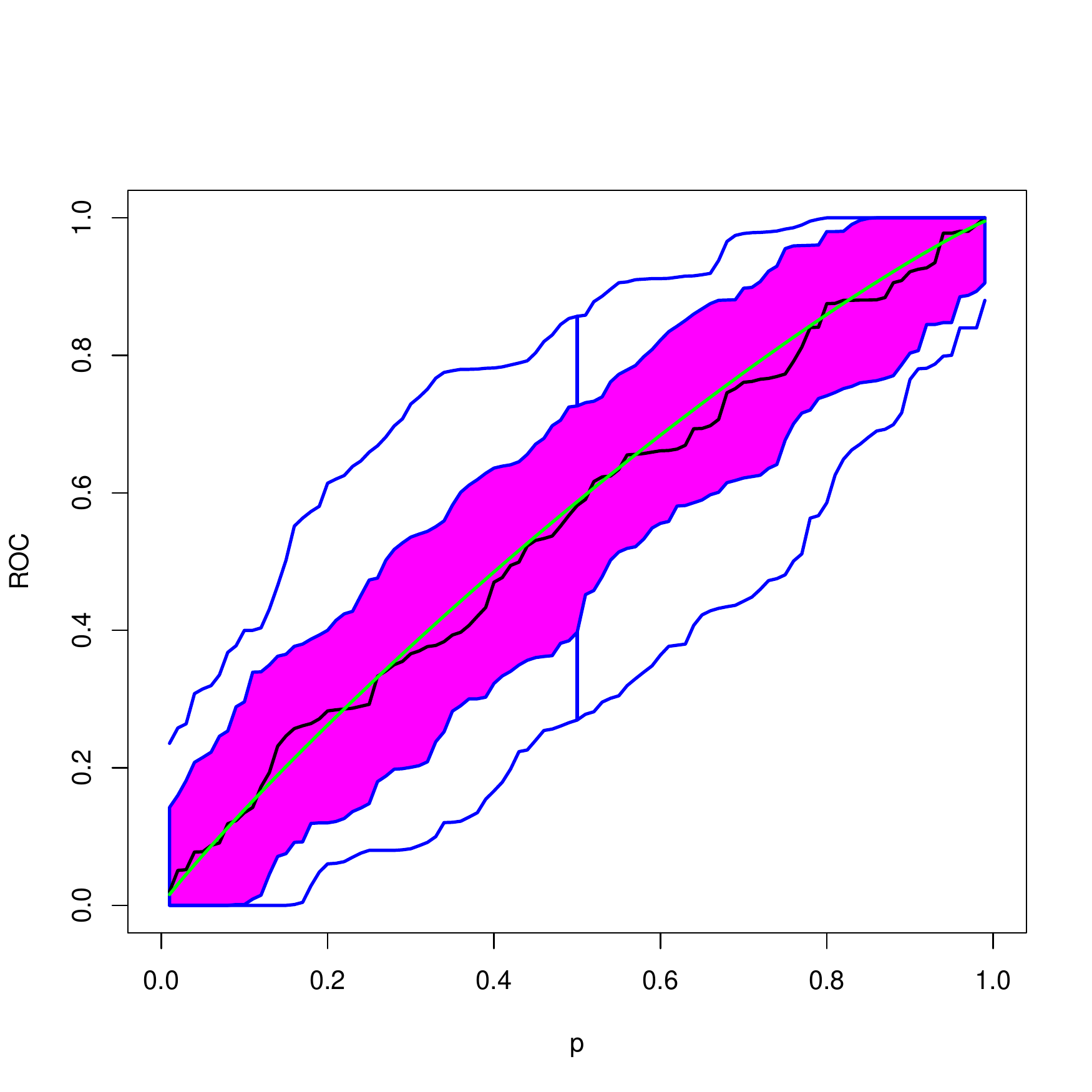}
   & \includegraphics[scale=0.25]{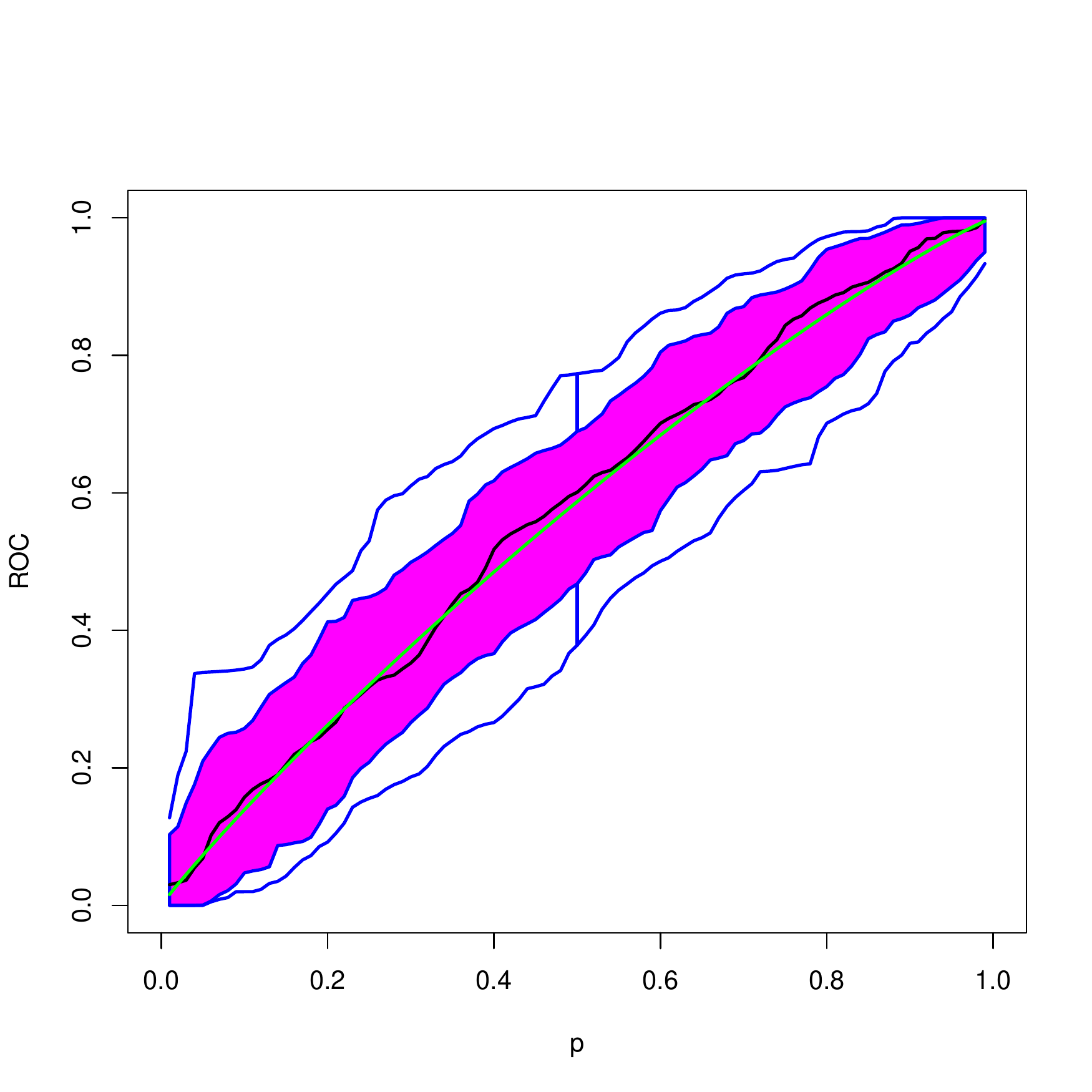} \\[-0.1in]
 (d)  
  &  \includegraphics[scale=0.25]{ROC_MUL_n20_H_preal_D_preallineallineal_mis_H_0_mis_D_0_dim2_caso2.pdf}  
  &  \includegraphics[scale=0.25]{ROC_MUL_n50_H_preal_D_preallineallineal_mis_H_0_mis_D_0_dim2_caso2.pdf}  
  &  \includegraphics[scale=0.25]{ROC_MUL_n100_H_preal_D_preallineallineal_mis_H_0_mis_D_0_dim2_caso2.pdf}  
 \end{tabular}
\vskip-0.1in  
\caption{\label{fig:fbx_dim2_caso2_00_solox1cuad}\small Functional boxplots of  $\wROC_{\conv}(p)$   and $\bX\in \real^2$, when the correct model is fitted (d) and   under misspecification of the regression which is assumed to be a linear model depending only on the square of the first component of $\bX$: (a) in both populations, (b) only in the diseased population and (c) only in the  healthy one.}
\end{center} 
\end{figure}

\subsection{Numerical study for data sets with  missing biomarkers}{\label{sec:montemiss}}
For the situation in which  missing data arise, we only report the results corresponding to $n=100$.
 
\subsubsection{The case of $\bx\in \real$}{\label{sec:dim1}}
We consider an homoscedastic  regression models for each populations, such that
\begin{eqnarray}
Z_{D,i} &=& 2+ 4 X_{D,i} +  \epsilon_{D,i} \label{trued}\\
Z_{H,i} &=&  0.5+   X_{H,i} + \sqrt{\frac{24}{9}} \epsilon_{H,i} \label{trueh}\; ,
\end{eqnarray}
where for all $i=1,\dots,n$ $\epsilon_{j,i} \sim N(0,1)$ are independent and independent from   $X_{j,i} \sim N(0,1/9)$, for $j=D,H$. Besides, the sample from one population was generated independently from the other one.

Missing data are generated according to two models denoted $\itM_1$ and $\itM_2$ which correspond to two logistic models for generating the missing probability. More precisely, under model $\itM_1$, $ \pi(x)=\pi_{\itM_1}(x)= 1/(1+\exp(2(x-0.5))$ while under $\itM_2$,  $ \pi(x)= \pi_{\itM_2}(x)=1/(1+\exp(\,-\,(x-0.5)/2))$. 
To generate the missing biomarkers, we first generate $\delta_{j,i} \sim Bi(1, \pi( X_{j,i} ))$  and then, we  define $Y_{j,i}=Z_{j,i}$, if $\delta_{j,i}=1$, and missing otherwise.

To construct the estimators $\wROC_{\ipw}$ or $\wROC_{\kernel}$, estimators of the propensity are needed.  We considered the situation in which the propensity is assumed to be known and equal to the true one denoted as $\wpi_D=\pi$, $\wpi_H=\pi$ in all Tables and Figures as well as a situation in which the parameters of the true logistic model are estimated from the data. This last case is labelled as $\wpi_D=\pi_{\log}$, $\wpi_H=\pi_{\log}$ in all Tables. The  results  for the three estimators $\wROC_{\ipw}$, $\wROC_{\kernel}$ and $\wROC_{\conv}$ of the ROC curve as well as the corresponding estimates of the area under the curve, are given in Table  \ref{tab:sum_dim1_ROC_AUC},  when the correct regression model is correctly fitted and the propensity is assumed to be  known or estimated as mentioned above. 

Two misspecification settings are considered. In the first one, which will affect only $\wROC_{\ipw}$ or $\wROC_{\kernel}$,  the propensity is estimated assuming a completely at random (\textsc{mcar}) model, either for one or for both populations.   The results are reported in Table \ref{tab:sum_mal_pi_ROC_AUC} where we  label  as $\pi_c$ the situation when the propensity is assumed to be constant. In the second misspecification setting whose goal is to analyse the sensitivity of  $\wROC_{\conv}$ to regression misspecification,  the regression is estimated using an incorrect model since the intercept is omitted. The summary measures for this situation are given in Table  \ref{tab:sum_malregre_ROC_AUC}.

We also provide  the functional boxplots of $\wROC(p)$  in Figures \ref{fig:fbx_11_22} and  \ref{fig:fbx_21_12} for the situation in which both the regression and propensity are correctly specified. On the other hand, Figure \ref{fig:fbx_mal_11_22}  illustrates the functional boxplots for misspecification of the propensity, while Figure  \ref{fig:fbx_11_22_nointer} corresponds to misspecification of the regression. 

The existence of missing biomarkers affect the good performance of the kernel estimator described for complete data sets. As it can be observed in Table \ref{tab:sum_dim1_ROC_AUC}, the best performance with respect to both the $MSE$ and $KS$ measures is now attained by  $\wROC_{\conv}$ in all cases, except when $\pi_D=\pi_{\itM_2}$ and $\pi_H=\pi_{\itM_1}$ and the propensity is estimated using a logistic model. The  advantage of $\wROC_{\conv}$ over $\wROC_{\ipw}$ and $\wROC_{\kernel}$ is also reflected in the functional boxplots given in Figures \ref{fig:fbx_11_22} and  \ref{fig:fbx_21_12}, since narrow central bands (containing the 50\% of the curves) are obtained. Regarding the estimation of the area under the curve, again in most situations the convolution based estimator leads to smaller biases and mean squared errors.

The misspecification of the propensity affects the estimators $\wROC_{\ipw}$ and $\wROC_{\kernel}$, but in  a small extent, since  the central areas of the functional boxplots still contain  the true ROC curve. The same behaviour is observed for  $\wROC_{\conv}$ under misspecification of the regression function.
As shown in  Table \ref{tab:sum_mal_pi_ROC_AUC}, the propensity misspecification also affects the inverse probability weighting and the smoothed kernel estimates of the $\AUC$ which present large biases, even though the $RB$ and mean squared error are less affected.  The same behaviour is observed for $\wAUC_{\conv}$ under misspecification of the regression model. 

\small
   
\begin{table}[ht!]
\centering
\setlength{\tabcolsep}{4pt}
 \renewcommand{\arraystretch}{0.8}
\begin{tabular}{c|rrr|rrr|}
  \hline
 $1000\times$ & $\wROC_{\ipw} $ & $\wROC_{\kernel}$ &  $\wROC_{\conv} $ &  $\wROC_{\ipw} $ & $\wROC_{\kernel}$ &  $\wROC_{\conv} $ \\
  \hline
 &   \multicolumn{3}{c|}{$\wpi_D=\pi$, $\wpi_H=\pi$} & \multicolumn{3}{c|}{$\wpi_D=\pi_{\log}$, $\wpi_H=\pi_{\log}$}\\\hline
&   \multicolumn{6}{c|}{$\pi_{H}=\pi_{\itM_1}$, $\pi_{D}=\pi_{\itM_1}$}\\\hline
$MSE$  &  4.44 & 3.67 & 2.97 & 4.22 & 3.45 & 2.97  \\ 
 $KS$ &   158.60 & 110.06 & 104.26 & 155.90 & 107.35 & 104.26 
 \\
   \hline

&   \multicolumn{6}{c|}{$\pi_{H}=\pi_{\itM_2}$, $\pi_{D}=\pi_{\itM_2}$}\\\hline
  $MSE$ & 7.05 & 5.79 & 4.57 & 6.54 & 5.30 & 4.57 \\ 
$KS$ & 194.76 & 136.44 & 129.25 & 190.46 & 132.22 & 129.25 \\ 
   \hline

&   \multicolumn{6}{c|}{$\pi_{H}=\pi_{\itM_2}$, $\pi_{D}=\pi_{\itM_1}$}\\\hline
$MSE$ & 5.73 & 4.72 & 4.13 & 5.46 & 4.45 & 4.13 \\ 
$KS$ & 182.73 & 126.95 & 126.46 & 180.12 & 123.95 & 126.46 \\  
   \hline

&   \multicolumn{6}{c|}{$\pi_{H}=\pi_{\itM_1}$, $\pi_{D}=\pi_{\itM_2}$}\\\hline
 $MSE$ & 5.71 & 4.70 & 3.39 & 5.26 & 4.26 & 3.39 \\ 
 $KS$ & 173.41 & 121.86 & 108.19 & 169.09 & 117.43 & 108.19 \\  
   \hline

 $1000\times$ & $\wAUC_{\ipw}$ & $\wAUC_{\kernel}$ &  $\wAUC_{\conv}$ & $\wAUC_{\ipw}$ & $\wAUC_{\kernel}$ &  $\wAUC_{\conv}$ \\
 \hline 
 &   \multicolumn{3}{c|}{$\wpi_D=\pi$, $\wpi_H=\pi$} & \multicolumn{3}{c|}{$\wpi_D=\pi_{\log}$, $\wpi_H=\pi_{\log}$}\\\hline
&   \multicolumn{6}{c|}{$\pi_{H}=\pi_{\itM_1}$, $\pi_{D}=\pi_{\itM_1}$}\\\hline
Bias & 0.09 & -3.14 & -0.34 & -0.68 & -3.90 & -0.34 \\ 
 $RB$ & 47.76 & 47.04 & 43.90 & 45.56 & 44.91 & 43.90 \\ 
$MSE$ & 1.93 & 1.89 & 1.64 & 1.76 & 1.73 & 1.64 \\  
  
\hline
&   \multicolumn{6}{c|}{$\pi_{H}=\pi_{\itM_2}$, $\pi_{D}=\pi_{\itM_2}$}\\\hline
Bias & 2.69 & -0.53 & 0.28 & 0.53 & -2.68 & 0.28 \\ 
$RB$ & 61.18 & 60.11 & 54.58 & 57.00 & 56.23 & 54.58 \\ 
$MSE$ & 3.17 & 3.08 & 2.57 & 2.75 & 2.68 & 2.57 \\ 
   \hline

&   \multicolumn{6}{c|}{$\pi_{H}=\pi_{\itM_2}$, $\pi_{D}=\pi_{\itM_1}$}\\\hline
Bias & 0.01 & -3.18 & -0.16 & -0.55 & -3.74 & -0.16 \\ 
$RB$ & 55.01 & 54.21 & 51.25 & 52.26 & 51.58 & 51.25 \\ 
$MSE$ & 2.51 & 2.46 & 2.22 & 2.31 & 2.26 & 2.22 \\ 
   \hline

&   \multicolumn{6}{c|}{$\pi_{H}=\pi_{\itM_1}$, $\pi_{D}=\pi_{\itM_2}$}\\\hline
Bias  & 2.76 & -0.50 & 0.13 & 0.40 & -2.84 & 0.13 \\ 
$RB$ & 54.74 & 53.72 & 47.55 & 50.13 & 49.44 & 47.55 \\ 
$MSE$ & 2.56 & 2.48 & 1.96 & 2.16 & 2.11 & 1.96 \\ 
   \hline
    
   \end{tabular}
\caption{\label{tab:sum_dim1_ROC_AUC}Summary measures for the ROC curve and the  $\AUC$.} 
\end{table}
 

\begin{table}[ht!]
\centering
\setlength{\tabcolsep}{4pt}
 \renewcommand{\arraystretch}{0.8}
\begin{tabular}{c|rrr|rrr|}
  \hline
 
  $1000\times$ & $\wROC_{\ipw} $ & $\wROC_{\kernel}$ &  $\wROC_{\conv} $ &  $\wROC_{\ipw} $ & $\wROC_{\kernel}$ &  $\wROC_{\conv} $ \\
  \hline
  &   \multicolumn{3}{c|}{$\wpi_D=\pi_c$, $\wpi_H=\pi_c$} & \multicolumn{3}{c|}{$\wpi_D=\pi_{\log}$, $\wpi_H=\pi_{c}$}\\\hline
&   \multicolumn{6}{c|}{$\pi_{H}=\pi_{\itM_1}$, $\pi_{D}=\pi_{\itM_1}$}\\\hline
  
$MSE$  & 5.01 & 4.43 & 2.97 & 3.98 & 3.17 & 2.97 \\ 
 $KS$ & 156.14 & 113.24 & 104.26 & 151.69 & 103.35 & 104.26 \\ 

   \hline
 
&   \multicolumn{6}{c|}{$\pi_{H}=\pi_{\itM_2}$, $\pi_{D}=\pi_{\itM_2}$}\\\hline
$MSE$  & 7.11 & 5.77 & 4.57 & 6.57 & 5.39 & 4.57 \\ 
$KS$ & 199.16 & 139.19 & 129.25 & 189.51 & 132.63 & 129.25 \\ 

   \hline
  &   \multicolumn{6}{c|}{$\pi_{H}=\pi_{\itM_2}$, $\pi_{D}=\pi_{\itM_1}$}\\\hline
$MSE$ & 7.61 & 6.84 & 4.13 & 5.49 & 4.54 & 4.13 \\ 
$KS $ & 187.91 & 138.12 & 126.46 & 178.94 & 124.64 & 126.46 \\ 
   \hline
 
 &   \multicolumn{6}{c|}{$\pi_{H}=\pi_{\itM_1}$, $\pi_{D}=\pi_{\itM_2}$}\\\hline
$MSE$ & 6.15 & 4.95 & 3.39 & 5.05 & 4.02 & 3.39 \\ 
 $KS$ & 179.24 & 124.99 & 108.19 & 165.92 & 114.51 & 108.19 \\ 
   \hline 
 
$1000\times$ & $\wAUC_{\ipw}$ & $\wAUC_{\kernel}$ &  $\wAUC_{\conv}$ & $\wAUC_{\ipw}$ & $\wAUC_{\kernel}$ &  $\wAUC_{\conv}$ \\
 \hline 
   &   \multicolumn{3}{c|}{$\wpi_D=\pi_c$, $\wpi_H=\pi_c$} & \multicolumn{3}{c|}{$\wpi_D=\pi_{\log}$, $\wpi_H=\pi_{c}$}\\\hline
&   \multicolumn{6}{c|}{$\pi_{H}=\pi_{\itM_1}$, $\pi_{D}=\pi_{\itM_1}$}\\\hline
 
 Bias  & -22.11 & -25.19 & -0.34 & 7.87 & 4.41 & -0.34 \\ 
$RB$ & 53.00 & 53.94 & 43.90 & 44.52 & 43.28 & 43.90 \\ 
$MSE$   & 2.46 & 2.55 & 1.64 & 1.65 & 1.57 & 1.64 \\ 
   \hline

&   \multicolumn{6}{c|}{$\pi_{H}=\pi_{\itM_2}$, $\pi_{D}=\pi_{\itM_2}$}\\\hline
Bias & 15.31 & 11.88 & 0.28 & -3.22 & -6.42 & 0.28 \\ 
$RB$ & 61.87 & 60.09 & 54.58 & 57.63 & 57.10 & 54.58 \\ 
$MSE$  & 3.18 & 3.02 & 2.57 & 2.79 & 2.76 & 2.57 \\ 
   \hline
  
  &   \multicolumn{6}{c|}{$\pi_{H}=\pi_{\itM_2}$, $\pi_{D}=\pi_{\itM_1}$}\\\hline 

Bias & -35.06 & -37.89 & -0.16 & -4.28 & -7.46 & -0.16 \\ 
$RB$ & 68.49 & 69.77 & 51.25 & 53.01 & 52.61 & 51.25 \\ 
$MSE$  & 4.05 & 4.19 & 2.22 & 2.36 & 2.34 & 2.22 \\ 
   \hline
 
 &   \multicolumn{6}{c|}{$\pi_{H}=\pi_{\itM_1}$, $\pi_{D}=\pi_{\itM_2}$}\\
 \hline

   Bias & 27.07 & 23.36 & 0.13 & 8.95 & 5.47 & 0.13 \\ 
$RB$ & 59.50 & 56.80 & 47.55 & 49.72 & 48.31 & 47.55 \\ 
$MSE$  & 2.92 & 2.67 & 1.96 & 2.09 & 1.98 & 1.96 \\ 
   \hline
 
\end{tabular}
\caption{\label{tab:sum_mal_pi_ROC_AUC}Summary measures for the ROC curve and the  AUC, under misspecification  of the propensity.} 
\end{table}


\begin{table}[ht!]
\centering
\setlength{\tabcolsep}{4pt}
 \renewcommand{\arraystretch}{0.8}
\begin{tabular}{c|rrr|rrr|}
  \hline
 $1000\times$ & $\wROC_{\ipw} $ & $\wROC_{\kernel}$ &  $\wROC_{\conv} $ &  $\wROC_{\ipw} $ & $\wROC_{\kernel}$ &  $\wROC_{\conv} $ \\
  \hline
   &   \multicolumn{3}{c|}{$\wpi_D=\pi$, $\wpi_H=\pi$} & \multicolumn{3}{c|}{$\wpi_D=\pi_{\log}$, $\wpi_H=\pi_{\log}$}\\\hline
 
 &   \multicolumn{6}{c|}{$\pi_{H}=\pi_{\itM_1}$, $\pi_{D}=\pi_{\itM_1}$}\\\hline 
$MSE$  & 4.44 & 3.67 & 3.88 & 4.22 & 3.45 & 3.88 \\ 
$KS$ & 158.60 & 110.06 & 121.36 & 155.90 & 107.35 & 121.36 \\ 

  \hline 
  
 &   \multicolumn{6}{c|}{$\pi_{H}=\pi_{\itM_2}$, $\pi_{D}=\pi_{\itM_2}$}\\\hline 
 
$MSE$  & 7.05 & 5.79 & 5.01 & 6.54 & 5.30 & 5.01 \\ 
 $KS$  & 194.76 & 136.44 & 132.55 & 190.46 & 132.22 & 132.55 \\ 
   \hline
  
 &   \multicolumn{6}{c|}{$\pi_{H}=\pi_{\itM_2}$, $\pi_{D}=\pi_{\itM_1}$}\\\hline

$MSE$ & 5.73 & 4.72 & 5.08 & 5.46 & 4.45 & 5.08 \\ 
 $KS$  & 182.73 & 126.95 & 135.78 & 180.12 & 123.95 & 135.78 \\ 
   \hline

 &   \multicolumn{6}{c|}{$\pi_{H}=\pi_{\itM_1}$, $\pi_{D}=\pi_{\itM_2}$}\\\hline
$MSE$ & 5.71 & 4.70 & 3.99 & 5.26 & 4.26 & 3.99 \\ 
$KS$ & 173.41 & 121.86 & 120.78 & 169.09 & 117.43 & 120.78 \\ 
 
   \hline

 $1000\times$ & $\wAUC_{\ipw}$ & $\wAUC_{\kernel}$ &  $\wAUC_{\conv}$ & $\wAUC_{\ipw}$ & $\wAUC_{\kernel}$ &  $\wAUC_{\conv}$ \\
 \hline 
  &   \multicolumn{3}{c|}{$\wpi_D=\pi$, $\wpi_H=\pi$} & \multicolumn{3}{c|}{$\wpi_D=\pi_{\log}$, $\wpi_H=\pi_{\log}$}\\\hline
 
 &   \multicolumn{6}{c|}{$\pi_{H}=\pi_{\itM_1}$, $\pi_{D}=\pi_{\itM_1}$}\\\hline

   Bias & 0.09 & -3.14 & 5.08 & -0.68 & -3.90 & 5.08 \\ 
$RB$ & 47.76 & 47.04 & 46.47 & 45.56 & 44.91 & 46.47 \\ 
$MSE$  & 1.93 & 1.89 & 1.81 & 1.76 & 1.73 & 1.81 \\ 
   \hline

 &   \multicolumn{6}{c|}{$\pi_{H}=\pi_{\itM_2}$, $\pi_{D}=\pi_{\itM_2}$}\\\hline 

 Bias & 2.69 & -0.53 & -16.87 & 0.53 & -2.68 & -16.87 \\ 
 $RB$  & 61.18 & 60.11 & 55.45 & 57.00 & 56.23 & 55.45 \\ 
$MSE$  & 3.17 & 3.08 & 2.67 & 2.75 & 2.68 & 2.67 \\ 

   \hline
 
 &   \multicolumn{6}{c|}{$\pi_{H}=\pi_{\itM_2}$, $\pi_{D}=\pi_{\itM_1}$}\\\hline
  
  Bias & 0.01 & -3.18 & 2.64 & -0.55 & -3.74 & 2.64 \\ 
$RB$ & 55.01 & 54.21 & 54.00 & 52.26 & 51.58 & 54.00 \\ 
$MSE$ & 2.51 & 2.46 & 2.44 & 2.31 & 2.26 & 2.44 \\ 
   \hline

 &   \multicolumn{6}{c|}{$\pi_{H}=\pi_{\itM_1}$, $\pi_{D}=\pi_{\itM_2}$}\\\hline

Bias & 2.76 & -0.50 & -14.90 & 0.40 & -2.84 & -14.90 \\ 
$RB$  & 54.74 & 53.72 & 48.14 & 50.13 & 49.44 & 48.14 \\ 
$MSE$ & 2.56 & 2.48 & 2.06 & 2.16 & 2.11 & 2.06 \\ 
   \hline
 
\end{tabular}
\caption{\label{tab:sum_malregre_ROC_AUC}Summary measures for the ROC curve and the  AUC under misspecification of the regression model, the regression is estimated without intercept.} 
\end{table}

\begin{figure}[ht!]
 \begin{center}
 \footnotesize
 \renewcommand{\arraystretch}{0.4}
 \newcolumntype{M}{>{\centering\arraybackslash}m{\dimexpr.15 \linewidth-1\tabcolsep}}
   \newcolumntype{G}{>{\centering\arraybackslash}m{\dimexpr.3\linewidth-1\tabcolsep}}
\begin{tabular}{M G G G}\\
& $\wROC_{\ipw}$ & $\wROC_{\kernel}$ & $\wROC_{\conv}$\\[0.1in]
& \multicolumn{3}{c}{$\pi_{H}=\pi_{\itM_1}$, $\pi_{D}=\pi_{\itM_1}$}\\
$\wpi_D=\pi$, $\wpi_H=\pi$   & 
\includegraphics[scale=0.25]{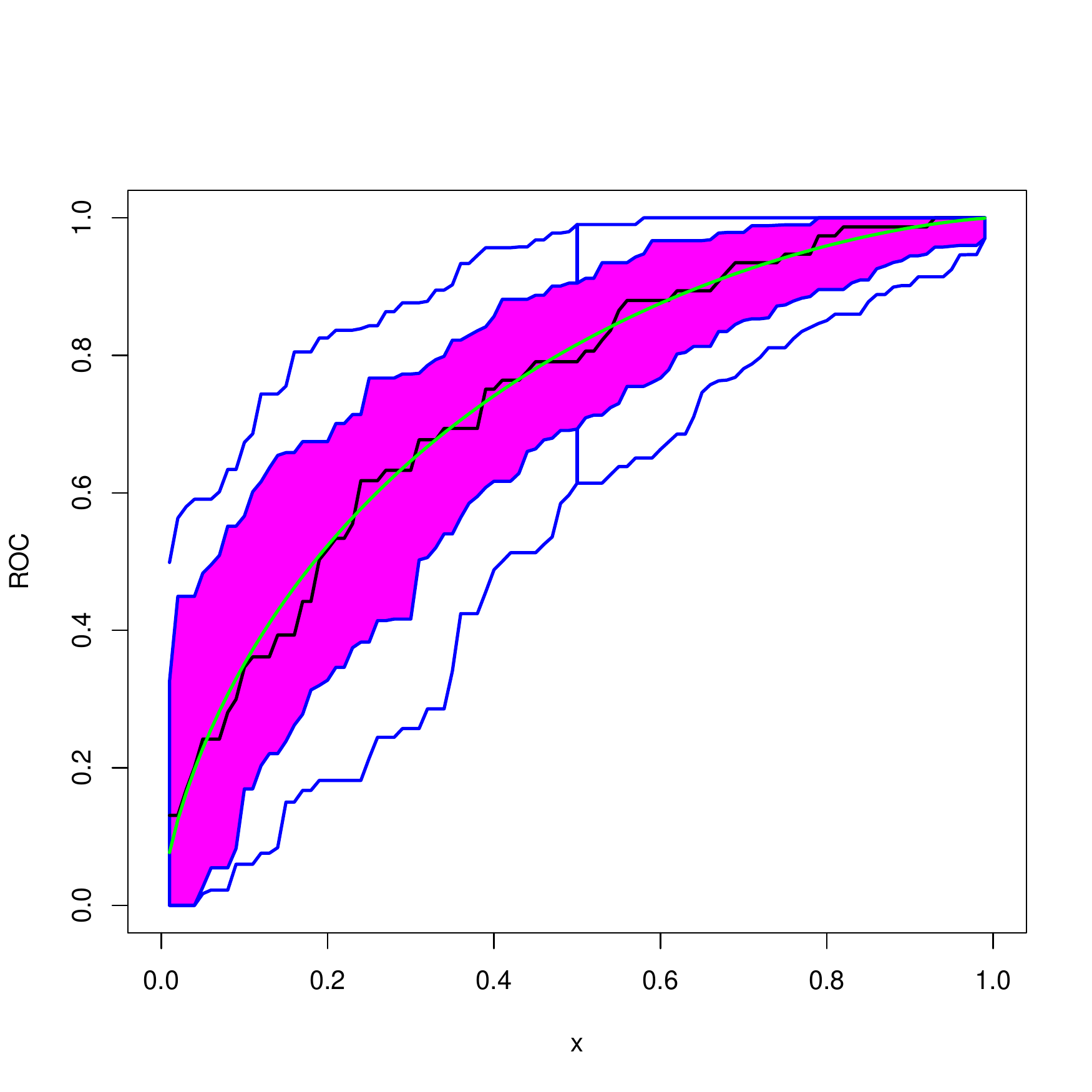} &
\includegraphics[scale=0.25]{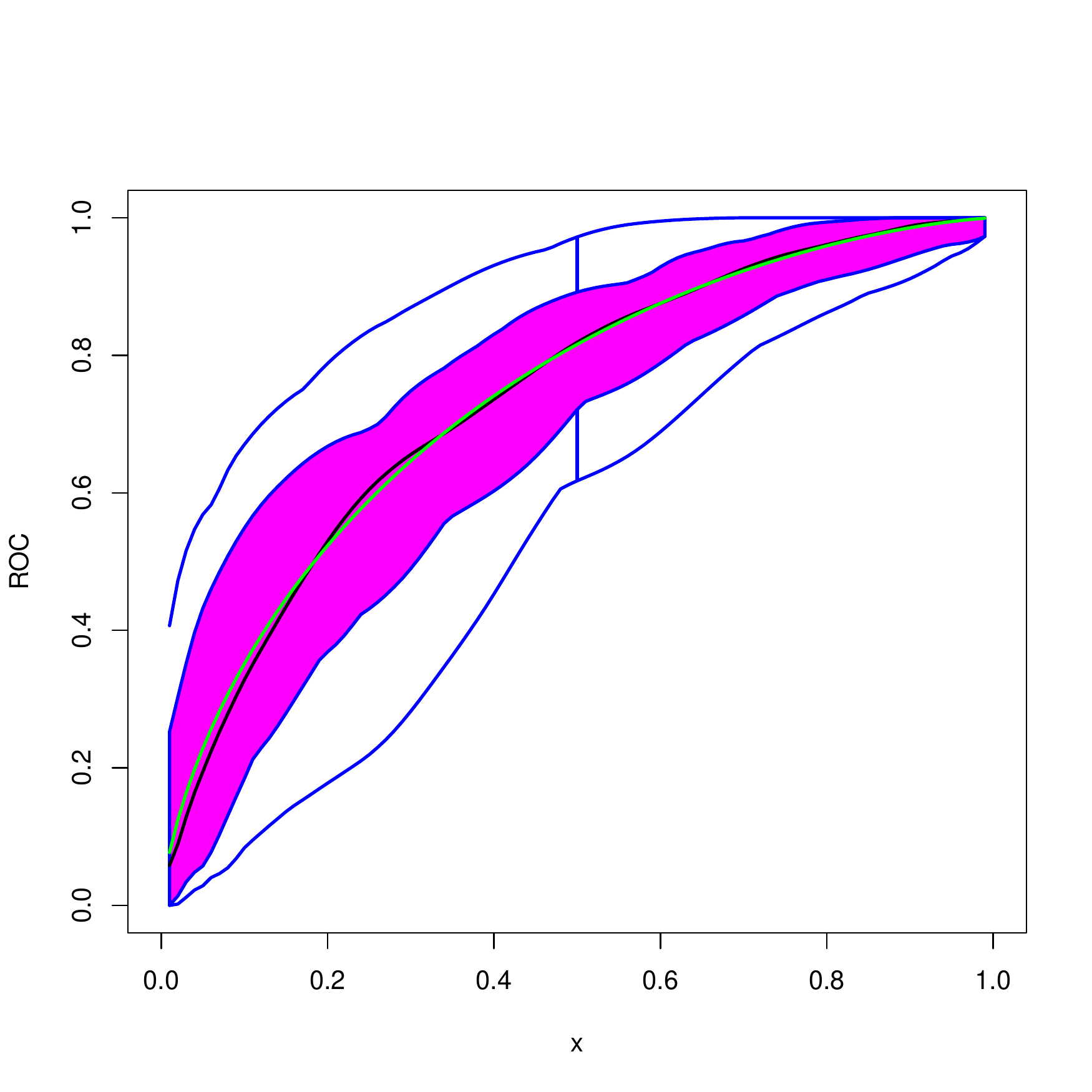} &
\includegraphics[scale=0.25]{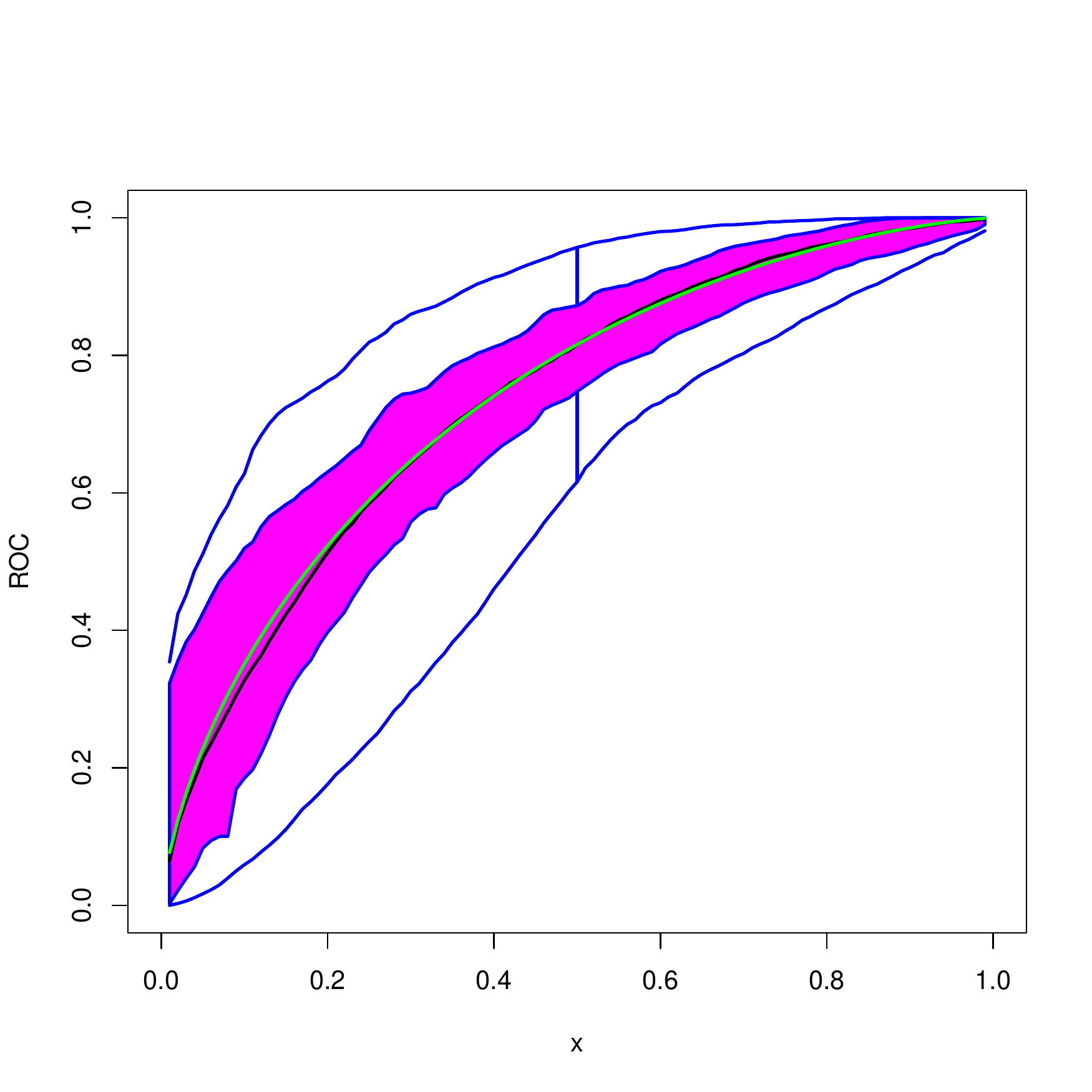}\\
 $\wpi_D=\pi_{\log}$, $\wpi_H=\pi_{\log}$ & 
\includegraphics[scale=0.25]{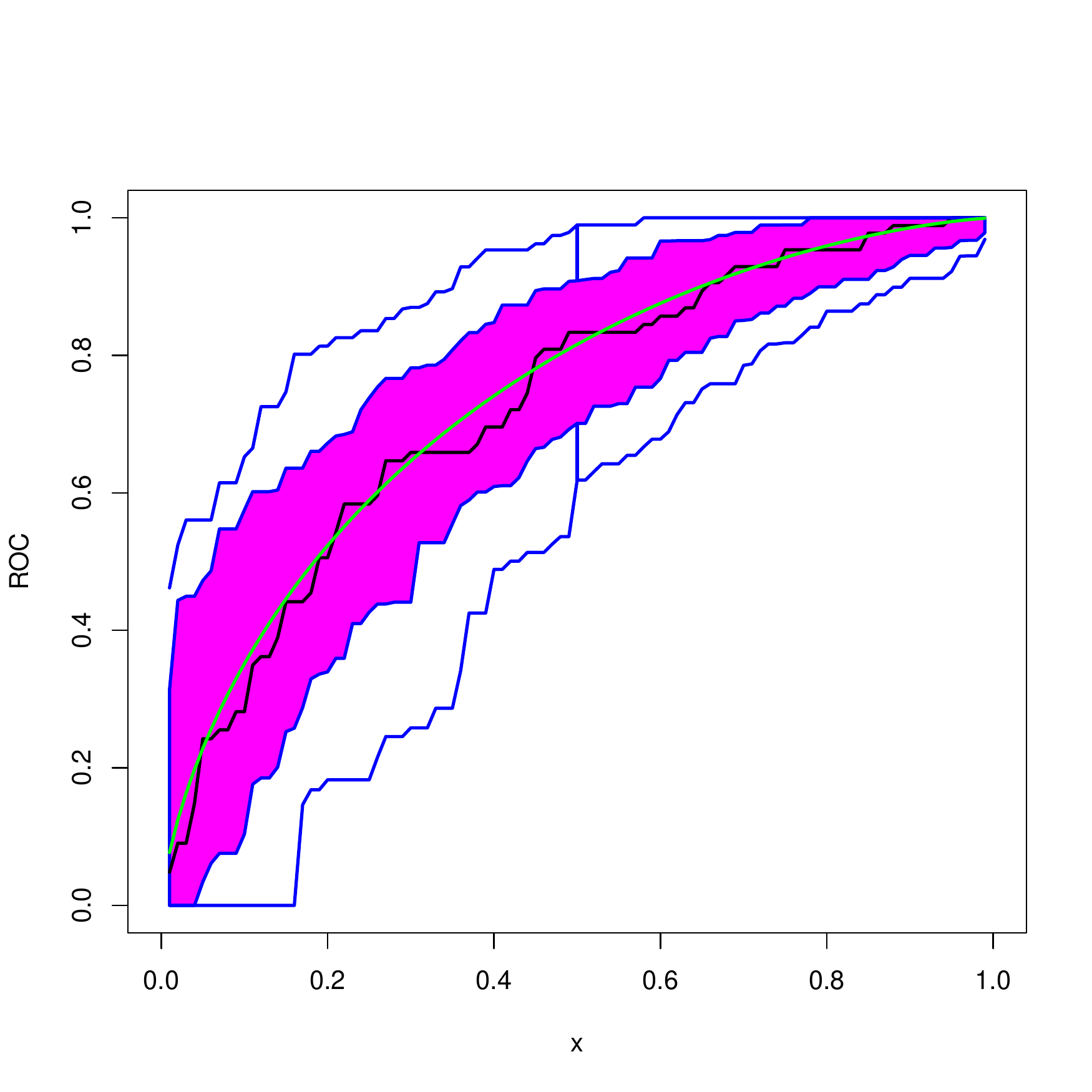} &   
\includegraphics[scale=0.25]{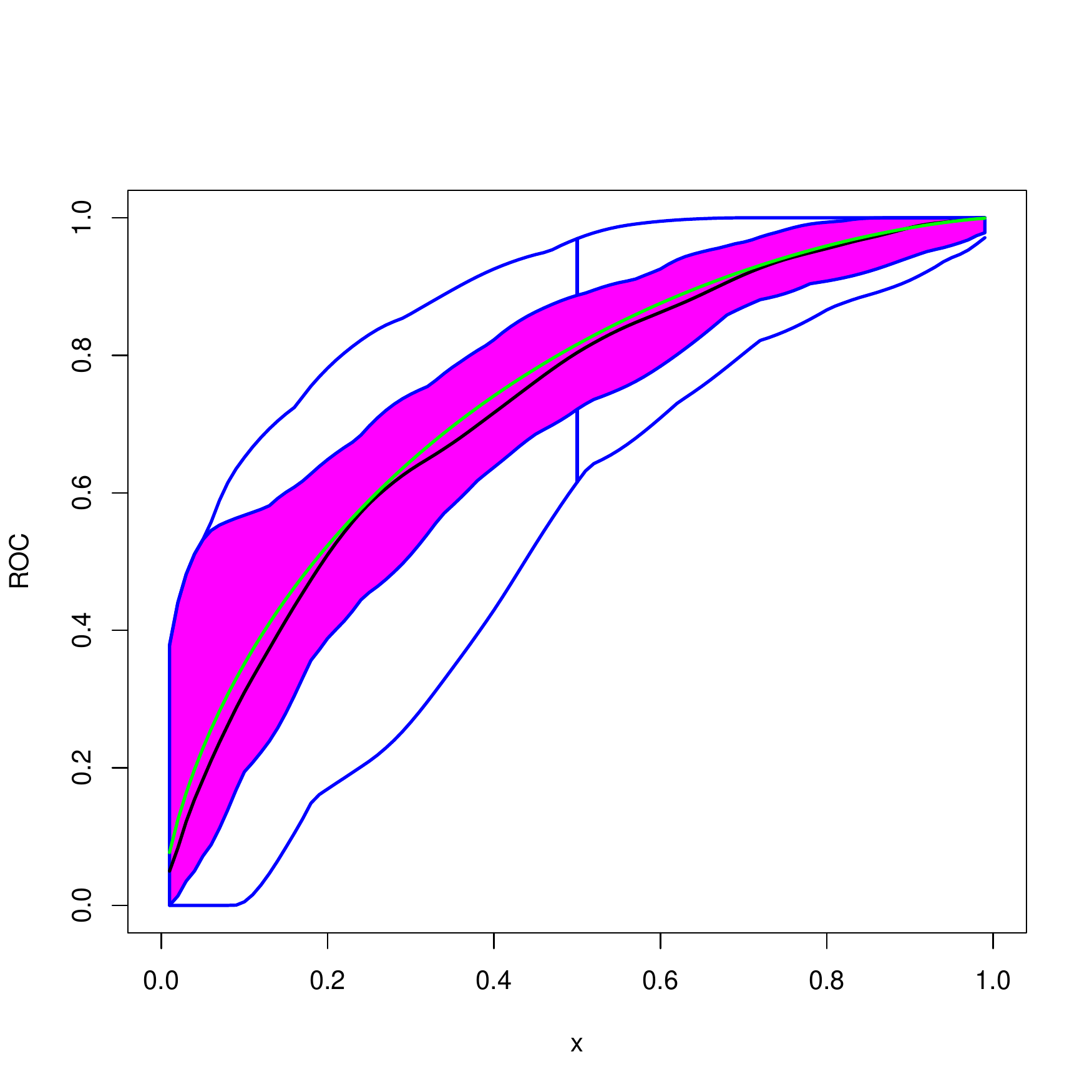}   &  
\includegraphics[scale=0.25]{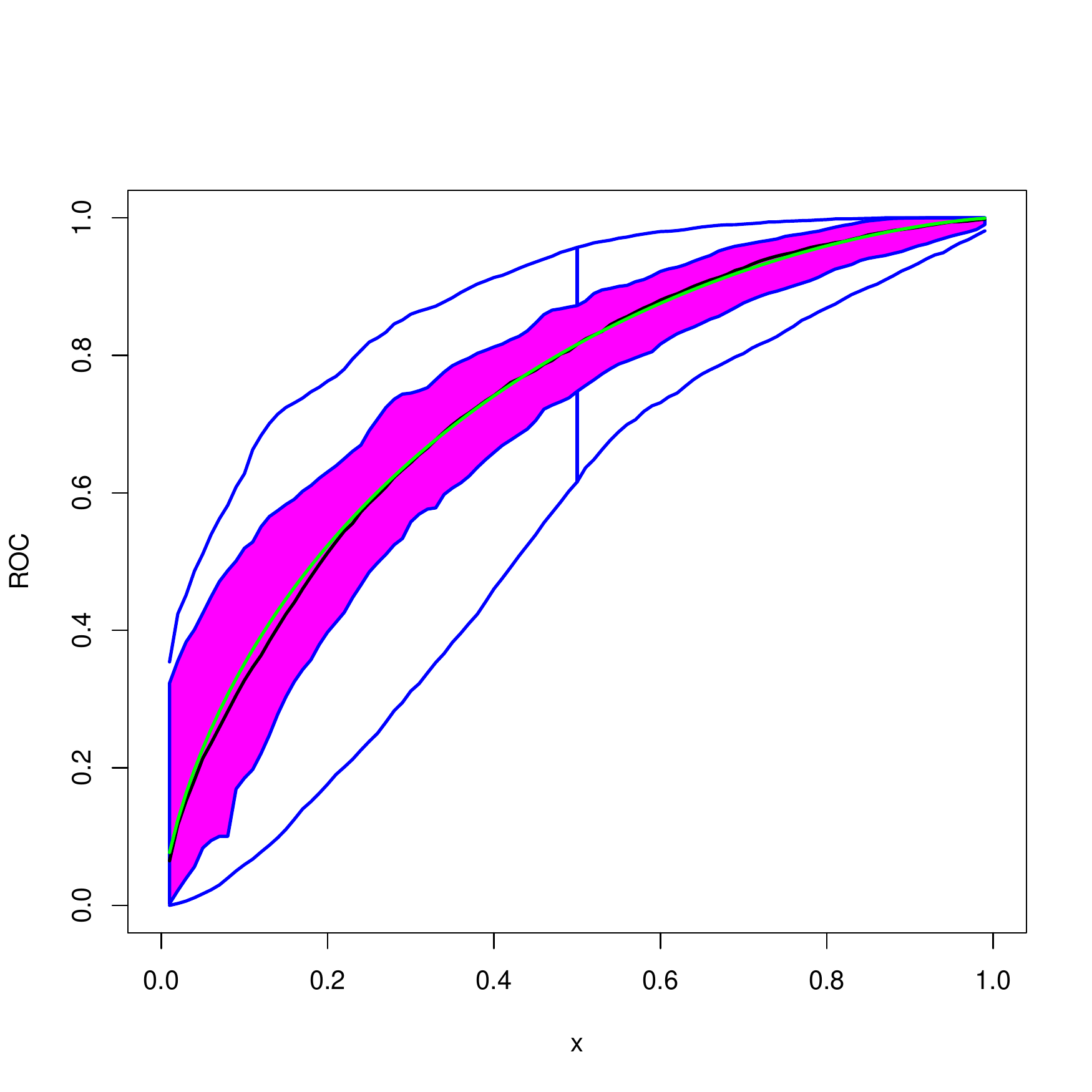}\\
 & \multicolumn{3}{c}{$\pi_{H}=\pi_{\itM_2}$, $\pi_{D}=\pi_{\itM_2}$}\\
$\wpi_D=\pi$, $\wpi_H=\pi$   & 
\includegraphics[scale=0.25]{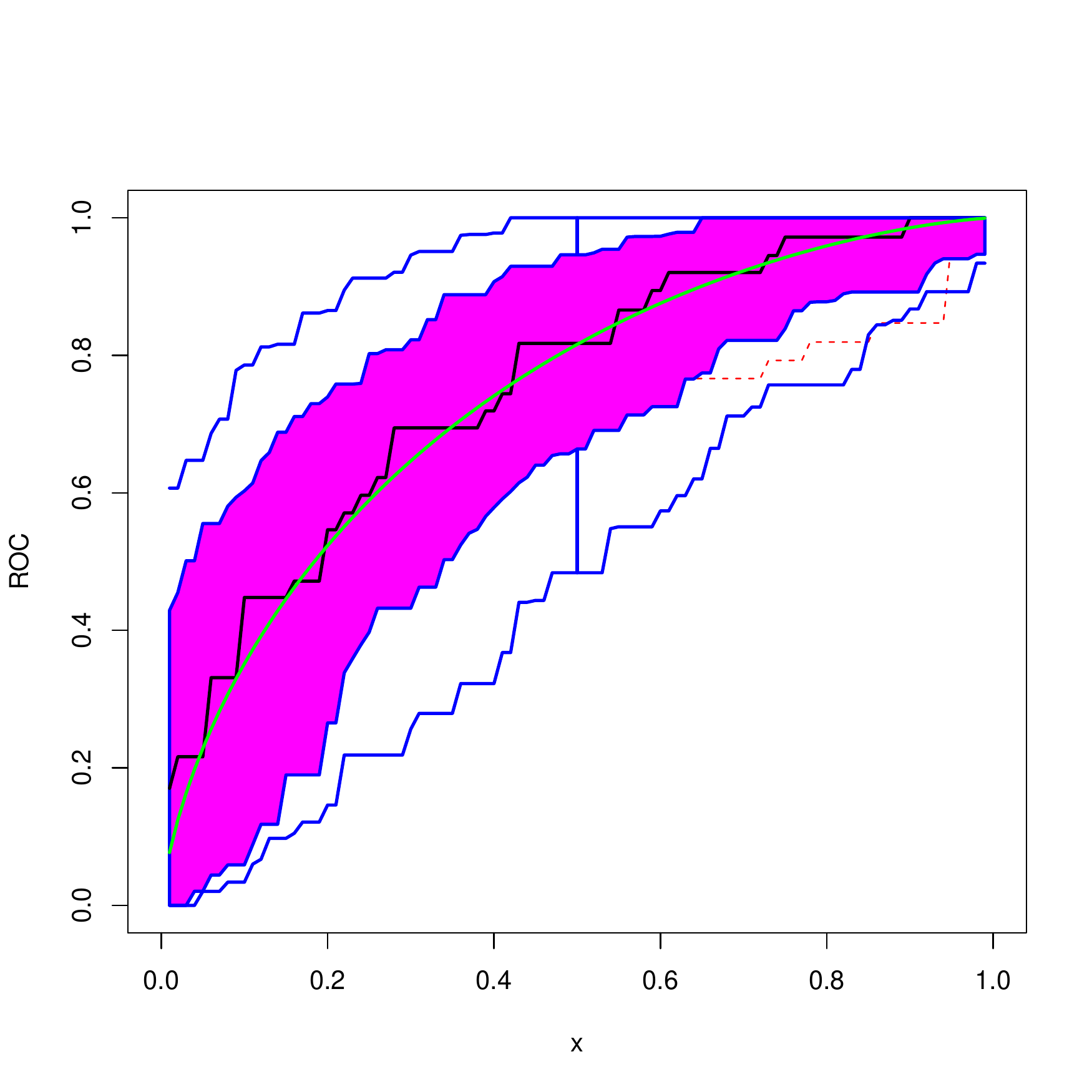} &
\includegraphics[scale=0.25]{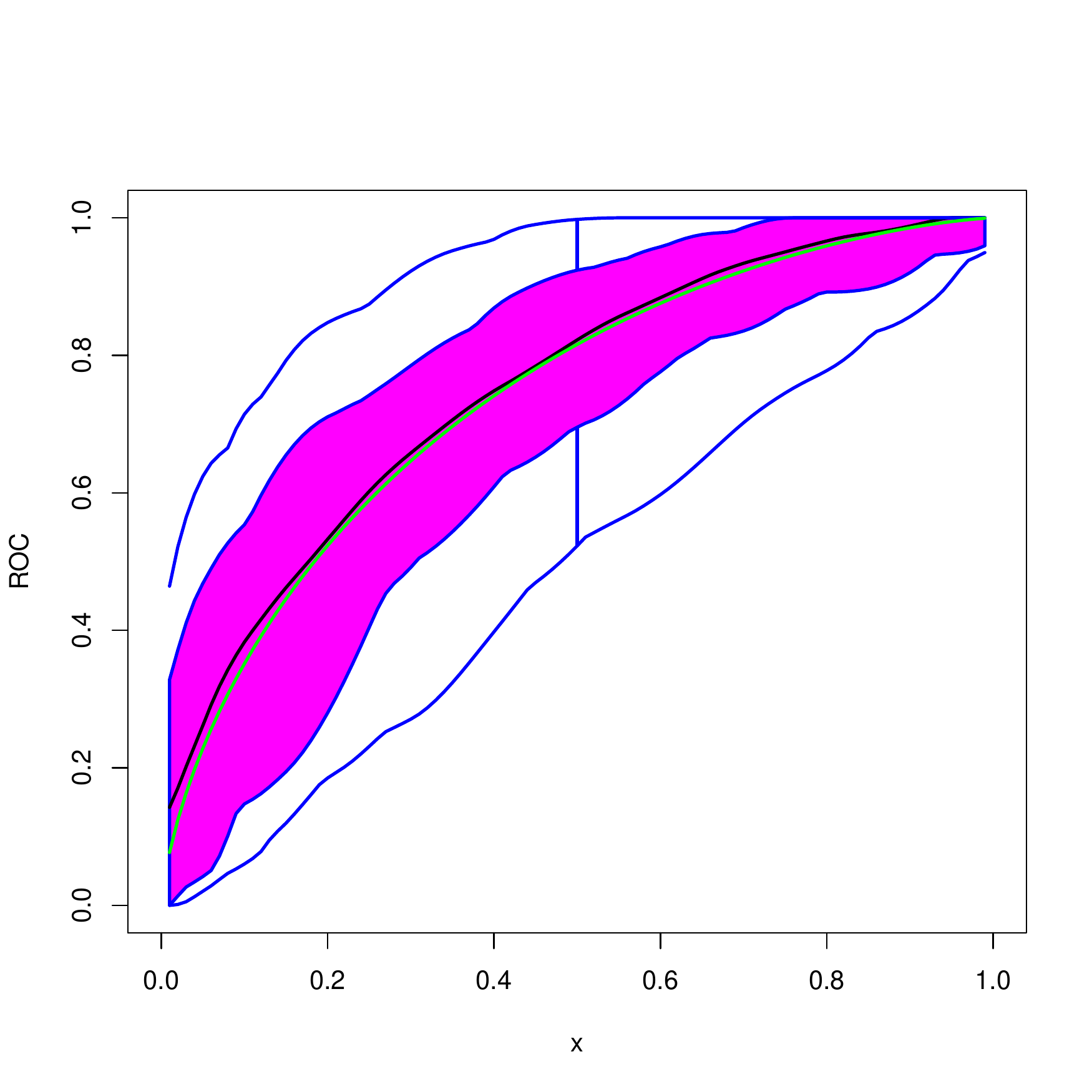} &
\includegraphics[scale=0.25]{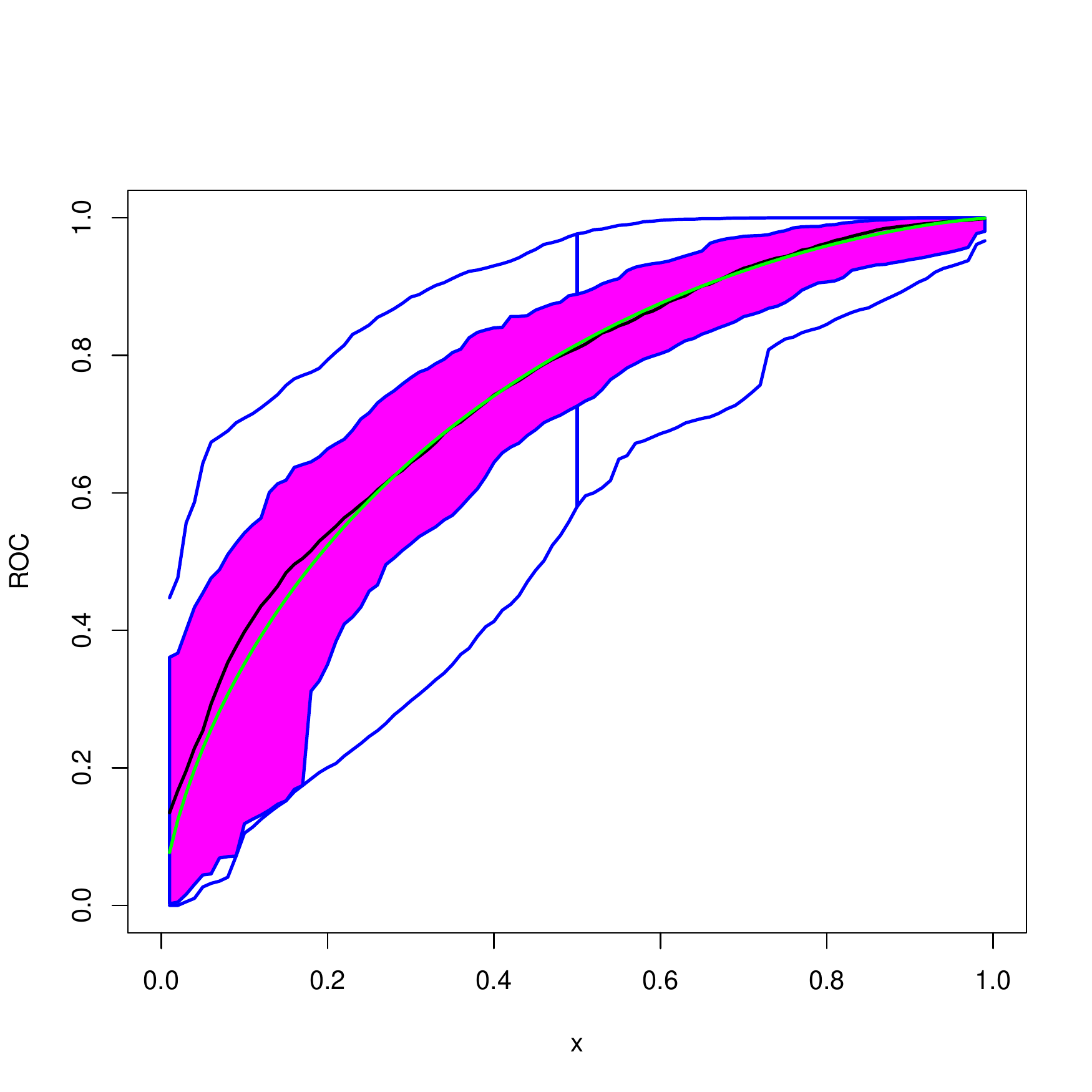}\\
 $\wpi_D=\pi_{\log}$, $\wpi_H=\pi_{\log}$ & 
\includegraphics[scale=0.25]{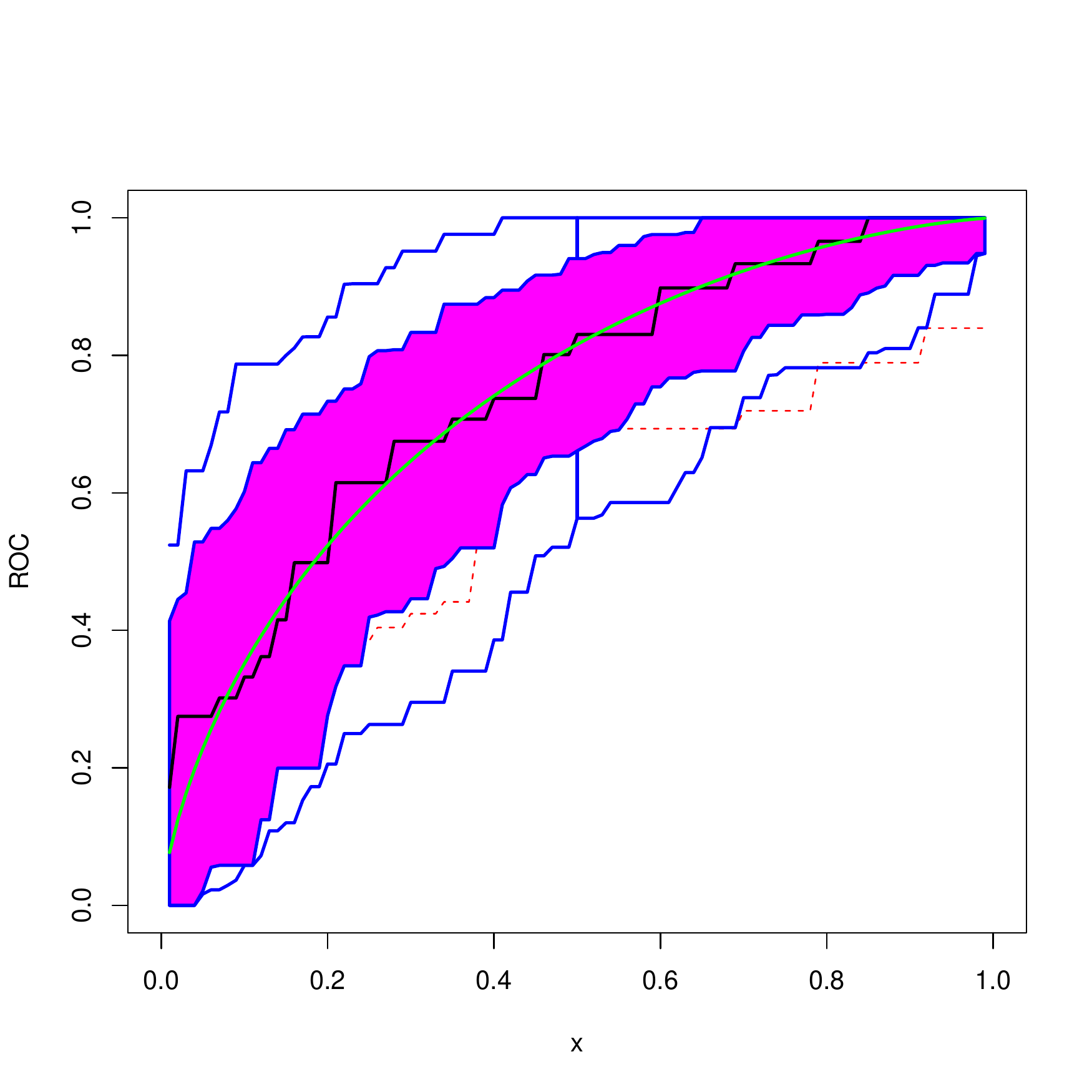} &   
\includegraphics[scale=0.25]{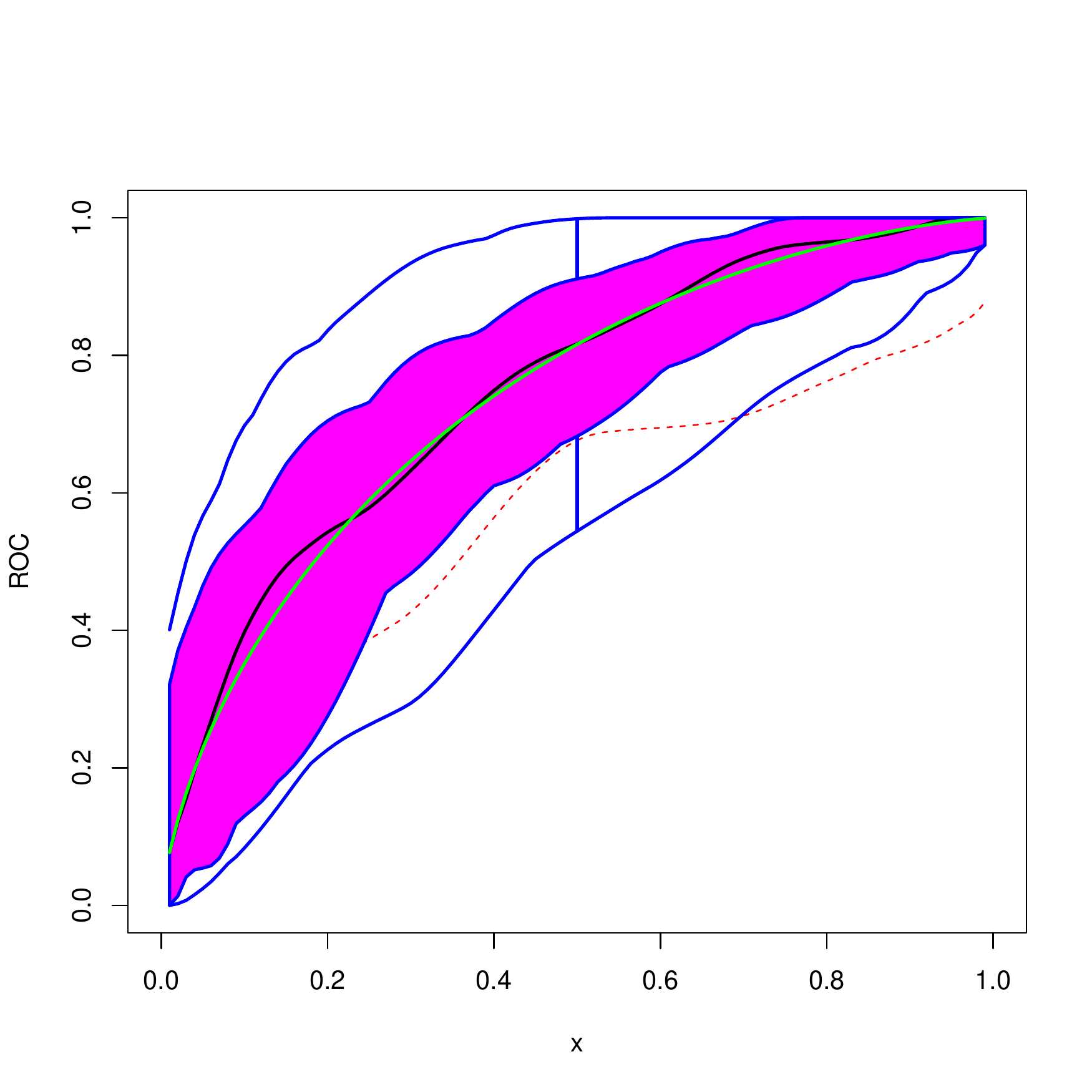}   &  
\includegraphics[scale=0.25]{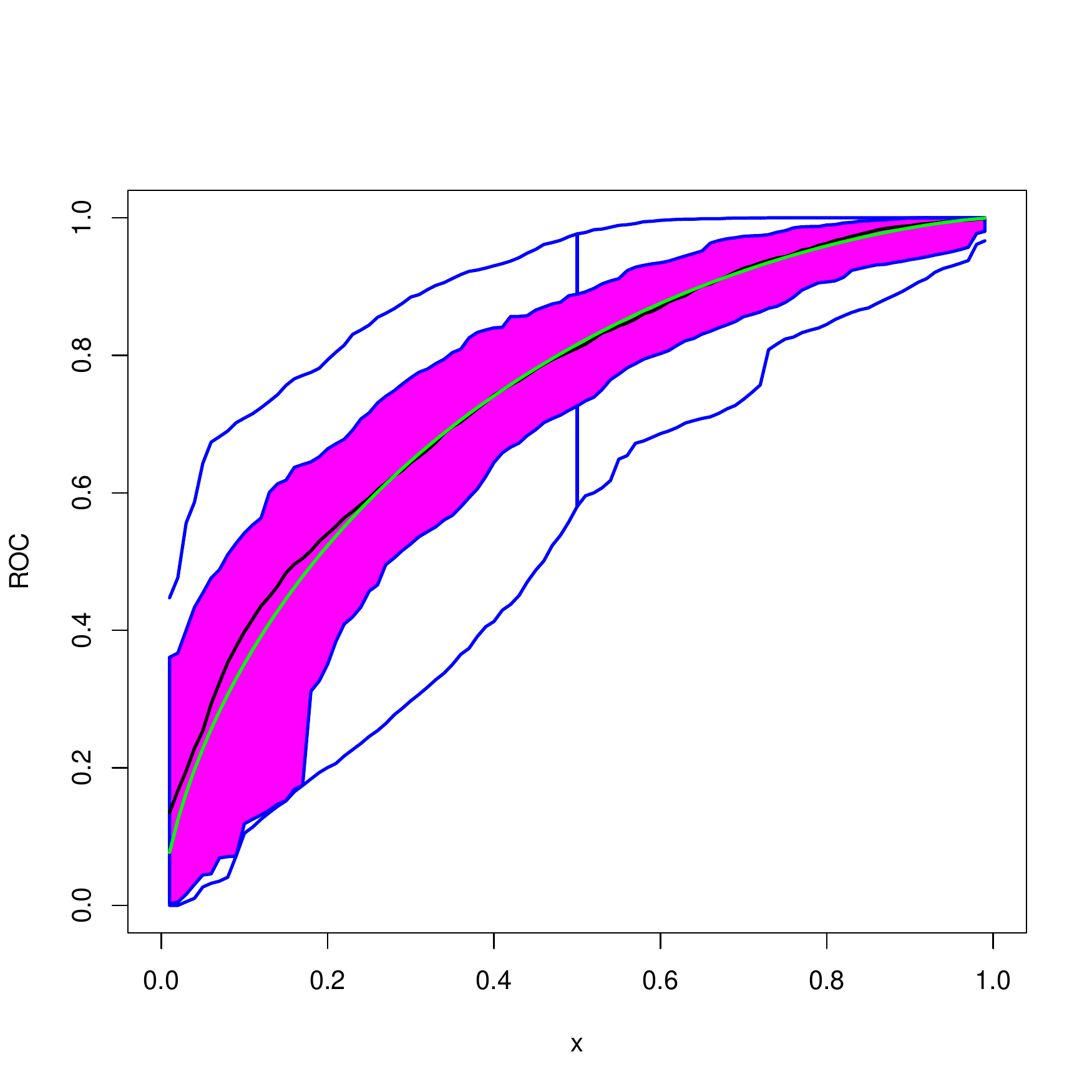}
   \end{tabular}
\vskip-0.1in  
\caption{\label{fig:fbx_11_22}\small Functional boxplots of  $\wROC(p)$. The green line corresponds to the true $ROC(p)$ and the dotted red lines to the outlying curves detected by the functional boxplot.}
\end{center} 
\end{figure}
\normalsize
 
\begin{figure}[ht!]
 \begin{center}
 \footnotesize
 \renewcommand{\arraystretch}{0.4}
 \newcolumntype{M}{>{\centering\arraybackslash}m{\dimexpr.15 \linewidth-1\tabcolsep}}
   \newcolumntype{G}{>{\centering\arraybackslash}m{\dimexpr.3\linewidth-1\tabcolsep}}
\begin{tabular}{M G G G}\\
& $\wROC_{\ipw}$ & $\wROC_{\kernel}$ & $\wROC_{\conv}$\\[0.1in]
& \multicolumn{3}{c}{$\pi_{H}=\pi_{\itM_2}$, $\pi_{D}=\pi_{\itM_1}$}\\
$\wpi_D=\pi$, $\wpi_H=\pi$   & 
\includegraphics[scale=0.25]{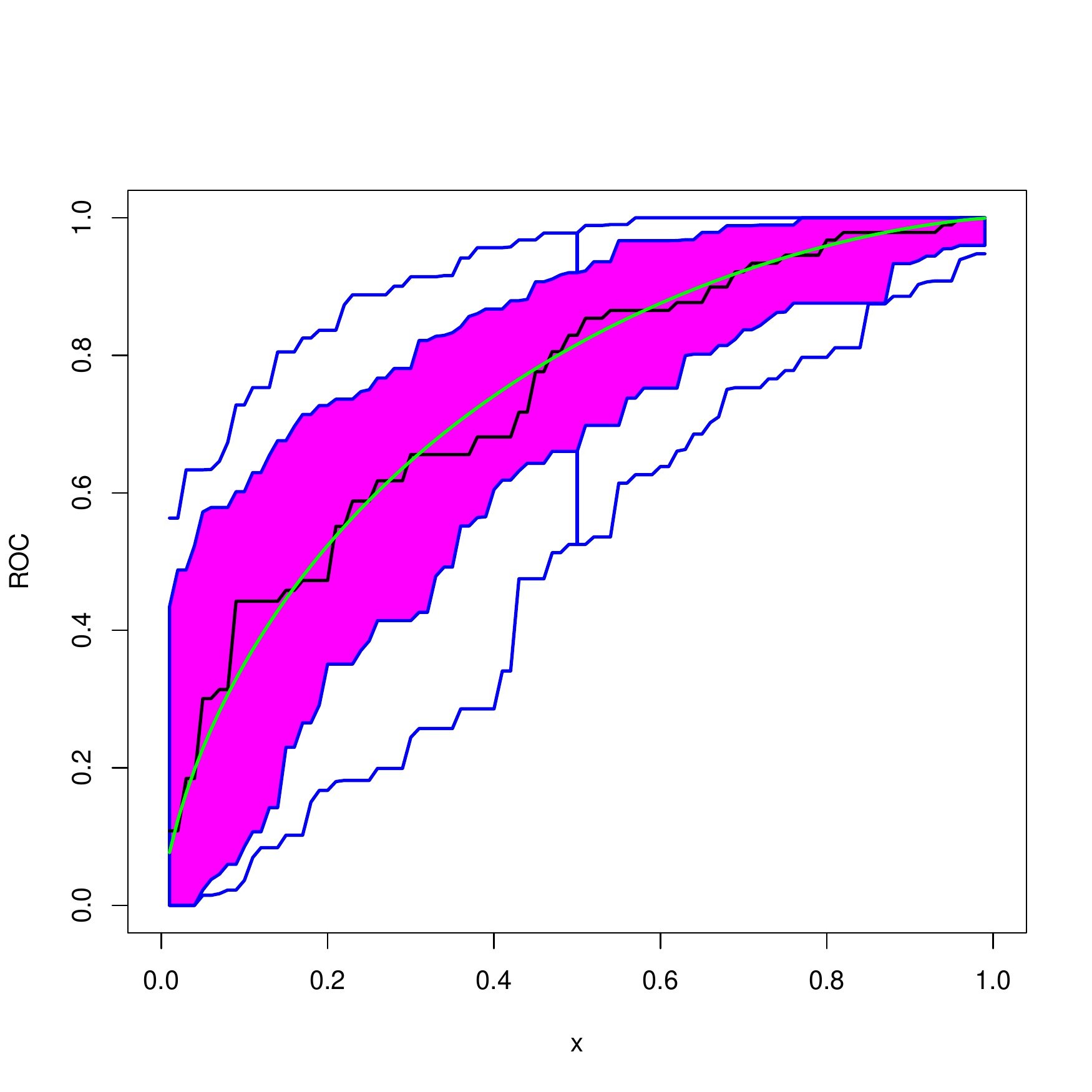} &
\includegraphics[scale=0.25]{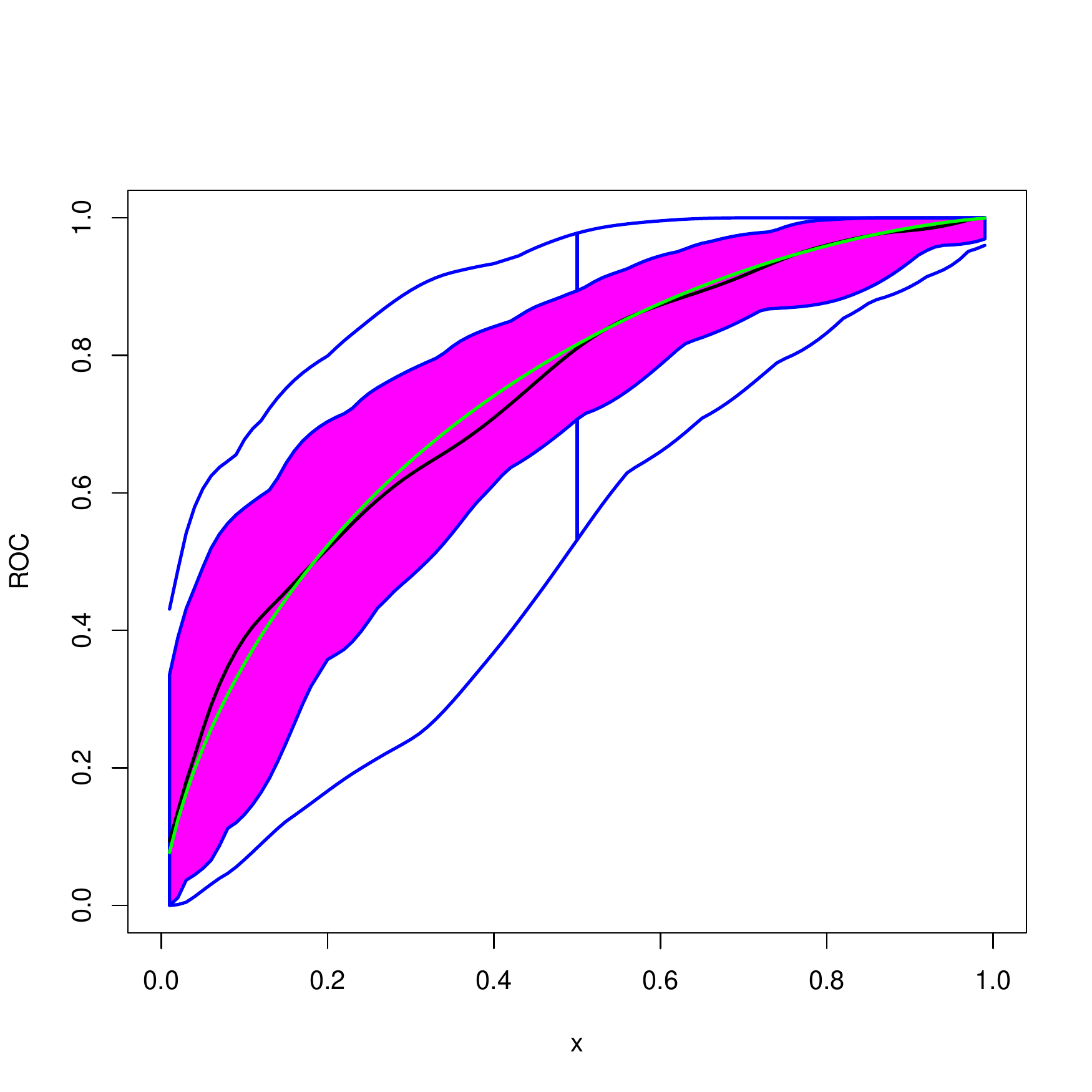} &
\includegraphics[scale=0.25]{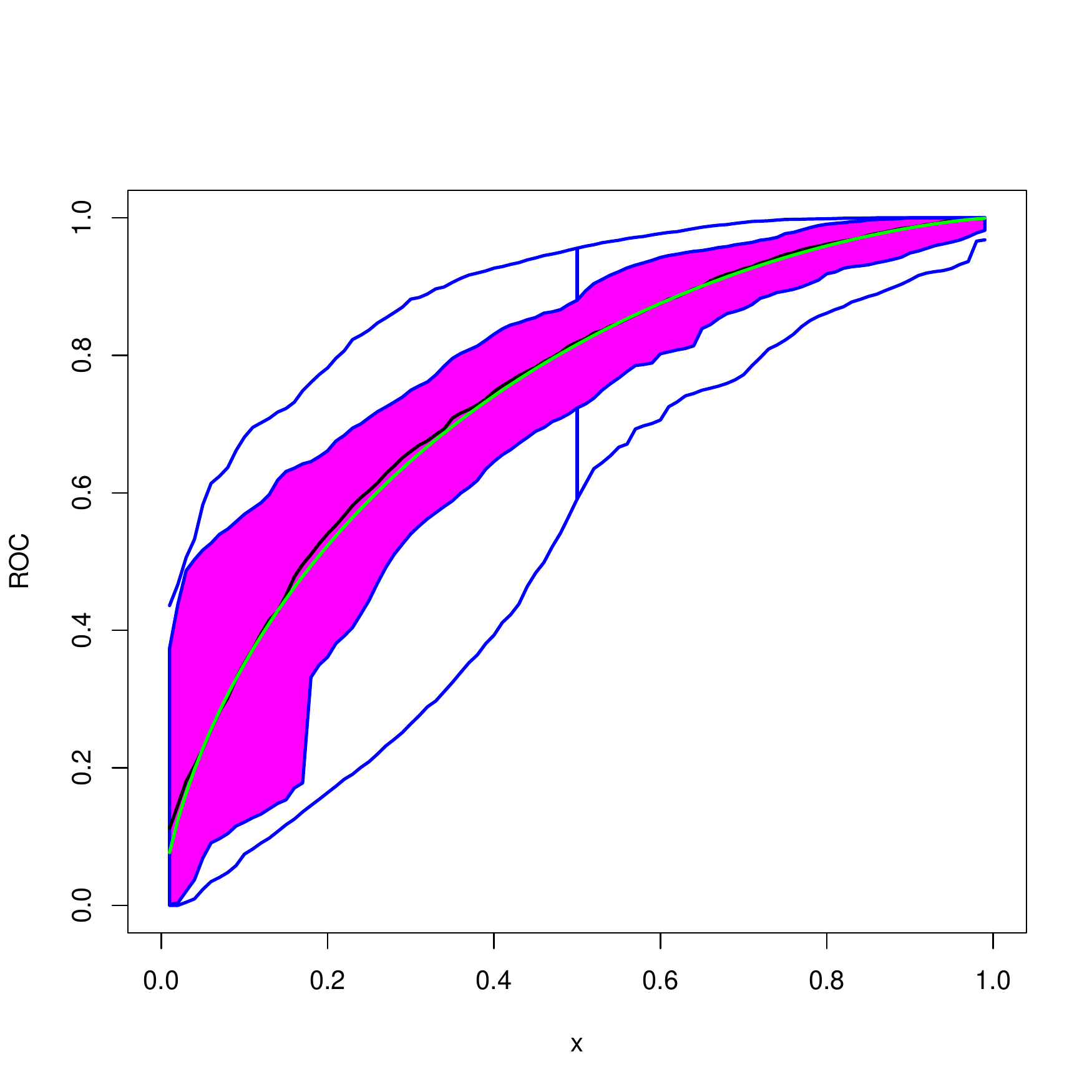}\\
 $\wpi_D=\pi_{\log}$, $\wpi_H=\pi_{\log}$ & 
\includegraphics[scale=0.25]{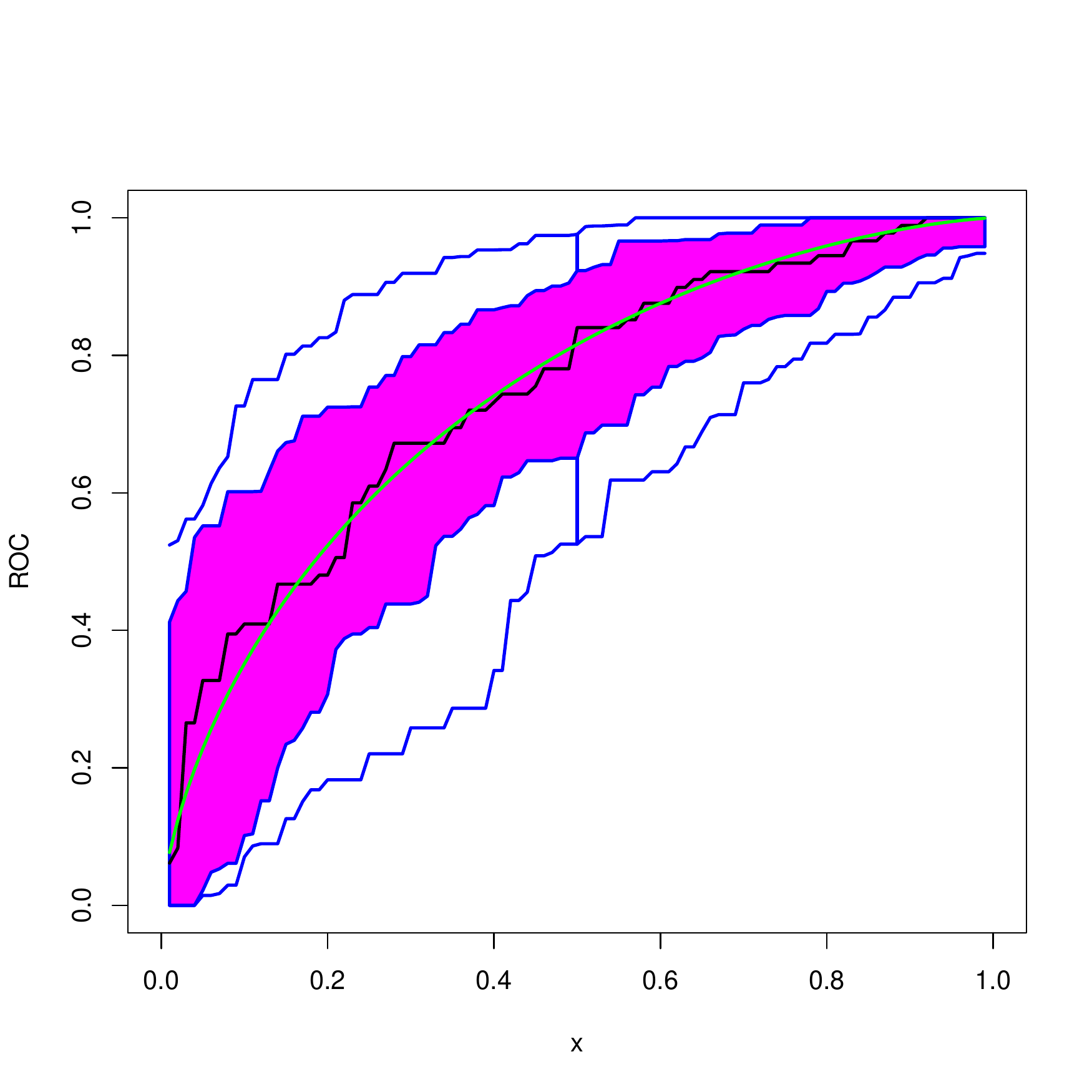} &   
\includegraphics[scale=0.25]{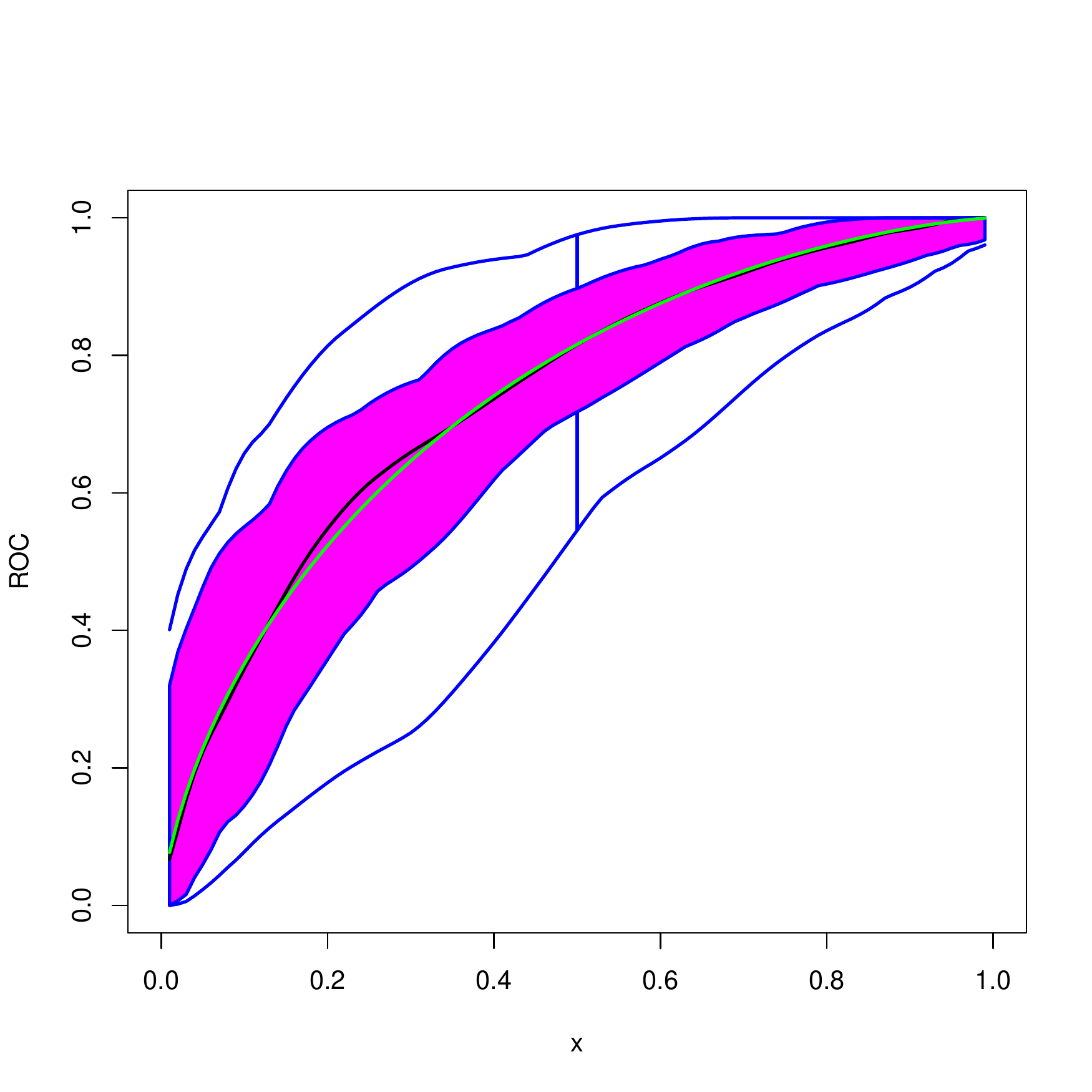}   &  
\includegraphics[scale=0.25]{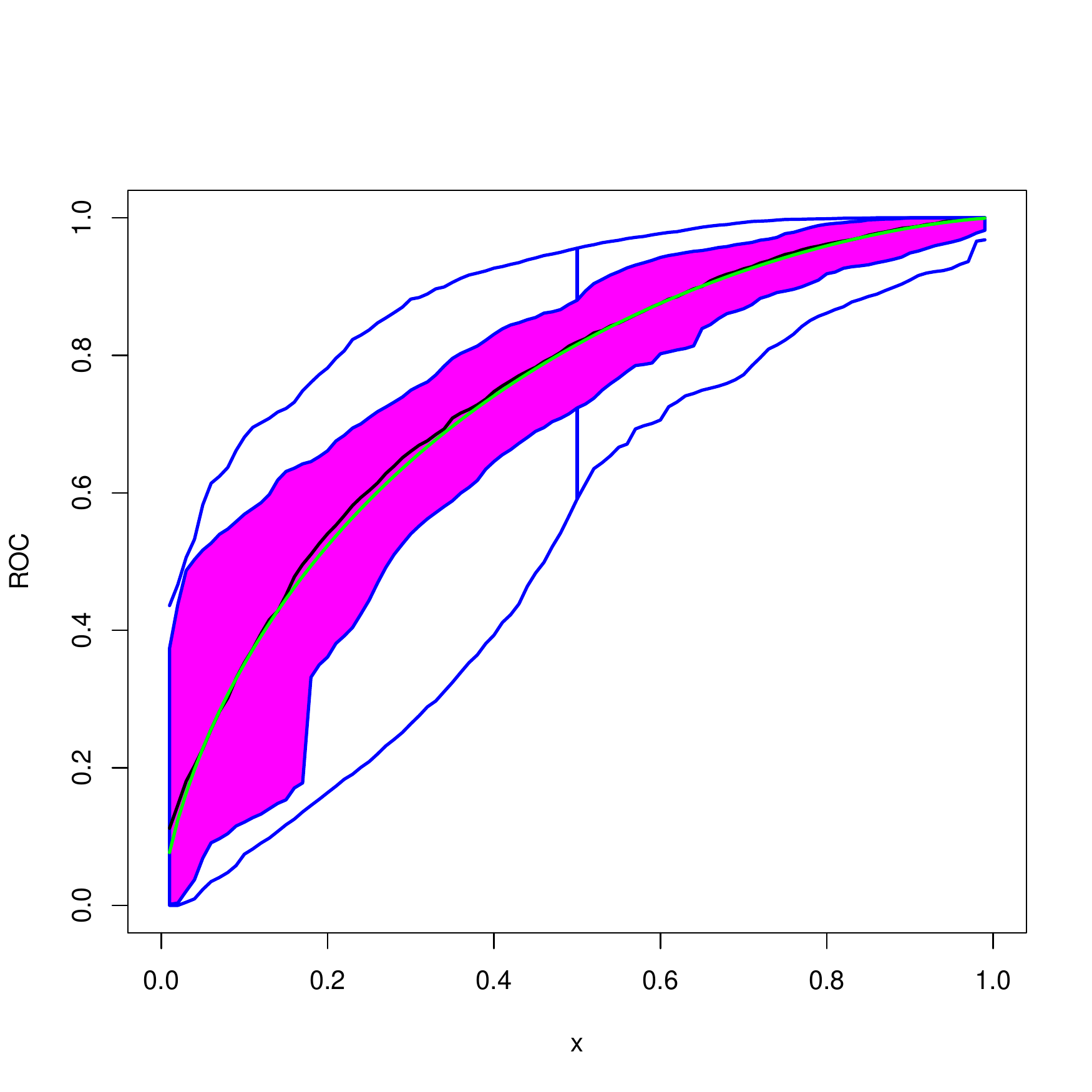}\\
 & \multicolumn{3}{c}{$\pi_{H}=\pi_{\itM_1}$, $\pi_{D}=\pi_{\itM_2}$}\\
$\wpi_D=\pi$, $\wpi_H=\pi$   & 
\includegraphics[scale=0.25]{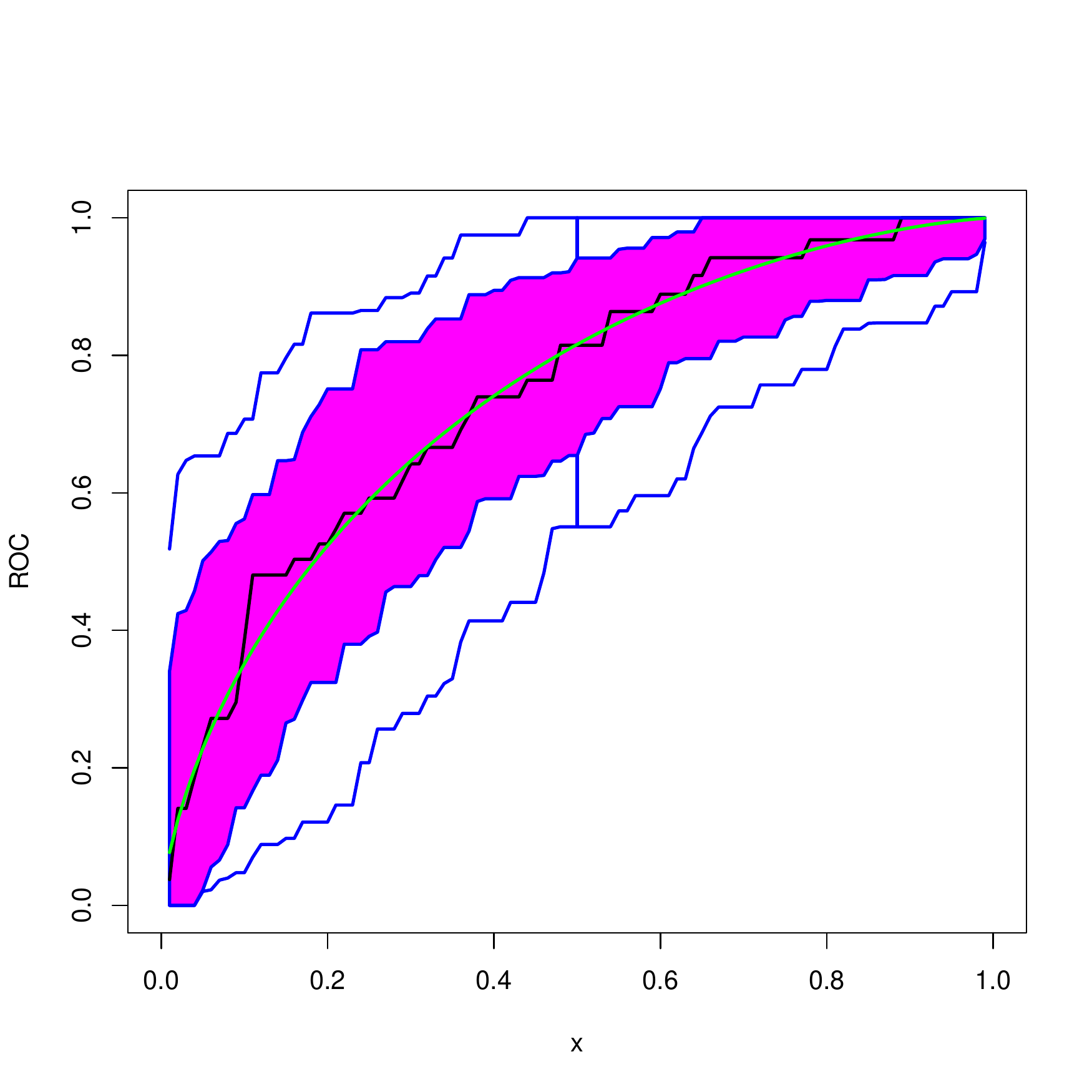} &
\includegraphics[scale=0.25]{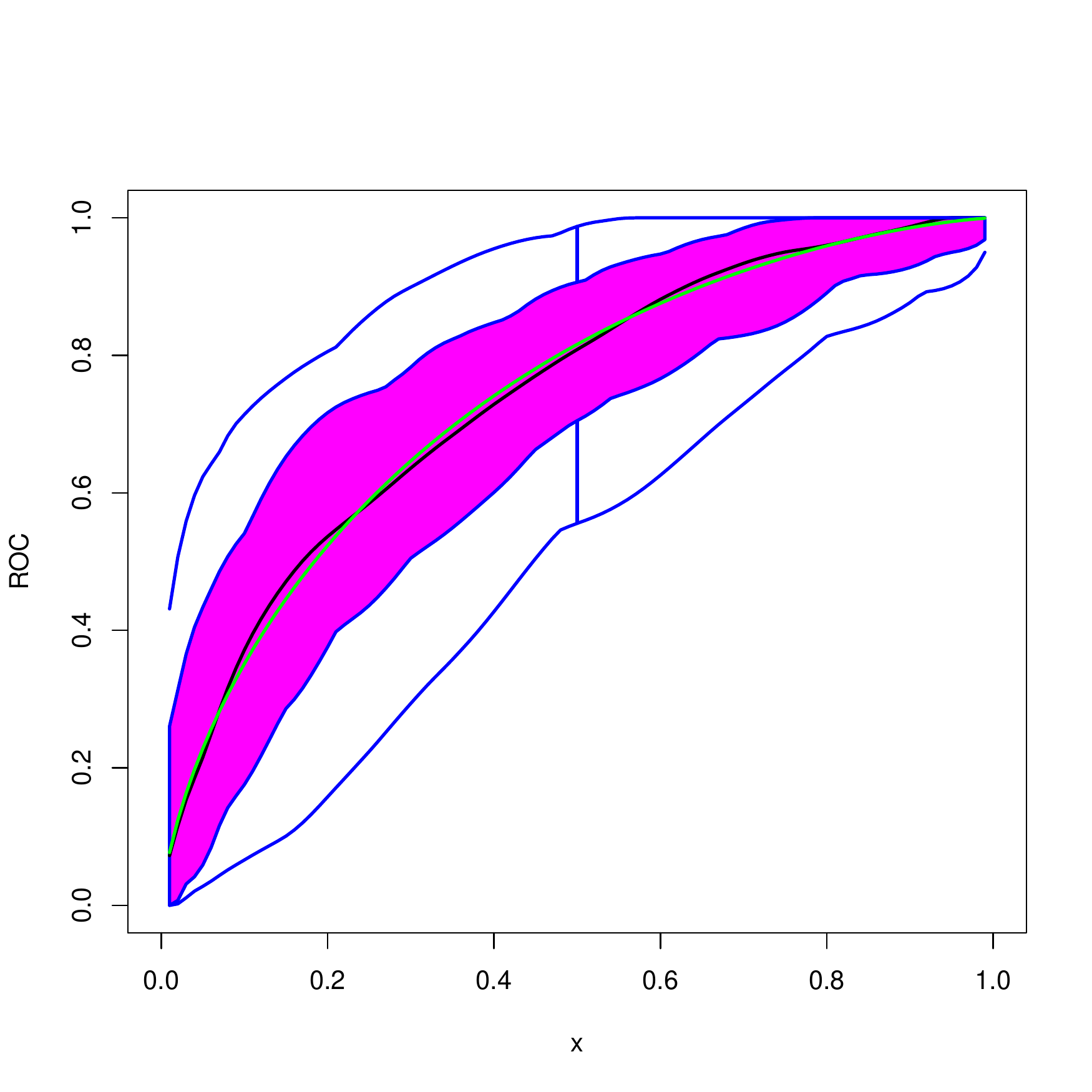} &
\includegraphics[scale=0.25]{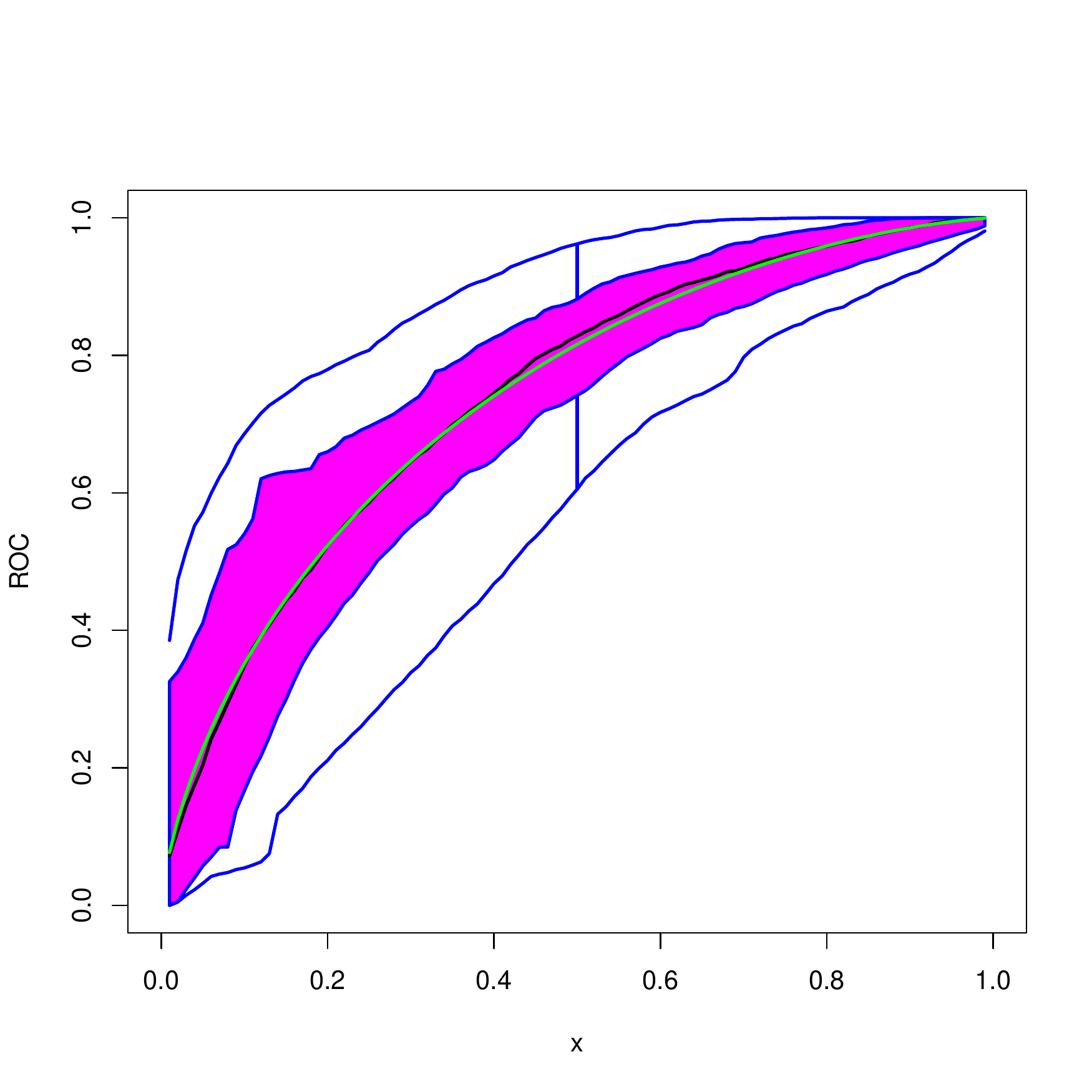}\\
 $\wpi_D=\pi_{\log}$, $\wpi_H=\pi_{\log}$ & 
\includegraphics[scale=0.25]{ROC_IPW_n100_H_plogit_D_plogitlineallineal_mis_H_3_mis_D_3.pdf} &   
\includegraphics[scale=0.25]{ROC_KER_n100_H_plogit_D_plogit_mis_H_3_mis_D_3_dim1.pdf}   &  
\includegraphics[scale=0.25]{ROC_MUL_n100_H_plogit_D_plogitlineallineal_mis_H_3_mis_D_3.pdf}
   \end{tabular}
\caption{\label{fig:fbx_21_12}\small Functional boxplots of  $\wROC(p)$. The green line corresponds to the true $ROC(p)$ and the dotted red lines to the outlying curves detected by the functional boxplot. }
\end{center} 
\end{figure}
\normalsize


\begin{figure}[ht!]
 \begin{center}
 \footnotesize
 \renewcommand{\arraystretch}{0.4}
 \newcolumntype{M}{>{\centering\arraybackslash}m{\dimexpr.12 \linewidth-1\tabcolsep}}
   \newcolumntype{G}{>{\centering\arraybackslash}m{\dimexpr.24\linewidth-1\tabcolsep}}
\begin{tabular}{M G G G G}\\
& $\wROC_{\ipw}$ & $\wROC_{\kernel}$ & $\wROC_{\ipw}$ & $\wROC_{\kernel}$\\[0.1in]
& \multicolumn{2}{c}{$\pi_{H}=\pi_{\itM_1}$, $\pi_{D}=\pi_{\itM_1}$} & \multicolumn{2}{c}{$\pi_{H}=\pi_{\itM_2}$, $\pi_{D}=\pi_{\itM_2}$}\\
$\wpi_D=\pi_c$, $\wpi_H=\pi_c$   & 
\includegraphics[scale=0.23]{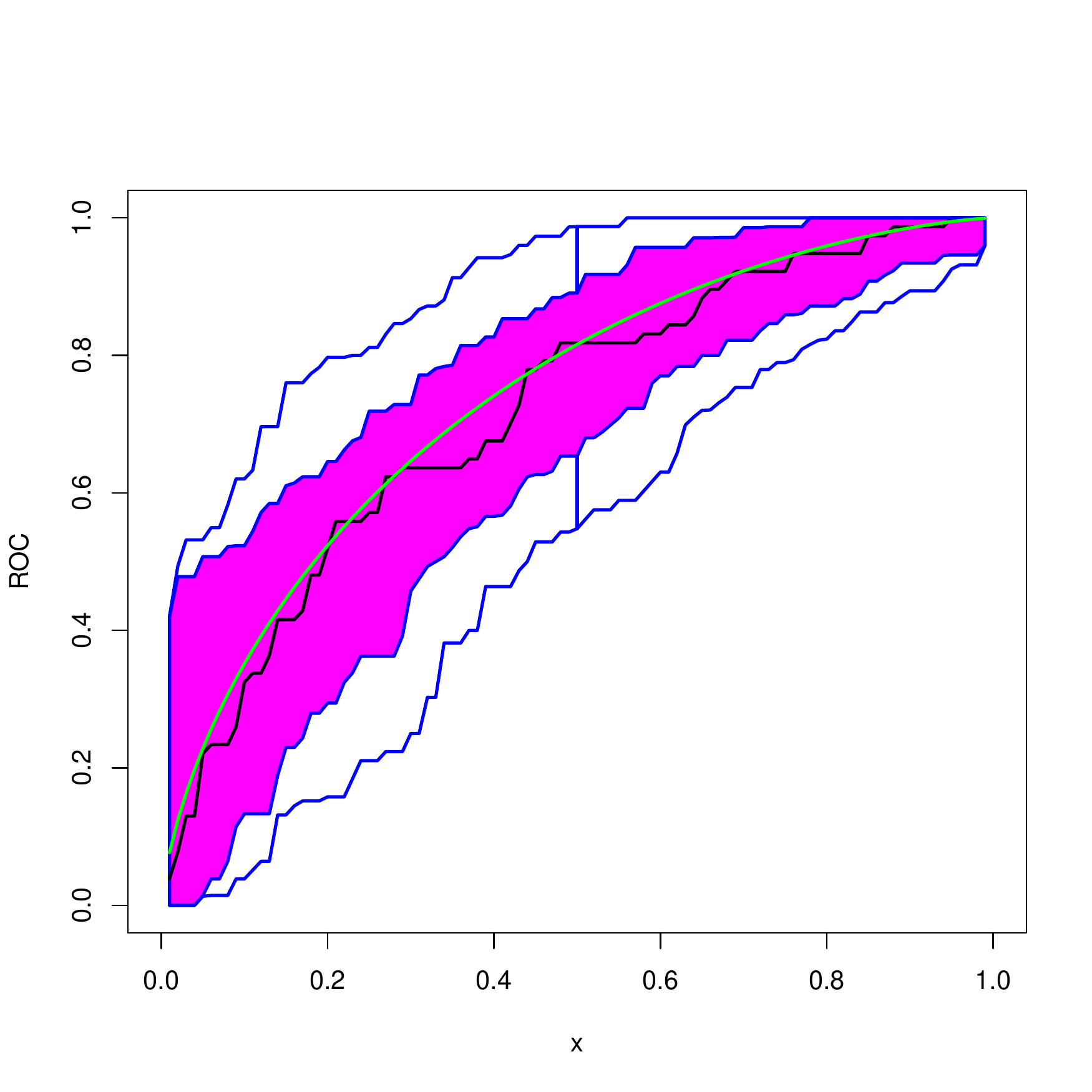} &
\includegraphics[scale=0.23]{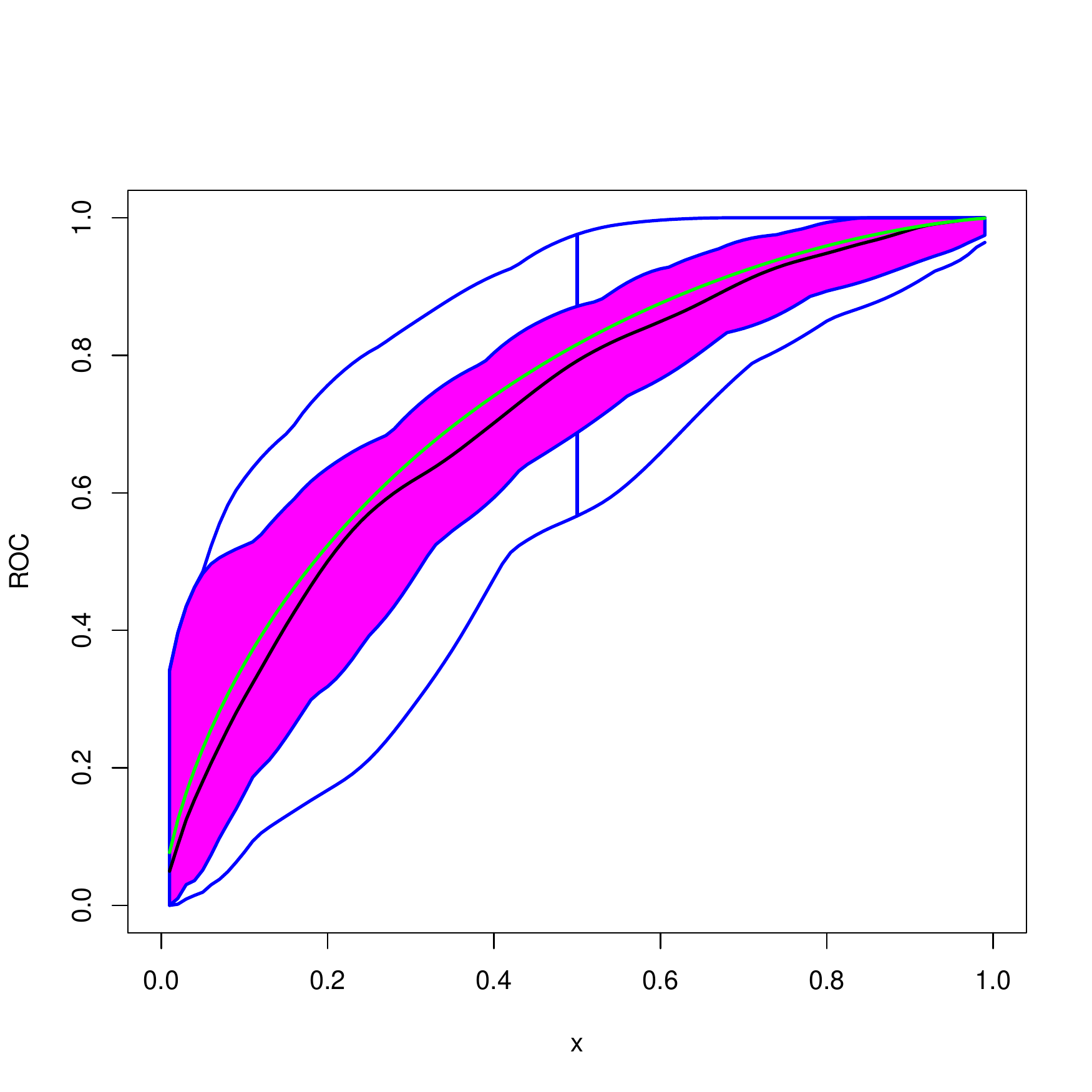} & 
\includegraphics[scale=0.23]{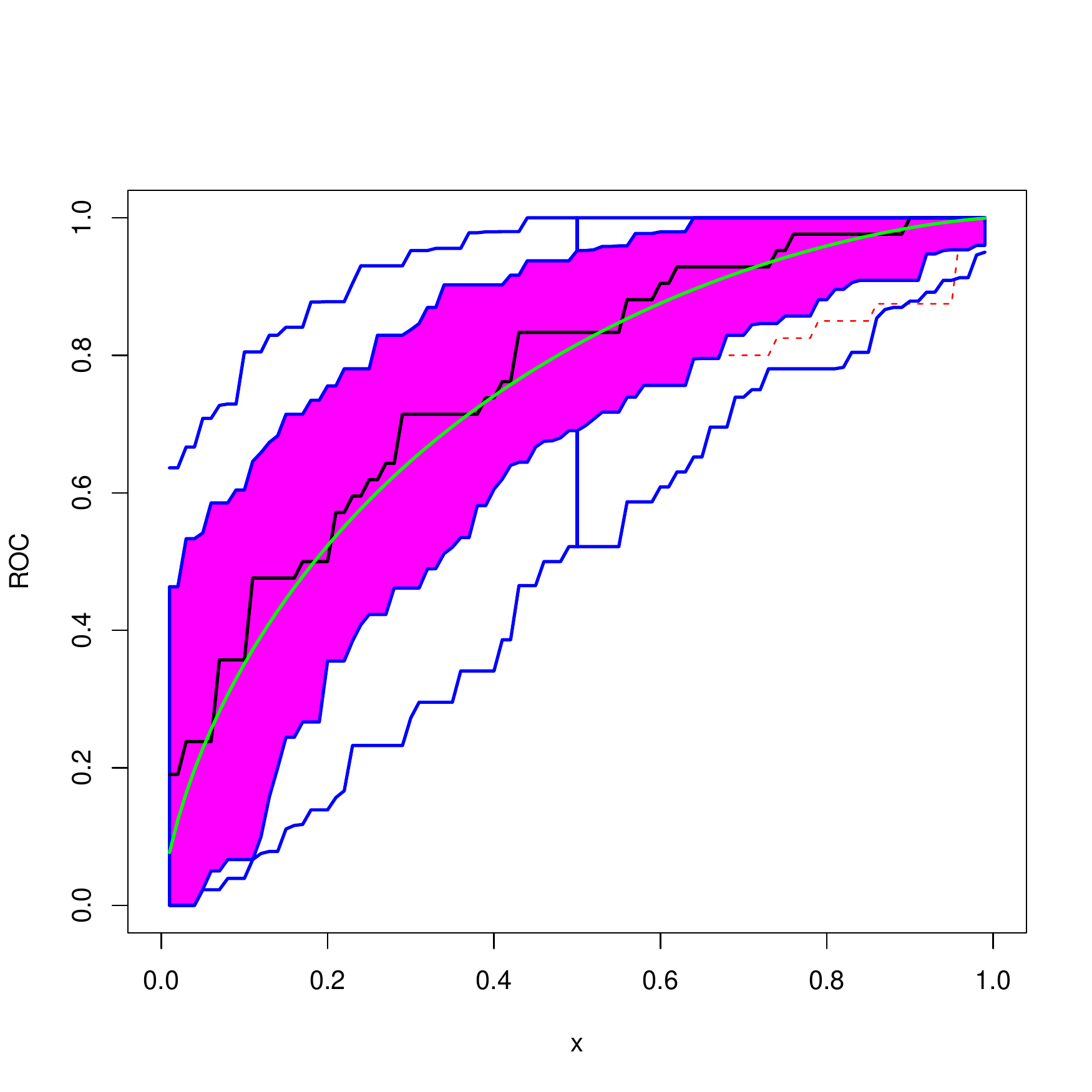} &
\includegraphics[scale=0.23]{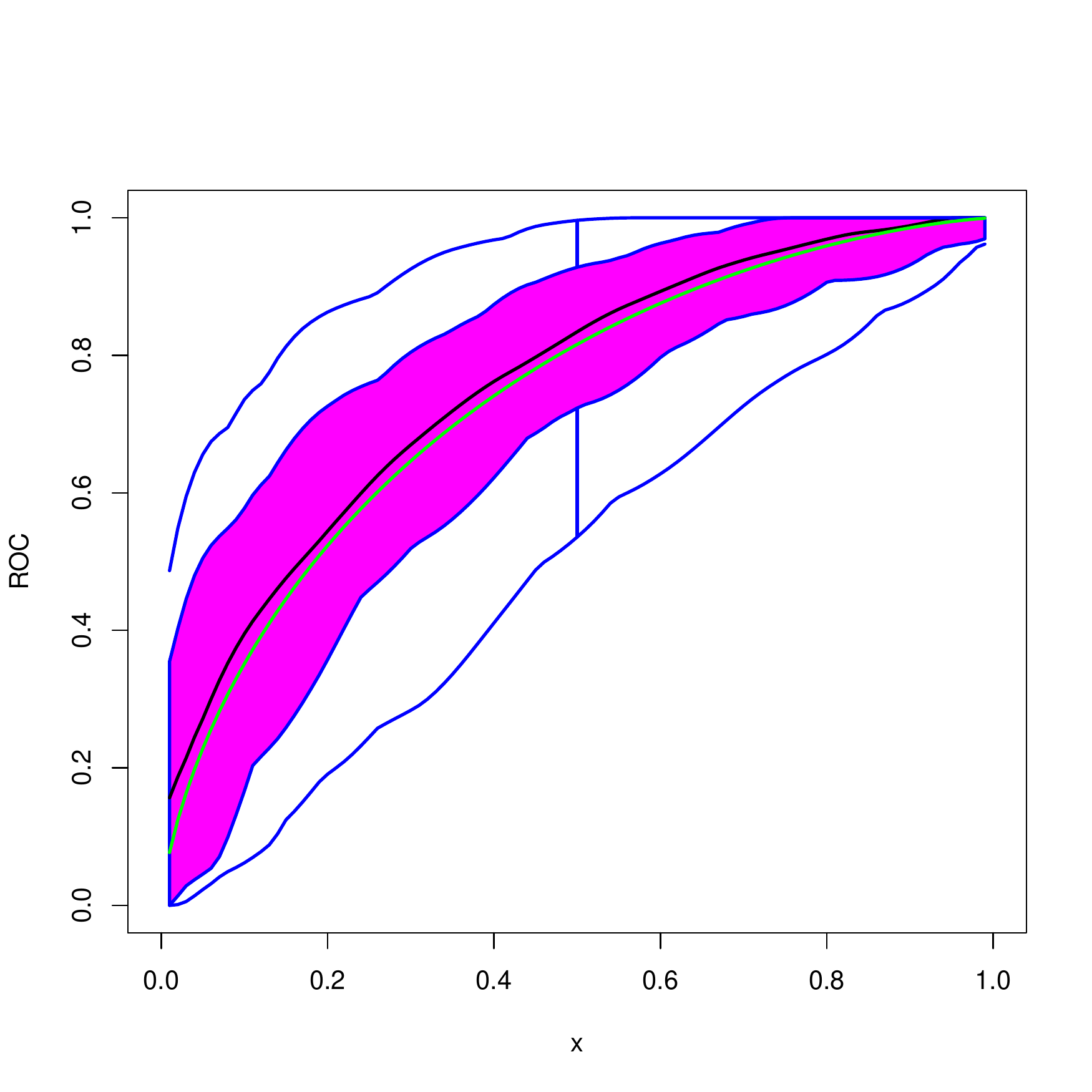} \\
 $\wpi_D=\pi_{c}$, $\wpi_H=\pi_{\log}$ & 
\includegraphics[scale=0.23]{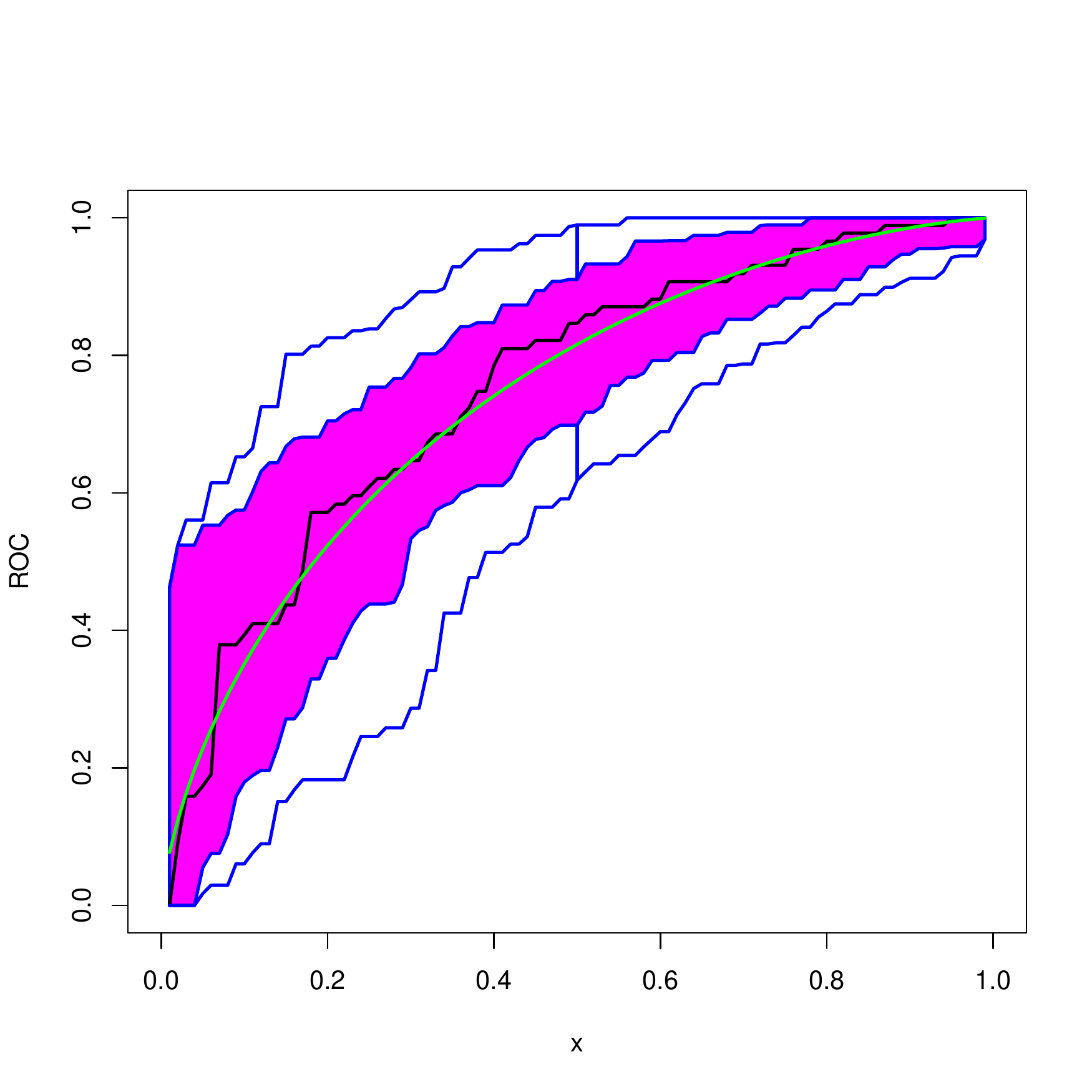} &   
\includegraphics[scale=0.23]{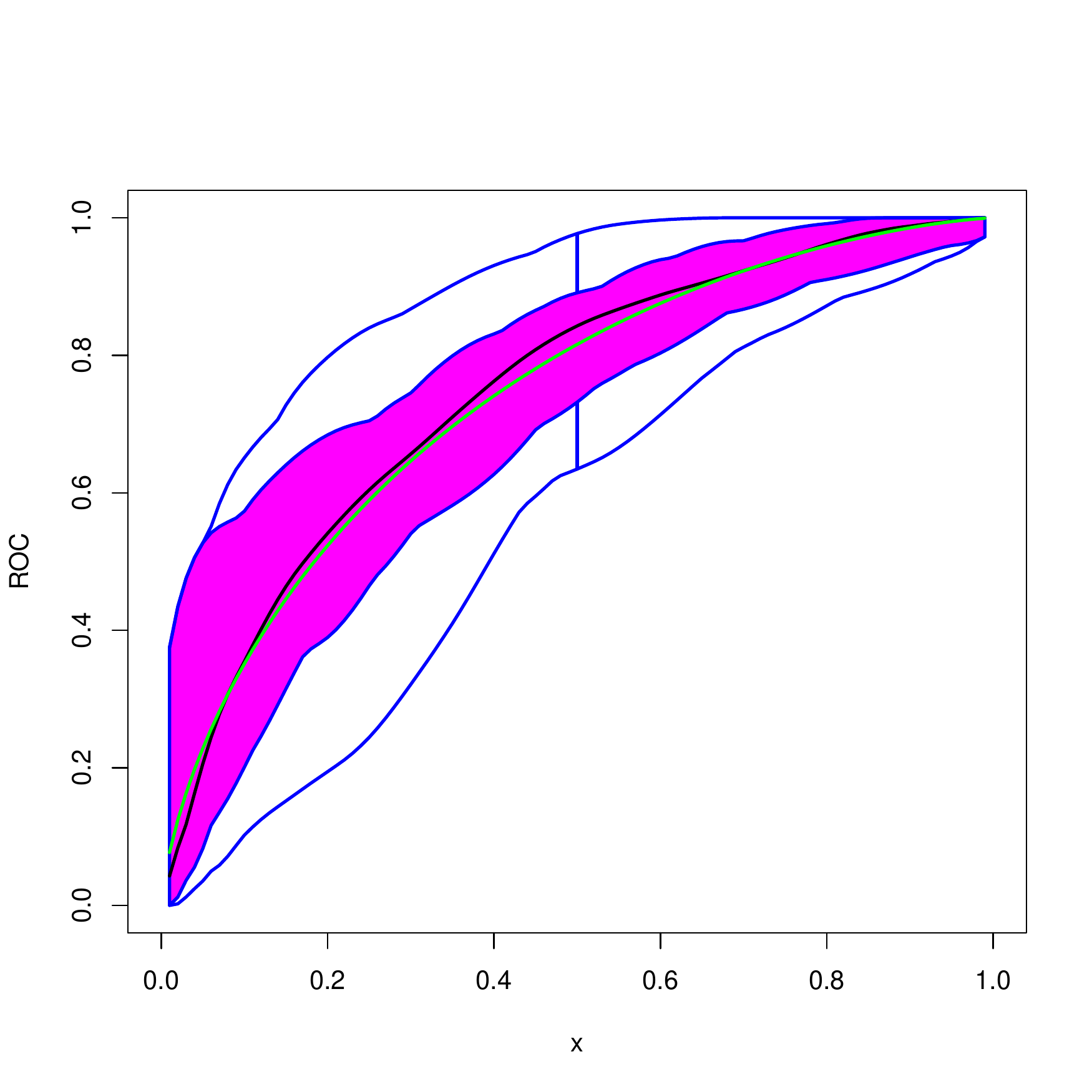}   &  
\includegraphics[scale=0.23]{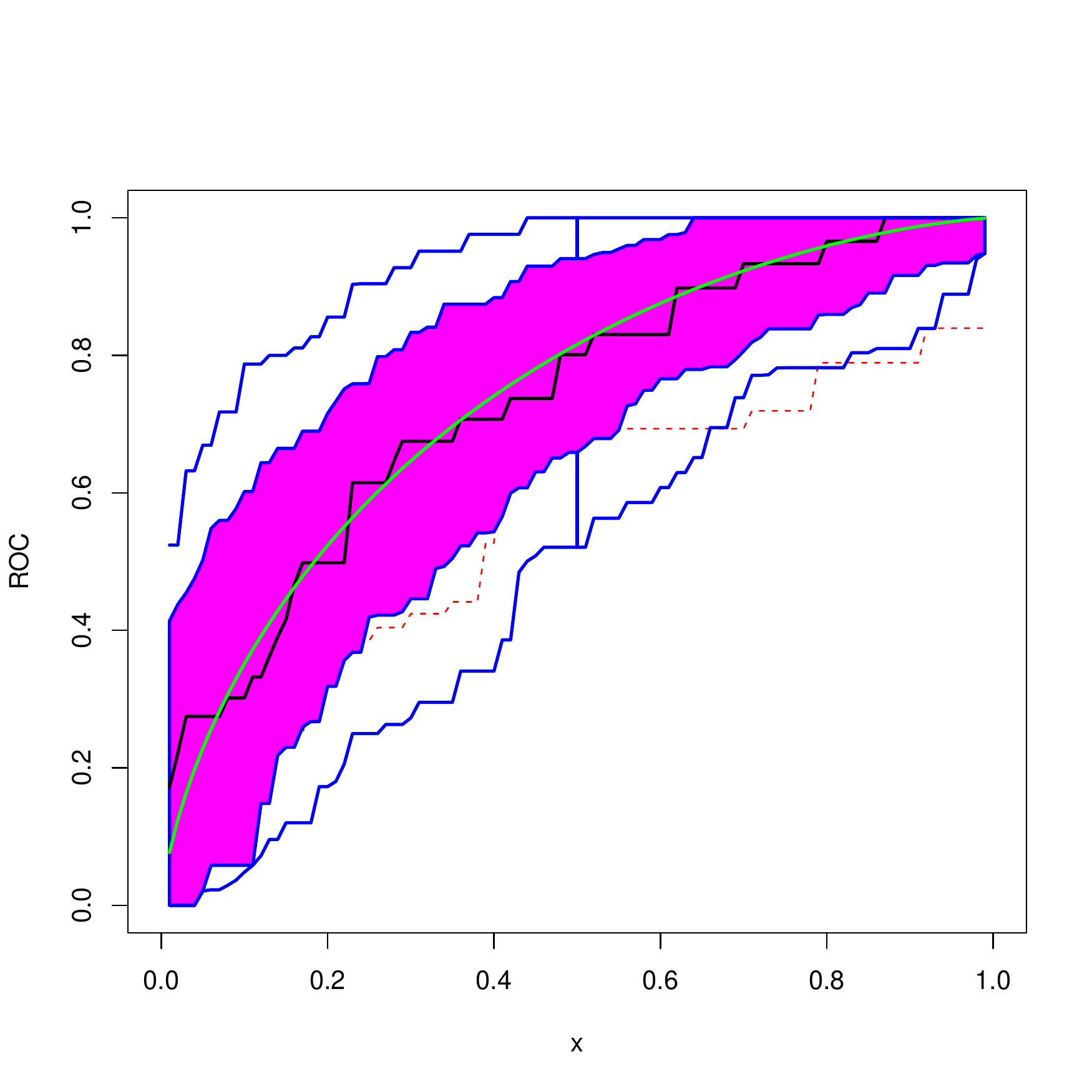} &   
\includegraphics[scale=0.23]{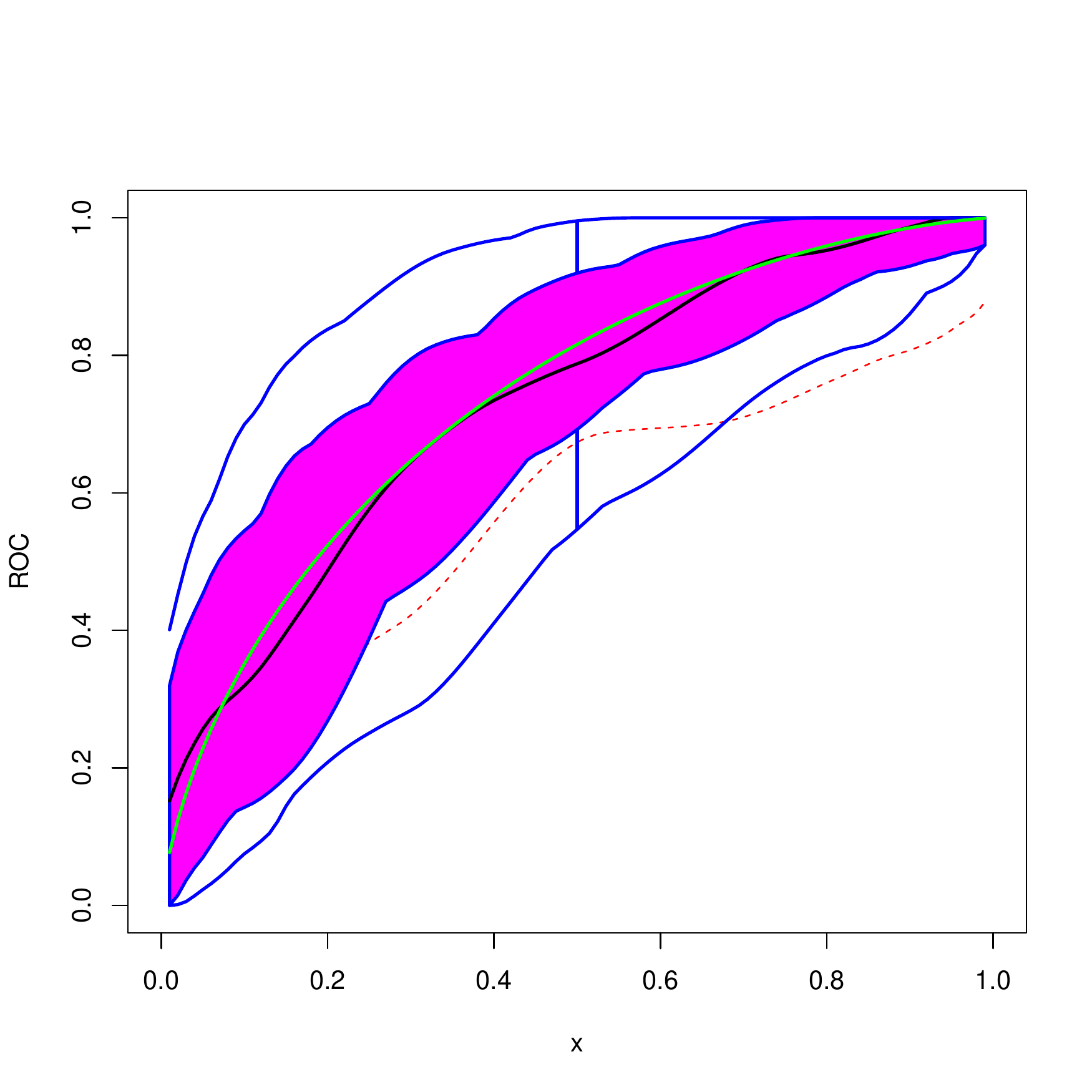}    \\
& \multicolumn{2}{c}{$\pi_{H}=\pi_{\itM_2}$, $\pi_{D}=\pi_{\itM_1}$} & \multicolumn{2}{c}{$\pi_{H}=\pi_{\itM_1}$, $\pi_{D}=\pi_{\itM_2}$}\\
$\wpi_D=\pi_c$, $\wpi_H=\pi_c$   & 
\includegraphics[scale=0.23]{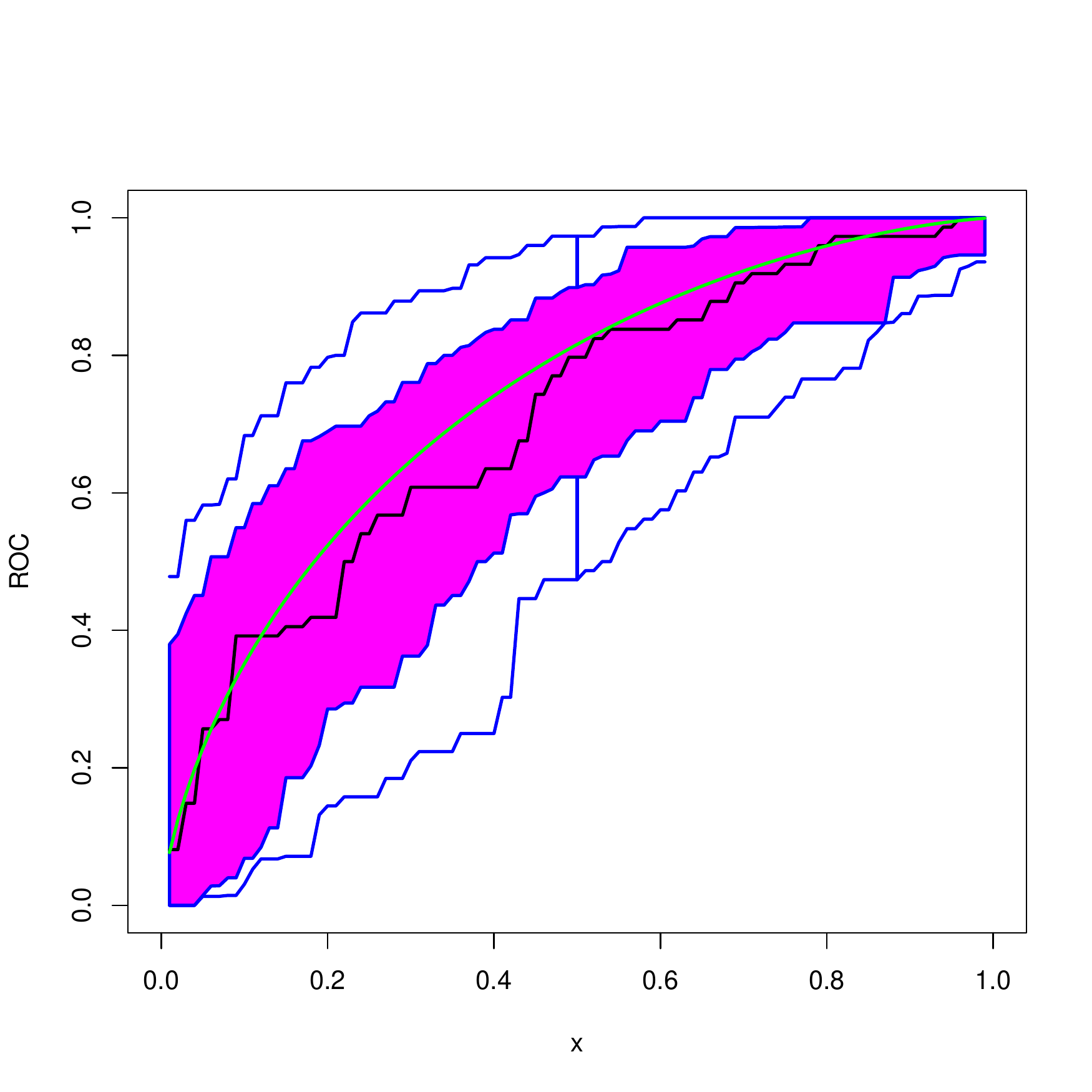} &
\includegraphics[scale=0.23]{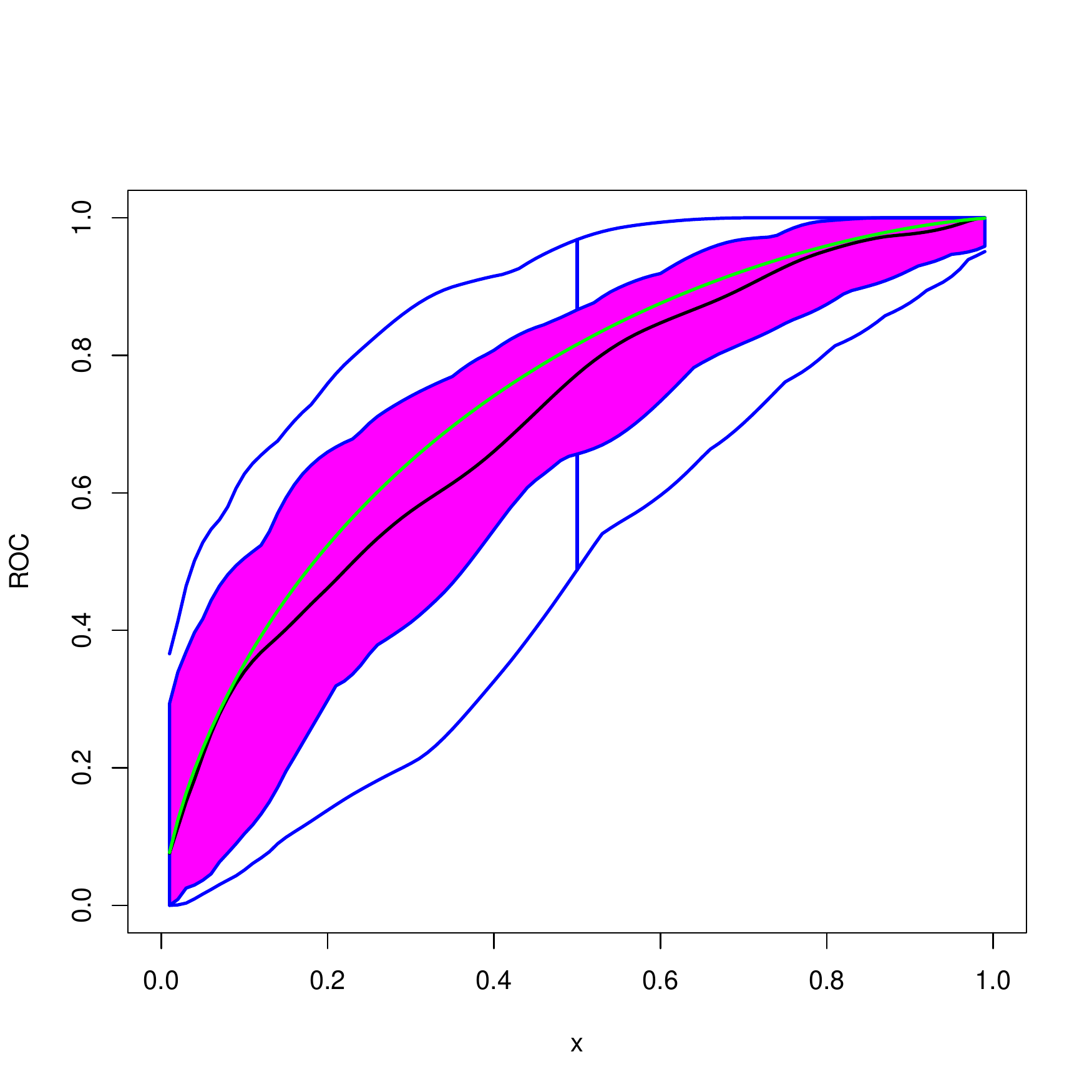} &
\includegraphics[scale=0.23]{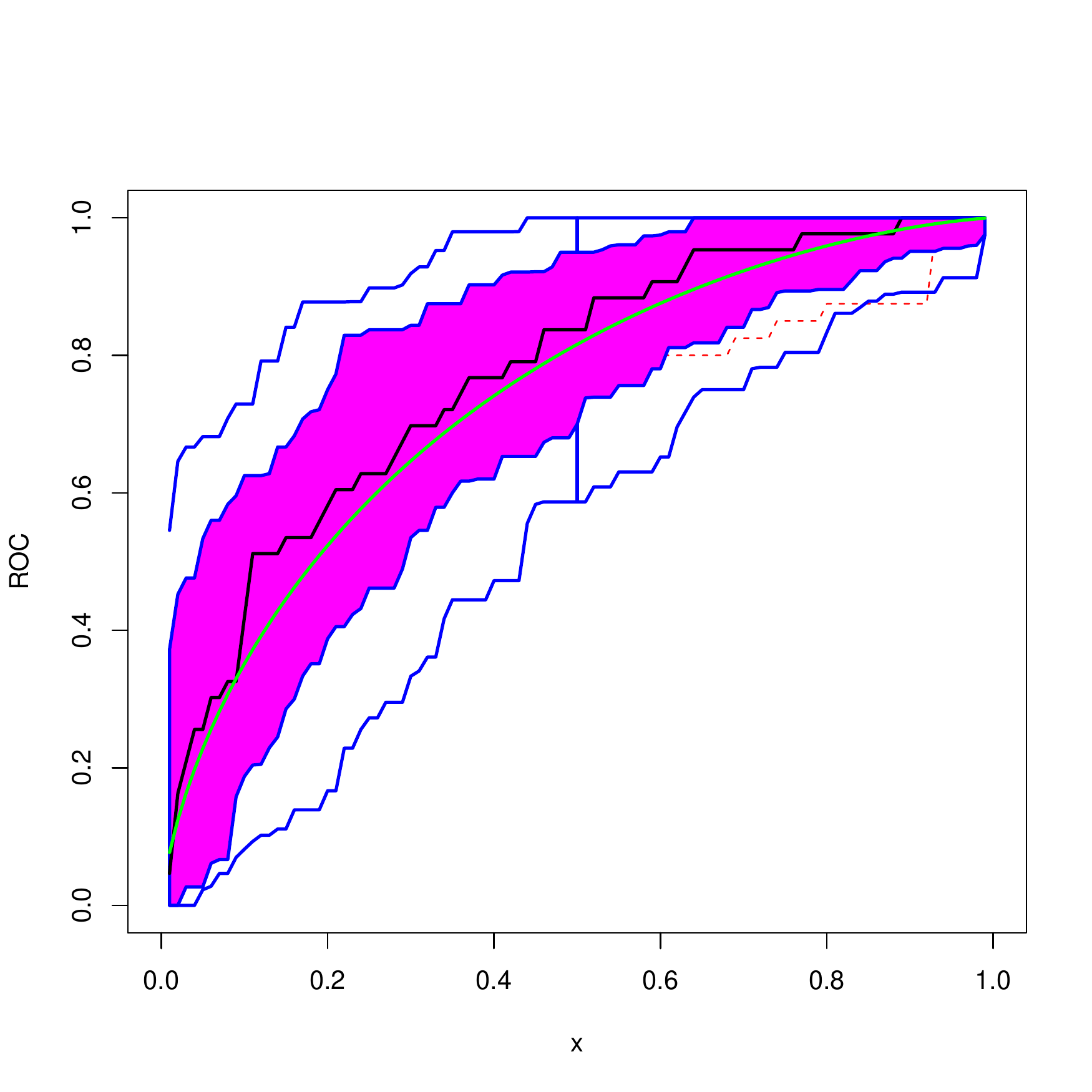} &
\includegraphics[scale=0.23]{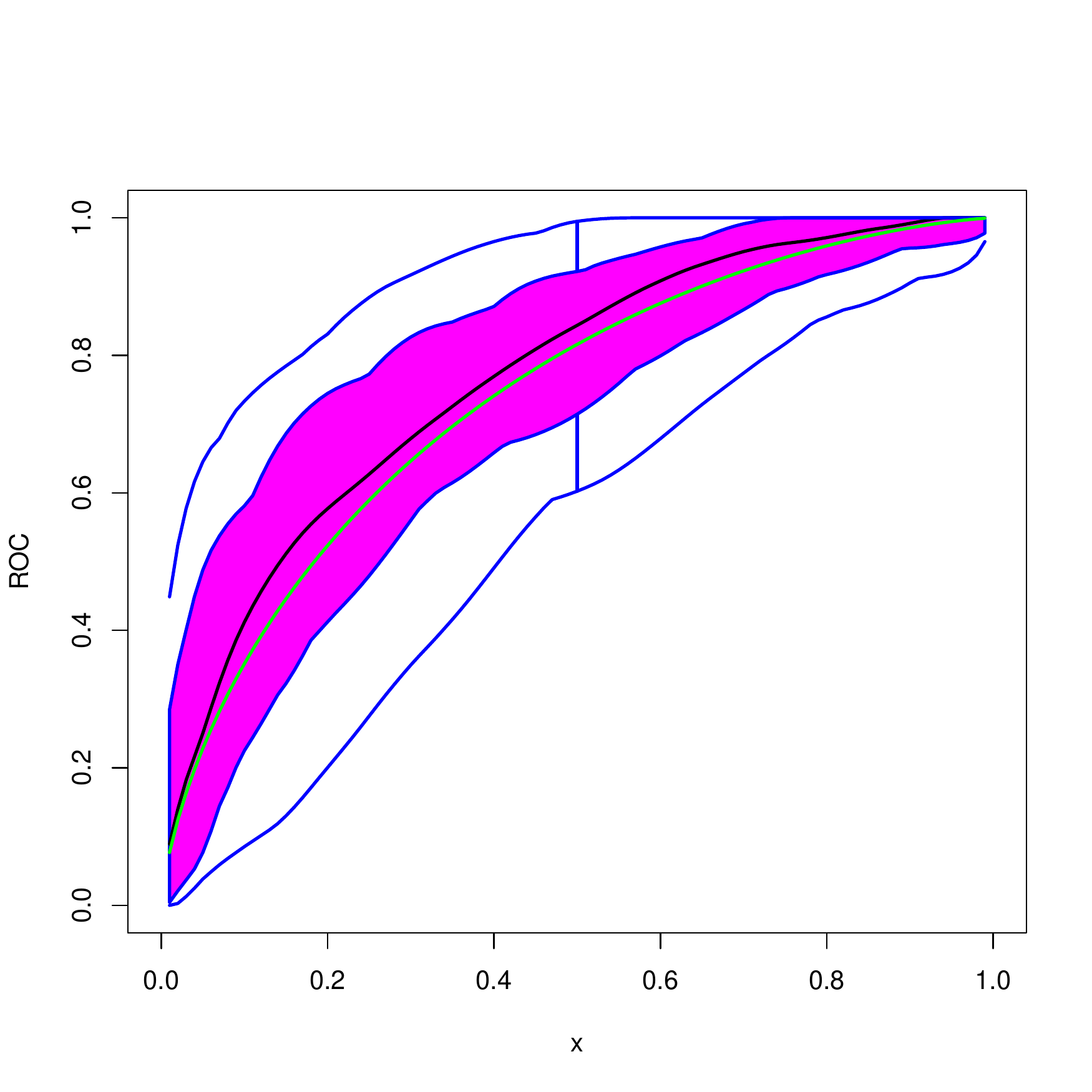}
\\

 $\wpi_D=\pi_{c}$, $\wpi_H=\pi_{\log}$ & 
\includegraphics[scale=0.23]{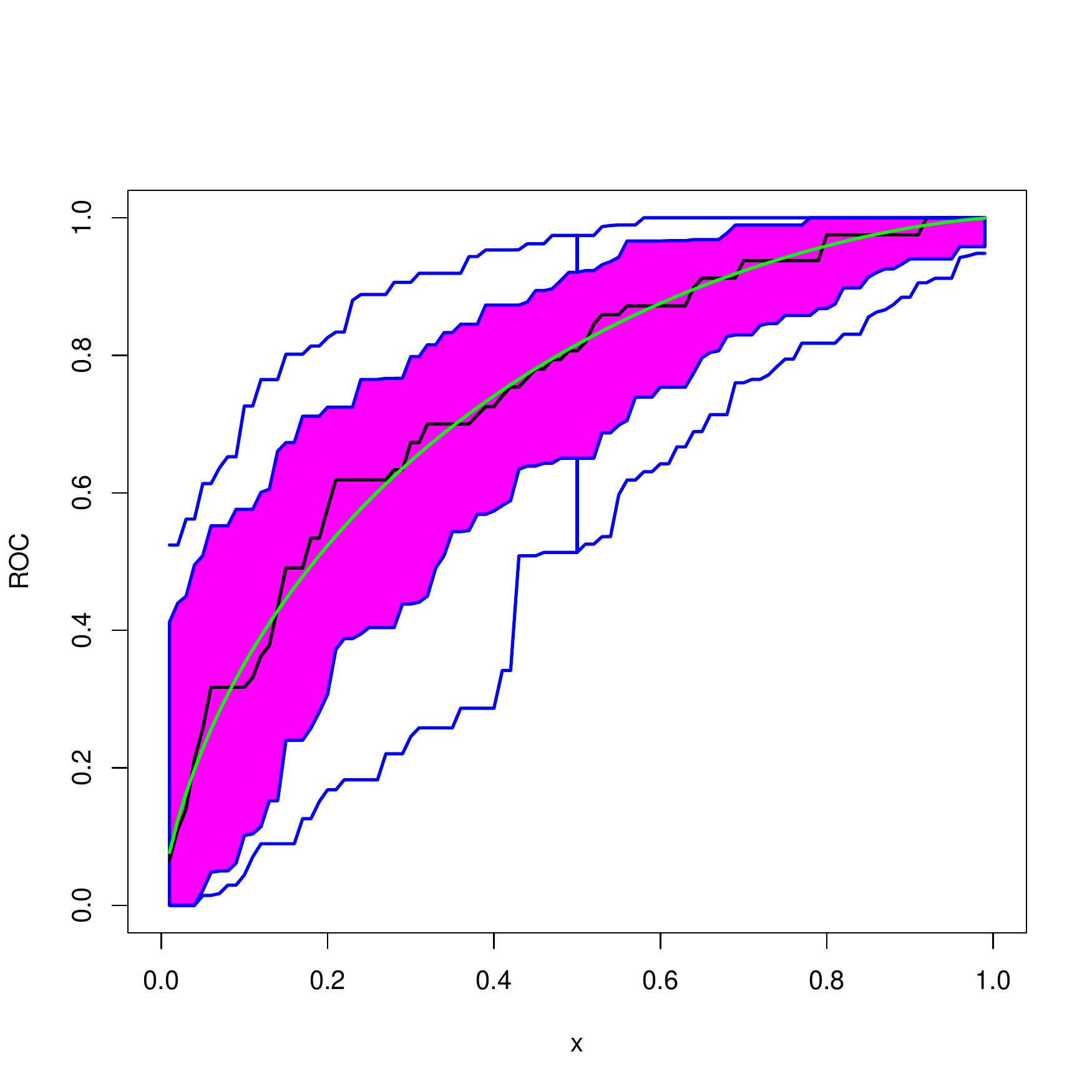} &   
\includegraphics[scale=0.23]{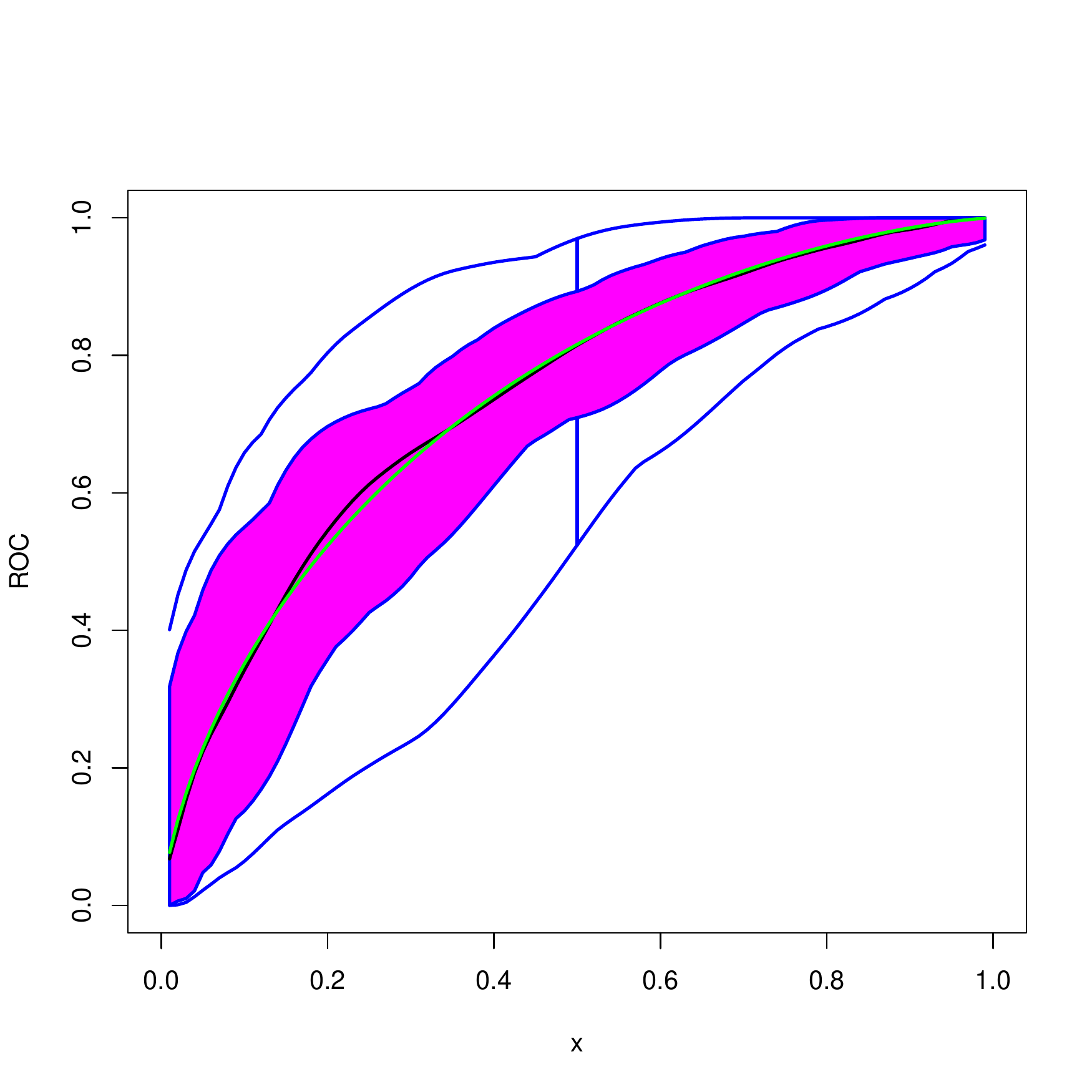}   & 
\includegraphics[scale=0.23]{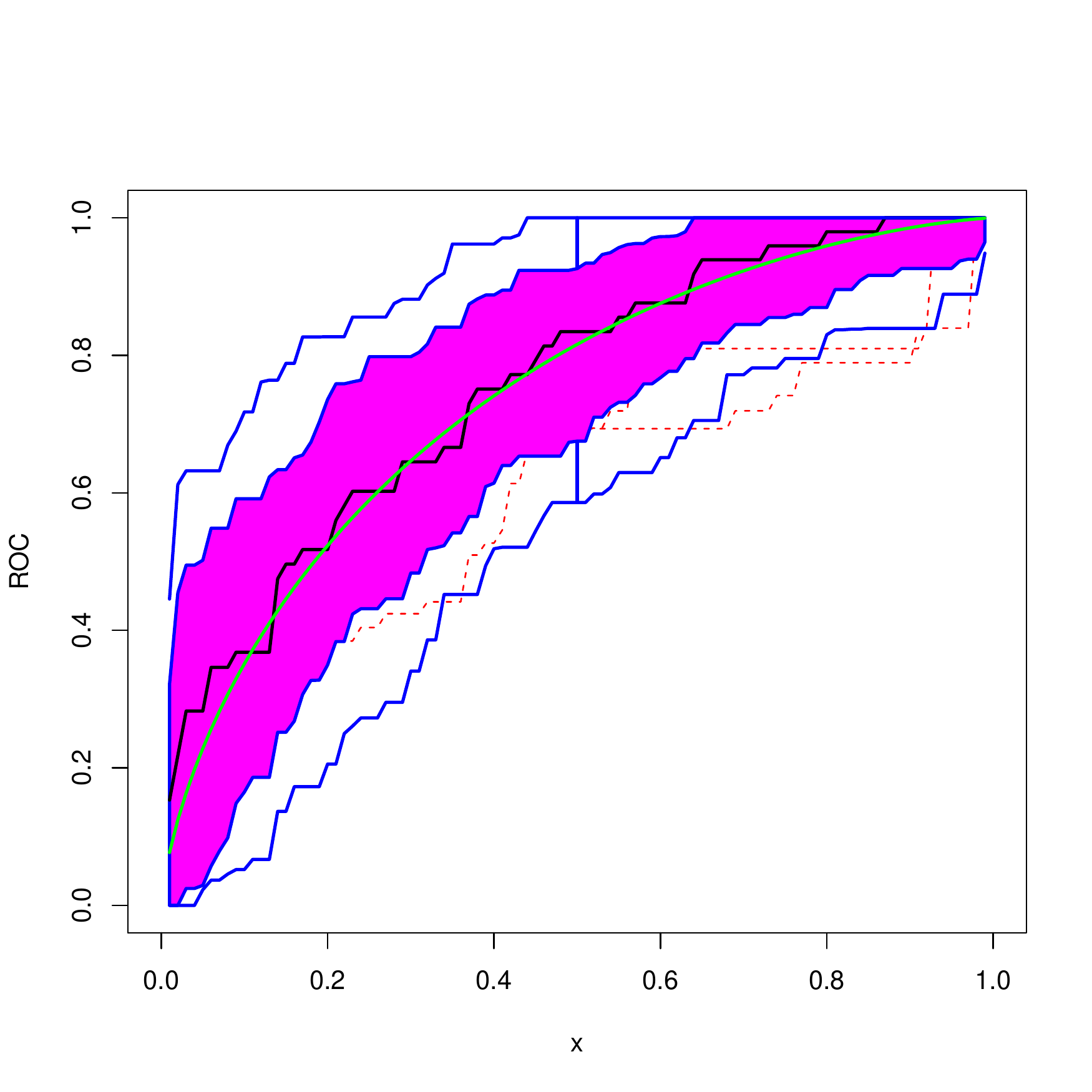} &   
\includegraphics[scale=0.23]{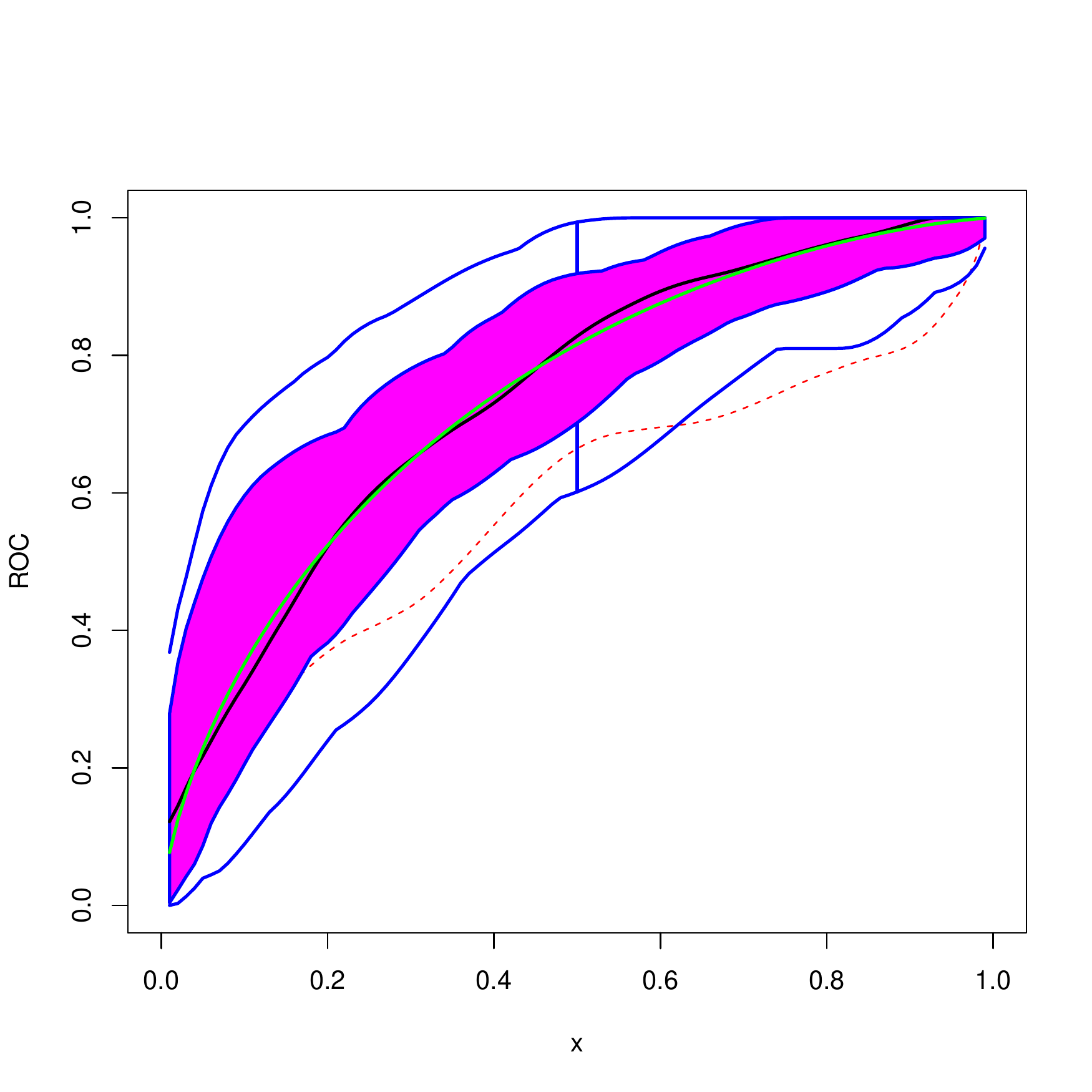}   
   \end{tabular}
\vskip-0.1in  
\caption{\label{fig:fbx_mal_11_22}\small Functional boxplots of  $\wROC(p)$   under misspecification of the propensity. The green line corresponds to the true $ROC(p)$ and the dotted red lines to the outlying curves detected by the functional boxplot. }
\end{center} 
\end{figure}
\normalsize

\begin{figure}[ht!]
 \begin{center}
 \footnotesize
 \renewcommand{\arraystretch}{0.4}
 \newcolumntype{M}{>{\centering\arraybackslash}m{\dimexpr.12 \linewidth-1\tabcolsep}}
   \newcolumntype{G}{>{\centering\arraybackslash}m{\dimexpr.24\linewidth-1\tabcolsep}}
\begin{tabular}{M G G G G}\\
&  $\pi_{H}=\pi_{\itM_1}$  & $\pi_{H}=\pi_{\itM_2}$ & 
$\pi_{H}=\pi_{\itM_2}$  &
$\pi_{H}=\pi_{\itM_1}$  \\
&   $\pi_{D}=\pi_{\itM_1}$ &   $\pi_{D}=\pi_{\itM_2}$ & 
 $\pi_{D}=\pi_{\itM_1}$ &
 $\pi_{D}=\pi_{\itM_2}$ \\

$\wpi_D=\pi$, $\wpi_H=\pi$   & 
\includegraphics[scale=0.23]{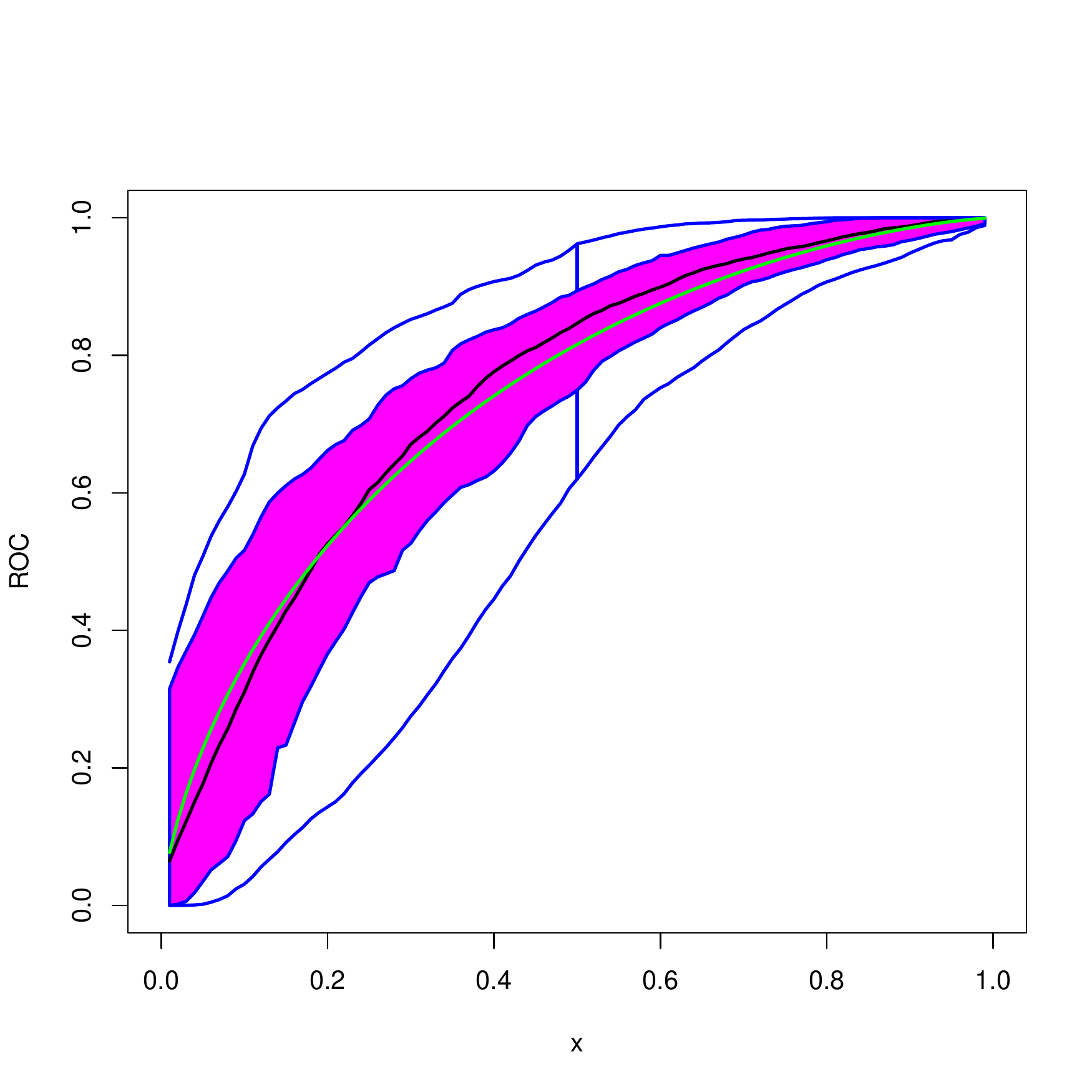} &
\includegraphics[scale=0.23]{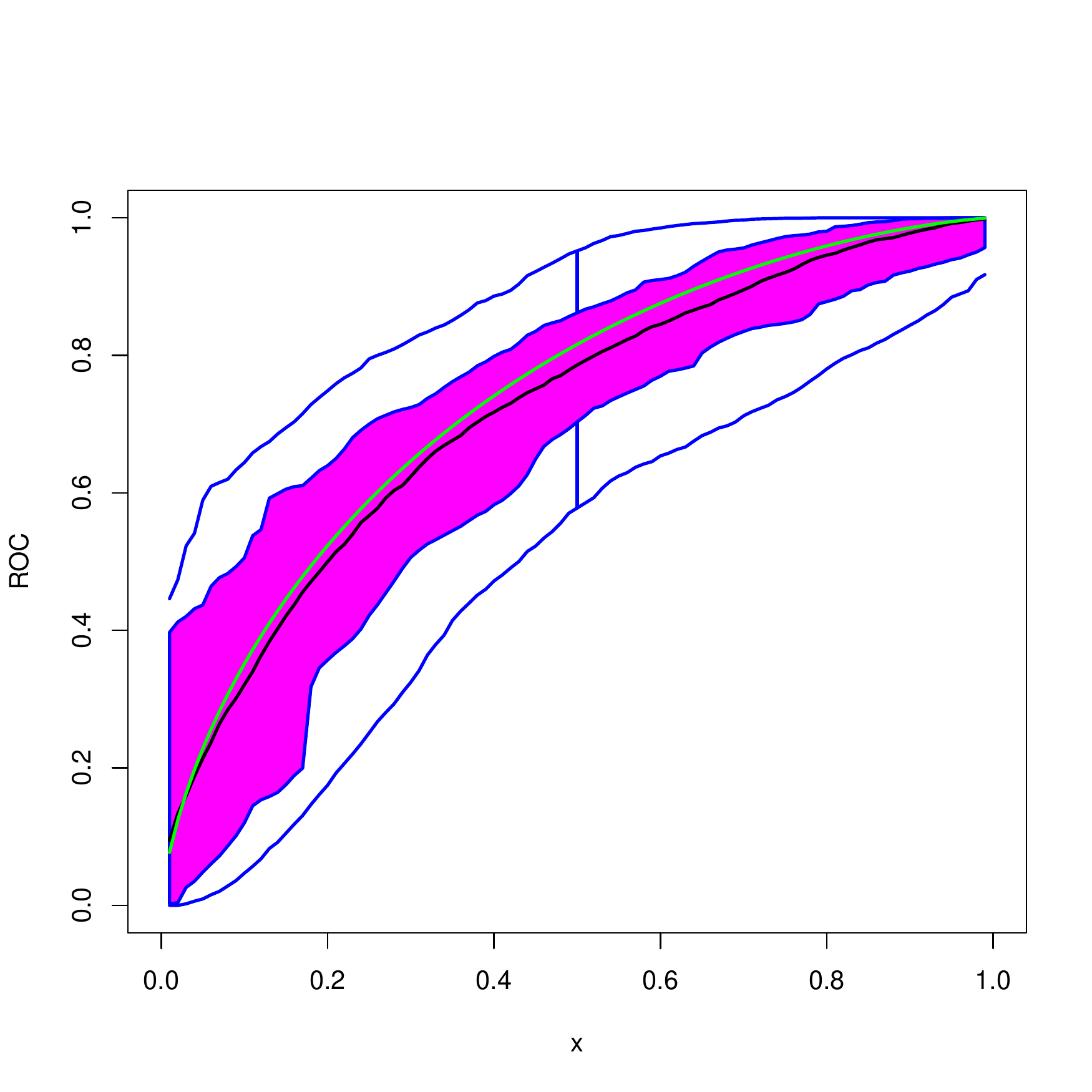} &
\includegraphics[scale=0.23]{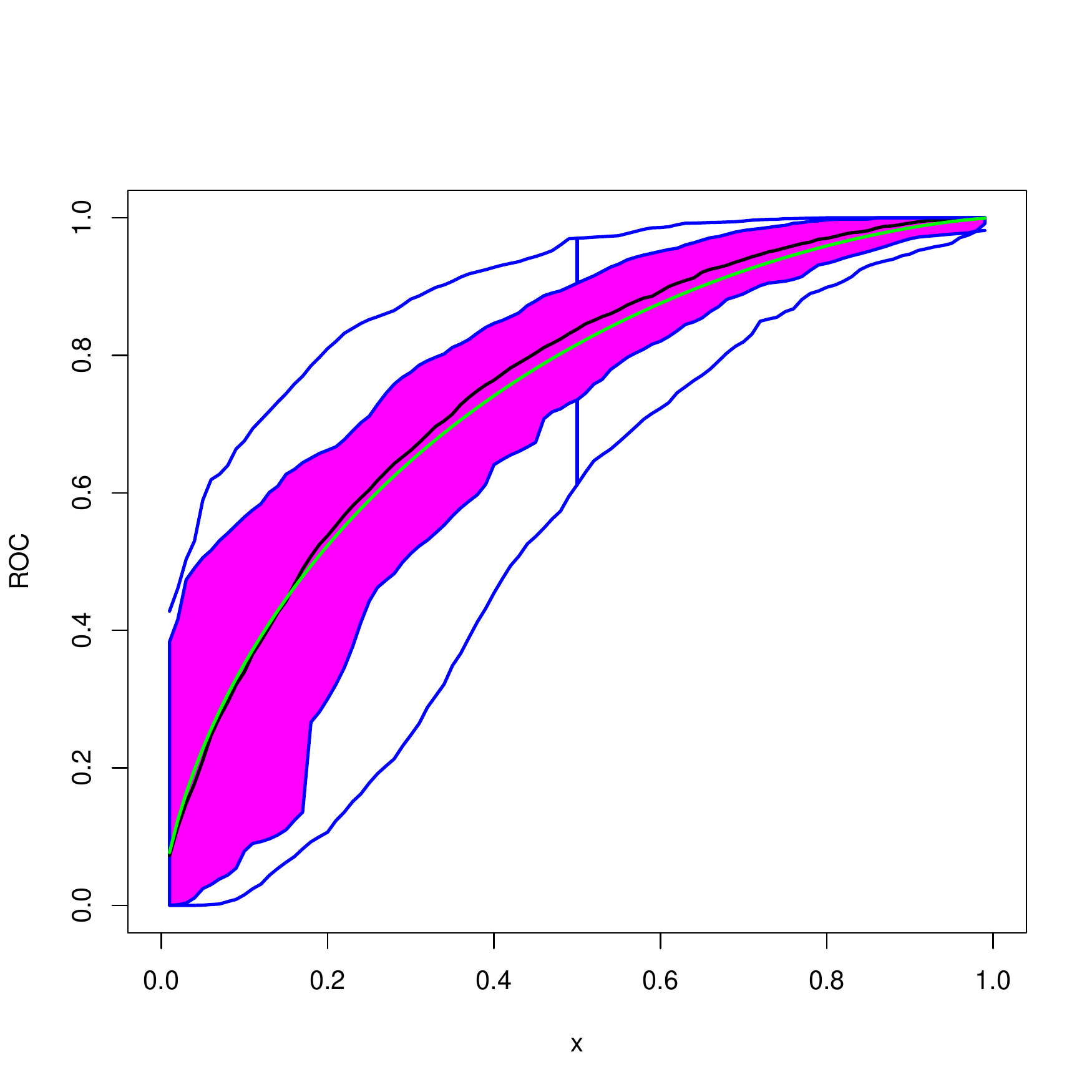} &
\includegraphics[scale=0.23]{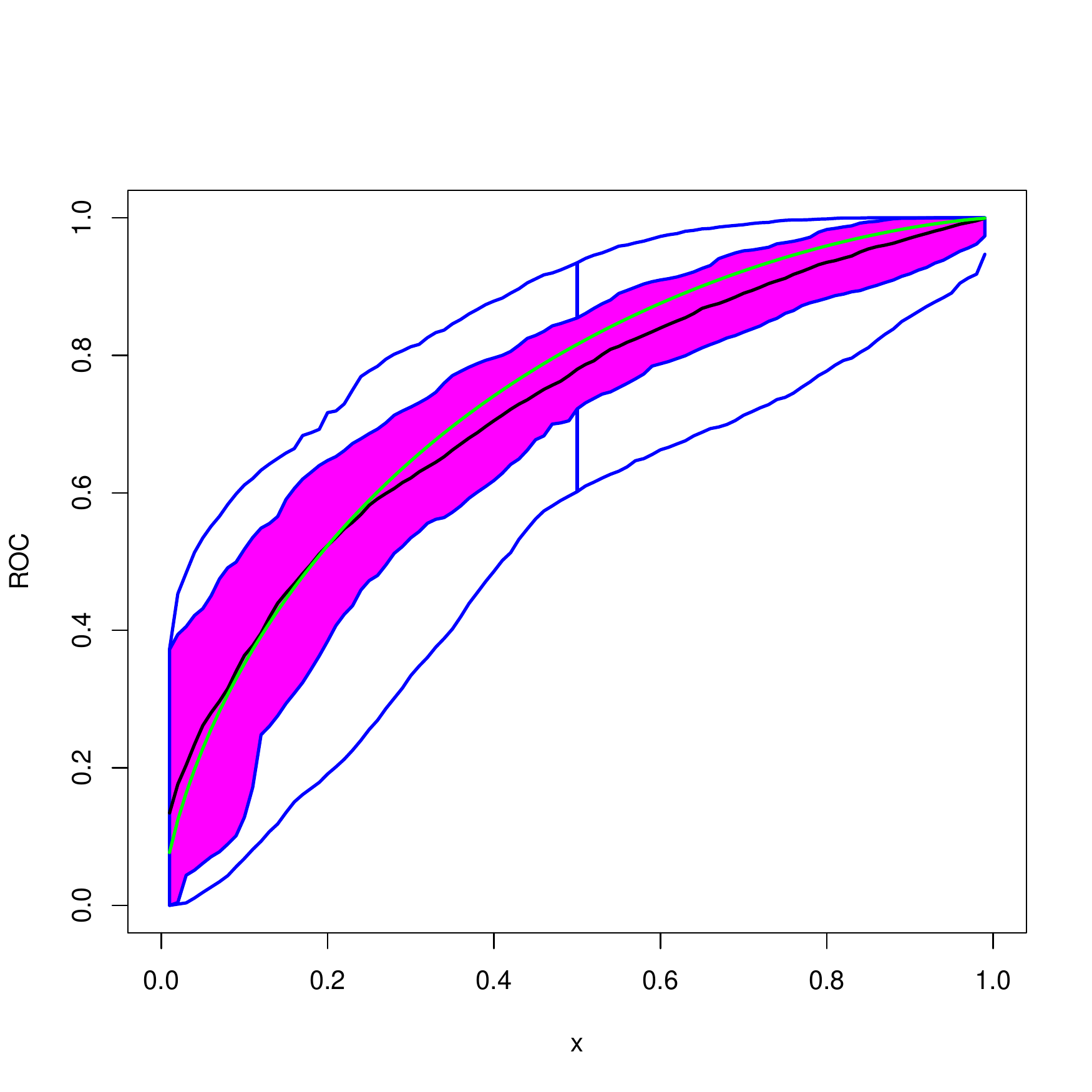}\\
\\
 $\wpi_D=\pi_{\log}$, $\wpi_H=\pi_{\log}$ & 
 
\includegraphics[scale=0.23]{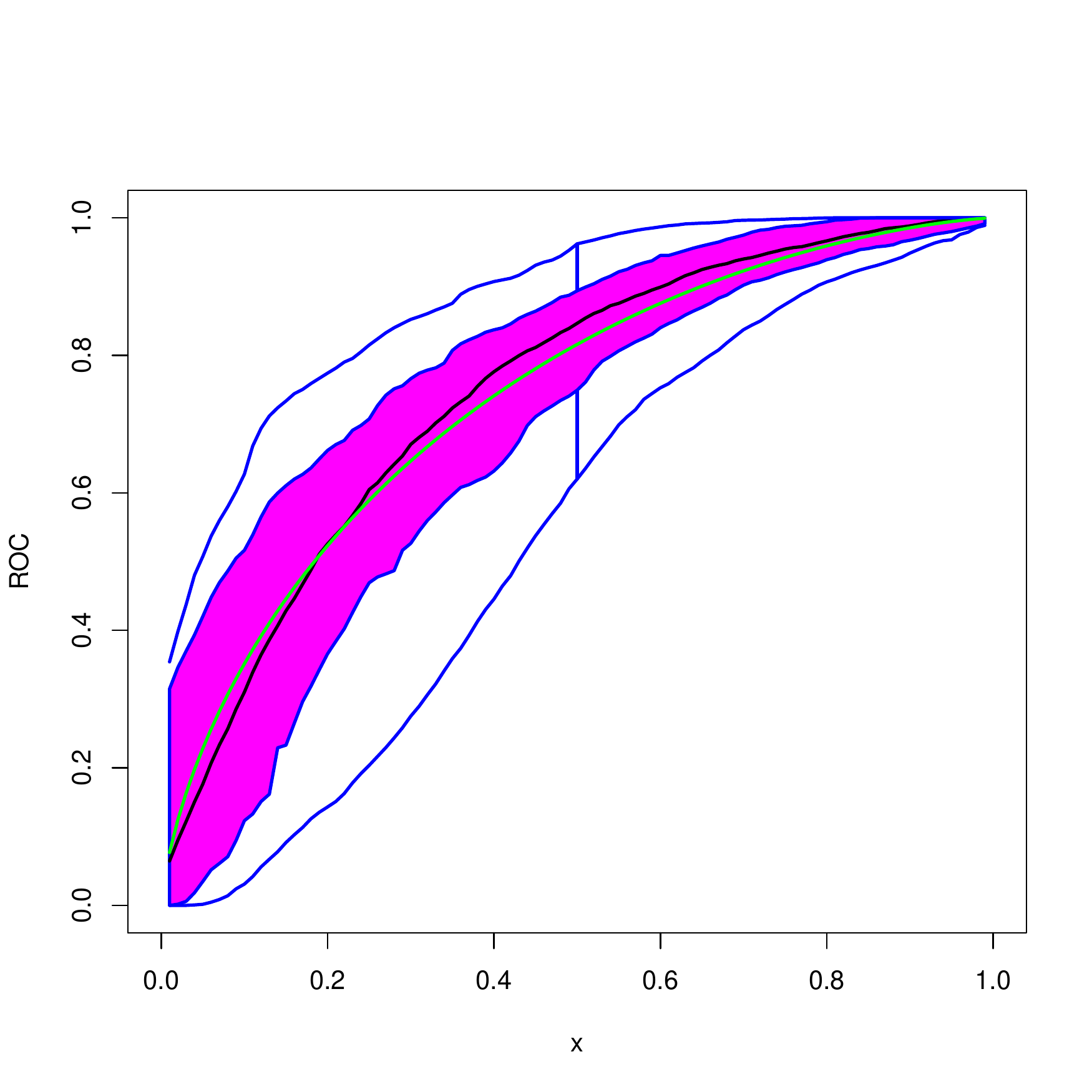} &
\includegraphics[scale=0.23]{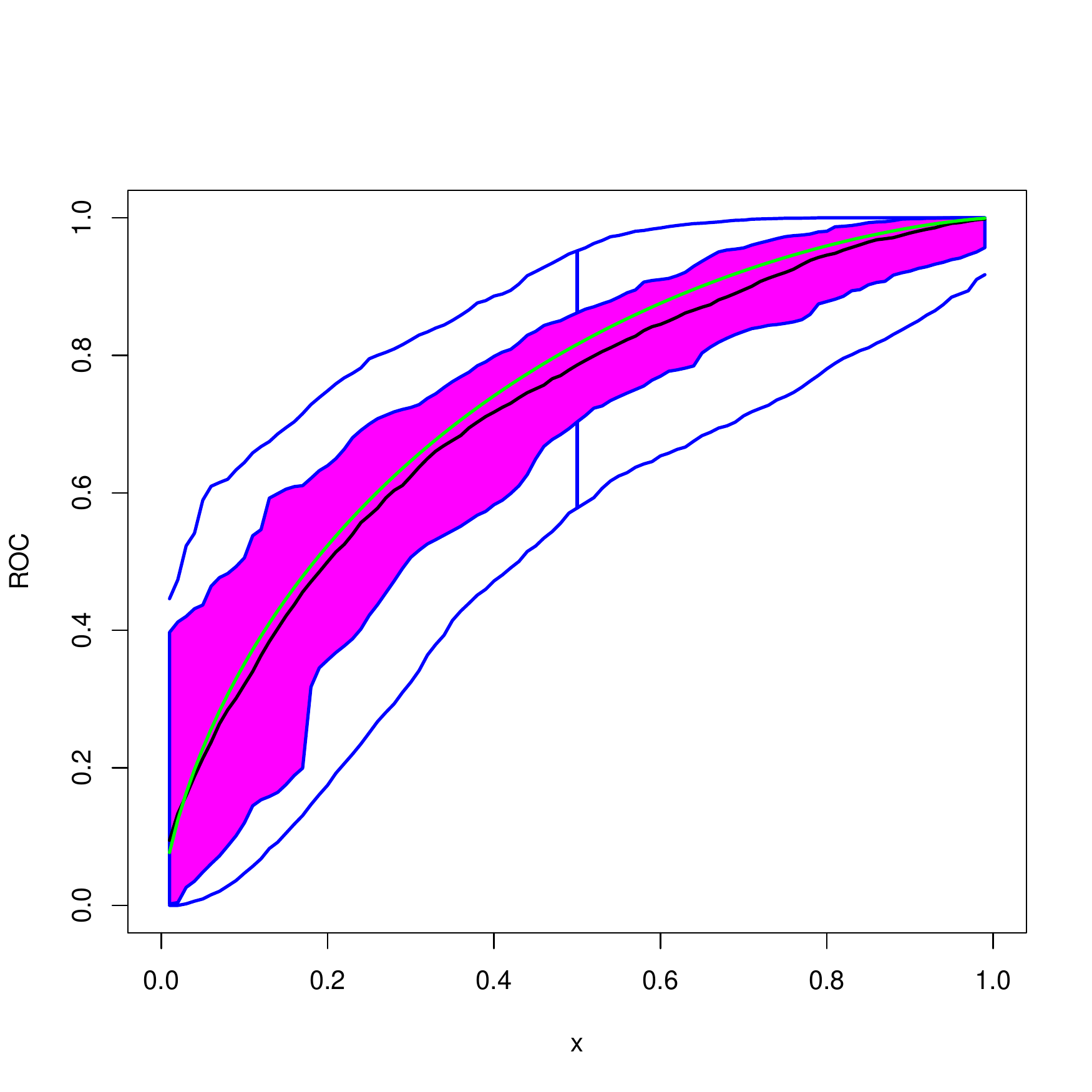} &
\includegraphics[scale=0.23]{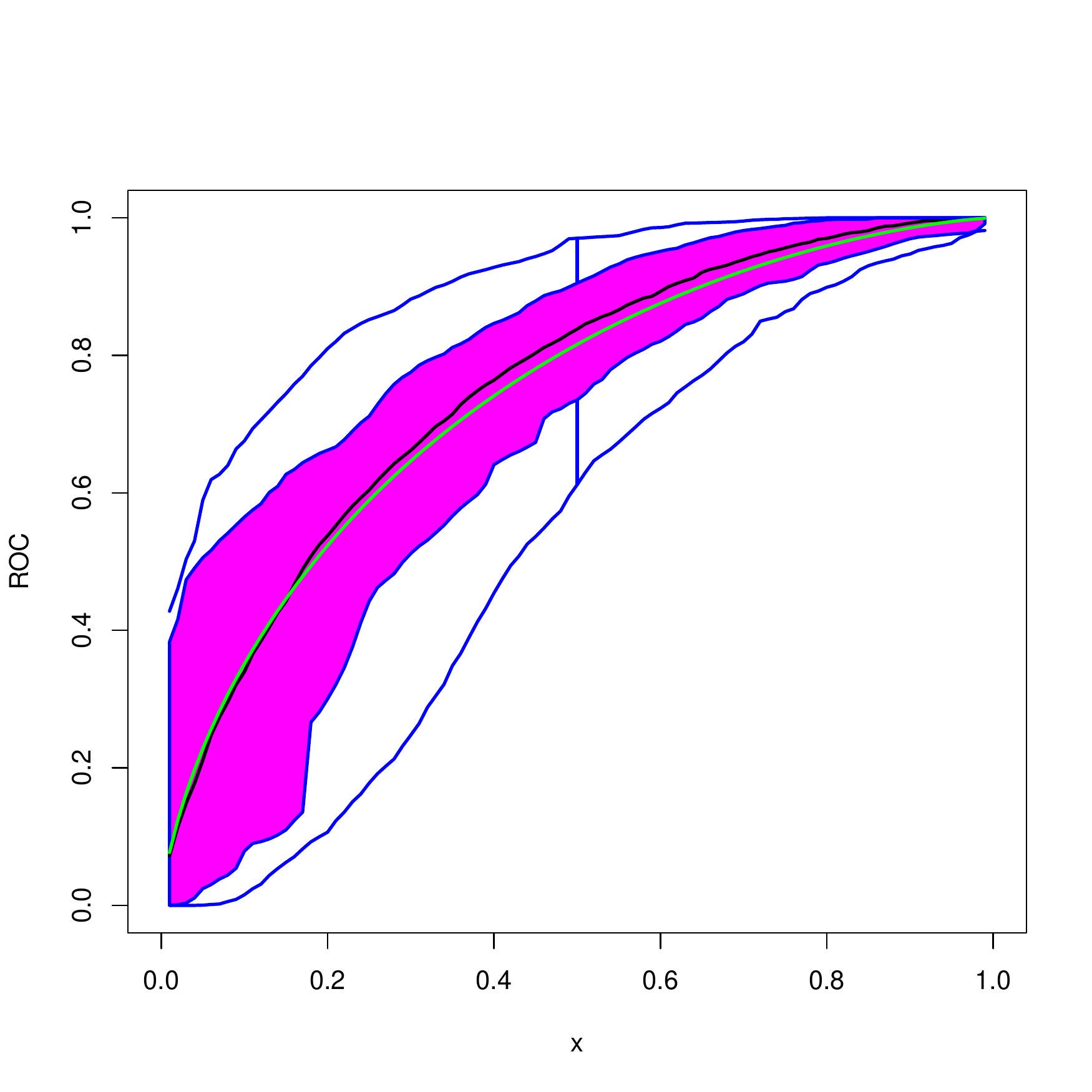}  &  
\includegraphics[scale=0.23]{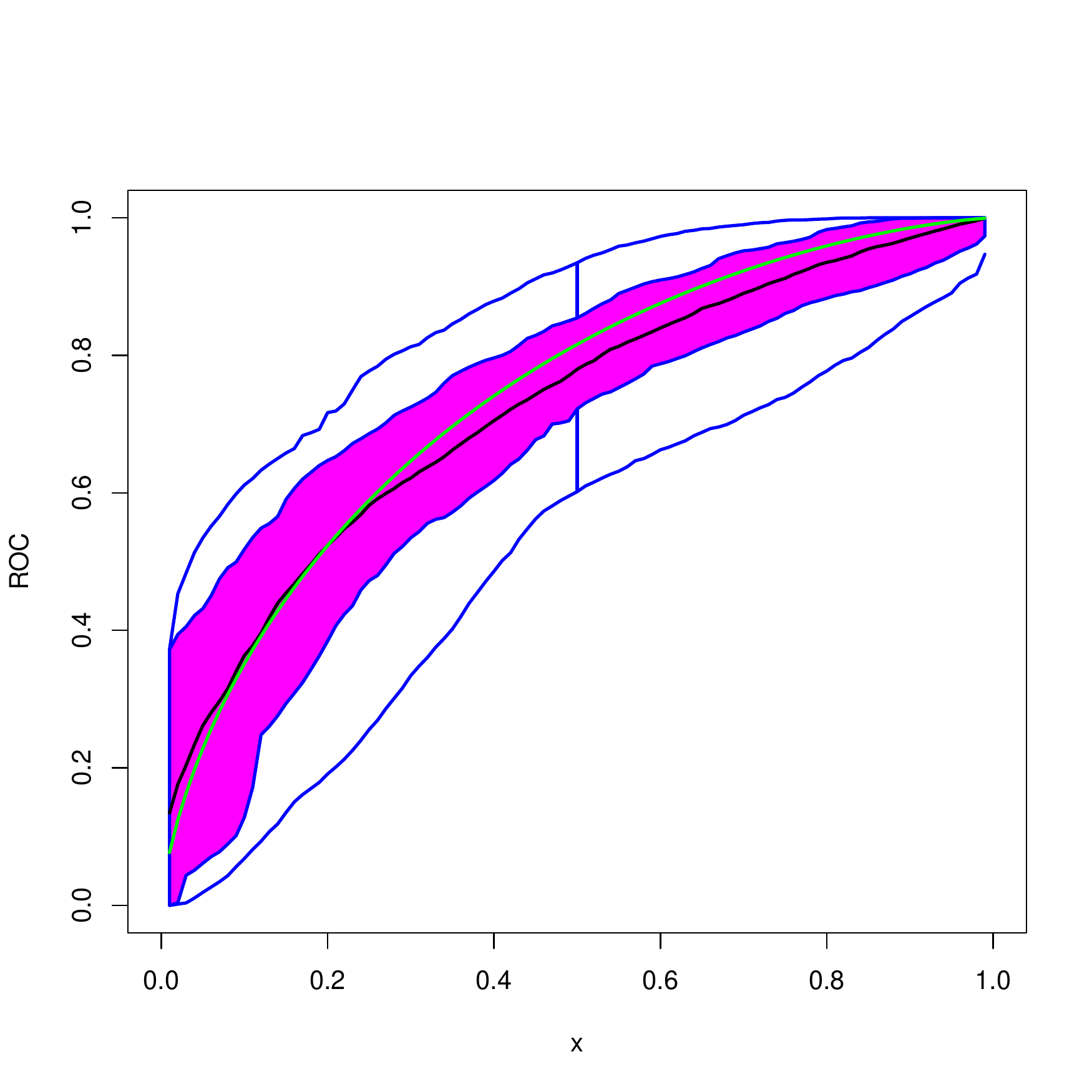}
   \end{tabular}
\vskip-0.1in  
\caption{\label{fig:fbx_11_22_nointer}\small Functional boxplots of  $\wROC_{\conv}(p)$,  under misspecification of the regression.  The green line corresponds to the true $ROC(p)$ and the dotted red lines to the outlying curves detected by the functional boxplot.}
\end{center} 
\end{figure}
\normalsize

\subsubsection{The case where $\bx\in \real^2$}{\label{sec:dim2-caso2-miss}}
As in Section \ref{sec:dim2-caso2}, we consider   homoscedastic  regression models for both populations and  generate   
missing data  according to a model  which depends only on the first component of the covariates.  More precisely, we  define $Y_{j,i}=Z_{j,i}$, if $\delta_{j,i}=1$, and missing otherwise, where $Z_{D,i}$  and $Z_{H,i}$  fulfil \eqref{trued} and \eqref{trueh}, respectively and $\delta_{j,i}\sim Bi(1, \pi( X_{j,i,1} ))$ with     $ \pi(x)=  1/(1+\exp(2x))$.

We considered the situation in which the propensity is assumed to be known and equal to the true one as well as a situation in which the parameters of the true logistic model are estimated from the data. Two misspecification settings are considered. In the first one, the propensity is estimated assuming a completely at random (\textsc{mcar}) model, either for one or for both populations. In the second one, the regression is misspecified and   we considered a regression  misspecification where   the intercept is omitted and the situation in which the model is estimated as a linear one depending only on the first component of the covariates. Tables \ref{tab:dim2_caso2_ROC} and \ref{tab:dim2_caso2_AUC} report the   results for the estimates of the  ROC curve and  AUC, respectively, while Figures \ref{fig:fbx_dim2_caso2_mal_44} and \ref{fig:fbx_dim2_caso2_44_malregre} present the functional boxplots of the ROC estimates when  the propensity and regression models are correctly estimated  and  under misspecification. In Figure \ref{fig:fbx_dim2_caso2_mal_44} when the propensity is misspecified we do not plot the boxplots of $\wROC_{\conv}$, since these estimators do not depend on the propensity estimates.

 Table \ref{tab:dim2_caso2_ROC}  reveals  that the convolution based estimator seems to have a stable performance even when a misspecified regression model is fitted.  It becomes evident that it still outperforms the two other competitors in most cases, even when the $MSE$ is duplicated when a regression model depending only on the first component of the covariates is fitted.  The bias of $\wAUC_{\conv}$ seems to be more affected  by the regression misspecification. It should be noticed that the biases and mean squared errors of the estimates  $\wAUC_{\ipw}$ and $\wAUC_{\kernel}$ are larger when the true propensity is used than when it is estimated using the true logistic model, a fact that has been observed also when estimating the marginal parameter in some regression models, see for instance, Wang \textsl{et al.} (1997).

 \small

\begin{table}[ht!]
\centering
\setlength{\tabcolsep}{4pt}
\renewcommand{\arraystretch}{0.9}
\begin{tabular}{c|rrr|rrr|}
  \hline
 $1000\times$ & $\wROC_{\ipw} $ & $\wROC_{\kernel}$ &  $\wROC_{\conv} $ &  $\wROC_{\ipw} $ & $\wROC_{\kernel}$ &  $\wROC_{\conv} $ \\
  \hline
   &   \multicolumn{3}{c|}{$\wpi_D=\pi$, $\wpi_H=\pi$} & \multicolumn{3}{c|}{$\wpi_D=\pi_{\log}$, $\wpi_H=\pi_{\log}$}\\\hline
 
$MSE$  & 7.32 & 6.12 & 2.93 & 5.68 & 4.48 & 2.93 \\ 
 $KS$ & 169.95 & 124.14 & 93.74 & 158.50 & 112.01 & 93.74 \\ 
   \hline 
 & \multicolumn{6}{c|}{Misspecified propensity}\\
\hline
 &   \multicolumn{3}{c|}{$\wpi_D=\pi_c$, $\wpi_H=\pi_c$} & \multicolumn{3}{c|}{$\wpi_D=\pi_{\log}$, $\wpi_H=\pi_{c}$}\\\hline

$MSE $ & 6.61 & 5.54 & 2.93 & 6.26 & 5.10 & 2.93 \\ 
$KS$ & 160.11 & 117.06 & 93.74 & 161.46 & 116.33 & 93.74 \\
   \hline 
  
 
  & \multicolumn{6}{c|}{Misspecified regression}\\
 \hline
   & \multicolumn{6}{c|}{Regression model estimated without intercept}\\
 \hline
 &   \multicolumn{3}{c|}{$\wpi_D=\pi$, $\wpi_H=\pi$} & \multicolumn{3}{c|}{$\wpi_D=\pi_{\log}$, $\wpi_H=\pi_{\log}$}\\ 
 \hline 
 $MSE$ & 7.32 & 6.12 & 2.96 & 5.68 & 4.48 & 2.96 \\ 
$KS$ & 169.95 & 124.14 & 90.08 & 158.50 & 112.01 & 90.08 \\ 
 \hline
 & \multicolumn{6}{c|}{Misspecified regression}\\
 \hline
   & \multicolumn{6}{c|}{Regression model estimated depending only on $X_{i,1}$}\\
 \hline
 &   \multicolumn{3}{c|}{$\wpi_D=\pi$, $\wpi_H=\pi$} & \multicolumn{3}{c|}{$\wpi_D=\pi_{\log}$, $\wpi_H=\pi_{\log}$}\\ 
 \hline 

$MSE $ & 7.32 & 6.12 & 5.92 & 5.68 & 4.48 & 5.92 \\ 
$KS$ & 169.95 & 124.14 & 126.76 & 158.50 & 112.01 & 126.76 \\ 
   \hline 
 \end{tabular}
\caption{\label{tab:dim2_caso2_ROC}Summary measures for ROC curve when $\bX\in \real^2$.} 
\end{table}

\begin{table}[ht!]
\centering
\setlength{\tabcolsep}{4pt}
\renewcommand{\arraystretch}{0.9}
\begin{tabular}{c|rrr|rrr|}
  \hline
& $\wAUC_{\ipw}$ & $\wAUC_{\kernel}$ &  $\wAUC_{\conv}$ & $\wAUC_{\ipw}$ & $\wAUC_{\kernel}$ &  $\wAUC_{\conv}$ \\
 \hline 
 $1000\times$ &    \multicolumn{3}{c|}{$\wpi_D=\pi$, $\wpi_H=\pi$} & \multicolumn{3}{c|}{$\wpi_D=\pi_{\log}$, $\wpi_H=\pi_{\log}$}\\
 \hline
Bias & 3.31 & 2.46 & 0.27 & 0.97 & 0.15 & 0.27 \\ 
 $RB$ & 85.46 & 84.17 & 56.65 & 63.59 & 62.69 & 56.65 \\ 
$MSE$  & 3.59 & 3.49 & 1.61 & 2.04 & 1.98 & 1.61 \\ 
   \hline
 
& \multicolumn{6}{c|}{Misspecified propensity}\\
\hline
 &   \multicolumn{3}{c|}{$\wpi_D=\pi_c$, $\wpi_H=\pi_c$} & \multicolumn{3}{c|}{$\wpi_D=\pi_{\log}$, $\wpi_H=\pi_{c}$}\\
 \hline
 
Bias & -1.99 & -2.95 & 0.27 & 13.72 & 12.56 & 0.27 \\ 
 $RB$ & 81.70 & 80.58 & 56.65 & 76.24 & 74.80 & 56.65 \\ 
$MSE $ & 3.27 & 3.19 & 1.61 & 2.79 & 2.68 & 1.61 \\ 
   \hline
 
 & \multicolumn{6}{c|}{Misspecified regression}\\
 \hline
   & \multicolumn{6}{c|}{Regression model estimated without intercept}\\
   \hline
 &   \multicolumn{3}{c|}{$\wpi_D=\pi$, $\wpi_H=\pi$} & \multicolumn{3}{c|}{$\wpi_D=\pi_{\log}$, $\wpi_H=\pi_{\log}$}\\
 \hline
Bias & 3.31 & 2.46 & -6.49 & 0.97 & 0.15 & -6.49 \\ 
$RB$ & 85.46 & 84.17 & 58.22 & 63.59 & 62.69 & 58.22 \\ 
 $MSE$ & 3.59 & 3.49 & 1.65 & 2.04 & 1.98 & 1.65 \\ 
   \hline
  & \multicolumn{6}{c|}{Misspecified regression}\\
 \hline
   & \multicolumn{6}{c|}{Regression model estimated depending only on $X_{i,1}$}\\
 \hline
 &   \multicolumn{3}{c|}{$\wpi_D=\pi$, $\wpi_H=\pi$} & \multicolumn{3}{c|}{$\wpi_D=\pi_{\log}$, $\wpi_H=\pi_{\log}$}\\ 
 \hline 
  
Bias & 3.31 & 2.46 & 2.88 & 0.97 & 0.15 & 2.88 \\ 
$RB$ & 85.46 & 84.17 & 83.68 & 63.59 & 62.69 & 83.68 \\ 
$MSE $ & 3.59 & 3.49 & 3.44 & 2.04 & 1.98 & 3.44 \\ 
   \hline
\end{tabular}
\caption{\label{tab:dim2_caso2_AUC}Summary measures for the area under the curve ($\AUC$) when $\bX\in \real^2$.} 
\end{table}

\begin{figure}[ht!]
 \begin{center}
 \footnotesize
 \renewcommand{\arraystretch}{0.4}
 \newcolumntype{M}{>{\centering\arraybackslash}m{\dimexpr.12 \linewidth-1\tabcolsep}}
   \newcolumntype{G}{>{\centering\arraybackslash}m{\dimexpr.3\linewidth-1\tabcolsep}}
\begin{tabular}{M G G G}\\
& $\wROC_{\ipw}$ & $\wROC_{\kernel}$ & $\wROC_{\conv}$\\ [0.15in]
 & \multicolumn{3}{c}{(a) Properly specified propensity and regression model}\\
 $\wpi_D=\pi$, $\wpi_H=\pi$   & 
 \includegraphics[scale=0.25]{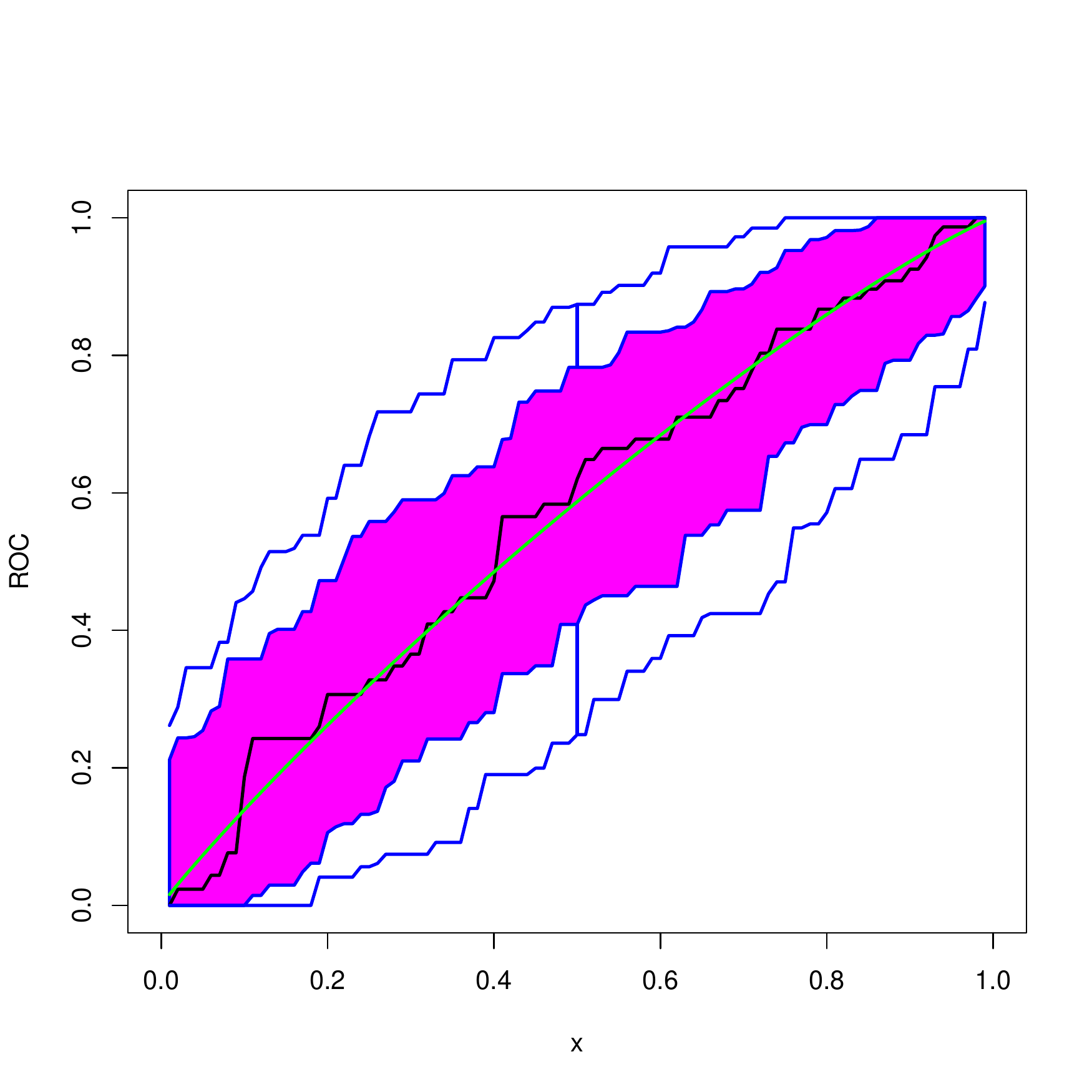} & 
\includegraphics[scale=0.25]{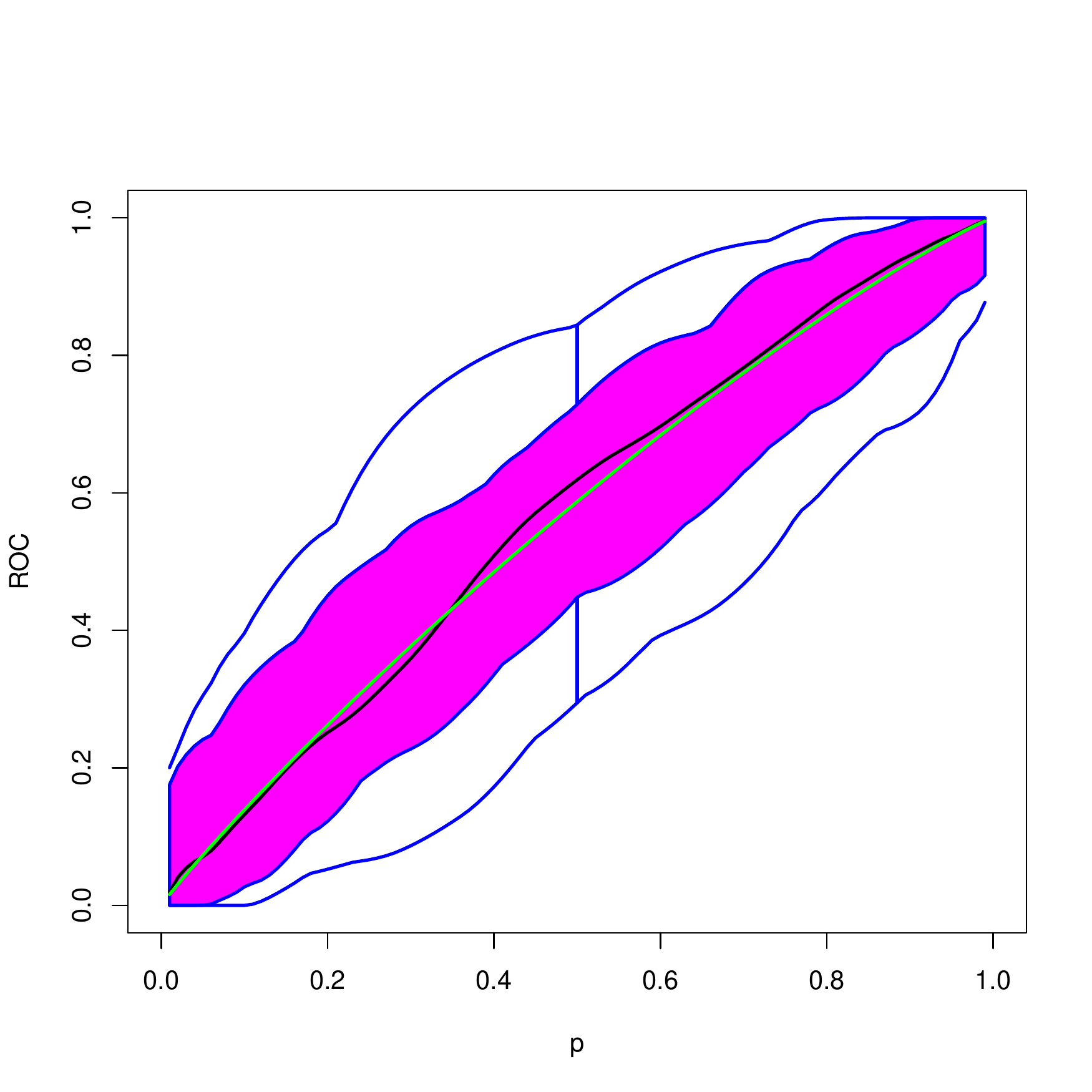} &
\includegraphics[scale=0.25]{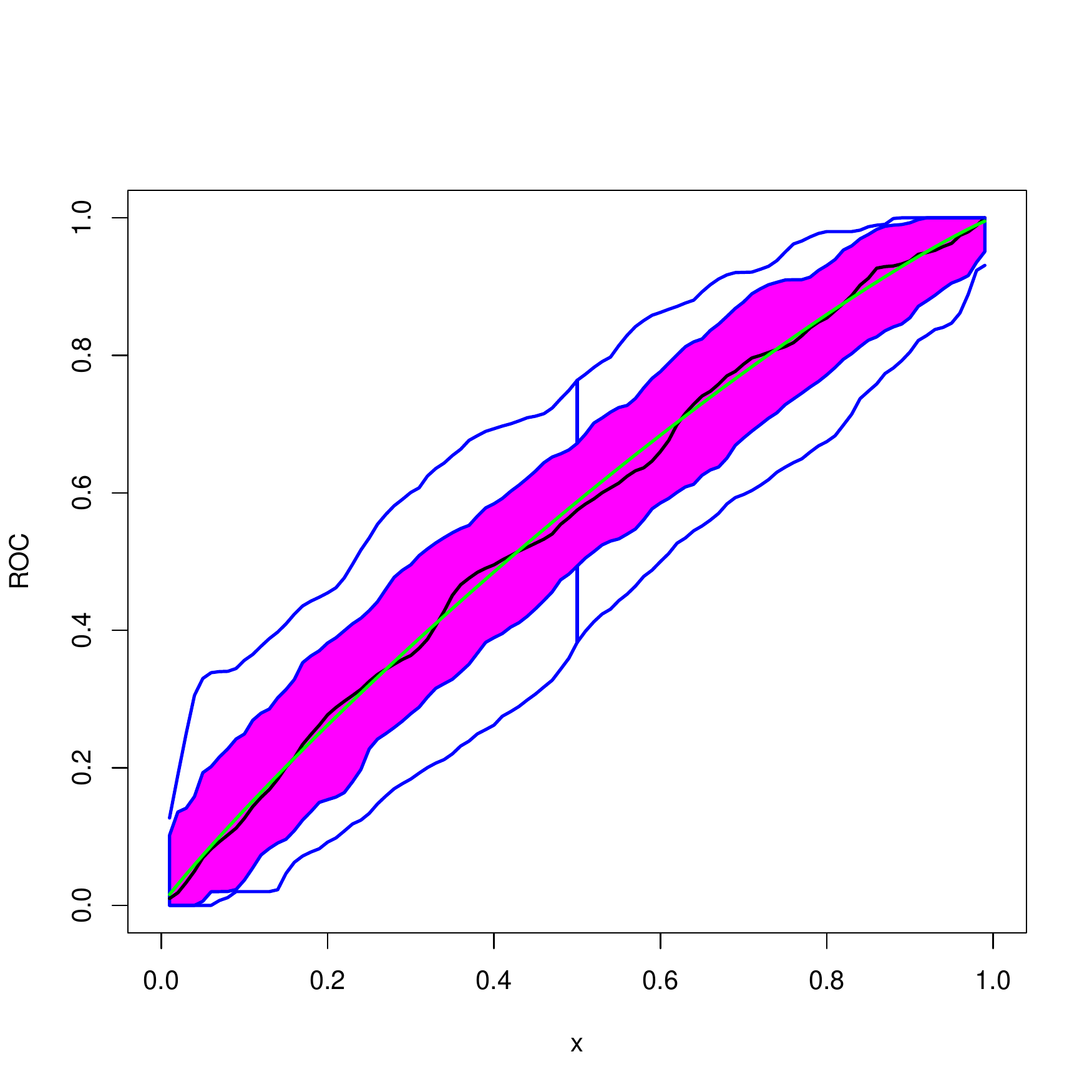}\\
 $\wpi_D=\pi_{\log}$, $\wpi_H=\pi_{\log}$ & 
 \includegraphics[scale=0.25]{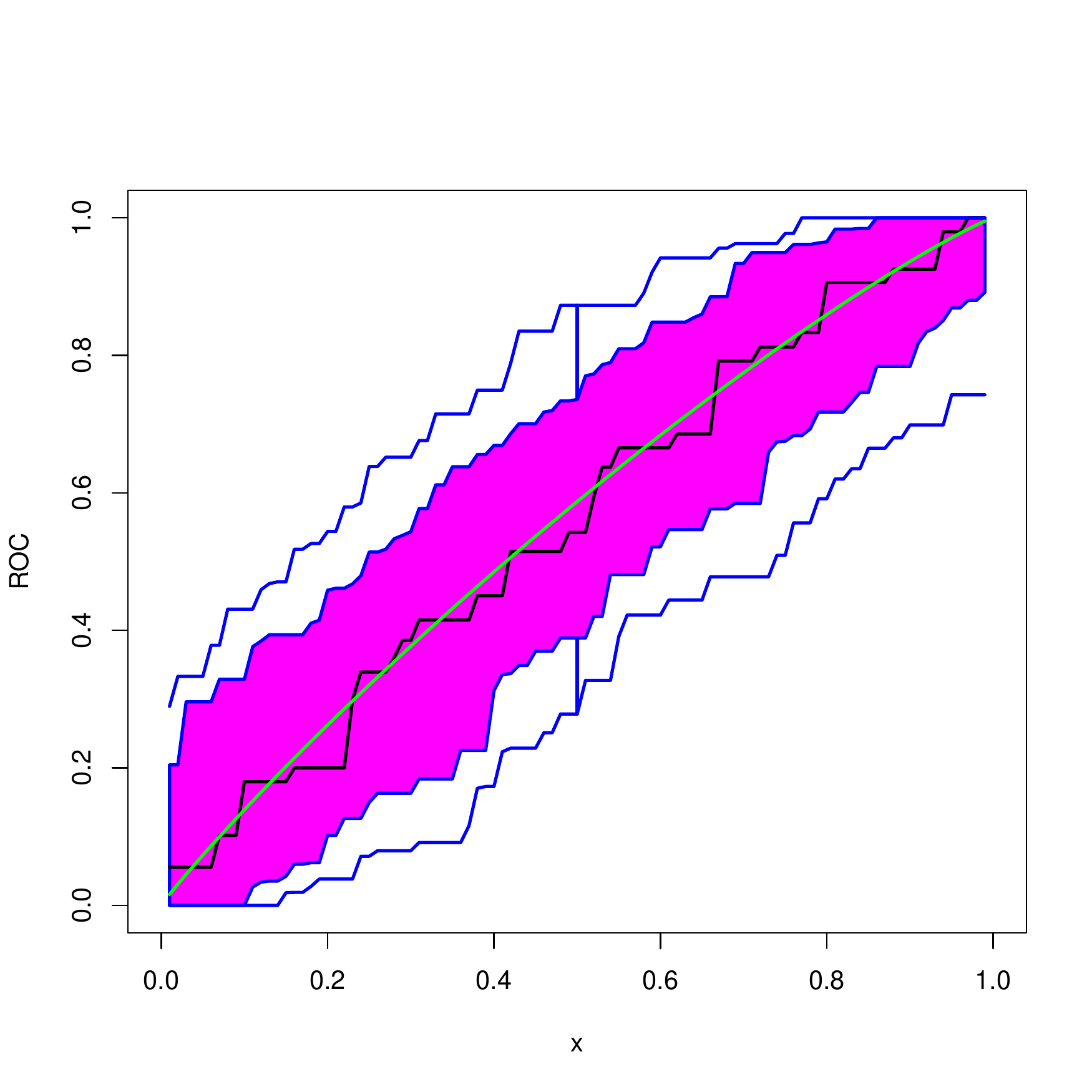}  & 
\includegraphics[scale=0.25]{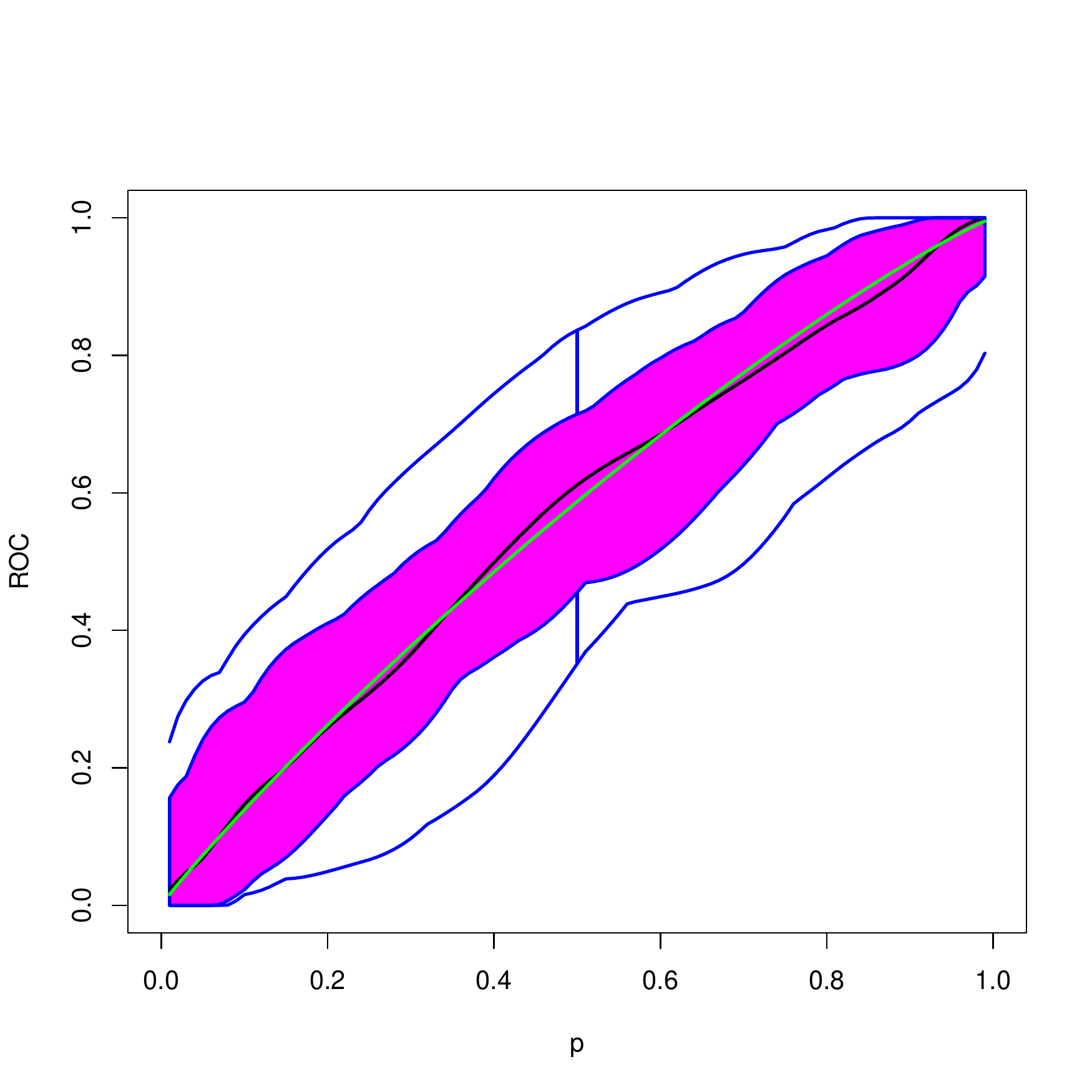} & 
\includegraphics[scale=0.25]{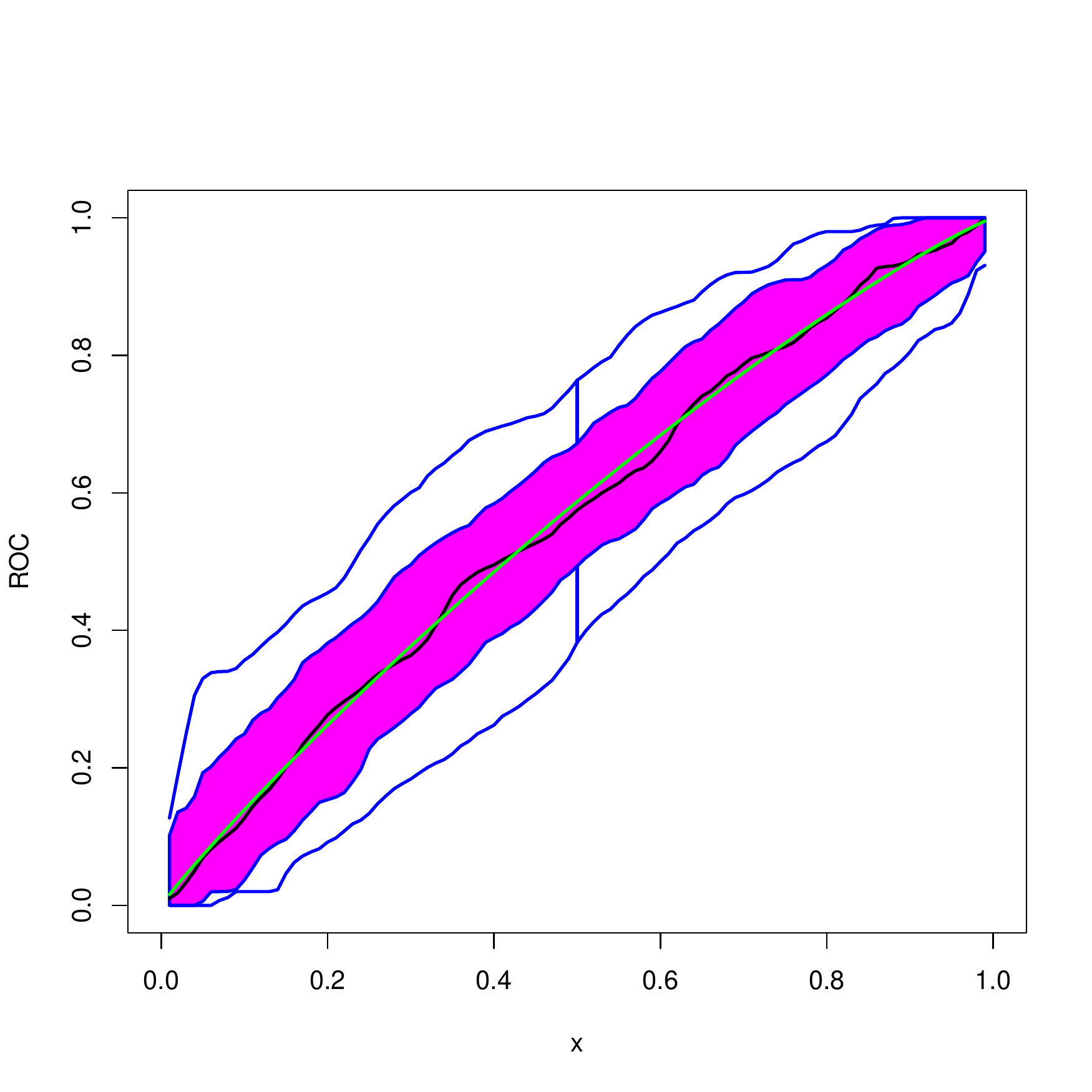} \\
 & \multicolumn{3}{c}{(b) Misspecified propensity}\\
 $\wpi_D=\pi_c$, $\wpi_H=\pi_c$   & 
\includegraphics[scale=0.25]{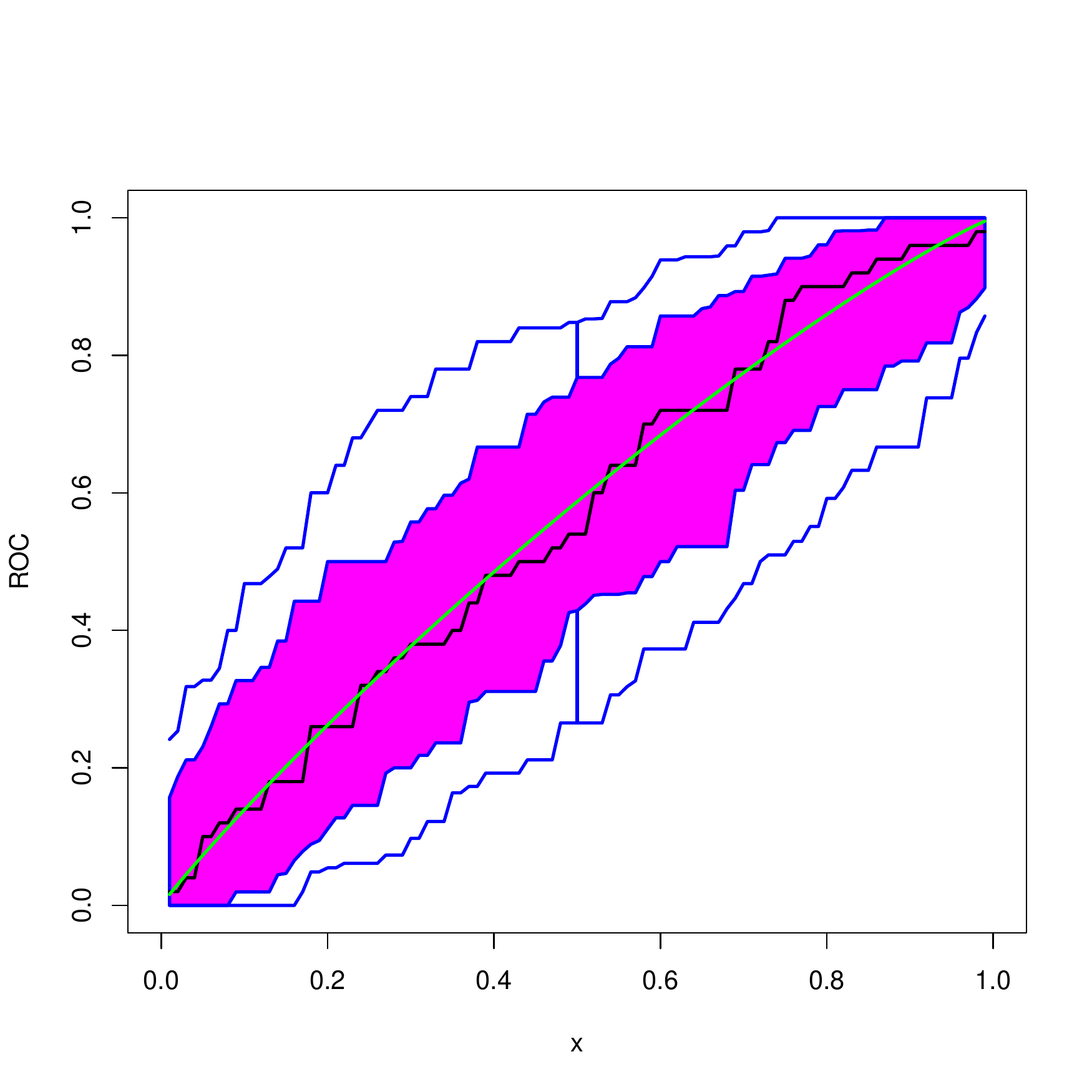} &
\includegraphics[scale=0.25]{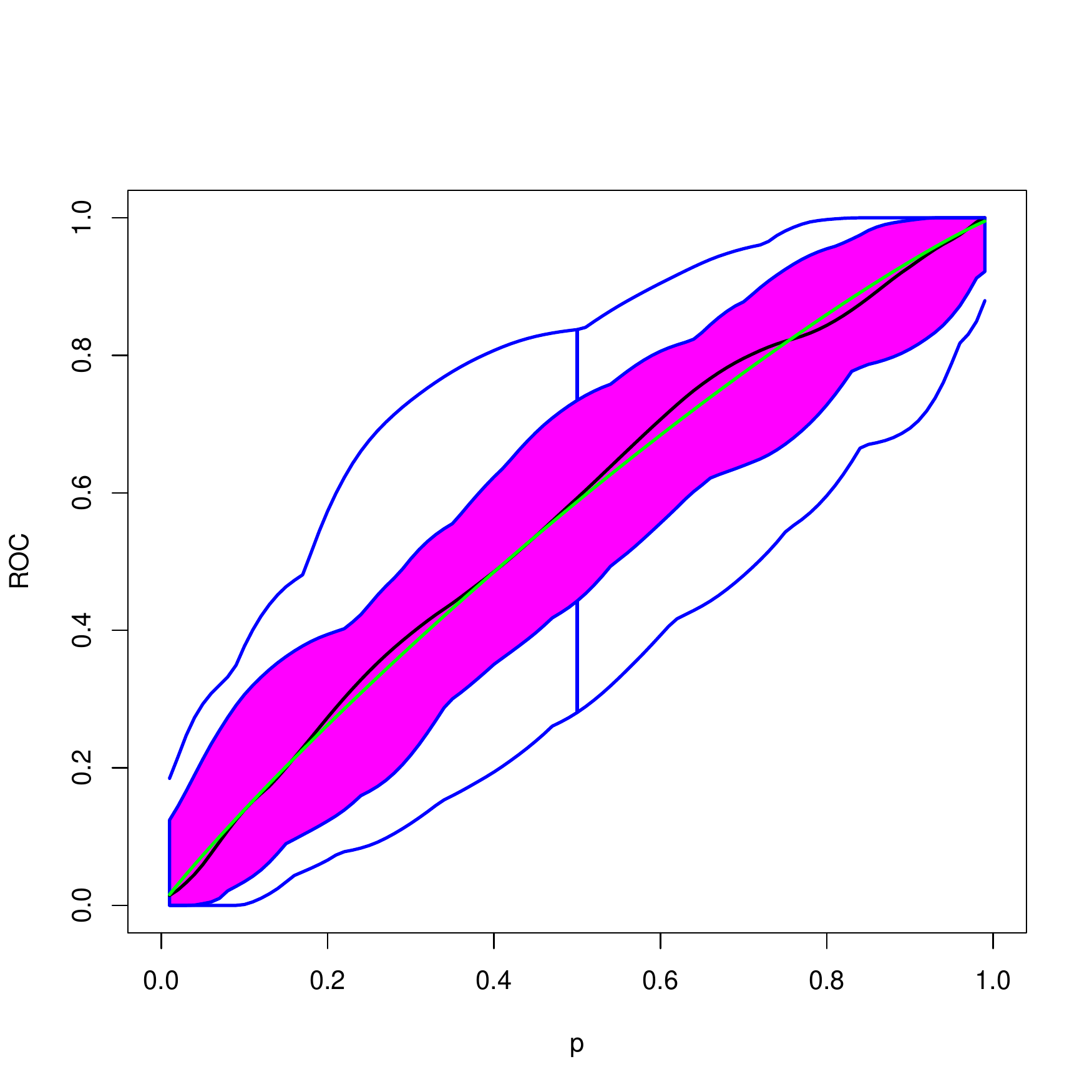} & \\
 $\wpi_H=\pi_{c}$, $\wpi_D=\pi_{\log}$ &
\includegraphics[scale=0.25]{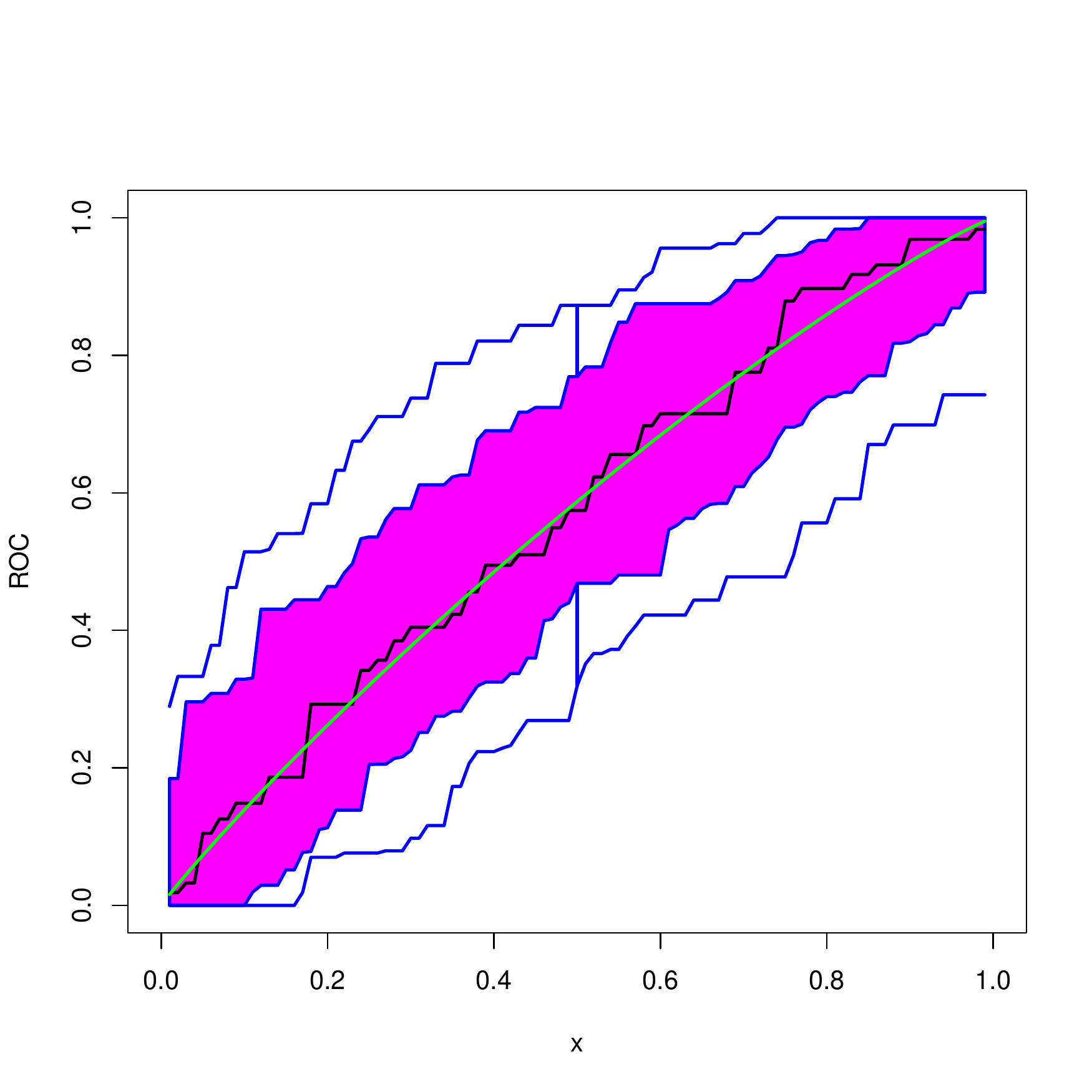} &
\includegraphics[scale=0.25]{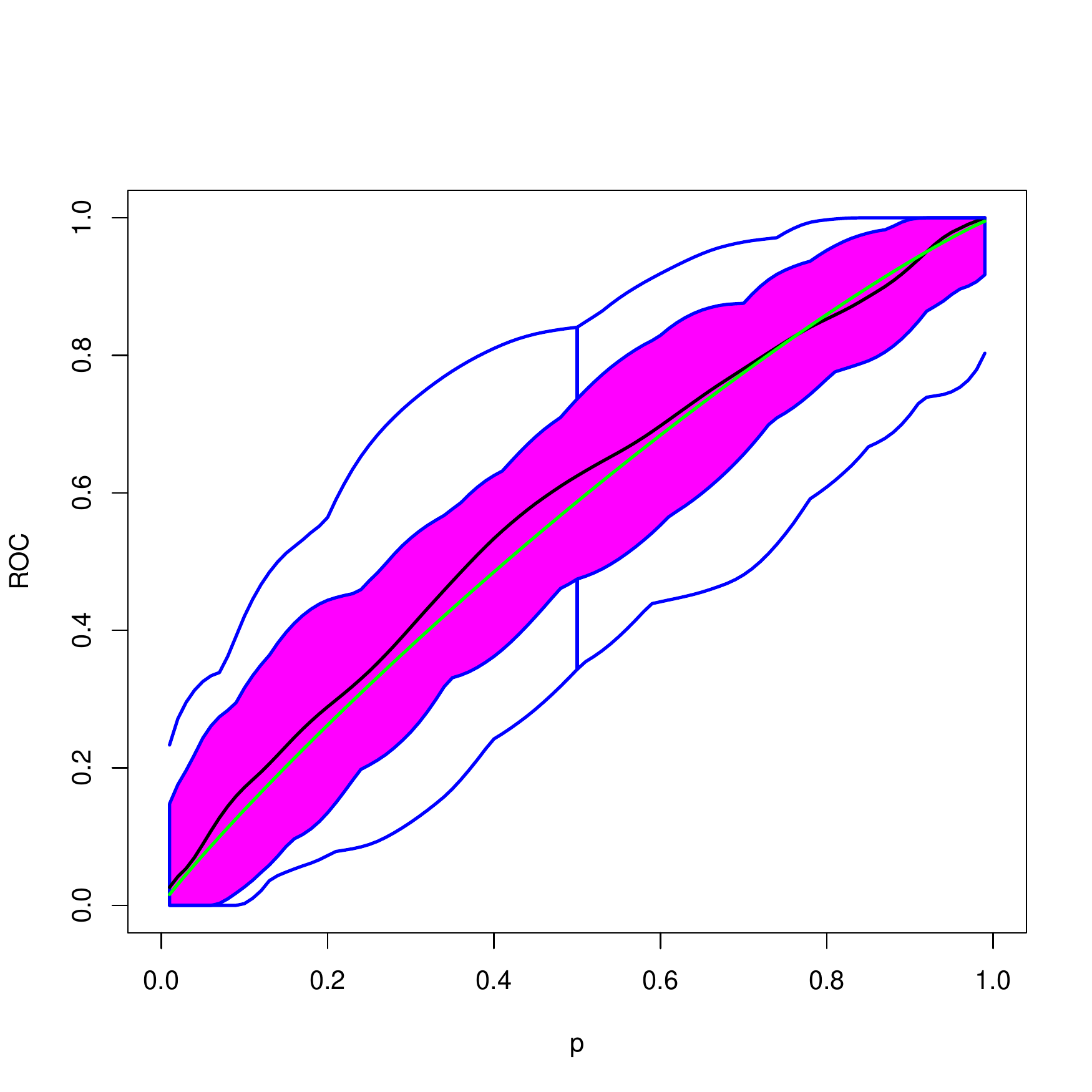} & 
 \end{tabular}
\vskip-0.1in  
\caption{\label{fig:fbx_dim2_caso2_mal_44}\small Functional boxplots of  $\wROC(p)$  under the assumed models and  under misspecification of the propensity for $\bX\in \real^2$. The green line corresponds to the true $\ROC(p)$ and the dotted red lines to the outlying curves detected by the functional boxplot.}
\end{center} 
\end{figure}
\normalsize
 

\begin{figure}[ht!]
 \begin{center}
\footnotesize
 \renewcommand{\arraystretch}{0.4}
 \newcolumntype{M}{>{\centering\arraybackslash}m{\dimexpr.12 \linewidth-1\tabcolsep}}
   \newcolumntype{G}{>{\centering\arraybackslash}m{\dimexpr.3\linewidth-1\tabcolsep}}
\begin{tabular}{M G G G}\\
   &  (a) & (b) & (c) \\
$\wpi_D=\pi$, $\wpi_H=\pi$   
&  \includegraphics[scale=0.25]{ROC_MUL_n100_H_preal_D_preallineallineal_mis_H_4_mis_D_4_dim2_caso2.pdf}
& \includegraphics[scale=0.25]{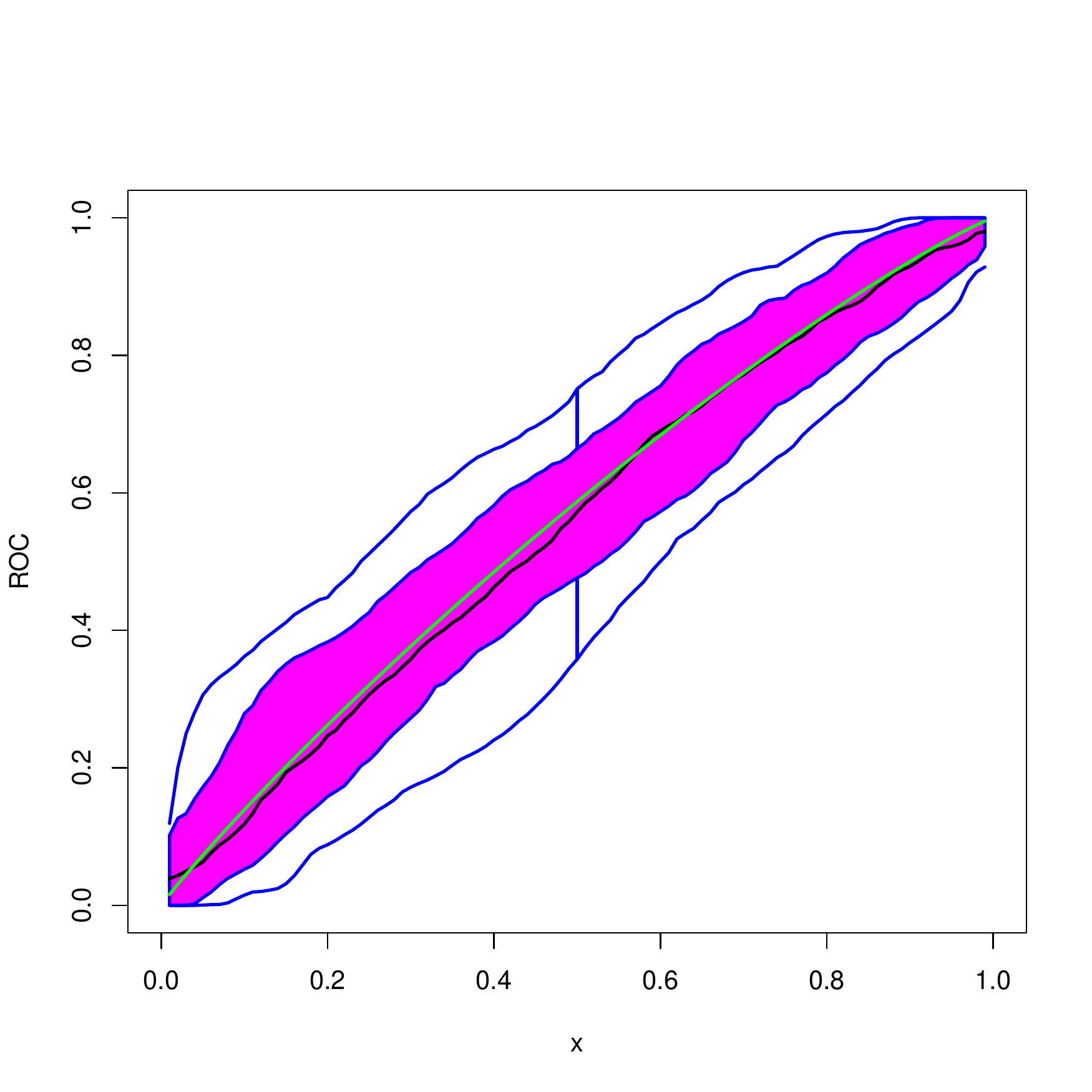} 
&  \includegraphics[scale=0.25]{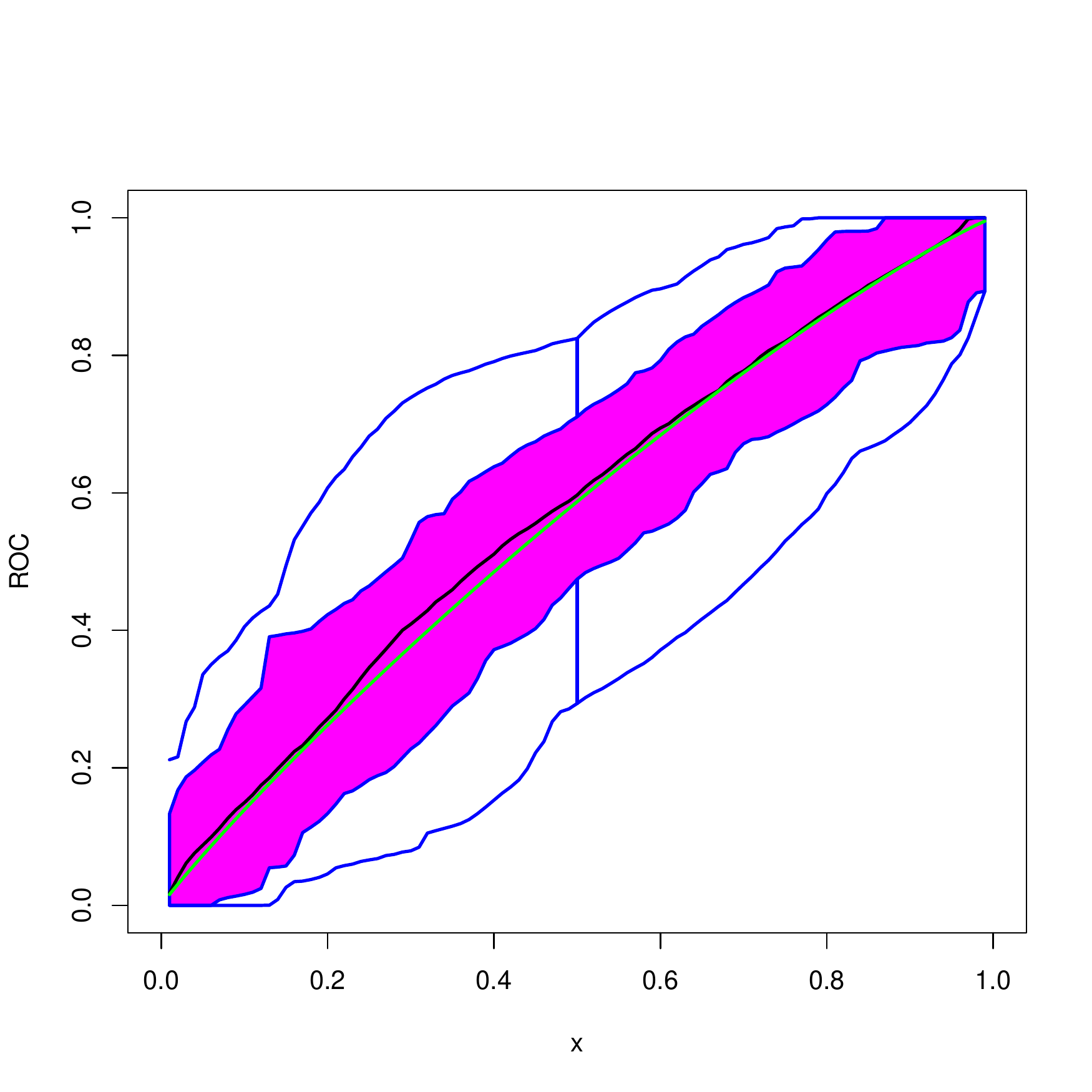} \\
 $\wpi_D=\pi_{\log}$, $\wpi_H=\pi_{\log}$ 
&  \includegraphics[scale=0.25]{ROC_MUL_n100_H_plogit_D_plogitlineallineal_mis_H_4_mis_D_4_dim2_caso2.pdf} 
& \includegraphics[scale=0.25]{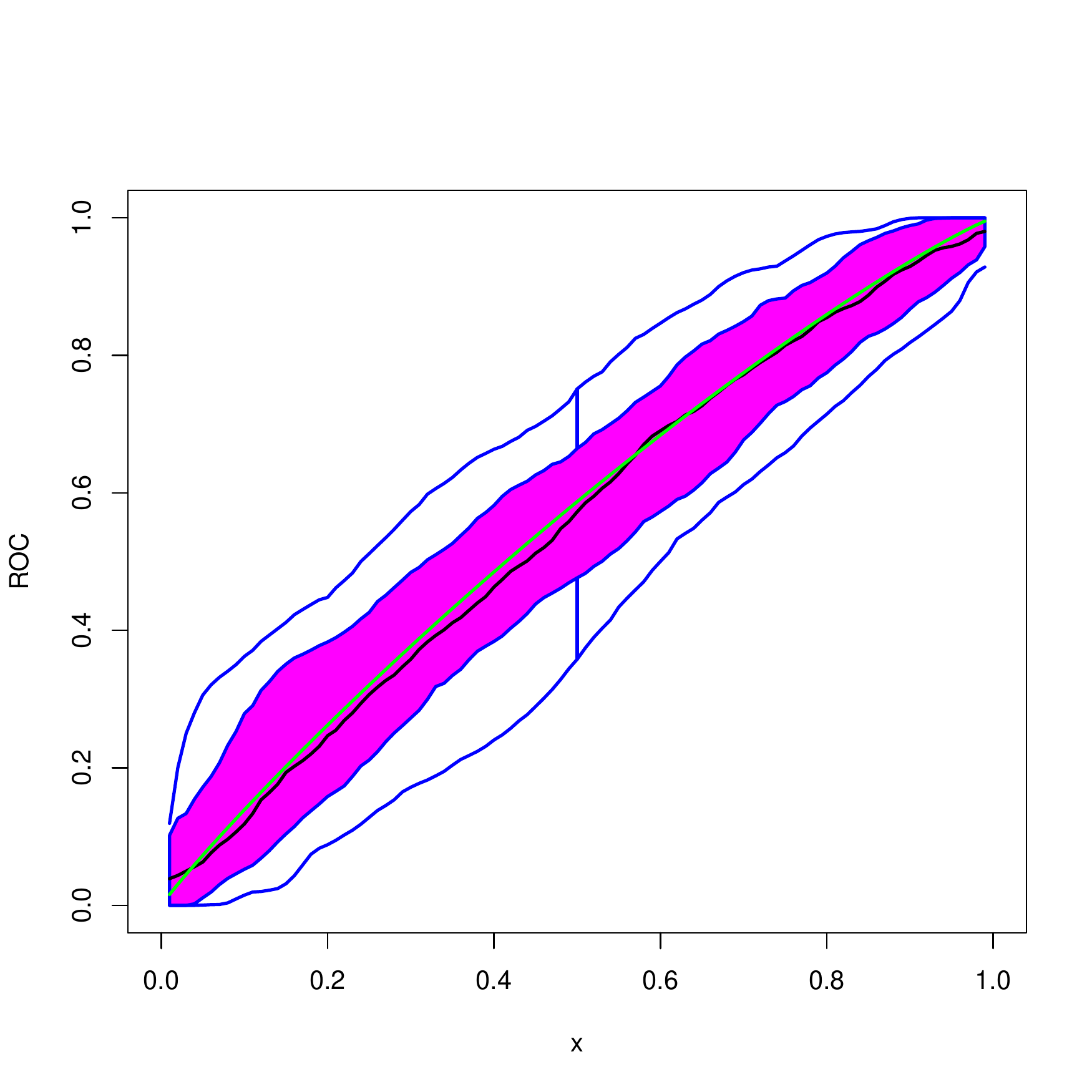}
&  \includegraphics[scale=0.25]{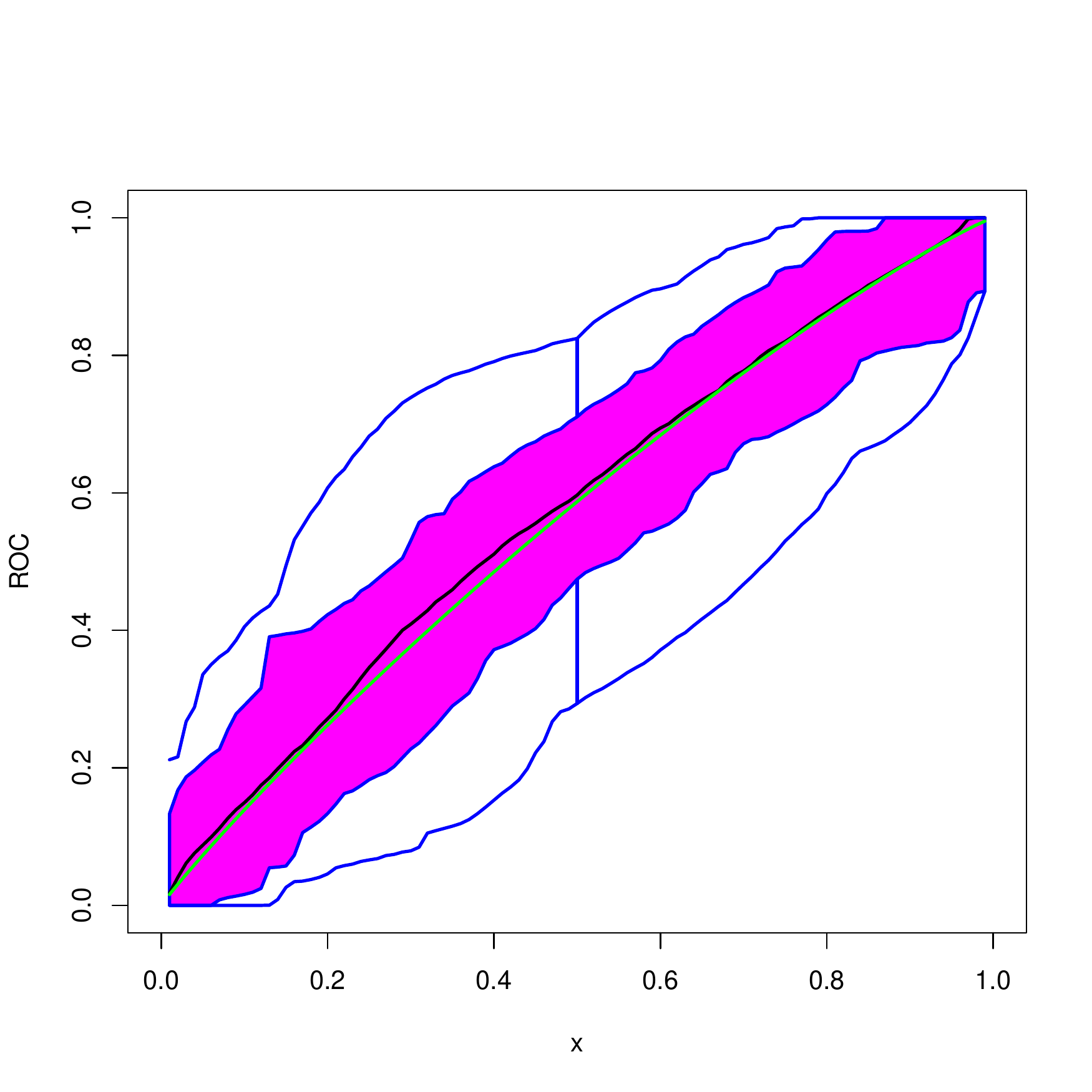}
  \end{tabular}
\vskip-0.1in  
\caption{\label{fig:fbx_dim2_caso2_44_malregre}\small Functional boxplots of  $\wROC_{\conv}(p)$   for $\bX\in \real^2$. In panel (a) the regression model is correctly fitted, while in (b) it is fitted without intercept and in (c) it is fitted as depending   only on $X_{i,1}$. The green line corresponds to the true $\ROC(p)$ and the dotted red lines to the outlying curves detected by the functional boxplot.}
\end{center} 
\end{figure}

\section{Data Analysis}{\label{sec:ejemplo}}
In this section, we analyse a data set available at \url{https://archive.ics.uci.edu/ml/datasets/automobile}.
The data set records the specification of a car in terms of various characteristics, its assigned insurance risk rating and its normalized losses in use as compared to other cars.  The two populations considered in our analysis correspond to cars with high or low risk according to its price. More precisely, initially a risk factor is assigned to each car according to its price. Then, if it is more risky (or less), this symbol is adjusted by moving it up (or down) the scale. A high positive value  indicates that the  automobile  is risky, while a negative with large absolute value indicates  that it is safe.
We then label as healthy population the cars with risk smaller or equal than 0 (\textsl{safe cars}) and as diseased those cars with positive risk (\textsl{risky cars}),  leading to  $n_H= 92$ and $n_D=113$.

The biomarker or response variable, $Y$, is chosen as the normalized losses which contains 41 missing observations among the 205 cars corresponding 21 of them to the  \textsl{safe cars}.  We fit the propensity using a logistic regression model based on the covariates  width, engine--size  and style of the car (with levels: hardtop, wagon, sedan, hatchback and convertible). Besides, to implement the convolution--based estimator a linear regression model was fitted to the biomarker using as   covariates   the width,  height and compression--ratio of the car.

 \begin{table}[ht!]
 \begin{center}
		\begin{tabular}{|cccc|}
	\hline
	  $\wAUC_{\simp}$ &  $\wAUC_{\ipw}$ & $\wAUC_{\kernel}$ & $\wAUC_{\conv}$\\\hline
			0.766 &  0.792 & 0.792 & 0.781
			\\
			\hline
		\end{tabular}
		\vskip-0.1in \caption{\label{tab:auc-ejemplo} \small Estimated AUC for the automobile data set.}
		\end{center}  
		\end{table}

The estimates for the area under the curve are given in Table \ref{tab:auc-ejemplo} where we report the   obtained  estimates of AUC  when using the inverse probability weighting procedure, the smoothed kernel  method as well as the  convolution--based one, denoted $\wAUC_{\ipw}$, $\wAUC_{\kernel}$ and $\wAUC_{\conv}$, respectively. For comparison purposes we also report the simplified estimate, denoted  $\wAUC_{\simp}$, that uses only the available biomarkers and which is known to be biased unless the missing is completely at random. In this data set,  the simplified estimator, that is based only on the complete available cases,  seems to underestimate the area under the curve.

\begin{figure}[ht!]
	\begin{center}
	   \newcolumntype{G}{>{\centering\arraybackslash}m{\dimexpr.3\linewidth-1\tabcolsep}}
 \begin{tabular}{G G G}\\
(a) & (b) & (c)\\[-0.1in]
\includegraphics[scale=0.3]{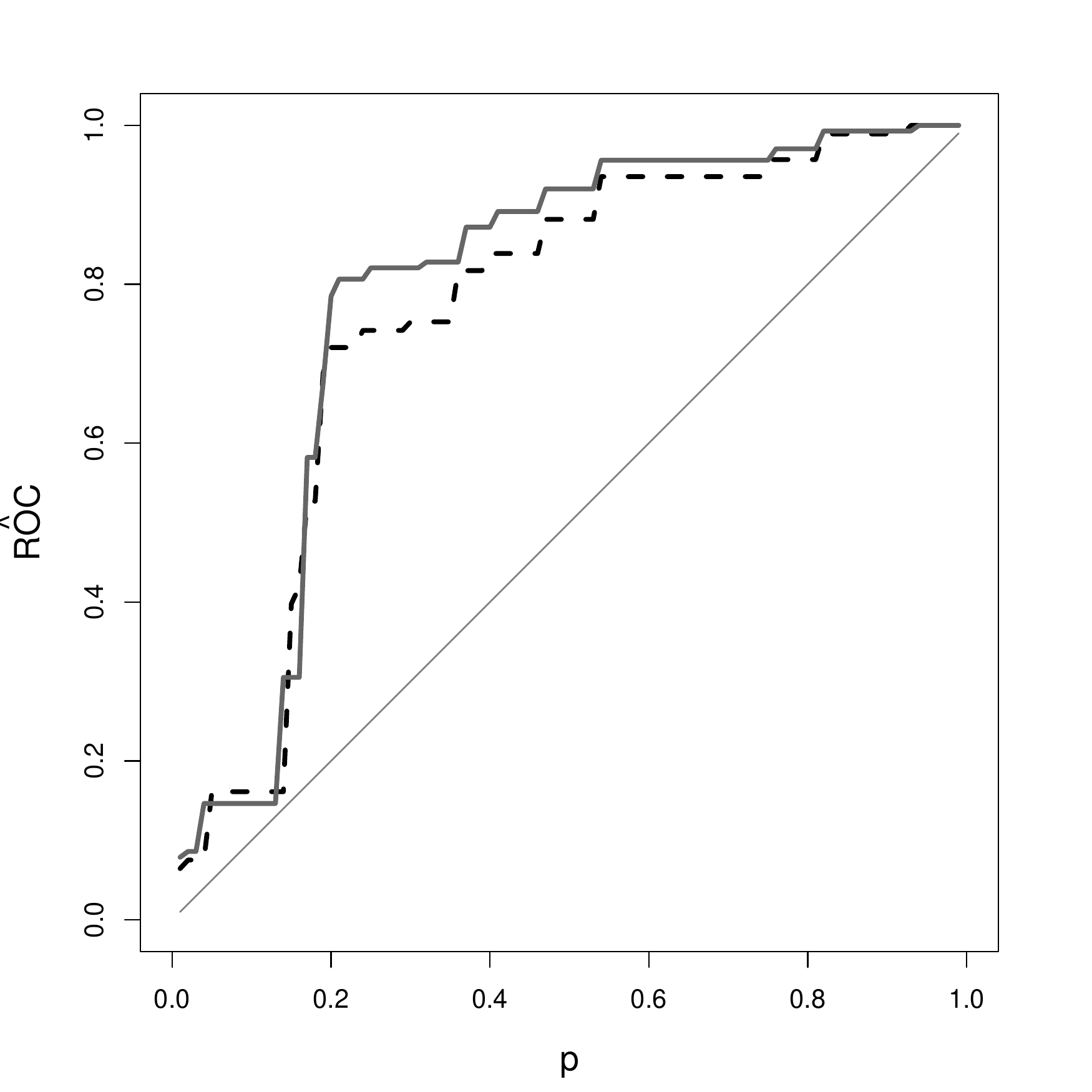} &
\includegraphics[scale=0.3]{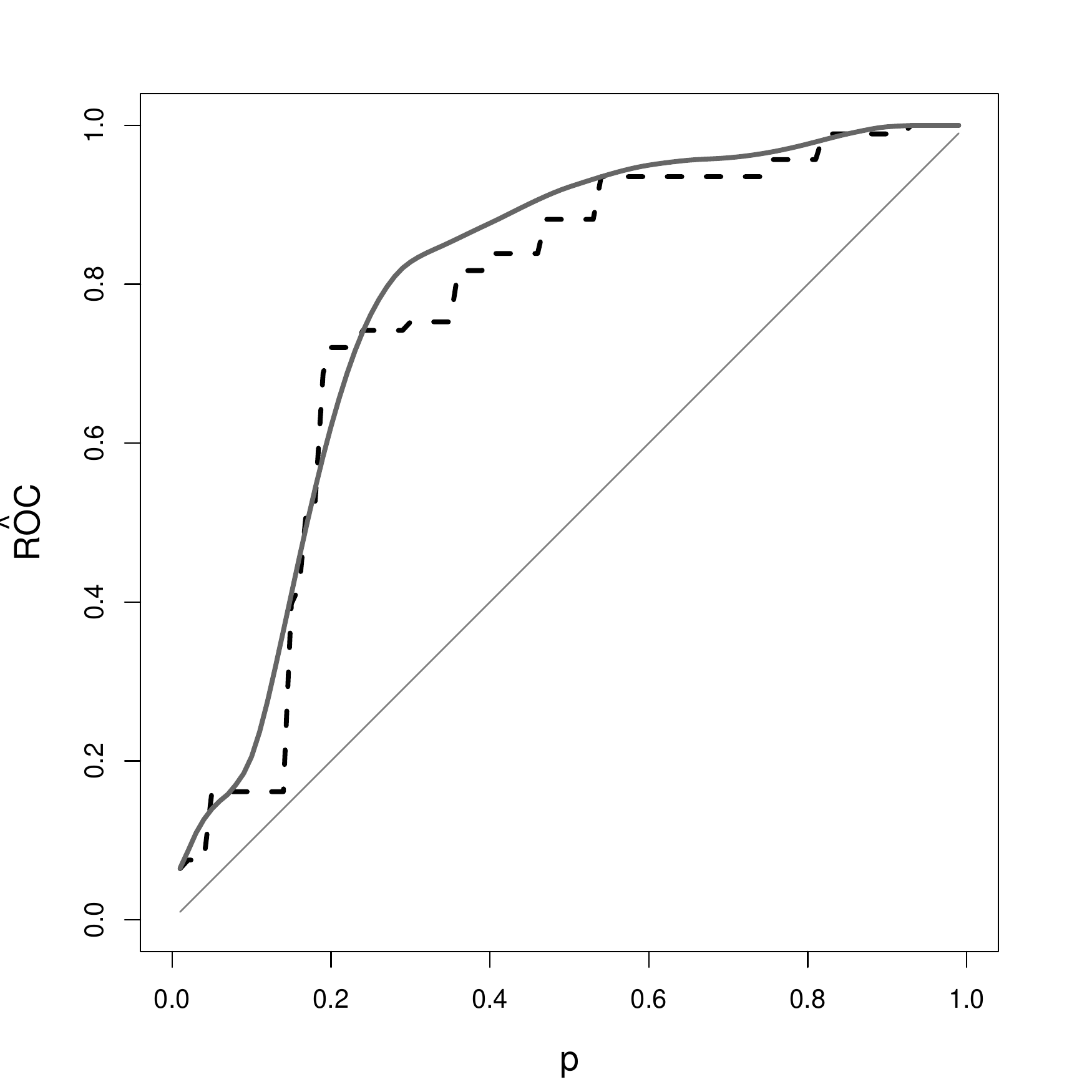} &
\includegraphics[scale=0.3]{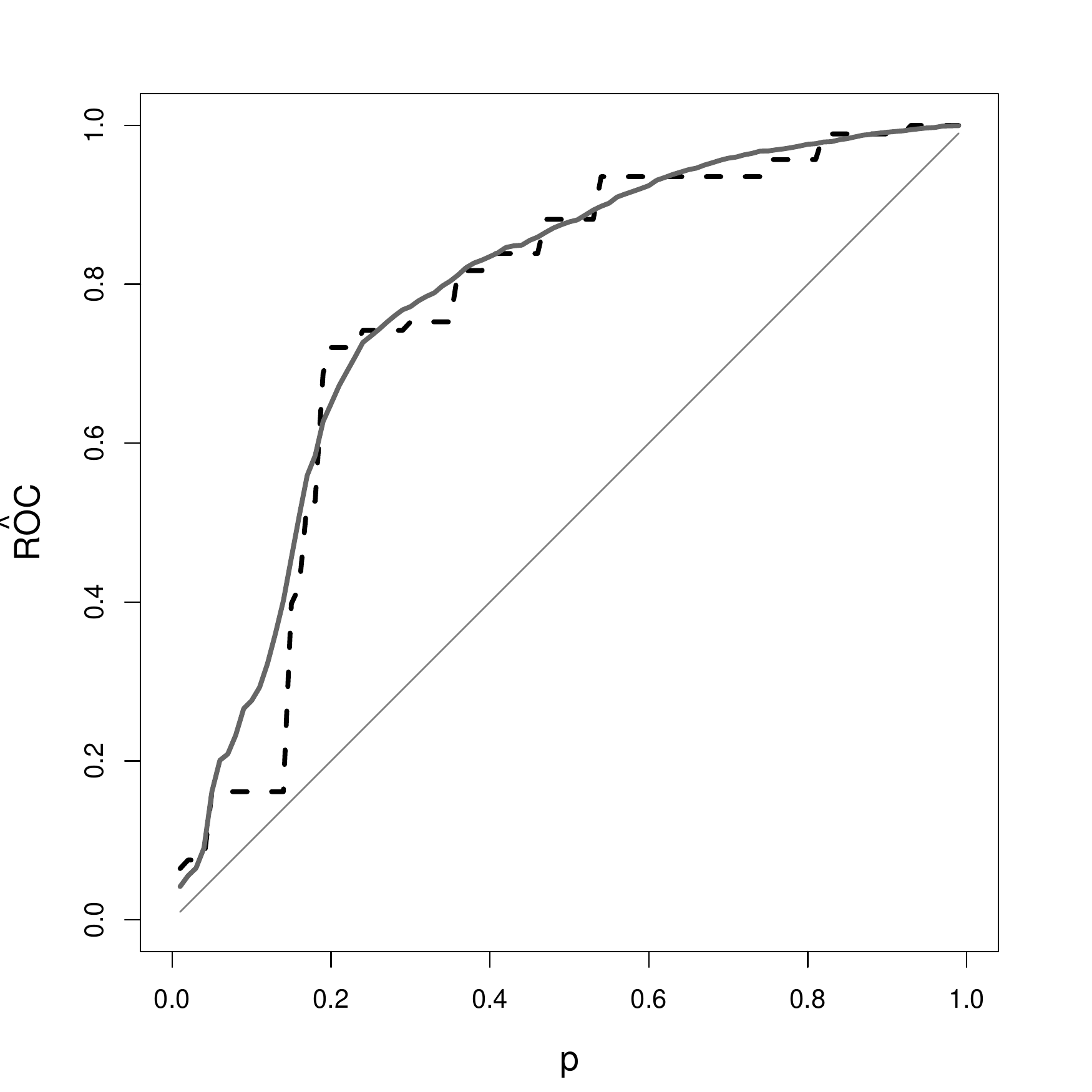}
\\
\multicolumn{3}{c}{(d)}\\[-0.1in]
\multicolumn{3}{c}{\includegraphics[scale=0.3]{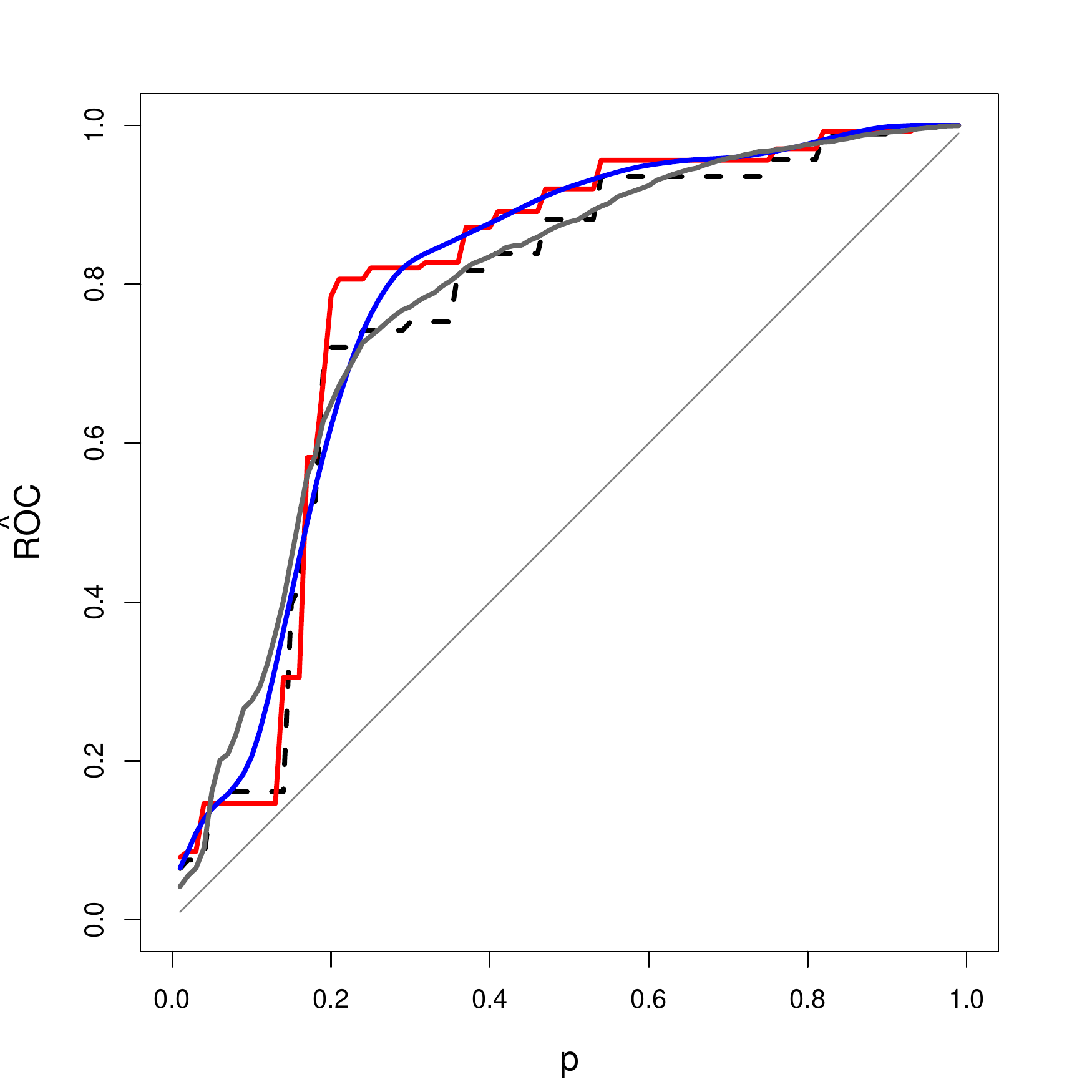}}\\
   \end{tabular}
\vskip-0.1in  
\caption{\label{fig:ROC_autos} \small Estimated ROC curves for the automobile data set. Panels   (a), (b) and (c) display  the simplified estimator (in dashed black lines) jointly with   $\wROC_{\ipw}$, $\wROC_{\kernel}$ and  $\wROC_{\conv}$ (in solid grey lines), respectively. The lower figure gives the plot of the four estimates, the red, blue and grey lines correspond to   $\wROC_{\ipw}$, $\wROC_{\kernel}$ and  $\wROC_{\conv}$, respectively. We also plot the line $y=x$.}
	\end{center} 
\end{figure}
 
 Figure \ref{fig:ROC_autos} displays the ROC curve estimates obtained with the four methods.  The simplified estimator, which uses only the observations at hand, is plotted with  black dashed lines, while in panels   (a), (b) and (c)    the estimates  $\wROC_{\ipw}$, $\wROC_{\kernel}$ and  $\wROC_{\conv}$ are given in solid grey lines. To compare the four estimators, panel (d) jointly represents all of them. In this case,  the red, blue and grey lines correspond to   $\wROC_{\ipw}$, $\wROC_{\kernel}$ and  $\wROC_{\conv}$, respectively. The largest differences between the inverse probability weighting and the simplified estimator are observed for values of $p$ between 0.2 and 0.6. In this region, the \textsc{ipw} estimate is close to the smoothed  kernel \textsc{ipw} ($\wROC_{\kernel}$). However, this last estimator as the convolution based one shows differences with the procedure that uses only the observations at hand for values of $p\in [0.1,0.2]$. 

 It is worth mentioning that any   analysis of a real data set involves the choice of   covariates with predictive capability on the propensity and also on  the biomarker. These covariates indirectly influence the estimation of the $\ROC$ curve improving the estimation if properly chosen.   In the considered data set, the recorded measurements allowed to construct   more flexible procedures that lead to $\ROC$  estimates with better performance than the simplified estimator that only uses the subsample containing the complete cases with available biomarkers values. Based on the obtained results, in presence of dropouts we recommend the use of $\wROC_{\kernel}$ since it gives better results in the range $[0.1 , 0.2]$ and $ [0.3 , 0.7]$.


\noi {\small\textbf{Acknowledgment.} This research was partially supported by Grants   \textsc{pict} 2018-00740 from \textsc{anpcyt} and   20020170100022BA from the Universidad de Buenos Aires, Argentina  and also by the Spanish Project {PID2020-116587GB-I00, CoDyNP}  from the Ministry of Economy and Competitiveness   (MINECO/AEI/FEDER, UE), Spain.}  

 \setcounter{section}{0}
\renewcommand{\thesection}{A.\arabic{section}}

\setcounter{equation}{0}
\renewcommand{\theequation}{A.\arabic{equation}}

\section{Appendix}{\label{sec:appen}}

\noi \textsc{Proof of Theorem \ref{sec:consist}.1.} We only derive a), since the proof of b) follows similarly. We begin by showing that for each fixed $p$, we have that $\left|\wROC_{\ipw}(p)-\ROC(p)\right| \convpp 0$. Note that $\left|\wROC_{\ipw}(p)-\ROC(p)\right|$ can be bounded as
\begin{eqnarray*}
 \left|\wROC_{\ipw}(p)-\ROC(p)\right| & \le & \left|\wF_{D, \ipw}\left(\wF_{H, \ipw}^{-1}\left(1-p\right)\right)-F_{D}\left(\wF_{H, \ipw}^{-1}\left(1-p\right)\right)\right| \\
 && +  \left|F_{D}\left(\wF_{H, \ipw}^{-1}\left(1-p\right)\right)-F_{D}\left(F_{H }^{-1}\left(1-p\right)\right)\right| \\
 & \le & \|\wF_{D,\ipw}-F_D\|_{\infty}+ \|f_D\|_{\infty}\left|\wF_{H, \ipw}^{-1}\left(1-p\right)- F_{H }^{-1}\left(1-p\right)\right|
\end{eqnarray*}
which together with Proposition \ref{sec:consist}.1 and Proposition \ref{sec:consist}.2 leads to $\left|\wROC_{\ipw}(p)-ROC(p)\right|\convpp 0$. The uniform convergence is a direct consequence of the fact that $\wROC_{\ipw}$ is a non--decreasing function of $p$ and  $\ROC:(0,1) \to [0,1]$ is a continuous non--decreasing function  such that $ \lim_{p\to 0} \ROC(p) = 0$ and  $ \lim_{p\to 1} \ROC(p) = 1$. \square
 
In order to prove Theorem  \ref{sec:consist}.2, we will need the following Lemma.

\noi \textbf{Lemma \ref{sec:appen}.1.} \textsl{Let  $\left(y_{{H},j},
\bx_{{H},j}\trasp, \delta_{{H},j}\right)$, $1\le j \le n_{H}$, be such that  (\ref{delta}) hold.  Assume that  \ref{ass:C1}, \ref{ass:C2}, \ref{ass:C4} to \ref{ass:C6} and \ref{ass:C8} to \ref{ass:C10} hold and that $\wpi_{H}(\bx )=G_{H}(\bx , \wbthe_{H})$, where $\sqrt{n_{H}}(  \wbthe_{H} -  \bthe_{H} )=O_{\prob}(1)$.
Then, we have that $\sqrt{n_H}\,\|\wF_{H, \ipw }-F_{H}\|_{\infty}=O_{\prob}(1)$.}

\noi \textsc{Proof.} First note that analogous arguments to those considered in   the proof of Theorem 4.1 in Bianco \textsl{et al.} (2010) allow to conclude that
$$
\frac{1}{\dst\frac{1}{n_H}\sum_{\ell=1}^{n_H}\frac{\delta_{H,\ell}}{\wpi_H(\bx_{H,\ell})}} \convpp 1\,.
$$
Hence, it will be enough to show that
\begin{equation}
\label{eq:aprobar}
\sqrt{n_H}\, \sup_{t\in \real} \left|\frac{1}{n_H} \sum_{j=1}^{n_H}\frac{\delta_{H,j}}{\wpi_H(\bx_{H,j})} \left\{\indica_{(-\infty,t]}(y_{H,j})-F_H(y)\right\}\right|=O_{\prob}(1)\,.
\end{equation}
Note that $(1/{n_H} )\sum_{j=1}^{n_H}\left({\delta_{H,j}}/{\wpi_H(\bx_{H,j})}\right) \left\{\indica_{(-\infty,t]}(y_{H,j})-F_H(y)\right\}  = S_{1,n_H}(t)-S_{2,n_H}(t)+S_{3,n_H}(t)$
where
\begin{align*}
S_{1,n_H}(t) &= 
\left\{\frac{1}{n_H} \sum_{j=1}^{n_H}\frac{\delta_{H,j}}{\pi_H(\bx_{H,j})} \indica_{(-\infty,t]}(y_{H,j})\right\}-F_H(t)
\\
S_{2,n_H}(t) & =\left\{\frac{1}{n_H} \sum_{j=1}^{n_H}\frac{\delta_{H,j}}{\pi_H(\bx_{H,j})} -1 \right\} F_H(t) 
\\
S_{3,n_H}(t)& = \frac{1}{n_H} \sum_{j=1}^{n_H}\left\{\frac{1}{\wpi_H(\bx_{H,j})}-\frac{1}{\pi_H(\bx_{H,j})}\right\}\;\delta_{H,j} \left\{\indica_{(-\infty,t]}(y_{H,j})-F_H(t)\right\}\,.
\end{align*}
The Central Limit Theorem and the fact that $F_H(t)\le 1$ entail  that $\sqrt{n_H}\,\sup_{t\in \real} \left|S_{2,n_H}(t)\right|=O_{\prob}(1)$.
On the other hand, taking into account that   the class 
$$\itG=\left\{g(\delta,\bx,y)= \frac{\delta }{\pi_H(\bx )} \indica_{(-\infty,t]}(y ) \qquad t\in \real \right\}$$ 
is a VC--class of functions, we immediately obtain that 
$$\sqrt{n_H}\,\sup_{t \in \real} \left|\frac{1}{n_H} \sum_{j=1}^{n_H}\frac{\delta_{H,j}}{\pi_H(\bx_{H,j})} \indica_{(-\infty,t]}(y_{H,j}) - \esp \frac{\delta_{H}}{\pi_H(\bx_{H})} \indica_{(-\infty,t]}(y_{H}) \right|=O_{\prob}(1)\,,$$
which, together with the fact that
$$\esp \frac{\delta_{H}}{\pi_H(\bx_{H})} \indica_{(-\infty,t]}(y_{H}) = F_H(t)\,,$$
implies that  $\sqrt{n_H}\,\sup_{t\in \real} \left|S_{1,n_H}(t)\right|=O_{\prob}(1)$. 

 It remains to show that  $\sqrt{n_H}\,\sup_{y\in \real} \left|S_{3,n_H}(t)\right|=O_{\prob}(1)$. For that purpose, define
	$$g_{\bthech,t}(\delta ,\bx ,y )= \left\{\frac{1}{G_H(\bx , \bthe)}-\frac{1}{G_H(\bx , \bthe_H)}\right\}\;\delta   \;\indica_{(-\infty,t]}(y )
	$$
	and the class of functions $\itG_H^{\star}=\{g_{\bthech,t}: \; \bthe \in \real^{q_H}, \, t\in \real\}$.  
	
	Note that $S_{3,n_H}(t)=S_{4,n_H}(t)-S_{5,n_H}(t)+M(t, \wbthe_H)$ with
	\begin{align*}
		S_{4,n_H}(t)  & = \left\{\frac{1}{n_H} \sum_{j=1}^{n_H}\left[\frac{1}{\wpi_H(\bx_{H,j})}-\frac{1}{\pi_H(\bx_{H,j})}\right]\;\delta_{H,j} \; \indica_{(-\infty,t]}(y_{H,j})\right\}-M(t, \wbthe_H)\\
		& =  \frac{1}{n_H} \sum_{j=1}^{n_H} g_{\wbthech,t}(\delta_{H,j},\bx_{H,j},y_{H,j})-M(t, \wbthe_H)\,,
	\end{align*}
	with
	$$M(t, \bthe )= \esp g_{\bthech,t} (\delta_H,\bx_H,y_H)= \esp\left\{\left[\frac{G(\bx , \bthe_H)}{G(\bx , \bthe)}-1\right] F_H(t|\bx_H)\right\}\, ,$$
	where $F_H(t|\bx_H)=\esp\left(\indica_{(-\infty,t]}(y_H ) | \bx_H\right)$ stands for the conditional distribution of $y_H$ given $\bx_H$ and
	\begin{align*}
		S_{5,n_H}(t) &=  \left\{\frac{1}{n_H} \sum_{j=1}^n\left[\frac{1}{\wpi_H(\bx_{H,j})}-\frac{1}{\pi_H(\bx_{H,j})}\right]\;\delta_{H,j} \right\}  F_H(t) \,.
	\end{align*}
	Assumption \ref{ass:C10}c) and the fact that $\{ \indica_{(-\infty,t]}(y )\; t\in \real\}$ is a VC-class implies that $\itG_H^{\star}$ is Donsker, so the uniform equicontinuity of the class and the fact that $\wbthe_H\convprob \bthe_H$ implies that $\sqrt{n_H} \sup_{t\in \real} \left| S_{4,n_H}(t)\right|=o_\prob(1)$. Using that the class $\itG_H$ defined in assumption \ref{ass:C10}c) has finite uniform--entropy, we immediately get that 
	$$ \frac{1}{\sqrt{n_H}} \sum_{j=1}^n\left[\frac{1}{\wpi_H(\bx_{H,j})}-\frac{1}{\pi_H(\bx_{H,j})}\right]\;\delta_{H,j} =o_{\prob}(1)$$
	leading to $\sqrt{n_H}  \sup_{t\in \real} |S_{5,n_H}(t)| = o_{\prob}(1)$. Hence, to conclude the proof it only remains to show that $\sqrt{n_H} \sup_{t\in \real} |M(t, \wbthe_H)|=O_{\prob}(1)$.
	We first note that 
	\begin{align*}
		\frac{G(\bx , \bthe_H)}{G(\bx , \wbthe_H)}-1 & =\frac{G(\bx , \bthe_H)-G(\bx , \wbthe_H)}{G(\bx , \wbthe_H)}\\
		& = -\;\frac{(\wbthe_H-\bthe_H)\trasp G^{\prime}(\bx , \bthe_H)}{G(\bx , \wbthe_H)} + \frac{(\wbthe_H-\bthe_H)\trasp  G^{\prime\,\prime}(\bx , \wbxi)(\wbthe_H-\bthe_H)}{G(\bx , \wbthe_H)}
	\end{align*}
	with $\wbxi$ and intermediate point between $\wbthe_H$ and $\bthe_H$. Hence, taking into account that  $F_H(t|\bx_H)\le 1$, we get that
	\begin{align*}
		\sqrt{n_H}  \sup_{t\in \real} |M(t, \wbthe_H)| & \le  \sqrt{n_H} \| \wbthe_H-\bthe_H \|\; \sup_{\bthech \in \itV} \esp \left(\frac{ \|G^{\prime}(\bx , \bthe_H)\|}{G(\bx , \bthe)}\right)\\
		& + \sqrt{n_H} \| \wbthe_H-\bthe_H \|^2 \; \sup_{\bthech \in \itV, \balfach \in \itV}\esp\left\{\frac{\lambda_1\left(G_{H}^{\prime\prime}(\bx_{H} ,\balfa)\right)}{G_{H}(\bx_H,  \bthe)}\right\}
	\end{align*}
	which together with the fact that $\sqrt{n_{H}}(  \wbthe_{H} -  \bthe_{H} )=O_{\prob}(1)$, entails that $\sqrt{n_H}  \sup_{t\in \real} |M(t, \wbthe_H)|=O_{\prob}(1)$, concluding the proof. \square 

\vskip0.1in
\noi \textsc{Proof of Theorem \ref{sec:consist}.2.} 
Recall that
$$\wROC_{\kernel}(p)=\frac{1}{\dst\sum_{\ell=1}^{n_D}\frac{\delta_{D,\ell}}{\wpi_D(\bx_{D,\ell})}} \sum_{j=1}^{n_D} \frac{\delta_{D,j}}{\wpi_D(\bx_{D,j})} \itK\left(\frac{p-\wZ_j}{h}\right)$$
where   $\wZ_j= 1-\wF_{H, \ipw}(y_{D,j})$ and $\itK(t)=\int_{-\infty}^t K(u)du$. Denote as $Z_j= 1- F_{H }(y_{D,j})$ and
$$\wROC_{\kernel}^{\star}(p)= \frac{1}{\dst\sum_{\ell=1}^{n_D}\frac{\delta_{D,\ell}}{\wpi_D(\bx_{D,\ell})}} \sum_{j=1}^{n_D} \frac{\delta_{D,j}}{\wpi_D(\bx_{D,j})} \itK\left(\frac{p-Z_j}{h}\right)\,.$$
As in   the proof of Theorem 4.1 in Bianco \textsl{et al.} (2010), standard arguments allow to show that
\begin{equation}
\frac{1}{\dst\frac{1}{n_D}\sum_{\ell=1}^{n_D}\frac{\delta_{D,\ell}}{\wpi_D(\bx_{D,\ell})}} \convpp 1\,.
\label{eq:deltawpi}
\end{equation}
Then, we have that $\wROC_{\kernel}^{\star}(p)\convpp \ROC(p)$, since $h\to 0$ and $n_D\, h\to \infty$.
Effectively, from  \ref{ass:C1} and \ref{ass:C2} and using that \ref{ass:C6} entails that $\itK$ is bounded, we easily get that 
$$\frac{1}{n_D}  \sum_{j=1}^{n_D} \left\{\frac{1}{\pi_D(\bx_{D,j})} -\frac{1}{\wpi_D(\bx_{D,j})}  \right\}\, \delta_{D,j}   \itK\left(\frac{p-Z_j}{h}\right) \convpp 0 \,,$$
which together with the fact that
$$\frac{1}{n_D}  \sum_{j=1}^{n_D} \frac{\delta_{D,j}}{\pi_D(\bx_{D,j})}   \itK\left(\frac{p-Z_j}{h}\right)  \convprob F_{Z}(p)=\ROC (p)\,,$$
entail that $\wROC_{\kernel}^{\star}(p)\convprob \ROC(p)$.

 On the other hand, if $\xi_j$ stands for an intermediate point between $\wZ_j$ and $Z_j$ we get that
\begin{align*}
\left|\wROC_{\kernel}(p)- \wROC_{\kernel}^{\star}(p) \right|& = \frac{1}{\dst\sum_{\ell=1}^{n_D}\frac{\delta_{D,\ell}}{\wpi_D(\bx_{D,\ell})}} \left| \sum_{j=1}^{n_D} \frac{\delta_{D,j}}{\wpi_D(\bx_{D,j})}  \left\{\itK\left(\frac{p-\wZ_j}{h}\right)-\itK\left(\frac{p-Z_j}{h}\right)\right\} \right|\\
& \le  \frac{1}{\dst\frac{1}{n_D}\sum_{\ell=1}^{n_D}\frac{\delta_{D,\ell}}{\wpi_D(\bx_{D,\ell})}} \frac{1}{n_D\; h} \sum_{j=1}^{n_D}  \frac{\delta_{D,j}}{\wpi_D(\bx_{D,j})}   K\left(\frac{p-Z_j}{h}\right)\left| \wZ_j-Z_j \right| \\
&+
\frac{1}{2\;\dst\frac{1}{n_D}\sum_{\ell=1}^{n_D}\frac{\delta_{D,\ell}}{\wpi_D(\bx_{D,\ell})}} \frac{1}{n_D\; h^2} \sum_{j=1}^{n_D}  \frac{\delta_{D,j}}{\wpi_D(\bx_{D,j})}  K^{\prime}\left(\frac{p-\xi_j}{h}\right)\left( \wZ_j-Z_j \right)^2\\
 & \le \frac{1}{\dst\frac{1}{n_D}\sum_{\ell=1}^{n_D}\frac{\delta_{D,\ell}}{\wpi_D(\bx_{D,\ell})}} \frac{1}{n_D\; h}  \sum_{j=1}^{n_D} \frac{\delta_{D,j}}{\wpi_D(\bx_{D,j})}   K\left(\frac{p-Z_j}{h}\right)  \; \|\wF_{H, \ipw }-F_{H }\|_{\infty}\\
&+
 \frac{1}{2\,n_D\; h^2} \;\|K^{\prime}\|_{\infty} \; n_D\,\|\wF_{H, \ipw }-F_{H }\|_{\infty}^2
\end{align*}
As above, using that  \ref{ass:C1} and \ref{ass:C2} hold, it is easy to see that
$$\frac{1}{n_D\; h}  \sum_{j=1}^{n_D} \left\{\frac{1}{\pi_D(\bx_{D,j})} -\frac{1}{\wpi_D(\bx_{D,j})}  \right\}\, \delta_{D,j}  K\left(\frac{p-Z_j}{h}\right)  \convpp 0 \,,$$
which together with the fact that
$$\frac{1}{n_D\; h}  \sum_{j=1}^{n_D} \frac{\delta_{D,j}}{\pi_D(\bx_{D,j})}   K\left(\frac{p-Z_j}{h}\right)  \convprob f_Z(p)=\ROC^{\prime}(p)\,,$$
entail that
$$\frac{1}{n_D\; h}  \sum_{j=1}^{n_D} \frac{\delta_{D,j}}{\wpi_D(\bx_{D,j})}   K\left(\frac{p-Z_j}{h}\right)  \convprob \ROC^{\prime}(p)\,.$$
Therefore,  from \ref{ass:C7} and \ref{ass:C8} we have that $n_D/n_H\to \tau/(1-\tau)$ with $0<\tau<1$ and $n_D h^2\to \infty$, so from \eqref{eq:deltawpi}  and   Lemma \ref{sec:appen}.1, we get that $\wROC_{\kernel}(p)- \wROC_{\kernel}^{\star}(p) \convprob 0$, concluding the proof. \square

\small
\section*{References}
\begin{description}
\item  An, Y. (2012). Smoothed empirical likelihood inference for ROC curves with missing data. \textsl{Open Journal of Statistics}, \textbf{2}, 21-27.

\item Bianco, A., Boente, G., Gonz\'alez--Manteiga, W. and P\'erez--Gonz\'alez, A. (2010). Estimation of the marginal location under a partially linear model with missing responses. \textsl{Computational Statistics and Data Analysis}, \textbf{54}, 546-564.

\item Bianco, A., Boente, G., Gonz\'alez--Manteiga, W. and P\'erez--Gonz\'alez, A. (2019). Plug--in marginal estimation  under a general regression model with missing responses and covariates.  \textsl{TEST}, \textbf{28}, 1-41. 

\item Ding, X. and Tang, N. (2018). Adjusted empirical likelihood estimation of distribution
function and quantile with nonignorable missing data. \textsl{Journal of Systems Science and Complexity}, \textbf{31}, 820-840.

\item Fahrmeir, L. and  Kaufmann, H. (1985). Consistency and asymptotic normality of the maximum likelihood estimator in generalized linear models. \textsl{Annals of Statistics}, \textbf{13}, 342-368. 

\item Gon{\c{c}}alves, L., Subtil, A., Oliveira, M. R. and  Bermudez, P. (2014) Roc Curve Estimation: An Overview. \textsl{REVSTAT-Statistical Journal}, \textbf{12}, 1-20.

\item  Horvitz, D. G.  and  Thompson, D. J.  (1952). A generalization of sampling without replacement from a finite universe. \textsl{Journal of the American Statistical Association}, \textbf{47}, 663-685.

\item Krzanowski, W. J. and Hand, D. J. (2009). \textsl{ROC curves for continuous data}. Chapman and Hall/CRC, Boca Raton.

\item Li, S. and Ning, Y. (2015).  Estimation of covariate-specific time-dependent ROC curves in the presence of missing biomarkers.  \textsl{Biometrics}, \textbf{71}, 666-676

\item  Liu, X. and Zhao,Y. (2012). Semi-empirical likelihood confidence intervals for ROC curves with missing data. \textsl{Journal of Statistical Planning and Inference}, \textbf{142}, 3123-3133.

\item Long Q., Zhan, X. and Shu, C. (2011a). Nonparametric multiple imputation for receiver operating characteristics analysis when some biomarker values are missing at random. \textsl{Statistis in Medicine}, \textbf{30}, 3149-3161.

\item Long Q., Zhan, X. and Johnson, B.A. (2011b). Robust estimation of area under ROC curve using auxiliary variables in the presence of missing biomarker values. \textsl{Biometrics}, \textbf{67}, 559-567.

  \item   M\"uller, U.  (2009). Estimating linear functionals in nonlinear regression with responses missing at random. \textsl{Annals of Statistics}, \textbf{37}, 2245-2277.
  
  \item Pardo-Fern\'andez, J. C., Rodr\'{\i}guez-Alvarez, M. X. and Van Keilegom, I. (2014). A review on ROC curves in the presence of covariates. \textsl{REVSTAT Statistical Journal}, \textbf{12}, 21-41.

\item Pepe, M. S. (2003). \textsl{The Statistical Evaluation of Medical Tests for Classification
and Prediction}, Oxford University Press, New York.

\item Pulit, M. (2016). A new method of kernel--smoothing estimation of the ROC curve. \textsl{Metrika}, \textbf{79}, 603-634.
 
 \item  Sued, M.  and  Yohai, V.  (2013). Robust location estimation with missing data. \textsl{Canadian Journal of Statistics}, \textbf{41}, 111-132.

 \item Sun, Y. and Genton, M. G. (2011). Functional boxplots. \textsl{Journal of Computational and Graphical Statistics}, \textbf{20}, 316-334. 

\item Wang, C.; Wang, S.; Zhao, L.P. and Ou,S.T., 1997. Weighted semiparametric estimation in regression analysis regression
with missing covariates data. \textsl{Journal of the American Statistical Association}, \textbf{92}, 512-525.
 
\item Yang, H. and Zhao, Y. (2015). Smoothed jackknife empirical likelihood inference for ROC curves with missing data. \textsl{Journal of Multivariate Analysis}, \textbf{140}, 123-138.
\end{description}
\end{document}